\documentclass[final]{./raa06letRevEd}
\usepackage{graphicx,times}
\usepackage{hyperref}
\usepackage{natbib}
\usepackage{mathrsfs}
\usepackage{bbm}
\usepackage{upgreek}
\usepackage{chemformula}
\usepackage{tabularx,ragged2e}
\newcolumntype{C}{>{\Centering\arraybackslash}X}
\usepackage{amssymb,amsmath}
\bibpunct{(}{)}{;}{a}{}{,}
\usepackage{listings}
\usepackage{CJK}
\renewcommand{\thepage}{99--\arabic{page}}
\hypersetup{colorlinks = true, linkcolor = blue, anchorcolor = red,
citecolor = blue, filecolor = red, pagecolor = red, urlcolor = blue}

\begin{document}

\begin{CJK}{GBK}{song}
   \title{Atmospheric regimes and trends on exoplanets and brown dwarfs}

\volnopage{{\bf 2020} Vol.~{\bf 20} No.~{\bf 7},~99(92pp)~
   {\small  doi: 10.1088/1674--4527/20/7/99}}
   \setcounter{page}{1}

   \author{Xi Zhang}
   
   \institute{ Department of Earth and Planetary Sciences, University of California Santa Cruz, Santa Cruz, CA 95064, USA; {\it xiz@ucsc.edu}\\
\vs \no
   {\small Received~~2020 April 14; accepted~~2020 June 11}
}

\abstract{A planetary atmosphere is the outer gas layer of a
planet. Besides its scientific significance among the first and most
accessible planetary layers observed from space, it is closely
connected with planetary formation and evolution, surface and
interior processes, and habitability of planets. Current theories of
planetary atmosphere{s} were primarily obtained
through the studies of eight large planets, Pluto and three large
moons (Io, Titan, and Triton) in the Solar System. Outside the Solar
System, more than four thousand extrasolar planets
(exoplanets) and two thousand brown dwarfs have been confirmed in
our {G}alaxy, and their population is rapidly growing.
The rich information from these exotic bodies offers a database to
test, in a statistical sense, the fundamental theories of planetary
climates. Here we review the current knowledge o{n}
atmospheres of exoplanets and brown dwarfs from recent observations
and theories. This review highlights important regimes and
statistical trends in an ensemble of atmospheres as an initial step
towards fully characterizing diverse substellar atmospheres, that
illustrates the underlying principles and critical problems.
Insights are obtained through analysis of the dependence of
atmospheric characteristics on basic planetary parameters. Dominant
processes that influence atmospheric stability, energy transport,
temperature, composition and flow pattern are discussed and
elaborated with simple scaling laws. We dedicate this review to Dr.
Adam P. Showman (1968--2020) in recognition of his fundamental
contribution to the understanding of atmospheric dynamics on giant
planets, exoplanets and brown dwarfs.  }
  \authorrunning{{\it X. Zhang}: Atmospheres on Exoplanets and Brown Dwarfs}            
   \titlerunning{{\it X. Zhang}: Atmospheres on Exoplanets and Brown Dwarfs}  
   \maketitle
\end{CJK}
%
\section{Introduction}           
\label{sect:intro}

The discovery of planets outside the Solar System has greatly
expanded the horizon in planetary science since
\cite{mayorJupitermassCompanionSolartype1995} discovered the first
exoplanet around a {S}un-like star
(51~Pegasi\,{\,}b). Among more than 4200 exoplanets confirmed to
date, {the }majority of them are larger than our Earth (e.g.,
\citealt{boruckiCharacteristicsPlanetaryCandidates2011,boruckiCHARACTERISTICSOFKEPLERPLANETARYCANDIDATES2011};
\citealt{howardPlanetOccurrence252012};
\citealt{dressingOccurrenceRateSmall2013};
\citealt{fressinFalsePositiveRate2013};
\citealt{petiguraPlateauPlanetPopulation2013}). Statistical analysis
of the current samples and structural models (e.g.,
\citealt{weissMassradiusRelation652014};
\citealt{lopezUNDERSTANDINGMASSRADIUSRELATION2014};
\citealt{rogersMOST1EARTHRADIUSPLANETS2015}) found that planets with
radii larger than about 1.5--1.7 Earth radii are likely to have
thick gaseous envelopes made of hydrogen, helium and hydrogen
compounds, while the smaller planets are mostly composed of iron
and silicates like the Earth. Although the larger planets should
hold thick atmospheres, we cannot rule out the possibility of
atmospheric existence on smaller planets because Io ($\sim$30\% of
the Earth radius) and Pluto ($\sim$20\% of the Earth radius) in our
Solar System still possess thin atmospheres. To date, most
atmospheres that have been characterized are on planets close to the
central stars via transit observations or on young planets via the direct
imaging technique, meaning that most of these planets are much
hotter than the Solar System planets. The current samples of exoplanet atmospheres are considered to be a high-temperature
extension of the planetary atmospheres in the Solar System.

In the same year as the discovery of 51~Pegasi\,{\,}b, the first
two brown dwarfs, which can be roughly defined as astronomical
bodies with a mass between 13 and 80 Jupiter masses, were also
confirmed (Gliese\,{\,}229B in
\citealt{nakajimaDiscoveryCoolBrown1995} and Teide\,{\,}1 in
\citealt{reboloBrownDwarfsPleiades1996}). The currently known
$\sim$2000 brown dwarfs are mostly free-floating, but some are also
orbiting star companions. Traditionally, brown dwarfs are considered
to be the low-mass branch below M dwarf stars in the
substellar main sequence: L, T and Y sub-categories
following the decreasing order of the effective temperature.
Nevertheless, to some extent, brown dwarfs behave more like planets.
Because their masses are below the hydrogen-burning limit but above
the deuterium-burning limit, {a} brown dwarf{'s}
interior is in the degenerate state, similar to {that
of} gas giant planets, and their radii are around one Jupiter
radius. Unlike ordinary stars in which trace elements
(other than hydrogen and helium) are mainly in the atomic form, the
colder photospheres of brown dwarfs are mostly composed of molecules
such as \ch{H2}, He, \ch{H2O}, CO and \ch{CH4}. The emission
temperature of the coldest brown dwarf detected so far (WISE 0855)
is even below the freezing point{ of water}, and
water ice clouds could form there as on Jupiter
(\citealt{luhmanNewParallaxMeasurement2014};
\citealt{skemerFIRSTSPECTRUMCOLDEST2016};
\citealt{esplinPHOTOMETRICMONITORINGCOLDEST2016}). Thus current
samples of brown dwarf atmospheres can be categorized as a
high-gravity and high-temperature extension of the
hydrogen-dominated giant planet{ary} atmospheres in the Solar
System.

To date{,} observations of the atmospheres of exoplanets and
brown dwarfs (hereafter collectively ``substellar atmospheres'')
mainly focus on transmission, reflection and emission spectra; phase
curves as the planets are circling the stars; rotational light curves
as the planets spin; and Doppler-shifted atomic or molecular lines
by the atmospheric winds and orbital motion. From these data, we can
retrieve the distributions of temperature, atmospheric compositions
and abundances of gases and clouds, and wind patterns in the
atmospheres. Based on{ the} NASA Exoplanet Archive, to date,
almost a hundred exoplanets have (some sort of) atmospheric
detection (\citealt{madhusudhanExoplanetaryAtmospheresKey2019}),
among which we have obtained $\sim$50 transmission spectra and
$\sim$30 emission spectra. The data quality is, however, not always
satisfactory. Arguably, the quality of the observational data of exoplanetary atmospheres
is about 40\,yr behind its counterpart in
the Solar System. For example, spectral coverage, spectral
resolution, and noise levels of the dayside emission spectra of a
canonical hot Jupiter HD~189733\,{\,}b from the Spitzer
telescope (\citealt{grillmairStrongWaterAbsorption2008}) and Hubble
Space Telescope ({HST, }\citealt{swainWaterMethaneCarbon2009})
are comparable to that of Jupiter spectra in the early era by
\cite{gillett814MICRONSpectrumJupiter1969}. The typical resolving
power of exoplanet observation is $\lambda/\Delta\lambda\sim$10 to
100 where $\lambda$ is the wavelength
(\citealt{konopackyDetectionCarbonMonoxide2013a}).
Higher-spectral-resolution
($\lambda/\Delta\lambda\sim$10{\,}000 to
100{\,}000) spectra in visible and near-infrared{
(near-IR)} wavelengths from ground-based transmission
observations can be achieved to determine the atmospheric
composition and even the wind speed using cross-correlation and
line-shape analysis techniques (e.g.,
\citealt{snellenOrbitalMotionAbsolute2010};
\citealt{crossfieldHighresolutionDifferentialNearinfrared2011};
\citealt{konopackyDetectionCarbonMonoxide2013a}). However, it seems
more difficult to estimate the uncertainties of ground-based data
than {those} of space-based data
(\citealt{kreidbergExoplanetAtmosphereMeasurements2018}). We expect
a great leap in the spectral data quality with future large
telescopes such as the James Webb Space Telescope ({JWST, }e.g.,
$\lambda/\Delta\lambda\sim$ 100 to 1000 in space), Atmospheric
Remote-sensing Infrared Exoplanet Large-survey (ARIEL), European
Extremely Large Telescope (ELT), Giant Magellan Telescope
(GMT) and Thirty Meter Telescope (TMT). A spatially
resolved image of an exoplanet is expected to remain difficult to achieve in the
foreseeable future.

On the other hand, without photon contamination from the host star
companions, spectral observations on high-temperature field brown
dwarfs typically have much higher data quality. For example, the
observational spectrum of a typical T4.5 dwarf 2MASS 0559-14 can
achieve a spectral resolving power of
$\lambda/\Delta\lambda\sim$2000 and a signal-to-noise ratio
{(S/N)}$>50$. The ``hot methane'' lines in the
near-{IR} can be spectrally resolved and led to
the reevaluation of the existing opacity database
(\citealt{yurchenkoSpectrumHotMethane2014}). These spectra allow a
much better estimate of the atmospheric properties such as
temperature and chemical compositions on both L and T dwarfs (e.g.,
\citealt{lineUniformAtmosphericRetrieval2017}). The closest brown
dwarf detected so far (Luhman 16 B) can even be spatially resolved
(in low resolution) using the Doppler imaging technique
(\citealt{crossfieldGlobalCloudMap2014}).

Despite the current data quality and future challenges in the
observations of atmospheres, exoplanet observation has some
advantages compared to that in the Solar System. For example, for
planets around other stars, we can naturally detect the orbital
light curve, i.e., the flux changes at different orbital phases,
whereas this can only be done by an orbiting or flyby spacecraft for
Solar System planets outside Earth{'s} orbit. As
another example, the giant planets in our Solar System are so cold
that water vapor is all trapped below the water cloud layer, and
hence water abundances remain unknown. On the other hand, hot
exoplanets show clear water vapor signals in the spectra, allowing a
better derivation of the carbon-to-oxygen ratio, a crucial parameter
for constraining planetary formation and evolution.

Characterizing planetary atmospheres and unveiling the principles
underlying their diverse weather and climate---as we have learned
from the Solar System studies---is challenging. Atmospheres are
fundamentally complex with many interacting processes and a large
number of free parameters. A big dataset with sufficient samples is
required for comparative planetology to understand the role of each
factor. Undoubtedly this dataset could only come from the
atmospheres of exoplanets and brown dwarfs. In the dawn of the
``third era'' in planetary science
(\citealt{ingersollThreeErasPlanetary2017}), researchers have been
astonished by the diversity of atmospheres outside the Solar System. These
substellar atmospheres have provided a wealth of information
complementary to their counterparts in the Solar System
(\citealt{pierrehumbertStrangeNewsOther2013}).

Here are a few examples. Tidally locked planets are synchronized to the central
stars with permanent dayside and nightside, a configuration we do
not have in our system. Super-Earths and mini-Neptunes are planets
with size between the Earth and Neptune. They are a new type of
world that is not present in the Solar System but dominates the
current, confirmed exoplanetary population. Hot Earth-sized planets
are so close to their central stars that their surfaces might be
melted or partially melted (e.g., 55~Cancri\,{\,}e or
Kepler-10\,{\,}b). They might be exciting analogs of Jupiter's
moon Io or the early Earth with atmosphere-magma
interaction. Planets very close to the central stars are perfect
samples to understand how atmospheres are evaporated or blown off.
In contrast, worlds that are very far away from
the{ir} central stars are useful to explore how atmospheres
condense on the surfaces. For planets in the ``habitable zone''
where liquid water could exist on the surface, various
climate states are possible, depending on parameters such as
planetary rotation rate, central star type, atmospheric
composition and orbital configuration. Furthermore, planets
at different ages could tell us how planetary atmospheres and
climates evolve with time and under different environments.

A number of excellent reviews of atmospheres of exoplanet{s} and
brown dwarfs have been published. Some articles generally cover a
bit of every aspect (e.g., \citealt{seagerExoplanetAtmospheres2010};
\citealt{baileyDawesReviewAtmospheres2014a};
\citealt{madhusudhanExoplanetaryAtmospheres2014};
\citealt{fortneyModelingExoplanetaryAtmospheres2018};
\citealt{madhusudhanExoplanetaryAtmospheresKey2019}) but most of
them focus on specific topics such as atmospheric observations
(e.g., \citealt{tinettiSpectroscopyPlanetaryAtmospheres2013};
\citealt{burrowsSpectraWindowsExoplanet2014};
\citealt{encrenazInfraredSpectroscopyExoplanets2014};
\citealt{pepeInstrumentationDetectionCharacterization2014};
\citealt{crossfieldObservationsExoplanetAtmospheres2015a};
\citealt{demingIllusionRealityAtmospheres2016};
\citealt{kreidbergExoplanetAtmosphereMeasurements2018};
\citealt{parmentierExoplanetPhaseCurves2018a};
\citealt{singObservationalTechniquesTransiting2018}), atmospheric
escape (e.g., \citealt{lammerAtmosphericEscapeEvolution2008};
\citealt{tianAtmosphericEscapeSolar2015};
\citealt{owenAtmosphericEscapeEvolution2019b}), atmospheric
radiation (e.g., \citealt{marleyCoolSideModeling2015};
\citealt{hengRadiativeTransferExoplanet2017}), atmospheric chemistry
(e.g., \citealt{loddersExoplanetChemistry2010};
\citealt{marleyCloudsHazesExoplanet2013};
\citealt{mosesChemicalKineticsExtrasolar2014};
\citealt{madhusudhanExoplanetaryAtmospheresChemistryFormation2016}),
atmospheric dynamics (e.g.,
\citealt{showmanAtmosphericCirculationExoplanets2010};
\citealt{showmanAtmosphericCirculationTerrestrial2013};
\citealt{hengAtmosphericDynamicsHot2015};
\citealt{pierrehumbertAtmosphericCirculationTideLocked2019};
\citealt{showmanATMOSPHERICDYNAMICSHOT2020}), space weather (e.g.,
\citealt{airapetianImpactSpaceWeather2020}), terrestrial climate
(e.g., \citealt{forgetPossibleClimatesTerrestrial2014}), giant
planets (e.g., \citealt{marleyAtmospheresExtrasolarGiant2007};
\citealt{fletcherExploringDiversityJupiterclass2014}), brown dwarfs
(e.g., \citealt{basriObservationsBrownDwarfs2000};
\citealt{kirkpatrickNewSpectralTypes2005};
\citealt{hellingModellingFormationAtmospheric2013};
\citealt{hellingAtmospheresBrownDwarfs2014};
\citealt{artigauVariabilityBrownDwarfs2018};
\citealt{billerTimeDomainBrown2017}) and habitability (e.g.,
\citealt{kastingEvolutionHabitablePlanet2003};
\citealt{madhusudhanExoplanetaryAtmospheresChemistryFormation2016};
\citealt{shieldsHabitabilityPlanetsOrbiting2016};
\citealt{kopparapuCharacterizingExoplanetHabitability2019}).
{However,} previous reviews focused less on statistical
properties in the emergent ensemble of substellar atmospheres,
motivating this article.

In this review, we consider these diverse atmospheres as a
systematic test bed for our current understanding of planetary
climates. We summarize the statistical ``trends'' discovered in
recent years and discuss various aspects to classify the atmospheres
into different climate ``regimes{.}" To be specific,
``regimes" and ``trends" refer to the dependence of the atmospheric
characteristics on the basic planetary parameters. Here ``basic
planetary parameters" refer to planetary parameters such as
the mass, radius, age, gravity, self-rotation rate, escape velocity,
semi-major axis, orbital period, eccentricity, obliquity,
metallicity (including elemental ratios such as
{carbon-to-oxygen (}C/O{)} ratio), surface albedo, internal
heat flux (internal luminosity) and equilibrium temperature. They
could also include host star parameters such as host star
type, stellar luminosity, and stellar irradiation spectra.
``Atmospheric characteristics" stand for the observed
properties of substellar atmospheres such as directly measured
broadband photometric fluxes and all kinds of spectral and
polarization signatures. It also includes the derived atmospheric
properties such as atmospheric existence, atmospheric pressure and
mass, bulk luminosity (or effective temperature), albedo,
distributions of temperature, gas and particle compositions, wind
and waves, and the time variability of those properties from
time-domain observations.

We are just beginning to discover and understand those trends and
regimes. There are dangers with this approach
because of the assumption, as pointed out by
\cite{stevensonPlanetaryDiversity2004}, that common processes are at
work on Solar System planets, including the Earth, exoplanets and
brown dwarfs but they yield different and diverse outcomes. If a
single fundamental mechanism controls an observable across the
sampled planets, we might observe a trend with a typical varying
parameter. A typical example is the Hertzsprung-Russell diagram for
stars. If a few fundamental mechanisms govern the observables, we
might expect a regime shift from one dominant mechanism to another
in the parameter space. However, if many factors could lead to
similar, almost indistinguishable observable phenomena, the trends
or regimes are washed out in a large sample. Given the current data
quality for substellar atmospheres, the statistical significance of
the trends and regimes in this review will be preliminary and
somewhat debatable. However, from a theoretical perspective, this is
also a good way to summarize our understanding of substellar
atmospheres, highlight fundamental principles underlying essential
processes and link back to our knowledge obtained from the
Solar System. We will also try to outline some simple analytical
scaling laws to help illuminate fundamental processes more
intuitively.

This comprehensive review is organized as follows. First, we will
make some general remarks on atmospheres. In
Section~\ref{sect:remarks}, {w}e start with the
fundamental equations in planetary atmospheres and elaborate on
vital processes and their complex interactions
in Section~\ref{sect:process}. We then summarize the difference between
the traditional ``cold" planetary atmospheres in the Solar System
and the currently characterized ``hot" atmospheres on exoplanets and
brown dwarfs in Section~\ref{sect:coldhot}. Then we feature several
important spectral and photometric observations to date for
characterizing substellar bodies in Section~\ref{sect:char}. That will
help lead into discussion on statistical trends and
regimes, summarized in several sub-fields. In each sub-field
section, we first introduce the fundamentals and then feature
several important regimes and trends. In
Section~\ref{sect:stability}, we discuss atmospheric stability with
a focus on the atmospheric escape from planets. We highlight the
``cosmic shoreline" in Section~\ref{sect:cos} and ``planet desert and
radius gap" in Section~\ref{sect:radgap} in recent observations and
underlying mechanisms. In Section~\ref{sect:temp}, we discuss the
thermal structure and radiative energy transport, with an emphasis
o{n} the radiative-convective boundary{ (RCB)},
vertical temperature inversion and mid-{infrared (mid-}IR{)}
brightness temperature trend on exoplanets in
Section~\ref{sect:verttemp}, thermal phase curves on tidally locked
exoplanets in Section~\ref{sect:horitemp}, and rotational light
curves on brown dwarfs and directly imaged planets in
Section~\ref{sect:rotlight}. We talk about atmospheric composition
and chemistry in Section~\ref{sect:chem}. In
Section~\ref{sect:gaschem}, we discuss gas chemistry, including both
thermochemistry and disequilibrium chemistry, followed by a review
of hazes and clouds in Section~\ref{sect:cloud}. In
Section~\ref{sect:dyn}, we concentrate on the atmospheric dynamics
and important regimes classified using non-dimensional numbers. We
describe three categories: highly irradiated planets such as the
tidally locked planets in Section~\ref{sect:hjdyn}, convective
atmospheres on directly imaged planets and brown dwarfs in
Section~\ref{sect:bddyn}. We only briefly review the terrestrial
climates in the habitable zone in Section~\ref{sect:hpdyn} because
of its complexity and the lack of data to reveal detailed trends
on extrasolar terrestrial atmospheres. We conclude
this review with prospects in Section~\ref{sect:prospects}.

\section{General Remarks}
\label{sect:remarks}

\subsection{Overview of Important Processes}
\label{sect:process}

Atmospheres in and out of the Solar System share similar fundamental
physical and chemical processes that should be understood in a
self-consistent mathematical framework. The fundamental equation set
is composed of a continuity equation, a momentum equation,  an
energy equation, an equation of state, an equation of radiative
transfer, and a series of transport equations for chemical species,
including both gas and particles. Equation set~(\ref{basiceq}) lists
the governing equations for a three-dimensional (3D), collisional,
neutral, inviscid, ideal-gas atmosphere with necessary assumptions.

 \begin{subequations} \label{basiceq}
 \begin{align}
& \frac{\partial\rho}{\partial t}+\nabla \cdot(\rho \vec{\bf{u}})=F_{\rho}\,,\\
& \frac{\partial \rho \vec{\bf{u}}}{\partial t}+\nabla \cdot(\rho \vec{\bf{u}} \vec{\bf{u}})+\nabla p + 2\bf{\Omega} \times \vec{\bf{u}}\nonumber\\
 &\quad\quad+ \bf{\Omega} \times (\bf{\Omega}  \times \vec{\bf{r}})-\rho \vec{\bf{g}} =  \vec{\bf{F_u}}\,,\\
& \frac{\partial}{\partial t} (\frac{p}{\gamma-1}+\frac{1}{2}\rho  \vec{\bf{u}}\cdot\vec{\bf{u}}+\rho\Phi)\nonumber\\
&\quad\quad+\nabla \cdot [(\frac{\gamma p}{\gamma-1}+\frac{1}{2}\rho \vec{\bf{u}}\cdot\vec{\bf{u}}+\rho\Phi)\vec{\bf{u}}-K_T\nabla T]\nonumber\\
&\quad\quad + Q(I_{\nu}) =F_{e}\,, \\
& p=\frac{\rho k_BT}{m}\,, \\
& \frac{dI_{\nu}}{d\tau_{\nu}}-I_{\nu}-\omega_{\nu}I^s_{\nu} e^{-\tau_{\nu}}P_{\nu}-(1-\omega_{\nu})J_{\nu}\nonumber\\
&\quad\quad-\omega_{\nu}S_{\nu}(\tau_{\nu}, P_{\nu}, I_{\nu})=0\,, \\
& \frac{\partial\rho\chi}{\partial t}+\nabla \cdot[\rho \chi \vec{\bf{u}}+ \rho D_{\chi} \nabla(\frac{\chi}{\chi_e})]=P_{\chi}-L_{\chi}\,.
\end{align}
\end{subequations}

Here bold represent{s} vector form. $t$, $\rho$, $p${
and} $T$ are time, density, pressure and temperature,
respectively. $\vec{\bf{u}}$, $\vec{\bf{\Omega}}$,
$\vec{\bf{r}}${ and} $\vec{\bf{g}}$ are the
{3D} velocity vector, rotational rate
vector, radial vector and gravitational acceleration
vector, respectively. $\Phi$ is the gravitational potential energy
by mass defined as $\vec{\bf{g}}=-\nabla\Phi$. $K_T$ is thermal
conductivity. $k_B$ is the Boltzmann constant. $m$ is the mean mass
of {an} air molecule. $\gamma$ is the adiabatic index,
i.e., the ratio of the specific heats $c_p/c_v$. $Q$ is the
radiative heating and cooling terms, $F_{\rho}, \vec{\bf{F_u}}, F_e$
are the external forcing terms of density, momentum and
energy, respectively. In the radiative transfer
Equation~(\ref{basiceq}e), $\nu$ is the spectral grid (wavelength or
frequency), $\tau_{\nu}$ is slant optical depth, $I$ is the light
intensity{ and} $I^s_{\nu} $ is the incoming stellar
intensity. $J_{\nu}$ is the self-emission source function, which is
the Planck function under Local Thermodynamic
Equilibrium (LTE).
$S_{\nu}$ is the scattering source function. $\omega_{\nu}$ is the
single scattering albedo, and $P_{\nu}$ is the scattering phase
function. In the chemical transport Equation~(\ref{basiceq}f),
$\chi$ is the mass mixing ratio of a specific species (either gas or
particle). $D_{\chi}$ is the molecular diffusivity that relaxes the
mass mixing ratio towards the equilibrium mass mixing ratio
$\chi_e$. Note that there is no eddy mixing term because the 3D
advection term by $\vec{\bf{u}}$ includes the eddy transport. $P$
and $L$ are the chemical/microphysical production and loss terms,
respectively.

The continuity{ equation,} Equation~(\ref{basiceq}a){,}
describes the bulk atmosphere {a}s a compressible fluid
and the external forcing term $F_{\rho}$ includes a mass loss to
space at the top of the atmosphere, mass exchange with the interior
(such as volcanism),{ and} surface/ocean and clouds through the
condensation and evaporation. The momentum{ equation,}
Equation~(\ref{basiceq}b){,} is a simplified form of the
Navier-Stokes equation in fluid mechanics neglecting the molecular
and dynamic viscous terms. The external forces include {the
}pressure gradient, Coriolis force, centrifugal force
and gravitational force. The latter could spatially vary
due to the oblateness of the body. Other external forces in the
$F_u$ term include the drag force from surface friction,
magnetic interaction, momentum gain or loss due to the mass
gain or loss, phase change and gravitational particle
settling. The energy equation,
Equation~(\ref{basiceq}c){,} describes the evolution of the
atmospheric energy flux, including internal energy, kinetic
energy{ (KE)} and gravitational potential energy. The $Q$
term represents diabatic heating and cooling from
atmospheric radiation. Thermal conduction via collisions is
described in the $K\nabla T$ term. The other energy forcing term
$F_e$ includes latent heat and energy exchange during the phase
transition, such as cloud formation, Ohmic heating through
interaction with the magnetic field, viscous heating due
to frictional drag and even the chemical potential energy change
during chemical reactions. The equation of
state~(\ref{basiceq}d) of the atmosphere approximately follows the
ideal gas law, which is valid in most photospheres. The equation of
state needs to be treated carefully in {a}
multi-component atmosphere, especially where clouds form
(\citealt{liSimulatingNonhydrostaticAtmospheres2019}).

The radiative transfer equation~(Eq.~(\ref{basiceq}e)) solves the photon
intensity distribution in the atmosphere at each wavelength and
angle. The radiative flux divergence is used in the energy equation
(the $Q$ term in Eq.~(\ref{basiceq}c)). Also, the actinic flux
derived from the intensity is applied to the photochemical
calculations. Multiple scattering from the gas (Rayleigh) and
particles needs to be considered. Chemical transport
Equation~(\ref{basiceq}f) includes advection and molecular diffusion
of the chemical species and the chemical production and loss terms.
The production and loss come from gas chemistry such as
photochemistry, neutral chemistry, ion chemistry and particle
microphysics in the haze and cloud formation such as nucleation,
coagulation, and condensation aggregation and coalescence processes.
The chemical equations are coupled together by the chemical reaction
network. Usually, the continuity Equation~(\ref{basiceq}a) would not
be altered by the mass-conserved gas chemistry, except that the gas
density could change in the condensation and evaporation processes.
Note that the chemical reactions do not conserve the total number of
molecules. Therefore{,} the mean molecular mass ($m$ in the
equation of state~(\ref{basiceq}d)) could be altered in the chemical
and microphysical processes.

In the system described by the equation set~(\ref{basiceq}), the
total momentum, mass and energy of the atmosphere do not have
to be conserved with time. They depend on the boundary conditions
(e.g., whether the atmosphere is escaping to space or condensing at
the surface) and internal processes (e.g., cloud formation
converting vapor to particles). In most cases, we assume the
observed planetary atmospheres have reached a steady state with
internal oscillations. In this situation, solving the statistically
averaged climate state is a boundary value problem, although setting
an appropriate boundary condition is not trivial. In the case of
short perturbations, such as {comet}
{Shoemaker-Levy 9} impact{ing} Jupiter's
atmosphere in 1994, the giant storm in Saturn's atmosphere in 2011,
dust storm evolution on Mars or climate change in
the{ atmosphere of} modern Earth, the above
equations could be solved as an initial value problem to understand
the evolution of the atmosphere under perturbations.

In this ``minimum recipe{,}" there are several unknown
parameters: temperature, pressure, density, wind velocity vector,
light intensity (and associated radiative heating and cooling rate
and actinic flux), and abundances of chemical tracers including
dust, haze and cloud particles. Complexity emerges because of the
coupling of parameters and interaction among processes, leading to a
high nonlinearity in this system. Realistic atmospheres could
only be much more complicated. For example, the equation
set~(\ref{basiceq}) does not explicitly include the magnetic field,
which becomes important when the atmosphere is so hot that it could
be partially ionized. In the high-temperature regime, magnetic field
might play a significant role (e.g.,
\citealt{batyginNonaxisymmetricFlowsHot2014};
\citealt{rogersConstraintsMagneticField2017}). Once the magnetic
field is coupled with the atmospheric flow, magnetohydrodynamics
(MHD) becomes complicated, especially if there is ion chemistry.
Maxwell's equations will need to be solved. We also did not include
the collisionless region in the upper atmosphere where the atoms and
molecules escape from the planet. In that case, the Boltzmann
equation needs to be solved. Interaction between the stellar wind
and the atmosphere is{ also} complicated. The
near-surface (boundary layer) physics that describes how the lower
atmosphere interacts with the surface is not detailed. If one is
interested in the deep atmosphere which does not obey the ideal gas
law, different equations of state also need to be adopted
in the high pressure and high-temperature regime although the
available data are sparse.

The {c}limate system contains a wide range of length
scales and timescales. Take Earth's atmosphere as an example. The
length scale spans from interactions between electromagnetic waves
and atoms/molecules at atomic/molecular scale
($\sim10^{-10}${\,}m), to aerosol and cloud microphysics
($10^{-8}-10^{-3}${\,}m {with} particle
size), to regional turbulence ($10^{-2}-10${\,}m), to
convective systems ($10^3${\,}m for tornados to
$10^6${\,}m for hurricanes), to synoptic weather systems
(e.g., $\sim 10^6${\,}m for baroclinic instability), to
planetary-scale dynamics ($\sim 10^6${\,}m for zonal jets
and overturning circulations), to finally more than the planetary
scale ($\sim 10^7${\,}m, such as planetary hydrodynamic
outflows). The timescale varies from molecular collisions (e.g.,
$\sim 10^{-10}${\,}s for the near-surface air), to
quantum state lifetime in radiation (e.g., $\sim
10^{-9}${\,}s for some electronically excited states), to
chemical reactions (from $\sim 10^{-8}${\,}s in radical
reactions to $\sim 10^{5}${\,}yr in
silicate weathering), to turbulent flow near the surface (seconds to
hours), to molecular and eddy diffusion (hours to weeks), to
hydrodynamical flow (hours to days), to radiative cooling (several
days at the surface), to seasonal variability (months), to
interannual variability (years to decades, e.g., ENSO), to ocean
dynamics ($>10^3${\,}yr), to orbital
change of the planet ($10^4-10^5${\,}yr),
to atmospheric escape ($>10^6${\,}yr), to
geological and interior processes ($10^6-10^8${\,}yr), to the secular variation of the host
star ($>10^9${\,}yr).

Tackling all of these length scales and timescales together is
impossible, and often investigations need to be simplified and
isolated. Also, breaking the system down to many scales with various
levels of complexities is the pathway for not only making models or
theories viable but {to} guarantee
understanding. Based on the ``minimum recipe" equation
set~(\ref{basiceq}) and using common simplifications such as
hydrostatic balance, large aspect ratio and small density
variation, one can formulate simpler equations to describe the
behavior of the atmosphere. Some famous forms include the
quasi-geostrophic equations, shallow water equations, primitive
equations, Boussinesq equations and anelastic equations. See
textbooks such as \cite{vallisAtmosphericOceanicFluid2006},
\cite{pedloskyGeophysicalFluidDynamics2013} and
\cite{holtonDynamicMeteorologyStratosphere2016} for details.

\subsection{Cold versus Hot Regimes}
\label{sect:coldhot}

To first order, we highlight ``cold" versus ``hot"
atmospheres enlightened by the emerging ensemble of exoplanets and
brown dwarfs across a broad range of temperatures. The regime
boundary between cold and hot is vaguely defined as the temperature
for water vapor-liquid phase transition at 1 bar
($\sim$373{\,}K). Traditional studies on planetary
atmospheres in the Solar System mostly focus on the ``cold regime"
except a few studies such as on the lower atmosphere of Venus and
deep atmospheres of giant planets. On the other hand, most
characterized exoplanets and brown dwarfs to date would fall in the
``hot regime{.}" Examples include ultra-hot
Jupiters with equilibrium temperatures higher than
2200{\,}K, including WASP-121\,{\,}b,
WASP-12\,{\,}b, WASP-103\,{\,}b, WASP-33\,{\,}b,
Kepler-13A\,{\,}b, WASP-18\,{\,}b and
HAT-P-7\,{\,}b (see spectra compiled in
\citealt{parmentierThermalDissociationCondensation2018}), and
scorching ones KELT-1\,{\,}b and KELT-9\,{\,}b. The latter
is the hottest known exoplanet to date with a dayside temperature of
$\sim$4600{\,}K
(\citealt{gaudiGiantPlanetUndergoing2017}). In the past two decades,
observations on substellar atmospheres gave birth to a new sub-field
in atmospheric science to study
``high-temperature atmospheres{.}" Conventional theories
of cold atmospheres in the Solar System might have
neglected critical processes in hot substellar atmospheres. In
Table~1, we highlight several possible essential differences
in the physical and chemical processes between the two regimes.

Compared with low-temperature
atmospheres, high-temperature atmospheres become more active {so that} processes generally operate faster. In atmospheric radiation,
the electron states in the atoms and molecules are easier to be
excited at {a} higher temperature. Numerous weak energy
transitions in the molecular electronic, vibrational and
rotational s{t}ates---usually negligible in the low-temperature
regime---have become significantly stronger to increase the opacity
of the atmosphere. The population of quantum states is
prone to deviate from the Boltzmann distribution under high
temperature, leading to Non-Local Thermodynamic
Equilibrium (non-LTE)
effect where the gas emission does not obey the simple Planck law
anymore. Instead, a complicated vibrational state ``chemistry"
impacts the atmospheric absorption and emission properties. Third,
the radiative timescale is shorter at {a} higher
temperature, implying a faster dissipation rate f{or} the
atmospheric heat.

\begin{table*}  
\bc
\begin{minipage}[]{120mm}
\caption[]{Atmospheric processes in low and high temperature regimes. \label{tab1}}\end{minipage}
\fns\renewcommand\baselinestretch{1.4}
\begin{tabular}{p{2.2cm}|p{6.2cm}|p{6.2cm}}
\hline\noalign{\smallskip}
Process & Low Temperature Regime & High Temperature Regime \\
\hline\noalign{\smallskip}
Radiation & LTE, less spectral lines &  non-LTE, more spectral lines from excited energy levels \\
\hline\noalign{\smallskip}
Gas chemistry    & one-way reactions dominate  &  forward and backward reactions, thermal ion chemistry\\
\hline\noalign{\smallskip}
Condensed phase  & molecular solid/liquid (e.g., \ch{H_2O}, \ch{CH_4}, \ch{NH_3}, \ch{N_2}, \ch{CO_2})  &     covalent/ionic/metallic refractive solids (e.g., silicate, Fe, KCl)\\
\hline\noalign{\smallskip}
Dynamics     & low-speed waves and wind, moist and dry convection, negligible magnetic coupling  &  high-speed waves and wind, dry convection, MHD effect \\
\hline\noalign{\smallskip}
Escape   & Jeans escape, non-thermal processes   &  hydrodynamic escape, non-thermal processes\\
\hline\noalign{\smallskip}
Surface interaction &  condensation/collapse on surface ocean or ice&   gas exchange with magma ocean or melted surface\\
\hline
\end{tabular}
\ec
\end{table*}

Given sufficient time, chemical reactions {proceed}
in both{ the} forward and backward directions towards
thermodynamical equilibrium---the minimum Gibbs free energy state.
In reality, because the reaction rates of the forward and backward
reactions usually have different temperature dependence, they
typically proceed at different speeds. In the cold regime, one
direction (namely the ``forward reaction") will proceed much faster
than the other direction. Other fast atmospheric processes, such as
wind transport, if more rapid than the backward reaction, lead to
chemical disequilibrium. In the high-temperature regime, both the
forward and backward reactions speed up, and species more easily
reach thermochemical equilibrium. Nevertheless, disequilibrium
chemistry is still essential because wind transport might also
become more potent at a higher temperature. Chemical models seem to
support that colder atmospheres show more substantial signs of
disequilibrium than hotter atmospheres, but more observations are
needed to confirm this hypothesis (e.g.,
\citealt{lineSystematicRetrievalAnalysis2013}).

Furthermore, in the cold regime, chemistry in the infrared{
(IR)} emission layers is mostly neutral chemistry among molecules.
Photoionization and thermal ionization could only be important in
the upper thermosphere and the auroral region. In the
high-temperature atmosphere, ionization more readily occurs, and
atomic neutrals and ions are common (e.g.,
\citealt{lavvasELECTRONDENSITIESALKALI2014}). For example, to date
about 15 atomic species have been detected in the atmosphere of a
very hot Jupiter KELT-9\,{\,}b (e.g.,
\citealt{yanExtendedHydrogenEnvelope2018};
\citealt{cauleyMagneticFieldStrengths2019};
\citealt{jenshoeijmakersAtomicIronTitanium2018};
\citealt{hoeijmakersSpectralSurveyUltrahot2019};
\citealt{pinoNeutralIronEmission2020}; \citealt{wyttenbachMasslossRateLocal2020}). Photoionization could also
be important in the photosphere if the planet is very close to the
central star. In these situations, ion-chemistry is also important
to understand the observed spectra.

Particles in the atmosphere could also be drastically different
between low and high-temperature regimes. The clouds on
Solar System planets are mostly molecular solids/liquids maintained
by intermolecular forces, such as sulfuric acid on Venus, water on
Earth and Mars, \ch{CO2} on Mars, ammonia and water on giant
planets, methane and hydrogen sulfide on Uranus and Neptune,
methane, ethane and hydrogen cyanide on Titan, as well as nitrogen
on Pluto and Triton. Observational spectra of hot atmospheres also
imply the existence of particles. {However,} all the
above cloud species in the Solar System will remain in the vapor
phase in the hot atmospheres. Instead, we expect different compounds
with much higher melting temperature, for example, refractive solids
maintained by network covalent bonds (e.g., silicate), metallic
bonds (e.g., iron) or ionic bonds (e.g., KCl). The formation
mechanisms of those mineral and iron clouds in the hot substellar
atmospheres are not well understood (see discussion in
Sect.~\ref{sect:cloud}). Atmospheric chemistry will also form
organic haze particles such as on Earth, Titan, Pluto, Triton
and giant planets. Experiments have shown that organic hazes are
able to form in various environment from 300--1500{\,}K
(\citealt{horstHazeProductionRates2018};
\citealt{heLaboratorySimulationsHaze2018,hePhotochemicalHazeFormation2018,heSulfurdrivenHazeFormation2020};
\citealt{fleuryPhotochemistryHotH2dominated2019};
\citealt{moranChemistryTemperateSuperEarth2020}). Whether the
detected particles in the hot substellar atmospheres are organic
haze particles or condensational dust clouds is still an open
question.

In atmospheric dynamics, a higher-temperature atmosphere usually has
a faster speed of {a} sound wave and other waves, and
perhaps higher wind speed too, depending on the rotation and other
parameters. While moist convection ubiquitously exists in thick
planetary atmospheres in the Solar System due to latent heat release
from the cloud condensation, it is less important in the
high-temperature regime than dry convection flux
(Sect.~\ref{sect:bddyn}). Moreover, because of partial
ionization of the atmosphere, the magnetic field will be more easily
coupled with the atmosphere. It exerts significant MHD drag
o{n} the atmospheric flow or causes significant Ohmic (or
Joule) heating. Strong magnetic fields ($\sim$20--120{\,}G) on several hot Jupiters have recently been
inferred through magnetic star-planet interactions
(\citealt{cauleyMagneticFieldStrengths2019}).

Atmospheric loss mechanisms could also be different between the two
regimes. In terms of atmospheric escape, most planetary atmospheres
in the Solar System are close to the hydrostatic state with a
moderate or weak Jeans escape. Hydrodynamic escape (i.e.,
atmospheric blow-off) could also occur in some cases such as the
solar wind and Earth's polar wind. Pluto's atmosphere was
{regarded} as a good candidate for ongoing
hydrodynamic escape (e.g.,
\citealt{zhuDensityThermalStructure2014}). However, the New Horizons
flyby in 2015 discovered a much colder atmosphere on Pluto, and thus
the atmospheric loss rate is much smaller
(\citealt{gladstoneAtmospherePlutoObserved2016};
\citealt{zhangHazeHeatsPluto2017}). On the other
hand, for a hot atmosphere close to the central star, the strong
stellar flux, and X-ray or {ultraviolet (}UV{)} heating in
the upper atmosphere could lead to{ outward} hydrodynamic
blow{-}off like a planetary wind. This flow has been
detected in recent observations of some exoplanets (e.g.,
GJ~436\,{\,}b, \citealt{ehrenreichGiantCometlikeCloud2015};
\citealt{lavieLongEgressGJ2017}. See
Sect.~\ref{sect:stabilityfund}).  The atmosphere can also be lost
to the surface or interior. In a very cold atmosphere, the bulk
atmospheric component, such as \ch{CO2} and \ch{N2}, could condense
onto the surface or even lead to total atmospheric
collapse (e.g., Mars, Io, Pluto and Triton). The condensation
does not readily occur in hot atmospheres. One exception could be
very hot rocky planets tidally locked to central stars.
The bulk composition on the dayside might be enriched in silicate
vapor such as on 51~Cancri\,{\,}e (e.g.,
\citealt{demoryVariabilitySuperEarth552016}), Kepler~1520\,{\,}b
(\citealt{rappaportPossibleDisintegratingShortperiod2012};
\citealt{perez-beckerCatastrophicEvaporationRocky2013}) and K2-22\,{\,}b
(\citealt{sanchis-ojedaK2ESPRINTProjectDiscovery2015}). The vapor
could condense to dust clouds when transport{ed} to the
nightside and collapse. A hot atmosphere could also melt its rocky
surface that leads to interesting interactions (such as ingassing)
between the atmosphere and {a} magma lake/ocean by
analogy with Jupiter's moon Io.

\subsection{Spectral and Photometric Characterization}
\label{sect:char}

Towards a more detailed classification of the substellar
atmospheres, photometry and spectroscopy play a central role. Both the atmospheric composition and temperature directly
control the broadband magnitudes and colors
{as well as} detailed spectral features in
transmission, emission and reflection spectra. Following conventional
stellar classification in{ the} Morgan-Keenan (MK) system
(\citealt{morganSpectralClassification1973}), brown dwarfs are
classified into several categories according to their spectral
colors in the optical and near-IR. The spectral types include L
dwarfs (\citealt{kirkpatrickImprovedOpticalSpectrum1999}), T dwarfs
(\citealt{burgasserSpectraDwarfsNearinfrared2002,
burgasserHubbleSpaceTelescope2006}, another classification scheme
from \citealt{geballeSpectralClassificationDwarfs2002} yielded the
similar results) and the Y class
(\citealt{cushingDISCOVERYDWARFSUSING2011} and
\citealt{kirkpatrickFurtherDefiningSpectral2012}). See the detailed
distinction between the M, L, T and Y spectral classes in the
review by \cite{kirkpatrickNewSpectralTypes2005}. In
optical wavelengths, the early-L dwarfs are characterized by
multiple atomic and molecular lines such as the neutral alkali
metals (e.g., Na\,I, K\,I, Rb\,I, Cs\,I), oxides (TiO and VO) and
hydride (e.g., FeH). Both alkali lines and hydrides increase
strength in the mid-L, but the oxides TiO and VO disappeared.  As
the dwarfs become colder such as in late-L and early-T, the spectra
show strong water features and alkali lines, whereas hydrides are
less important. In the late-T, water dominates the absorption and
the line widths of Na\,I and K\,I spread widely. Finally, in the
cold and faint Y-class, the optical features almost disappeared. The
characterization of brown dwarfs in near-IR is also
similar. Early-L spectra show features of \ch{H2O}, FeH and
CO, and atomic metal lines such as Na, Fe and K. \ch{CH4}
appears in early-T. \ch{CH4} and \ch{H2O} dominate the entire T-type
spectra. The Y-dwarfs show up at the cold end of the spectral
sequence where the alkali resonance lines disappear and possibly
ammonia absorption bands emerge in the near-IR (e.g.,
\citealt{cushingDISCOVERYDWARFSUSING2011};
\citealt{kirkpatrickFurtherDefiningSpectral2012};
\citealt{lineUniformAtmosphericRetrieval2017};
\citealt{zaleskyUniformRetrievalAnalysis2019}). On the other hand,
the mid-IR classification has not been well established yet
(\citealt{kirkpatrickNewSpectralTypes2005}). The effective
temperature of L dwarfs ranges from 1300{\,}K to
2500{\,}K and T dwarfs are typically below
1500{\,}K. The coldest known Y dwarf detected so far is
WISE 0855 with an effective temperature of about
235--260{\,}K
(\citealt{luhmanNewParallaxMeasurement2014}) where water could
condense as clouds (e.g., \citealt{skemerFIRSTSPECTRUMCOLDEST2016};
\citealt{esplinPHOTOMETRICMONITORINGCOLDEST2016};
\citealt{morleyBandSpectrumColdest2018};
\citealt{milesObservationsDisequilibriumCO2020}).

Within the L dwarfs, spectroscopic diversity can be further
classified using gravity as in the MK system because both the
opacity distribution and vertical temperature profile in the
atmosphere are significantly influenced by gravity. For example, the
weak FeH absorption and weak Na\,I and K\,I doublets indicate low
gravity objects (\citealt{cruzYoungDwarfsIdentified2009}). For brown
dwarfs, gravity is also a good proxy of age.
\cite{cruzYoungDwarfsIdentified2009} proposed a gravity
classification scheme for the optical spectra: $\alpha$ for normal
gravity, $\beta$ for intermediate gravity and $\gamma$ for
very low gravity; th{ose} latter two correspond to ages
of $\sim$100 and $\sim$10{\,}Myr, respectively.
{Utilizing} equivalent widths for gravity-sensitive
features (VO, FeH, K\,I, Na\,I and the {$H$
}band continuum shape) in the near-IR spectra,
\cite{allersNearinfraredSpectroscopicStudy2013} classified the young
brown dwarfs {in}to three types. They the low-gravity (VL-G),
intermediate gravity (INT-G) and field (FLD-G), corresponding
to ages of $\sim$30, $\sim$30--200{\,}Myr and
$\sim$200~Myr, respectively. Note that the gravity types are still
very uncertain.

Do planets also {manifest} typical spectral types?
Planets are more diverse than brown dwarfs because their temperature
and compositions are affected by many factors, such as the distance
to the star, metallicity, gravity and internal heat. Despite their complex nature,
\cite{fortneyUnifiedTheoryAtmospheres2008} proposed that the dayside
atmospheres of hot Jupiters could be classified into two categories:
the hotter ``pM" and the cooler ``pL" classes, by analogy to M and L
brown dwarfs{ respectively}. The ``pM" planets with an effective temperature
greater than 2000{\,}K will {exhibit}
strong thermal inversion (i.e., temperature increases with altitude)
in the upper atmosphere caused by the TiO and VO opacity sources and
high irradiation from the parent stars, as well as a large day-night
temperature difference due to the shorter radiative timescale than
the advective timescale. Their dayside spectra are expected to
{display} emission features in photospheres. Note that
the existence of TiO or VO in the upper atmosphere might require
strong vertical mixing (\citealt{spiegelCANTiOEXPLAIN2009}). On the
other hand, the cooler class, ``pL", could re-radiate away the
incoming stellar radiation more easily and show no thermal inversion
in the photosphere. Water absorption features in the near-IR instead
dominate their spectra. The search for evidence of the two classes
and thermal inversion is still ongoing. Some recently characterized
ultra-hot Jupiters with equilibrium temperature greater than
2200{\,}K have been confirmed with temperature inversion
and emission features detected, including WASP-121\,{\,}b
(\citealt{evansUltrahotGasgiantExoplanet2017}), WASP-18\,{\,}b
(\citealt{sheppardEvidenceDaysideThermal2017};
\citealt{arcangeliOpacityWaterDissociation2018}) WASP-33\,{\,}b
(\citealt{haynesSPECTROSCOPICEVIDENCETEMPERATURE2015};
\citealt{nugrohoHighresolutionSpectroscopicDetection2017}). Even
though the detailed mechanisms of thermal inversion might
not exactly be due to the previously proposed TiO/VO opacity, and
the transition between planets with and without inversion might not
coincide with 2000{\,}K as proposed in
\cite{fortneyUnifiedTheoryAtmospheres2008}, it seems the
exoplanetary atmospheres do show some typical spectral categories
that can be further characterized in future spectral observations.

In addition to temperature, gravity might also play a role.
\cite{parmentierThermalDissociationCondensation2018} classified the
hot Jupiter spectra at secondary eclipse using gravity and the
dayside temperature. In the higher gravity and/or lower temperature
regime, TiO is expected to rain out (e.g.,
\citealt{spiegelCANTiOEXPLAIN2009};
\citealt{parmentier3DMixingHot2013};
\citealt{parmentierTransitionsCloudComposition2016};
\citealt{beattyEvidenceAtmosphericColdtrap2017}). In the higher
temperature regime (such as on ultra-hot Jupiters with an
equilibrium temperature, $T_{\rm eq}>$2200{\,}K), most
spectrally relevant molecules, except some with very strong bonds
such as \ch{N2}, CO and SiO, tend to be thermally dissociated,
resulting in spectra with very weak features in general. \ch{H-}
opacity becomes an important opacity source in the high-temperature
regime as well. More discussions are in Section~\ref{sect:verttemp}.

Photometrically, the substellar bodies can be characterized using
color-magnitude diagrams (CMDs) similar to the Hertzsprung-Russell
diagram for stars. \cite{dupuyHAWAIIINFRAREDPARALLAX2012} compiled a
large number of brown dwarfs and illustrated their evolution
sequence in both near-IR and mid-IR. One typical diagram is the
{$J-H$} color versus {$J$}{ }band
magnitude in Figure~1, in which we convert the {$H$} band
flux to the HST channels for
comparison with data from hot Jupiters and ultra-hot Jupiters
(\citealt{manjavacasCloudAtlasHubble2019}). In the optical and
near-IR sequences, the spectral sequence of brown dwarfs from M, L,
T to Y types is evident. As the {$J$ }band flux
decreases, {$J-H$} color is gradually reddening from M
to late L, and suddenly shifts to blue in early-T within an
effective temperature range of only about 200{\,}K. It
continues bluer to the mid- and late-T but eventually turns back to
red in the Y-types. Also, some discrepancies between optical and
near-{IR} types for L dwarfs and the evolution
sequences exist (\citealt{kirkpatrickNewSpectralTypes2005}). In
terms of gravity, the low-gravity objects (VL-G) are systematically
redder and brighter than the field brown dwarfs. {For
m}ore photometric behaviors on the gravity dependence{,}
refer to \cite{liuHAWAIIINFRAREDPARALLAX2016}.

Those spectral trends are statistically robust, but underlying
mechanisms are not fully understood.
\cite{stephensJHKMagnitudesDwarfs2003} suggested that the optical
sequence primarily came from temperature, but the
near-{IR} diagram is influenced more by clouds or
possibly gravity. \cite{kirkpatrickNewSpectralTypes2005} also argued
that the main driver of spectral evolution is
temperature, but condensational clouds also play an important role
in the spectral change. The inclination angle of those bodies viewed
from Earth could also impact the color diversity (e.g.,
\citealt{kirkpatrickDiscoveriesNearinfraredProper2010};
\citealt{metchevWeatherOtherWorlds2015};
\citealt{vosViewingGeometryBrown2017}). The temperature and cloud
formation seem mainly driving the M-L sequence as the objects redden
as they cool. The observed temperature and spectral types are only
correlated well from early to mid-L. The correlation breaks down in
the very sharp transition from mid-L to mid-T as the
near-{IR} color changes blueward in a very narrow
effective temperature range ($\sim1400\pm200${\,}K)
(\citealt{kirkpatrickNewSpectralTypes2005}). This transition has
been observed for young, old and spectrally peculiar objects
in the near-IR (\citealt{liuHAWAIIINFRAREDPARALLAX2016}). The
underlying mechanisms of the so-called L/T transition problem are
not known yet. It was proposed to be relevant to the change of the
cloud properties in the atmospheres (e.g.,
\citealt{saumonEvolutionDwarfsColorMagnitude2008};
\citealt{marleyPatchyCloudModel2010}) but alternative mechanisms
have also been suggested, such as dynamical regime change driven by
gas composition change with temperature (e.g.
\citealt{tremblinFINGERINGCONVECTIONCLOUDLESS2015,
tremblinCLOUDLESSATMOSPHERESDWARFS2016,tremblinCloudlessAtmospheresYoung2017,tremblinThermocompositionalDiabaticConvection2019}).
We will discuss the details in Section~\ref{sect:bdcloud} and
Section~\ref{sect:bddyn}.

\begin{figure*}  
   \centering
   \includegraphics[width=0.85\textwidth, angle=0]{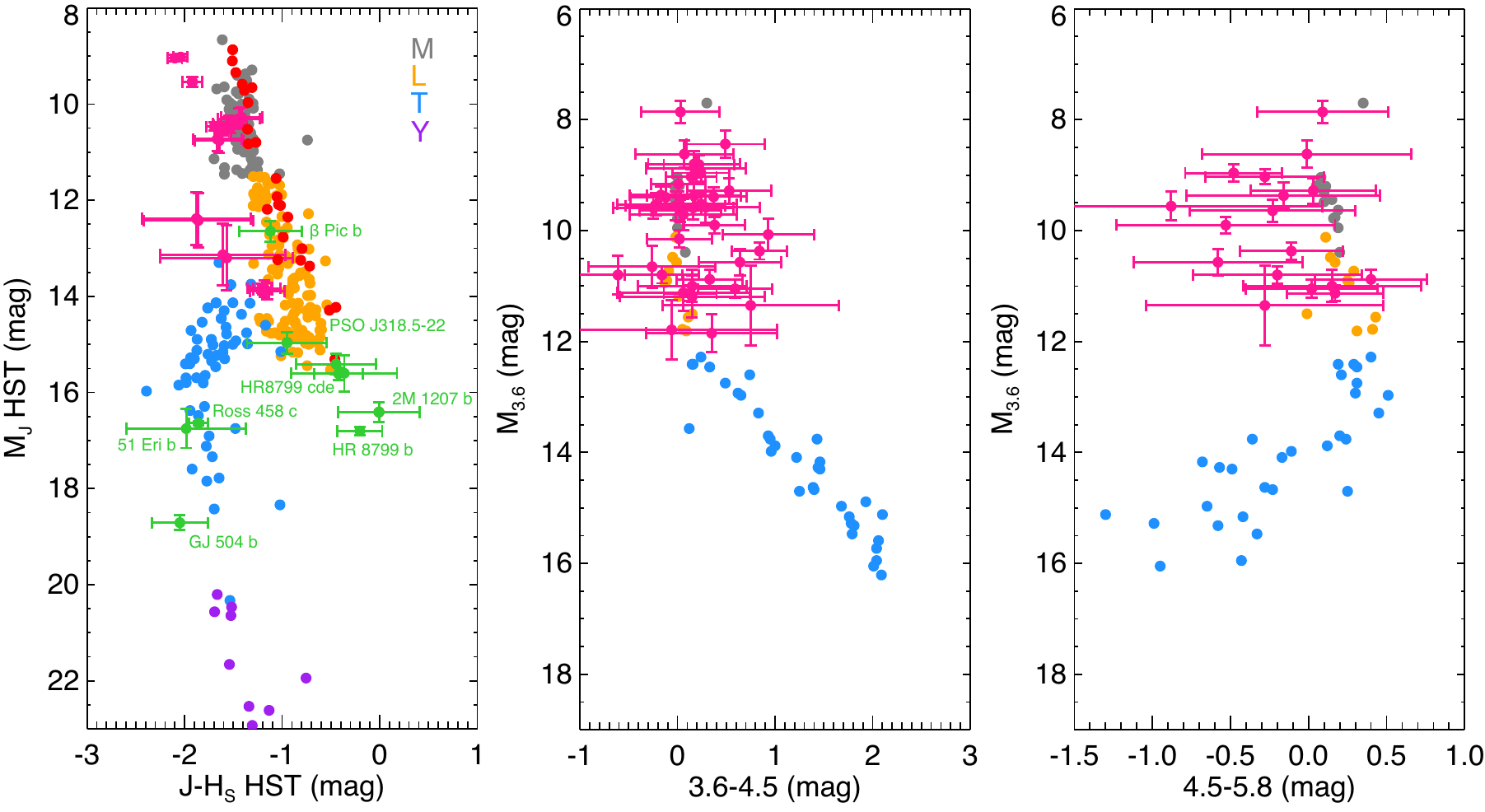}
   \caption{\baselineskip 3.8mm {CMDs} for hot Jupiters ({\it pink}), directly imaged planets ({\it green}) and brown dwarfs (\citealt{dupuyHAWAIIINFRAREDPARALLAX2012}). {\it Grey}, {\it orange}, {\it blue} and {\it purple} stand for M, L, T and Y dwarfs respectively. {\it Red dots} are very low gravity objects from \cite{liuHAWAIIINFRAREDPARALLAX2016}. For hot Jupiters, HST data ({{\it left panel}}) are from \cite{manjavacasCloudAtlasHubble2019} and thermal {IR} data (the {{\it middle}} and {{\it right panels}}) come from \cite{triaudColourMagnitudeDiagrams2014} and \cite{kammerSPITZERSECONDARYECLIPSE2015}. The directly imaged planet data are from various sources: Beta-pic\,{\,}b from \cite{bonnefoyPhysicalOrbitalProperties2014},
PSO~J318 from \cite{liuEXTREMELYREDYOUNG2013} and
\cite{liuHAWAIIINFRAREDPARALLAX2016}, 51~Eri\,{\,}b from
\cite{macintoshDiscoverySpectroscopyYoung2015}, GJ~504\,{\,}b
from \cite{liuHAWAIIINFRAREDPARALLAX2016}, Ross~458\,{\,}c from
\cite{cushingDISCOVERYDWARFSUSING2011}, the HR~8799 system (b, c, d
and e) from \cite{zurloFirstLightVLT2016}{ and}
2M~1207\,{\,}b from
\cite{allersNearinfraredSpectroscopicStudy2013}. In the {{\it left panel}}, both brown dwarfs and directly imaged
planets are converted into HST colors based on the scaling
relationship in \cite{manjavacasCloudAtlasHubble2019}. }
   \label{figcmd}
\end{figure*}

The mid-IR {CMD} in the
Spitzer/IRAC bands{ of} 3.6 and 4.5 microns is also
{depicted} in Figure~\ref{figcmd}
(\citealt{triaudColourMagnitudeDiagrams2014,triaudColourMagnitudeDiagrams2014a}).
The mid-IR {CMD} does not
{display} very distinct spectral types, and the
[3.6]--[4.5] color stays roughly the same in M and mid-L. At around
1400{\,}K, as the temperature drops, the [3.6]--[4.5]
color exhibits a clear redward shift. This transition is the mid-IR
L/T transition. The [4.5]--[5.8] {CMD} (also see Fig.~\ref{figcmd}) also shows the sharp L/T
transition but towards the blue end as the effective temperature
decreases (\citealt{triaudColourMagnitudeDiagrams2014a}). The
temperature-driven gas chemistry probably causes both the redward
turns in the [3.6]--[4.5] diagram and blueward turn in the
[4.5]--[5.8] diagram (\citealt{triaudColourMagnitudeDiagrams2014a}).
In the mid-IR, the vibrational-rotational bands of \ch{CH4},
CO and \ch{H2O} dominate the absorptions at 3.6, 4.5
and 5.8 microns, respectively. As the temperature drops below the
L/T transition temperature, the thermochemical reaction CO $+$
3\ch{H2} $\rightarrow$ \ch{CH4} $+$ \ch{H2O} favors the production
of \ch{CH4} and \ch{H2O}. Consequently, both the absorption at 3.6
and 5.8 microns increase, but CO absorption at{ the} 4.5-micron
band decreases, resulting in the color change in the mid-IR L/T
transition. Clouds might affect the sharp gradient as well, but it
has yet to be investigated in detail.

Do exoplanets also follow similar color-magnitude sequences? We
first consider close-in exoplanets. Because the emission
from close-in planets is mainly from the re-radiation of
external stellar energy rather than internal energy, this
might not be an apples-to-apples comparison. The dayside emissions
of close-in planets (mostly hot Jupiters) from HST near-IR (from
\citealt{manjavacasCloudAtlasHubble2019}) and Spitzer mid-IR
channels (from \citealt{triaudColourMagnitudeDiagrams2014} and
\citealt{kammerSPITZERSECONDARYECLIPSE2015}) are plotted on top of
the brown dwarf samples in Figure~\ref{figcmd} for
comparison. There is no well-characterized hot Jupiter in the
T-dwarf temperature range yet. It looks that the planets and brown
dwarfs might show similar trends, indicating their spectral sequence
might share some similar underlying mechanisms. Note that the
scattering of color indices in the exoplanet sample is much larger
than the brown dwarfs in both near-IR and mid-IR. Also, the radii of
hot Jupiters might change by a factor of several (from
0.5--2{\,}$R_J$), which could influence the magnitude but
are not likely to cause such a large diversity
{apparent} in Figure~\ref{figcmd}
(\citealt{triaudColourMagnitudeDiagrams2014a}). Instead, this
scattering in planetary samples suggests that the planets have a
larger diversity influenced by other parameters such as gravity,
host star irradiation, internal heat and metallicity.
{Implementing} a more physically based model,
\cite{adamsReassessingExoplanetLight2018} derived simple physical
model parameters based on the observed light curves. They found that
although there seem{ to be} statistical trends in the
{CMDs}, the trends in the
individual derived parameters are not obvious.

Young, directly imaged planets offer a more direct comparison to
brown dwarfs because of their similar self-luminous nature. One
would expect those directly imaged planets should{ be}
locate{d} within the low-mass brown dwarfs (such as the VL-G
sequence, \citealt{liuHAWAIIINFRAREDPARALLAX2016}). In the currently
limited samples, it seems that the near-IR photometric behaviors of
several characterized directly imaged planets follow the L dwarf
spectral sequence generally well (Fig.~\ref{figcmd}).
$\beta$-pic\,{\,}b and PSO~J318 resemble mid-L and late-T
types, respectively. Three (51~Eri\,{\,}b,
GJ~504\,{\,}b and Ross~458\,{\,}c) follow the T
sequence. However, the HR~8799 system (b, c, d and e) and
2M~1207\,{\,}b, which have a similar effective temperature
as T-dwarfs, continue the L dwarf sequence further towards
red. To date, whether the directly imaged exoplanets exhibit a clear
L/T transition is inconclusive.

Population studies have shed light on possible statistical
properties of an ensemble of exoplanetary spectra and light curves
(e.g., \citealt{stevensonQuantifyingPredictingPresence2016};
\citealt{singContinuumClearCloudy2016};
\citealt{crossfieldTrendsAtmosphericProperties2017};
\citealt{fuStatisticalAnalysisHubble2017};
\citealt{tsiarasPopulationStudyGaseous2018};
\citealt{fisherRetrievalAnalysis382018};
\citealt{wakefordExoplanetAtmosphereForecast2019};
\citealt{gaoAerosolCompositionHot2020}). For
example, spectral strengths of metals and water vapor in the
transmission spectra of transit{ing} exoplants can be
{employed} to quantify how cloudy the atmospheres
are. The
presence of high-altitude condensational clouds and photochemical
hazes could significantly weaken the spectral absorption features.
Such flattened transmission spectral features have been detected for
many hot Jupiters (e.g.,
\citealt{demingINFRAREDTRANSMISSIONSPECTROSCOPY2013};
\citealt{mandellEXOPLANETTRANSITSPECTROSCOPY2013};
\citealt{gibsonOpticalTransmissionSpectrum2013};
\citealt{singHSTHotJupiterTransmission2013};
\citealt{jordanGroundbasedOpticalTransmission2013};
\citealt{wilkinsEmergent11Mm2014};
\citealt{mallonnTransmissionSpectroscopyHATP32b2016};
\citealt{singContinuumClearCloudy2016};
\citealt{fuStatisticalAnalysisHubble2017}). Cooler and smaller
planets such as warm Neptunes and super-Earths are also inferred to
possess high altitude aerosols (GJ~1214\,{\,}b,
\citealt{kreidbergCloudsAtmosphereSuperEarth2014};
GJ~436\,{\,}b,
\citealt{knutsonFeaturelessTransmissionSpectrum2014};
HD~97658\,{\,}b, \citealt{knutsonHubbleSpaceTelescope2014};
GJ~3470\,{\,}b,
\citealt{ehrenreichNearinfraredTransmissionSpectrum2014};
HAT-P-26\,{\,}b,
\citealt{wakefordHATP26bNeptunemassExoplanet2017}; HD 106315\,c, \citealt{kreidbergTentativeEvidenceWater2020}).
HAT-P-11\,{\,}b is partially cloudy as water vapor can
be seen in the HST near IR band
(\citealt{fraineWaterVapourAbsorption2014}) and the nearly flat
optical transmission spectrum from HST STIS
(\citealt{chachanHubblePanCETStudy2019}). Also, two
super-puffs---planets with very low masses but large radii,
Kepler~51\,{\,}b and d---have been observed to show flat
transmission spectra in the near IR, indicating abundant atmospheric
hazes or dust particles
(\citealt{libby-robertsFeaturelessTransmissionSpectra2020};
\citealt{wangDustyOutflowsPlanetary2019};
\citealt{gaoDeflatingSuperpuffsImpact2020}). Two other cooler
sub-Neptunes{,} K2-18\,{\,}b
(\citealt{bennekeWaterVaporClouds2019};
\citealt{tsiarasWaterVapourAtmosphere2019}) and K2-25\,{\,}b
(\citealt{thaoZodiacalExoplanetsTime2020}){,} might also have
hazes or clouds in their atmospheres but the details are not certain
yet. A systematic analysis of the transmission spectra in
\cite{wakefordExoplanetAtmosphereForecast2019} showed that, on most
hot Jupiters, the amplitudes of the near-{IR}
water spectral features are $\sim$1/3 of that expected in clear-sky
models, indicating a ubiquitous presence of suspended particles
(such as clouds and hazes) on gas giants. A{n} HST campaign
(\citealt{singContinuumClearCloudy2016}) observed hot Jupiters
across a broad range of physical parameters and inferred the
cloudiness from the strength of water band signals. The spectral
strength of water is found to correlate well with the relative
absorption strength difference between optical and near-IR and also
with that between the near-IR and mid-IR. This provides strong
evidence that the clouds and hazes significantly shape both the
optical and {IR} color of transiting giant
exoplanets.

\cite{stevensonQuantifyingPredictingPresence2016} proposed a
cloudiness metric using the {$J$ }band water
feature amplitude ($A_H$) in transmission spectra. The larger the
$A_H$ is, the less cloudy the atmosphere is. In a sample of 14
exoplanets, there is a positive correlation between the cloudiness
proxy $A_H$ and the equilibrium temperature $T_{\rm eq}$ when
$T_{\rm eq} <$ 750{\,}K. A weak correlation between water
signal and gravity was also proposed.
\cite{fuStatisticalAnalysisHubble2017} generalized that study to
include 34 transiting exoplanets and found a positive correlation
between the cloudiness $A_H$ (similar to that in
\citealt{stevensonQuantifyingPredictingPresence2016}) and $T_{\rm
eq}$ between 500--2500{\,}K (also see a recent study in
\citealt{gaoAerosolCompositionHot2020}). For Neptune-size{d}
planets, \citealt{crossfieldTrendsAtmosphericProperties2017}
reported that the $A_H$ for smaller planets might also correlate
with the equilibrium temperature and bulk H/He mass fraction, which
they interpreted as a correlation between the metallicity and
cloudiness. We will discuss the theories of clouds and hazes on
exoplanets and their implications on the above observational trends
in Section~\ref{sect:cloud}.

The transmission spectra on close-in brown dwarfs are difficult to
obtain because of their high gravity and small scale height.
{However,} the emission spectra of directly imaged
planets and brown dwarfs have higher {S/N} compared with their close-in counterparts. As we
showed in Figure~\ref{figcmd}, their near IR
{CMD} indicate{s} the existence
of clouds in their atmospheres for (e.g., L dwarfs). The spectra of
many directly imaged planets might also need clouds or hazes to
explain (e.g., HR~8799 planets:
\citealt{barmanCloudsChemistryAtmosphere2011};
\citealt{madhusudhanMODELATMOSPHERESMASSIVE2011};
\citealt{marleyMASSESRADIICLOUD2012};
\citealt{currieDEEPTHERMALINFRARED2014};
\citealt{skemerDIRECTLYIMAGEDLT2014}; $\beta$-Pic\,{\,}b:
\citealt{chilcoteFirstHbandSpectrum2014},
\citealt{chilcote12MuNearIR2017}; 51~Eri\,{\,}b:
\citealt{macintoshDiscoverySpectroscopyYoung2015};
\citealt{zahnlePHOTOLYTICHAZESATMOSPHERE2016}).
{Moreover,} there are also many dusty brown dwarfs
typically in the L spectral type, some even
{displaying a} possible silicate feature in the
mid-IR (\citealt{cushingSpitzerInfraredSpectrograph2006}).

The data quality of emission spectra on close-in exoplanets is
generally lower than that on directly imaged planets because of
stellar contamination. Nevertheless, their mid-IR broadband emission
at 3.6 and 4.5 microns can be observed by Spitzer during their
secondary eclipses (e.g.,
\citealt{triaudColourMagnitudeDiagrams2014a}). Although their mid-IR
{CMDs} {manifest}
large scattering (Fig.~\ref{figcmd}), recent studies have
searched for an internal correlation of the mid-IR brightness
temperatures between the 3.6 and 4.5 micron data
(\citealt{kammerSPITZERSECONDARYECLIPSE2015};
\citealt{wallackInvestigatingTrendsAtmospheric2019};
\citealt{garhartStatisticalCharacterizationHot2020}). The
statistical analysis seems to suggest a systematic deviation of the
mid-IR spectra from the blackbodies. Moreover, there seems{ to
be} a statistically increasing trend of the observed brightness
temperatures between 4.5 and 3.6 microns with increasing
equilibrium temperature in the range of 800--2500{\,}K
(\citealt{garhartStatisticalCharacterizationHot2020}). This trend is
still a puzzle that no current theory can explain. We will discuss
it in detail in Section~\ref{sect:verttemp}.

Although it is difficult to resolve surface features o{n}
distant substellar bodies, time-domain observations provide clues on
their temporal and spatial variations.  Horizontal information
o{n} substellar atmospheres is primarily obtained from
light curve observations. In addition, eclipse
mapping (e.g., \citealt{rauscherEclipseMappingHot2007};
\citealt{dewitConsistentMappingDistant2012}) has been suggested to
be able to probe the spatial feature{s} in future
observations.
There are three kinds of light curves: reflection, transit
and emission. The stellar flux strongly contaminates reflection
light curves for close-in planets, and the signals are weak for
planets far away from their host stars. For transit{ing}
planets, transit light curves in principle could also be used to
probe the difference between east limb and west limb (e.g.,
\citealt{lineInfluenceNonuniformCloud2016};
\citealt{kemptonObservationalDiagnosticDistinguishing2017};
\citealt{powellTransitSignaturesInhomogeneous2019}) but the current
{S/N} is still not good enough.

Emission light curves originate from the time evolution of
hemisphere-averaged thermal flux emitted from the planets towards
the observer. There are two general types of emission light curves.
For close-in exoplanets and close-in brown dwarf companions with
self-rotations synchronized with their orbits around central stars
due to gravitational tides, emission light curves trace different
phases in the orbits and are also called thermal phase
curves. Most thermal phase curves are detected through the ``warm
Spitzer band" at 3.6 and 4.5 microns (see review in
\citealt{parmentierExoplanetPhaseCurves2018a}). For very hot
planets, it is also possible to observe emission light
curves from the visible band such as {Transiting Exoplanet
Survey Satellite }({TESS, }e.g.,
\citealt{shporerTESSFullOrbital2019}). On the other hand, if a
cooler planet is bright, the detected light curves in short
wavelengths (e.g., HST near-IR band, Kepler band) might include a strong
stellar reflection component (e.g.,
\citealt{parmentierTransitionsCloudComposition2016}). The shape of
the phase curves directly probes the photospheric inhomogeneity on
these synchronously rotating planets. For example,
\cite{knutsonMapDaynightContrast2007} detected a phase offset of the
light curve peak, suggesting that the heat redistribution due to
atmospheric jets and waves shifts the hot spot away from the
substellar point. The temporal variation of the phase curves between
different rotations also suggests complicated weather patterns on
these planets. One example is Kepler observations
{targeting} HAT-P-7\,{\,}b, on which the peak
brightness offset changes dramatically with time
(\citealt{armstrongVariabilityAtmosphereHot2016}). Another example
is Kepler~76\,{\,}b
(\citealt{jacksonVariabilityAtmosphereHot2019}). The phase curve
amplitude can vary by a factor of two in tens of days, associated
with the peak offset varying accordingly. Population studies have
also been performed to understand the statistical properties---such
as the albedo and heat redistribution---of an ensemble of
exoplanetary phase curves (e.g.,
\citealt{cowanStatisticsAlbedoHeat2011};
\citealt{schwartzPhaseOffsetsEnergy2017};
\citealt{zhangPhaseCurvesWASP33b2018};
\citealt{keatingUniformlyHotNightside2019}). Current data samples on
transiting planets might have revealed some possible interesting
trends of the dayside temperature, nightside temperature, day-night
temperature difference, and phase offset on various parameters such
as equilibrium temperature and rotation rate. Details will be
discussed in Section~\ref{sect:horitemp}.

The emission light curves observed on directly imaged planets and
brown dwarfs fall into a different category. On these bodies,
{IR} emission is modulated by planetary
self-rotation and in-and-out-of-view of the weather patterns in the
photospheres, producing rotational light curves. Photometric
variability has been monitored for brown dwarfs since their
discovery. Their rotational light curves unveil very active weather
associated with temperature and cloud patterns, especially around
the L/T transition. The short-term and long-term variations of the
rotational light curves can be used to retrieve the surface features
(e.g., \citealt{apaiZonesSpotsPlanetaryscale2017}) and even the wind
speed (\citealt{allersMeasurementWindSpeed2020}). Recent progress
have been summarized in a series of papers on brown dwarfs from the
``Weather on other Worlds" program
(\citealt{heinzeWeatherOtherWorlds2013,heinzeWeatherOtherWorlds2015,metchevWeatherOtherWorlds2011,metchevWeatherOtherWorlds2015,miles-paezWeatherOtherWorlds2017})
and on both low-gravity brown dwarfs and planetary-mass companions
from the ``Cloud Atlas" program
(\citealt{lewCloudAtlasDiscovery2016,lewCloudAtlasUnraveling2019,lewCloudAtlasWeak2020,manjavacasCloudAtlasDiscovery2017,manjavacasCloudAtlasHubble2019,manjavacasCloudAtlasRotational2019,miles-paezCloudAtlasVariability2019,zhouCloudAtlasHighcontrast2019,zhouCloudAtlasHighprecision2020,zhouCloudAtlasRotational2018}),
as well as the reviews in \cite{billerTimeDomainBrown2017} and
\cite{artigauVariabilityBrownDwarfs2018}. We will discuss the
rotational light curves, their variability, and the underlying
mechanisms in Section~\ref{sect:rotlight} and
Section~\ref{sect:bddyn}.

\section{Atmospheric Stability}
\label{sect:stability}
\subsection{Fundamentals}
\label{sect:stabilityfund}

The stability of a planetary atmosphere primarily depends on the
planetary mass, radius and atmospheric temperature. For
planets with surfaces, if the surface temperature drops below the
main constituents' saturation temperatures, the atmosphere will
collapse. Possible ice-albedo feedback---the condensed ices (e.g.,
water, \ch{CO2, \;N2, \;CO}, and \ch{CH4}) could reflect more stellar flux
to space and further cools down the surface---accelerates
the process. {A} planet with a collapsed atmosphere
enters {a} snowball climate, with the surface pressure
being in thermodynamical equilibrium with the surface ices.
{Such} atmospheric collapse could be common for the
terrestrial climate. The current atmospheres of Pluto and Triton are
in this state. Earth was in the snowball phase several times. The
atmospheres of Mars (\citealt{forget3DModellingEarly2013}) and Titan
(\citealt{lorenzPhotochemicallyDrivenCollapse1997}) might have
collapsed in the past. Atmospheric collapse and condensation will
also greatly change the compositions of the atmosphere (see
Sect.~\ref{sect:gaschem}). The atmosphere could also be absorbed
into the magma ocean or the interior in the early age (e.g.,
\citealt{olsonNebularAtmosphereMagma2019};
\citealt{kiteAtmosphereOriginsExoplanet2020}).

In this section{,} we will mainly focus on escape to space
(e.g., \citealt{jeansDynamicalTheoryGases1904};
\citealt{parkerDynamicsInterplanetaryGas1958}). The atmosphere
escapes via both thermal and non-thermal processes. In thermal
escape, if the upper atmosphere temperature is so high---either due
to strong stellar heating, gravitational energy released during the
accretion phase or other heating mechanisms---that the thermal
velocities of molecules or atoms exceed the escape velocity of the
planet, that volatiles {are} no longer gravitationally
bound. If the atmosphere remains in balance and the velocities of
the molecules or atoms still follow the Maxwellian distribution,
only a fraction of the molecules in the high-velocity tail of the
distribution will be able to escape. This scenario is the Jeans
escape. The particles will escape to space from the exobase, which
is the altitude above which the atmosphere is no longer collisional.
If the temperature of the upper atmosphere is very
high{,} the entire atmosphere {can}
escap{e} hydrodynamically, driven by the pressure
gradient. Hydrodynamic
escape can still be diabatic. In some situations such as giant
impact by incoming asteroids or comets, if the atmosphere has enough
internal and {KE} per unit mass to escape
isentropically, extreme escape could occur as a quick blow-off. The
division between the two regimes (Jeans and pressure-driven escape)
can be roughly characterized by the Jeans parameter
$\lambda=GM_pm/k_BTR_p$, a dimensionless number that describes the
ratio of the gravitational energy $GM_pm/R_p$ to the thermal energy
of the upper atmosphere $k_BT$, where $M_p$ and $R_p$ are the
planetary mass and radius (or more precisely, the exobase radius),
respectively. $G$ is the gravitational constant, and $k_B$ is the
Boltzmann constant. $T$ is the temperature at the exobase. $m$ is
the mass of the escaping species. The Jeans parameter is also the
ratio of the pressure scale height to the planet radius. Moreover,
the square root of $\lambda$ is roughly equal to the ratio of the
escape velocity $v_e=(2G M_p/R_p)^{1/2}$ to the adiabatic sound
speed $v_s=(\gamma k_B T/m)^{1/2}$ where $\gamma$ is the adiabatic
index. This expression includes three crucial parameters of the
planetary atmosphere: mass, radius and temperature at the
exobase and the mass of the escaping particle, which is usually H or
He atom{s}.

Although atmospheric thermal escape has been studied for at least
170\,yr (from J.~J.~Waterson
\citealt{watersonPhysicsMediaThat1851}, also see early works by
\citealt{jeansDynamicalTheoryGases1904};
\citealt{parkerDynamicsInterplanetaryGas1958};
\citealt{huntenThermalNonthermalEscape1982,
huntenMassFractionationHydrodynamic1987,
huntenKuiperPrizeLecture1990};
\citealt{zahnleMassFractionationNoble1990}), the theory and
especially numerical simulations are still incomplete. Crudely
speaking, the transition between the two end-members---hydrostatic
Jeans escape (large $\lambda$) to hydrodynamic escape (small
$\lambda$)---is found to occur at around $\lambda\sim1$ (e.g.,
\citealt{volkovTHERMALLYDRIVENATMOSPHERIC2011};
\citealt{volkovKineticSimulationsThermal2011},
\citealt{tianAtmosphericEscapeSolar2015}). The reality is, however,
much more complicated. For example, the behavior also depends on the
collisional property of the medium characterized by the ``Knudsen
number" $\mathrm{Kn}$---the ratio of the mean free path of the
escaping gas to the planetary radius. Usually, the transition from
hydrodynamic to free molecular flow at the exobase is difficult to
resolve without molecular dynamics or Boltzmann numerical
simulation. The direct simulation Monte Carlo (DSMC) results
(\citealt{volkovKineticSimulationsThermal2011}) show that, for a
single component atmosphere, evaluated using the Jeans parameter and
Knudsen number, thermal escape processes fall into different
regimes. In the collisional regime (small $\mathrm{Kn}$), an
analytical theory is also consistent with the DSMC results in
\cite{gruzinovRateThermalAtmospheric2011}. The thermal escape at the
top of the planetary atmosphere can occur in three regimes: Parker,
Fourier and Jeans. In the traditional Parker regime
(\citealt{parkerDynamicsInterplanetaryGas1958,parkerDYNAMICALPROPERTIESLAR1964,parkerDYNAMICALPROPERTIESSTELLAR1964})
where the Jeans parameter is small ($\lambda < 2$), outflow behaves
as a supersonic ideal hydrodynamic wind. Thermal conductivity is
negligible, and the temperature structure is controlled by
isentropic expansion starting from the sonic surface. When{ the}
Jeans parameter is large ($\lambda
> \sim4-6$), {the }atmosphere escapes in a molecule-by-molecule fashion.
The escape rates are not significantly different from the
traditional Jeans flux if $\lambda>\sim6$. Thermal conduction is
important in this regime, and temperature could be nearly
isothermal. In between the Parker and Jeans regimes, thermal
conduction balances the hydrodynamic expansion. This transition
(so-called Fourier regime) occurs in a very narrow range of
$\lambda\sim2-4$. If the atmosphere is not very collisional, such as
{a} low-density medium (large $\mathrm{Kn}$), thermal
conduction is also significant. Otherwise, the traditional Parker
wind solution can lead to inaccurate results
(\citealt{volkovHydrodynamicModelThermal2016}).

{An a}tmosphere could also be lost to space via many
non-thermal processes such as photochemically driven escape, ion
pick-up by the stellar wind, stellar wind stripping, charge exchange
and so on (e.g., \citealt{holmstroemEnergeticNeutralAtoms2008};
\citealt{kislyakovaXUVExposedNonHydrostaticHydrogenRich2013};
\citealt{kislyakovaStellarWindInteraction2014};
\citealt{kislyakovaStellarDrivenEvolution2015};
\citealt{dongProximaCentauriHabitable2017}). To understand the
non-thermal escape processes requires sophisticated photochemical
and ion-chemical calculations in the upper atmosphere and a coupled
magnetosphere-ionosphere-thermosphere{ (MIT})
simulation for the interaction between the atmosphere and the solar
wind. As noted, atmosphere could also be removed by surface
weathering and ingassing processes. Typical examples are{ the}
silicate-carbonate cycle on Earth (weathering) and helium rain
(maybe including neon) in the giant planets. We do not discuss those
processes in detail here. Atmospheric escape is not only
important for understanding atmospheric mass evolution, but also
strongly impacts the atmospheric composition via mass fractionation
(e.g., \citealt{zahnleMassFractionationTransonic1986,
zahnleMassFractionationNoble1990};
\citealt{huntenMassFractionationHydrodynamic1987}) and altering the
planetary redox state over time (e.g.,
\citealt{catlingBiogenicMethaneHydrogen2001}). See the reviews in
\cite{lammerAtmosphericEscapeEvolution2008},
\cite{tianAtmosphericEscapeClimate2013} and
\cite{kislyakovaStellarDrivenEvolution2015} for more details.

Atmospheric escape becomes relevant for exoplanets since multiple species
have been detected in their upper atmospheres thanks to the high-resolution
facilities in the ultraviolet and visible, such as HST, VLT/ESPRESSO,
TNG/HARPS, and GTC/OSIRIS. The first extended hydrogen cloud surrounding a canonical hot Jupiter
HD~209458\,{\,}b was discovered by the Lyman-$\alpha$ transit
technique (\citealt{vidal-madjarExtendedUpperAtmosphere2003}).
Lyman-$\alpha$ has also been detected on another hot Jupiter
HD~189733\,{\,}b (e.g.,
\citealt{desetangsEvaporationPlanetHD2010,desetangsTemporalVariationsEvaporating2012};
\citealt{bourrierAtmosphericEscapeHD2013}, two smaller planets
including a warm Neptune GJ~436\,{\,}b
(\citealt{kulowLYAlphaTRANSIT2014};
\citealt{ehrenreichGiantCometlikeCloud2015};
\citealt{lavieLongEgressGJ2017};
\citealt{dossantosHubblePanCETProgram2019}) and GJ~3470\,{\,}b
(\citealt{bourrierHubblePanCETExtended2018}) and possibly
TRAPPIST-1\,{\,}b and c
(\citealt{bourrierReconnaissanceTRAPPIST1Exoplanet2017}), Kepler-444
(\citealt{bourrierStrongLymanaVariations2017}) and K2-18\,{\,}b
(\citealt{dossantosHighenergyEnvironmentAtmospheric2020}),
suggesting strong hydrogen escape from these bodies. Other hydrogen lines
in the Balmer series in the visible such as H$\alpha$ (and H$\beta$ in some cases) have
also been detected on two hot Jupiters HD~189733\,{\,}b (e.g.,
\citealt{jensenDETECTIONAlphaEXOPLANETARY2012};
\citealt{cauleyVARIATIONPRETRANSITBALMER2016,cauleyDecadeUpalphaTransits2017,cauleyEvidenceAbnormalVariability2017})
and WASP-52\,{\,}b (\citealt{chenDetectionNaHa2020}), as well as
four ultra-hot Jupiters: MASCARA-2\,{\,}b (also known as
KELT-20\,{\,}b, \citealt{casasayas-barrisNaHaAbsorption2018}),
WASP-12\,{\,}b (\citealt{jensenHydrogenSodiumAbsorption2018}),
KELT-9\,{\,}b (\citealt{yanExtendedHydrogenEnvelope2018};
\citealt{cauleyAtmosphericDynamicsVariable2019};
\citealt{turnerDetectionIonizedCalcium2020}; \citealt{wyttenbachMasslossRateLocal2020}){ and}
WASP-121\,{\,}b (\citealt{cabotDetectionNeutralAtomic2020}).
These observations suggested extended hydrogen atmospheres that
might originate from the neutral hydrogen escape. Note that some
H$\alpha$ signals from the young, forming planets could instead come
from the ongoing accretion, for example, PDS~70\,{\,}b and
PDS~70\,{\,}c (e.g.,
\citealt{haffertTwoAccretingProtoplanets2019};
\citealt{aoyamaConstrainingPlanetaryGas2019};
\citealt{hashimotoAccretionPropertiesPDS2020}). On the other hand,
extended hydrogen exospheres were not detected on some other
planets, such as super-Earths 55~Cnc\,{\,}e
(\citealt{ehrenreichHintTransitingExtended2012}),
HD~97658\,{\,}b (\citealt{bourrierNoHydrogenExosphere2017}),
GJ~1132\,{\,}b (\citealt{waalkesLyaGJ11322019}) and
$\pi$~Men\,{\,}c
(\citealt{garciamunozMenAtmosphereHydrogendominated2020}).

In extended atmospheres{,} heavier species including helium and easily
ionized metals such as Na, K, Ca, Mg, Si, and Fe have also been
detected near or beyond the planetary Roche lobe, for example, on
HD~209458\,{\,}b
(\citealt{vidal-madjarDetectionOxygenCarbon2004};
\citealt{linskyOBSERVATIONSMASSLOSS2010};
\citealt{vidal-madjarMagnesiumAtmospherePlanet2013};
\citealt{singHubbleSpaceTelescope2019};
\citealt{cubillosNearultravioletTransmissionSpectroscopy2020}), and
other planets including ultra-hot Jupiters (e.g.,
\citealt{fossatiMETALSEXOSPHEREHIGHLY2010};
\citealt{jenshoeijmakersAtomicIronTitanium2018,hoeijmakersSpectralSurveyUltrahot2019};
\citealt{yanIonizedCalciumAtmospheres2019};
\citealt{cauleyAtmosphericDynamicsVariable2019};
\citealt{turnerDetectionIonizedCalcium2020};
\citealt{chenDetectionNaHa2020}). In particular, extended helium
atmospheres have also been recently observed on hot Jupiter
HD~209458\,{\,}b (\citealt{alonso-florianoHe108302019}), HD189733\,{\,}b (\citealt{salzDetectionHeLambda2018}), a
Jupiter-size{d} Neptune-mass planet WASP-107\,{\,}b
(\citealt{spakeHeliumErodingAtmosphere2018}; \citealt{kirkConfirmationWASP107bExtended2020}), Saturn-mass planet
WASP-69\,{\,}b
(\citealt{nortmannGroundbasedDetectionExtended2018}) and a
Neptune-size{d} planet HAT-P-11\,{\,}b
(\citealt{allartSpectrallyResolvedHelium2018};
\citealt{mansfieldDetectionHeliumAtmosphere2018}). Moreover, the
circumstellar gas replenished by mass loss from ablating low-mass
planets could absorb stellar chromospheric emission. The Dispersed
Matter Planet Project (e.g.,
\citealt{barnesAblatingPlanetEccentric2020,haswellDispersedMatterPlanet2020,staabCompactMultiplanetSystem2020})
has recently detected low stellar chromospheric emission around
about 40 out of 3000 nearby bright stars, indicating possible
existence of highly irradiated, mass-losing exoplanets in these
systems. Also, the observed high variability in the transit depths
of so-called ``super-comets" such as
Kepler~1520\,{\,}b
(\citealt{rappaportPossibleDisintegratingShortperiod2012};
\citealt{perez-beckerCatastrophicEvaporationRocky2013}) and
K2-22\,{\,}b
(\citealt{sanchis-ojedaK2ESPRINTProjectDiscovery2015}) suggests that
they might experience significant evaporation (e.g.,
\citealt{budajExtrasolarEnigmasDisintegrating2020}).

 The temperature structure of the upper atmosphere is crucial for
 determining the species escape rates and the resulting transit observations.
 Atmospheric layers at the very top should be very hot---so-called thermosphere.
 All thermospheres on Solar System planets are hot, especially on giant planets, but the cause
is still debatable (e.g., \citealt{yelleJupiterThermosphereIonosphere2004}). On exoplanets,
it was suggested that intense stellar heating and insufficient cooling---primarily due to
thermal dissociation of coolants (e.g., \citealt{mosesDisequilibriumCarbonOxygen2011};
\citealt{koskinenEscapeHeavyAtoms2013})---allows the upper atmosphere to reach{ a} temperature of
$\sim10{\,}000$\,K. High-resolution observations in the {UV} and visible provide unambiguous
evidence of the hot upper atmospheres {for} several exoplanets using the Lyman-$\alpha$
(e.g., HD 209458 b, GJ 436 b, GJ 3470 b), hydrogen Balmer series and metal lines (e.g.,
HD 189733 b, KELT-9 b, KELT-20 b, WASP-12 b, WASP-121 b, WASP-52 b), and Helium
line (e.g., HD 209458 b, WASP-107 b, WASP-69 b, HAT-P-11 b).
The temperature (and its gradient), density, associated mass loss rate,
and even the wind speed at the upper atmosphere can be derived from
the powerful high-resolution spectroscopy (e.g., \citealt{hengNONISOTHERMALTHEORYINTERPRETING2015}; \citealt{wyttenbachSpectrallyResolvedDetection2015,
wyttenbachHotExoplanetAtmospheres2017, wyttenbachMasslossRateLocal2020};
\citealt{fisherHowMuchInformation2019}; \citealt{welbanksDegeneraciesRetrievalsExoplanetary2019}; \citealt{seidelWindChangeRetrieving2020}). If hydrodynamic escape occurs, adiabatic
cooling might lead to the temperature decreasing with altitude again.
For more discussion on high-resolution spectroscopy, refer to \cite{birkbySpectroscopicDirectDetection2018}.

Observations have motivated many theoretical studies that
investigated the upper atmospheres and mass loss on hot Jupiters and
smaller planets (e.g.,
\citealt{lammerAtmosphericLossExoplanets2003};
\citealt{yelleAeronomyExtrasolarGiant2004};
\citealt{tianTransonicHydrodynamicEscape2005};
\citealt{erkaevRocheLobeEffects2007};
\citealt{garciamunozPhysicalChemicalAeronomy2007};
\citealt{koskinenStabilityLimitAtmospheres2007};
\citealt{schneiterThreedimensionalHydrodynamicalSimulation2007};
\citealt{holmstroemEnergeticNeutralAtoms2008};
\citealt{penzaMassLossHot2008};
\citealt{murray-clayAtmosphericEscapeHot2009};
\citealt{stoneAnisotropicWindsClosein2009};
\citealt{guoEscapingParticleFluxes2011};
\citealt{guoEscapingParticleFluxes2013};
\citealt{trammellHotJupiterMagnetospheres2011};
\citealt{lopezHowThermalEvolution2012};
\citealt{owenPlanetaryEvaporationUV2012};
\citealt{lopezRoleCoreMass2013};
\citealt{owenKeplerPlanetsTale2013};
\citealt{erkaevXUVExposedNonHydrostaticHydrogenRich2013};
\citealt{lammerOriginLossNebulacaptured2014};
\citealt{koskinenEscapeHeavyAtoms2013,koskinenEscapeHeavyAtoms2013a};
\citealt{tremblinCollidingPlanetaryStellar2013};
\citealt{bourrier3DModelHydrogen2013,bourrierAtmosphericEscapeHD2013,bourrierEvaporatingPlanetWind2016,bourrierModelingMagnesiumEscape2014};
\citealt{jinPlanetaryPopulationSynthesis2014};
\citealt{kurokawaMassLossEvolutionClosein2014}; \citealt{shaikhislamovAtmosphereExpansionMass2014}; \citealt{salzTPCIPLUTOCLOUDYInterface2015, salzHighenergyIrradiationMass2015, salzSimulatingEscapingAtmospheres2016};
\citealt{owenEvaporationValleyKepler2017};
\citealt{dongProximaCentauriHabitable2017};
\citealt{zahnleCosmicShorelineEvidence2017};
\citealt{wangEvaporationLowmassPlanet2018};
\citealt{wangDustyOutflowsPlanetary2019};
\citealt{jinCompositionalImprintsDensity2018};
\citealt{mordasiniPlanetaryEvolutionAtmospheric2020};
\citealt{lamponModellingHeTriplet2020}). For close-in
exoplanets around {S}un-like stars, hydrodynamic escape
of atomic hydrogen could occur inside an orbit of about
0.1{\,}AU (e.g.,
\citealt{lammerAtmosphericLossExoplanets2003};
\citealt{yelleAeronomyExtrasolarGiant2004}). It was found{ that}
the transition between {a} stable atmosphere and{
an} unstable atmosphere (i.e., escaping) is located around
0.14--0.16{\,}AU{,} around sub-like stars in 3D
simulations (\citealt{koskinenStabilityLimitAtmospheres2007}). In
the context of planetary formation and evolution, atmospheric escape
could greatly affect the evolution of close-in small planets,
especially their planetary size distribution (see reviews in
\citealt{tianAtmosphericEscapeSolar2015} and
\citealt{owenAtmosphericEscapeEvolution2019b}). Thus{,}
atmospheric escape has become essential in understanding the current
planetary data sample.

In general, there are two important regimes for the thermal escape
rate. The escape rate can be ``supply-limited' or
``energy-limited{.}'' In the supply-limited regime, the
``limiting flux principle''
(\citealt{huntenEscapeH2Titan1973,huntenEscapeLightGases1973})
states that the thermal escape flux might be limited by several
bottlenecks below the exobase such as the cold trap at the
tropopause, atmospheric chemistry, cloud formation and
vertical diffusion. Take hydrogen escap{ing}
Earth as an example. The hydrogen primarily comes from
tropospheric water, which condenses as clouds in the upper
troposphere before{ being} transported to the stratosphere. This
cold trap of water vapor in the tropopause region leads to a very
dry stratosphere, with water molar fraction of a few parts
per million (ppm). Stratospheric chemistry converts water vapor and
some other hydrogen-bearing species such as methane to hydrogen
atoms. The conversion rate depends on the chemical pathways and UV
photons in the stratosphere. The final bottleneck is
diffusion above the homopause, where the species are gravitationally
separated according to their molecular weights. Hydrogen atoms rise
through the heavier species to the exobase by molecular diffusion
and eventually escape {in}to space.  The cold trap could
effectively limit the hydrogen loss from a terrestrial planet. The
efficiency of the cold trap primarily depends on the ratio of latent heat of condensable species (e.g., \ch{H2O})
to the sensible heat of the non-condensable species (e.g., \ch{N2},
\;\ch{CO2}) at the surface
(\citealt{wordsworthWATERLOSSTERRESTRIAL2013}). If the partial
pressure of non-condensable species is small, the cold
trap is not efficient, and the upper atmosphere will be moist. In
that case, hydrogen escape will eventually lead to severe
oxidization of the entire atmosphere on exoplanets. Hydrogen escape
on terrestrial planets in the Solar System is generally limited by
diffusion (\citealt{kastingEvolutionHabitablePlanet2003}). It is
also thought that a canonical hot Jupiter HD~209458\,{\,}b is
possibly escaping at the diffusion limit (e.g.,
\citealt{vidal-madjarDetectionOxygenCarbon2004,vidal-madjarExtendedUpperAtmosphere2003};
\citealt{yelleAeronomyExtrasolarGiant2004};
\citealt{koskinenEscapeHeavyAtoms2013a};
\citealt{zahnleCosmicShorelineEvidence2017}).

In the second regime, hydrogen escape rate is limited by the energy
available for escape. This energy limit could come from the incoming
energy itself, but could also from the limiting steps converting the
incoming energy into available energy driving the escape
such as radiative processes---hydrogen radiative
recombination (\citealt{murray-clayAtmosphericEscapeHot2009}) or
ionization photons (\citealt{owenUVDrivenEvaporation2015}). In other
words, the conversion efficiency of incoming energy
to {KE} is crucial. The
energy-limited escape has been widely assumed in hydrogen escape on
warm and hot close-in exoplanets (e.g.,
\citealt{lammerAtmosphericLossExoplanets2003,lammerDeterminingMassLoss2009};
\citealt{vidal-madjarExtendedUpperAtmosphere2003};
\citealt{desetangsAtmosphericEscapeHot2004};
\citealt{baraffeEffectEvaporationEvolution2004};
\citealt{desetangsDiagramDetermineEvaporation2007};
\citealt{erkaevRocheLobeEffects2007a};
\citealt{hubbardEffectsMassLoss2007,hubbardMassFunctionConstraint2007};
\citealt{davisEvidenceLostPopulation2009};
\citealt{leitzingerCouldCoRoT7bKepler10b2011};
\citealt{owenPlanetaryEvaporationUV2012};
\citealt{lopezRoleCoreMass2013};
\citealt{owenKeplerPlanetsTale2013}). For most planets in the
cold-temperature regime (see Sect.~\ref{sect:coldhot}), hydrogen
escape is not violent and the energy supply from stellar
heating in the upper atmosphere is sufficient to drive
escape under{ the} hydrostatic situation. As temperature
increases, the atmospheric escape could rapidly transit from the
Jeans regime to the hydrodynamic regime in a rather narrow range of
Jeans parameter
(\citealt{volkovTHERMALLYDRIVENATMOSPHERIC2011,volkovKineticSimulationsThermal2011}).
For hotter hydrogen atmospheres, when hydrodynamic escape occurs, a
rapid blow-off of the main constituents requires a large amount of
heating from the stellar X-ray and extreme
ultraviolet{ (XUV}) or even softer near/{f}ar ultraviolet{ (NUV/FUV}) photons (e.g.,
\citealt{garciamunozRapidEscapeUltrahot2019}). The partitioning
between the two is not well understood at this moment and perhaps
varies case by case (\citealt{owenPlanetaryEvaporationUV2012};
\citealt{owenKeplerPlanetsTale2013}). The energy loss processes in
the upper atmosphere (i.e., the thermosphere) {are} also
complicated. Most of the energy could be radiated {in}to space, or thermally conducted to the lower atmosphere. The
energy used to drive the intensive planetary wind and atmospheric
mass loss is thus limited.

In the energy-limited regime, it is important to quantify both the
heating efficiency and wind efficiency. The former measures the
X-ray and UV heating and radiative cooling processes in the upper
atmosphere, for instance, \ch{CO2} cooling on terrestrial planets
(\citealt{tianConservationTotalEscape2013}), \ch{H3+} ion
cooling for giant planets
(\citealt{koskinenStabilityLimitAtmospheres2007}) or cooling
of hydrogen radiative recombination
(\citealt{murray-clayAtmosphericEscapeHot2009}). A careful treatment
of the radiative transfer and chemistry is needed. The wind
efficiency is a global measure of how much incoming energy is
converted to {KE} for the
blow-off. A simple but widely { applied}
energy-limited formula (\citealt{watsonDynamicsRapidlyEscaping1981})
for hydrodynamic escape is

 \begin{equation}\label{energylimit}  
\dot{M}\sim\eta\frac{L_{\mathrm{HE}}R_p^3}{4GM_pa^2},
\end{equation}
where $L_{\mathrm{HE}}$ is the high-energy portion of the stellar
luminosity. $a$ is the star-planet distance. Here we neglect the
potential energy reduction factor due to the Roche lobe effect
(\citealt{erkaevRocheLobeEffects2007}), as well as the difference
between the planetary radius (which is vague for giant planets) and
the level where the wind is launched. {E}nergy-limited escape is usually valid if the cooling is
dominated by adiabatic (and subsonic) expansion with a large escape
rate (\citealt{johnsonMOLECULARKINETICSIMULATIONSESCAPE2013}). The
wind efficiency ($\eta$) is usually treated as a constant for
simplicity, for instance, 10\%--20\% (e.g.,
\citealt{lopezHowThermalEvolution2012};
\citealt{owenPlanetaryEvaporationUV2012};
\citealt{lopezRoleCoreMass2013};
\citealt{owenKeplerPlanetsTale2013};
\citealt{kurokawaMassLossEvolutionClosein2014}). However, it would
be good to keep in mind that the wind efficiency in this simple
formula is usually not constant and needs to be
{utilized} with caution. Also, hydrodynamic escape is
essentially a self-limiting process because the rapid
non-hydrostatic expansion of the atmosphere will adiabatically cool
itself down and slow down the wind. The atmospheric structure,
including the temperature and density, might change dramatically and
thus the heating level and wind efficiency (e.g.,
\citealt{koskinenEscapeHeavyAtoms2013a}). Previous studies (e.g.,
\citealt{erkaevXUVExposedNonHydrostaticHydrogenRich2013};
\citealt{lammerProbingBlowoffCriteria2013}) indicate that the simple
energy-limited formula might only apply to high-gravity bodies like
hot Jupiters where thermospheric expansion is not extreme
rather than low-gravity bodies such as cooler
Earth-size{d} planets. Otherwise, escape rates could be
significantly overestimated. Moreover, recent hydrodynamic
simulations with thermochemistry and radiative transfer
(\citealt{wangEvaporationLowmassPlanet2018})
{demonstrated} that for small planets
($M_p<\sim10M_e$), the mass-loss rate scales with radius
square{d} ($R^2$) instead of radius cube{d} in the
conventional formula, leading towards a ``photo-limited" scenario
(e.g., \citealt{owenUVDrivenEvaporation2015}) where EUV
photoheating is strong and the gravitational potential
is shallow.

Here, instead of focusing on the detailed modeling and theories on
the escape of exoplanet atmospheres (see reviews in
\citealt{lammerAtmosphericEscapeEvolution2008};
\citealt{tianAtmosphericEscapeSolar2015};
\citealt{owenAtmosphericEscapeEvolution2019b}), we highlight two
important regime classifications of currently detected exoplanets
from observational statistics known as the ``cosmic shoreline" and
the ``planet desert and radius gap{.}"

\subsection{``Cosmic Shoreline'': Irradiation or Impact?}
\label{sect:cos}

\begin{figure*}

\vspace{-2.5cm} 
   \centering
   \includegraphics[width=0.8\textwidth, angle=0]{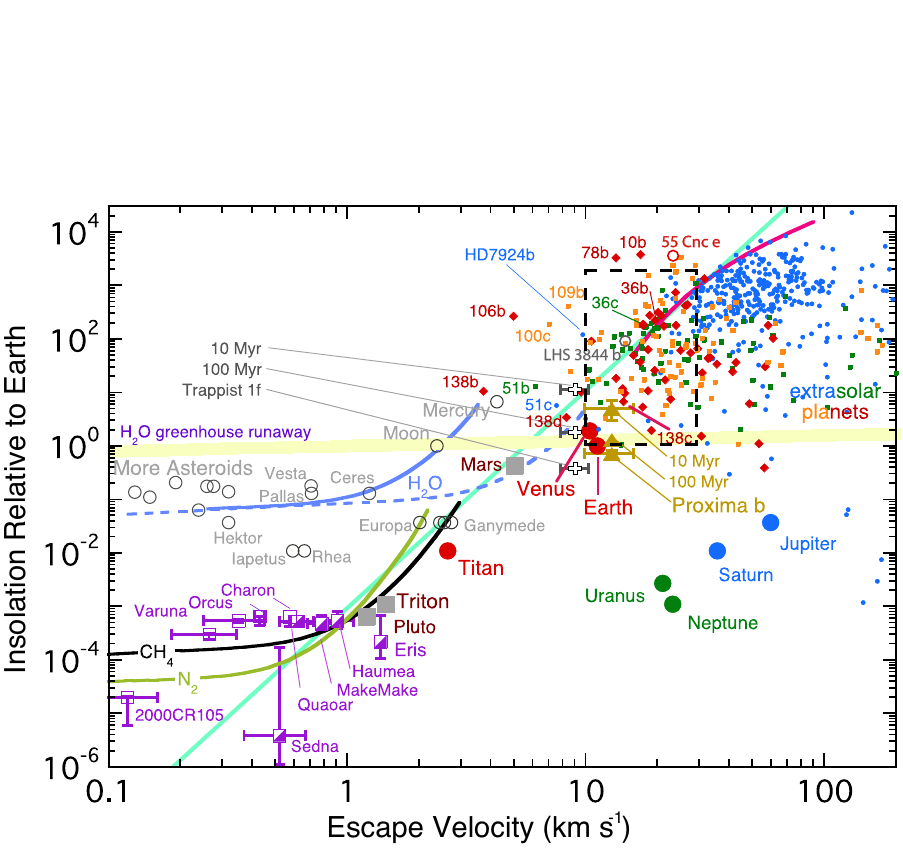}
   \caption{\baselineskip 3.8mm Diagram of insolation versus escape velocity slightly updated from the hypothetic cosmic shoreline ({\it cyan}) figure in \cite{zahnleCosmicShorelineEvidence2017} including 55~Cancri\,{\,}e and the recently detected airless TESS planet LHS~3844\,{\,}b (\citealt{kreidbergAbsenceThickAtmosphere2019}). We assumed the density of LHD~3844\,{\,}b {to be }the same as the Earth because its mass has not been measured yet. The presence or absence of an atmosphere on Solar System objects is indicated by {\it filled} or {\it open symbols}, respectively. The extrasolar planets are color-coded for Saturn-like ($R > 8{\,}R_E$, {\it blue}), Neptune-like ($3{\,}R_E < R < 8{\,}R_E$, {\it green}), Venus-like ($R < 1.6{\,}R_E$, {\it red}) and the rest ($1.6{\,}R_E < R < 3{\,}R_E$, {\it red}). Also {displayed} are hydrodynamic thermal escape curves for \ch{CH4}, \;\ch{N2} and \ch{H2O} ({\it solid} for \ch{H2O} and {\it dashed} if hydrogen escapes), the thermal stability limit for hot extrasolar giant planets ({\it magenta}) and the runaway greenhouse threshold for steam atmospheres ({\it yellow}). The {\it black rectangular box} approximately indicates the ``radius gap" region in Fig.~\ref{figrgap}. The escape velocity ranges from 10 to 30{\,}$\mathrm{km{\,}s^{-1}}$ and from the insolation ranges from 1$\times$ to 2000$\times$ Earth's insolation.}
   \label{figcs}
   \end{figure*}

After proto-atmosphere accretion, the long-term existence of
{an} atmosphere is controlled by the planet's ability
to hold its atmosphere. Knowing the fundamental processes such as
condensation and escape, one can predict whether a planet has an
atmosphere or not. The dominant mechanisms could be statistically
tested against existing data. \cite{zahnleOriginsAtmospheres1998}
first analyzed the Solar System data and put planets and large moons
in a diagram of solar insolation versus escape velocity. An
empirical division exists between those bodies with and without
apparent atmospheres. \cite{zahnleCosmicShorelineEvidence2017}
expanded this idea to include asteroids, Kuiper Belt Objects and
exoplanets in the same diagram (Fig.~\ref{figcs}). Although the escape velocity spans more than two orders of
magnitude and the stellar insolation changes about eight orders of
magnitude in these $\sim$600 samples, an empirical division between
atmospheric bodies and airless{ ones} is
relatively clear. The regime boundary seems to follow a straight line
in log-log space,{ the} so-called ``Cosmic
Shoreline{,}" The region below the shoreline is the
``atmospheric regime"--- planets tend to have atmospheres when insolation is low and gravity is high; planets above the
shoreline fall in{to} the ``airless regime"--- they do not seem
to harbor an apparent atmosphere.

The existence of the cosmic shoreline is intuitively understandable,
but the detailed mechanisms are not easy to decipher. To
first order, escape velocity measures the depth of the gravitational
potential on a planet. {S}tellar insolation
represents several external driving forces that lead to atmospheric
loss. For example, insolation itself affects the planetary
equilibrium temperature and might lead to a thermally unstable state
of the entire atmosphere. The high-energy portion of
stellar photons in the X-ray, XUV and FUV can directly
trigger hydrodynamic escape of the atmosphere. The stellar
wind is responsible for many non-thermal processes such as stellar
wind stripping, sputtering and ion pick-up. The empirical
cosmic shoreline in Figure~\ref{figcs} can be expressed as
 \begin{equation}\label{cosmic}
I\propto v_e^4,
\end{equation}
where $I$ is stellar insolation at the planetary body, and
$v_e$ is escape velocity.

The underlying principle of this simple scaling law is not obvious.
Here we restate the derivation in
\cite{zahnleCosmicShorelineEvidence2017} using the Jeans parameter
$\lambda$. Since the atmosphere will be lost rapidly through
hydrodynamic escape as lambda exceeds unity, we expect the cosmic
shoreline corresponds to $\lambda=1$. From{ the}
Stefan-Boltzmann law, $I$ scales with $T^4$.
\cite{zahnleCosmicShorelineEvidence2017} assumed the molecular
weight $m = T^{-1}$ in diverse planetary atmospheres
(which is also an empirical observation). Put together, we found
$\lambda\sim I^{-1/2}v_e^2/k_B $, and thus the $\lambda\sim1$
corresponds to $I\propto v_e^4$. An alternative but very similar
version to represent the cosmic shoreline is
{employing} the XUV flux as the vertical axis in
Figure~\ref{figcs} (see fig.~2 in
\citealt{zahnleCosmicShorelineEvidence2017}). One could also obtain
a scaling law by assuming that the X-ray and XUV heating primarily
drive the hydrodynamic escape. In this scenario, the total
fractional mass loss of the atmosphere is the time integral of the
energy-limited escape formula (Eq.~(\ref{energylimit}))
proportional to $I_{\mathrm{xuv}}t_{\mathrm{xuv}}R_p^3/M_p^2$ where
$t_{\mathrm{xuv}}$ is the Kelvin-Helmholtz timescale or cooling
timescale in which the planet is under high XUV exposure. Typically
the timescale is on the order of a few Myr (e.g.,
\citealt{jacksonCoronalXrayageRelation2012};
\citealt{tuExtremeUltravioletXray2015}). Assume the mass-radius
relationship is $M_p = \rho R_p^3$, which could be problematic
because the mass-radius relationship of the planet is not simple,
and one can achieve $I_{\mathrm{xuv}} \propto v_e^3\rho^{1/2}$,
which is also roughly consistent with the current sample.

In the low insolation regime (left-lower corner of
Fig.~\ref{figcs}), the collapsed atmospheres on Kuiper Belt
Objects (including Pluto and Triton) can be divided by another type
of cosmic shoreline, which does not follow the simple power-law but
{manifests} as curves in Figure~\ref{figcs}.
\cite{zahnleCosmicShorelineEvidence2017} proposed hydrodynamic
thermal escape models for \ch{CH4} and \ch{N2} assuming vapor
pressure equilibrium at the surface. Their models could explain the
regime division in those low-temperature bodies.

{T}hermal escape is not the sole explanation. There
has been a long-standing hypothesis that{ a} planetary
atmosphere can be entirely removed by impact erosion (e.g.,
\citealt{walkerImpactErosionPlanetary1986};
\citealt{meloshImpactErosionPrimordial1989};
\citealt{zahnleImpactgeneratedAtmospheresTitan1992};
\citealt{zahnleXenologicalConstraintsImpact1993};
\citealt{zahnleOriginsAtmospheres1998};
\citealt{griffithInfluxCometaryVolatiles1995};
\citealt{chenErosionTerrestrialPlanet1997};
\citealt{brainAtmosphericLossOnset1998};
\citealt{newmanImpactErosionPlanetary1999};
\citealt{gendaSurvivalProtoatmosphereStage2003};
\citealt{gendaEnhancedAtmosphericLoss2005};
\citealt{catlingPlanetaryAirLeak2009};
\citealt{shuvalovAtmosphericErosionInduced2009};
\citealt{shuvalovImpactInducedErosion2014};
\citealt{korycanskyTitanImpactsEscape2011};
\citealt{schlichtingAtmosphericMassLoss2015};
\citealt{zahnleCosmicShorelineEvidence2017};
\citealt{bierstekerAtmosphericMasslossDue2019};
\citealt{wyattSusceptibilityPlanetaryAtmospheres2019}). The impact
erosion scenario has been proposed to understand the
early atmosphere{ of Mars}
(\citealt{meloshImpactErosionPrimordial1989}) and the dichotomy
between gas-rich Titan and airless Ganymede/Callisto
(\citealt{zahnleImpactgeneratedAtmospheresTitan1992}). Although
large uncertainties still remain in evaluating the detailed
mechanisms, presumably a thinner atmosphere is easier to be eroded
away than a thicker atmosphere, meaning that the impact erosion is a
runaway process. \cite{zahnleCosmicShorelineEvidence2017} also
tested this hypothesis {utilizing} all planet samples
in Figure~\ref{figcs}. They simply assumed that impact velocities
are proportional to orbital velocities for close-in planets and
plotted against the escape velocities of the planets. It was found
that, again, there is a regime division between the bodies with and
without atmospheres (see their fig.~4). The regime
boundary follows $v_{\rm imp} = 4\sim5~v_e$ where the
$v_{\rm imp}$ is the impact velocity. Future investigations are
worth put{ting} forward in this direction and
pin{ning} down the uncertainties
(\citealt{wyattSusceptibilityPlanetaryAtmospheres2019}).

If the cosmic shoreline is real, this empirical law might predict
the existence of atmospheres on exoplanets. For example, the
recently detected airless body LHS 3844 b
(\citealt{kreidbergAbsenceThickAtmosphere2019}) lies above the
cosmic shoreline (Fig.~\ref{figcs}). However, there are some exceptions, such
as Kepler 51 b and c, very low-density bodies but located above the
empirical line, suggesting the cosmic shoreline
might also depend on the age of the planet. A more massive, older
planet{,} 55 Cancri e{,} is also an outlier. Both thermal
phase curve observations (\citealt{demoryMapLargeDaynight2016}) and
HST transmission spectra
(\citealt{tsiarasDetectionAtmosphereSuperEarth2016}) indicated a
substantial atmosphere on 55 Cancri e. It would also be interesting
to put the future yet-to-be-characterized habitable-zone terrestrial
planets such as the Trappist-1 system in the diagram. {C}urrent observations can rule out the existence of
significant hydrogen atmospheres on TRAPPIST-1\,{\,}b and
TRAPPIST-1\,{\,}c (\citealt{dewitCombinedTransmissionSpectrum2016}).
\cite{zahnleCosmicShorelineEvidence2017} found Proxima Centauri b
and Trappist-1f are both near the cosmic shoreline (``on the beach")
and thus we cannot conclude the existence of their atmospheres at
this moment. This coincidence is interesting because the known
terrestrial planets with atmospheres, including the Earth with life
on it, are all located close to the cosmic shoreline.

How did we detect an airless exoplanet? For tidally locked
terrestrial exoplanets, an airless body could possess a higher
amplitude in the thermal emission light curve due to little heat
redistribution between the dayside and the nightside
(\citealt{kreidbergAbsenceThickAtmosphere2019}). On the other hand,
the presence of an atmosphere could naturally reduce the dayside
thermal emission via cloud formation and heat redistribution
(\citealt{kollIdentifyingCandidateAtmospheres2019}) and also
increase the planetary albedo
(\citealt{mansfieldIdentifyingAtmospheresRocky2019}). Until
recently, we have found the first indirect evidence of an airless
exoplanet LHS 3844 b (\citealt{kreidbergAbsenceThickAtmosphere2019})
{applying} the thermal {IR} light
curves from Spitzer. Future observations will further narrow down
the cosmic shoreline region's width and profile the detailed shape
of the stability zone among extrasolar terrestrial
planets.

\subsection{Planet Desert and Radius Gap}
\label{sect:radgap}

If a thick gas envelope is lost via atmospheric escape, the observed
planetary size shrinks. If this process occurs commonly on a large
number of planets, atmospheric escape might imprint itself in
planetary size distributions as a function of insolation or orbital
distance. The fractional mass-loss rate on close-in hot Jupiters is
small---at around{ the} 1\% level (e.g.,
\citealt{hubbardEffectsMassLoss2007};
\citealt{owenKeplerPlanetsTale2013})---and thus the radius change is
tiny. On the other hand, smaller planets with lighter gravity could
have a significantly large fractional mass loss. In extreme cases,
the gas envelope can be completely {s}tripped off, and a bare solid core is left behind. For
planets smaller than Neptune, a few percent of hydrogen and helium
loss in mass will significantly reduce the planetary size---a radius
change that could be observable in the old planet population.
Intuitively, one can expect a planet closer to the central star is
smaller and denser, and that further away is larger and lighter.
Statistically, one might also expect that the occurrence rate of
short-period planets drop{s} as the star-planet separation (or
the orbital period) decreases.

\begin{figure*}
   \centering
   \includegraphics[width=0.8\textwidth, angle=0]{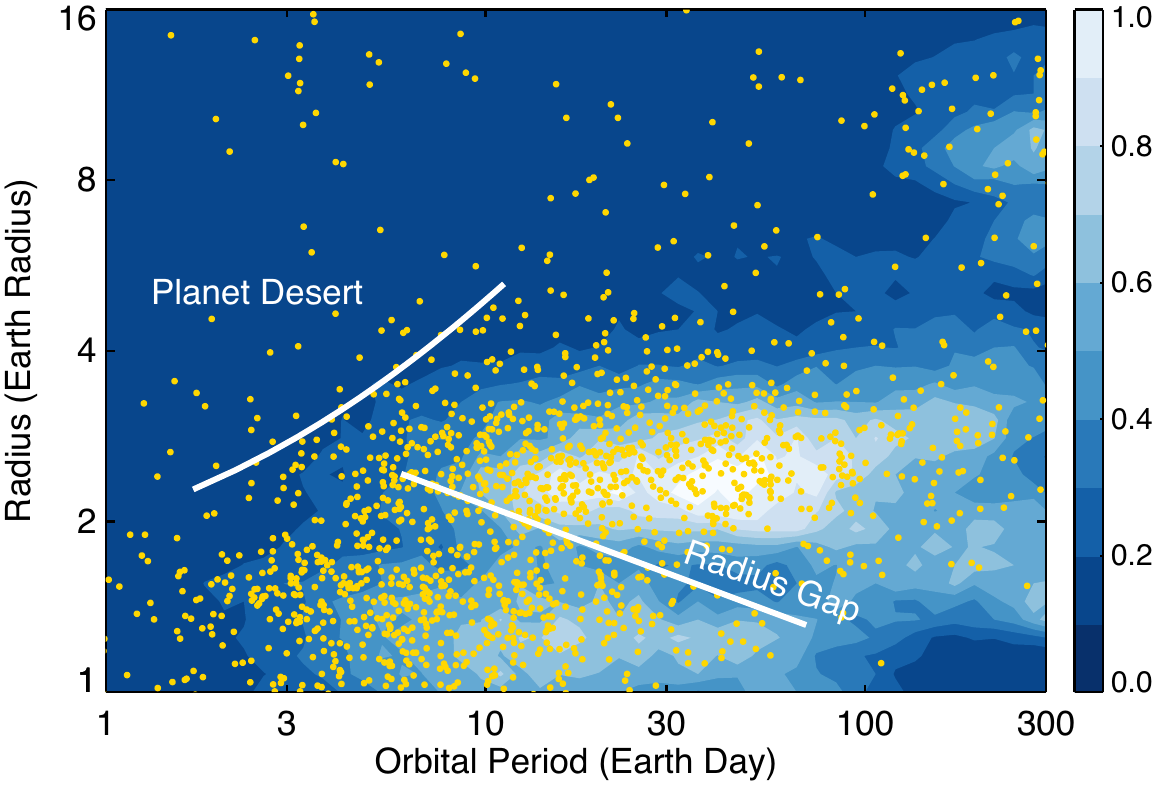}
   \caption{\baselineskip 3.8mm Size-period distributions of close-in small exoplanets (dots and contours from \citealt{fultonCaliforniaKeplerSurvey2018a} and \citealt{petiguraCaliforniaKeplerSurvey2018}) for all star types. The points represent the {California-Kepler Survey (}CKS{)} planets. Color indicates the relative planetary occurrence
rate (normalized by the maximum value) calculated from the simulated
110{\,}733 planets in a sample of
1{\,}000{\,}000 Sun-like stars in Table 9 in
\cite{petiguraCaliforniaKeplerSurvey2018}. Note that the number{
of} planets per 100 stars per bin depends on the interval size in
the period-radius plane. For reference, the maximum occurrence rate
is about {four} planets per 100 stars having periods
within 0.125{\,}dex of 40{\,}d and
radii within 0.05{\,}dex of 2.5\,$R_E$. The top white
line indicates an analytical curve of{ the} planet desert from
Eq.~(\ref{pldesert}) with $R_c=1.5\,R_E$, above which planetary
occurrence is rare. The lower white line
{signifies} the predicted photoevaporation gap
scaling from \cite{owenEvaporationValleyKepler2017}: $R_p\propto
P^{-0.25}$ (Eq.~(\ref{plgap})). Below the line{,} the planets are
assumed to be bare cores in theory, i.e., no H$_2$-He
atmosphere{,} {b}ut the theory seems to
overestimate the slope of the radius gap from the observations
(\citealt{vaneylenAsteroseismicViewRadius2018};
\citealt{martinezSpectroscopicAnalysisCaliforniaKepler2019}). The
ranges of escape velocity and insolation level of the planets around
the radius gap are {depicted with} the rectangular
box in Fig.~\ref{figcs}.}
   \label{figrgap}
   \end{figure*}

Early studies have noticed negative correlations between orbital
period and planetary mass or gravity for short-period planets and
pointed out thermal escape as a possible cause
(\citealt{mazehIntriguingCorrelationMasses2005};
\citealt{southworthMethodDirectDetermination2007}). With Kepler
data, a deficit in the occur{re}nce rate of close-in
small planets is discovered (e.g.,
\citealt{youdinExoplanetCensusGeneral2011};
\citealt{szaboShortPeriodCensorSubJupiter2011};
\citealt{howardPlanetOccurrence252012};
\citealt{beaugeEmergingTrendsPeriodRadius2013};
\citealt{petiguraPlateauPlanetPopulation2013};
\citealt{fressinFalsePositiveRate2013};
\citealt{burkePlanetaryCandidatesObserved2014};
\citealt{silburtStatisticalReconstructionPlanet2015};
\citealt{muldersStellarmassdependentDropPlanet2015};
\citealt{thompsonPlanetaryCandidatesObserved2018}). It is also found
that the inner planets are denser and smaller than the outer ones in
multi-planet systems (\citealt{wuDensityEccentricityKepler2013};
\citealt{ciardiRelativeSizesPlanets2013}). However, for transit
observations, uncertainties in star radii can greatly influence the
planetary size estimation. Recently, using CKS data together with Gaia
astrometry data, \cite{fultonCaliforniaKeplerSurvey2018a} and
\cite{fultonCaliforniaKeplerSurveyIII2017} measured the planetary
radii down to the 5\% precision level. Th{e}s{e} new
data of short-period small planets (within 100{\,}d and below 10 Earth radii $R_E$) clearly
{exhibit} two prominent features in the radius-orbit
diagram (Fig.~\ref{figrgap}, from
\citealt{fultonCaliforniaKeplerSurvey2018a} and
\citealt{petiguraCaliforniaKeplerSurvey2018}). The first one is the
``Planetary Desert" of short-period large planets in the upper
corner of the graph (e.g.,
\citealt{wuDensityEccentricityKepler2013};
\citealt{ciardiRelativeSizesPlanets2013}). The
second one is an intriguing ``radius gap" (or ``valley") of that
planetary occurrence rate that decline{s}
diagonally with increasing semi-major axis. The gap occurs at 1.5--2\,$R_E$,
separating the planetary population into two groups: super-Earths
(1--1.7\,$R_E$) and sub-Neptunes (1.7--4\,$R_E$).

The radius-period diagram with a ``desert" and a ``gap"---if the
planets were not born to be so ({for }alternatives{,} refer
to \citealt{zengGrowthModelInterpretation2019})---shows a strong
signature of the atmospheric mass loss after formation. One
possibility is the thermal escape via photoevaporation. In fact, the
evaporation desert and gap have been theoretically predicted by
atmospheric escape models in \cite{owenKeplerPlanetsTale2013} and
\cite{lopezRoleCoreMass2013} and also investigated by subsequent
photoevaporation studies (e.g.,
\citealt{jinPlanetaryPopulationSynthesis2014};
\citealt{howeEvolutionaryModelsSuperEarths2015};
\citealt{chenEvolutionaryAnalysisGaseous2016};
\citealt{owenEvaporationValleyKepler2017};
\citealt{jinCompositionalImprintsDensity2018};
\citealt{mordasiniPlanetaryEvolutionAtmospheric2020}). The planetary
desert is easy to understand as a result of photoevaporation. For a
given initial planetary mass and core mass, planets located closer
to the star experience faster erosion, resulting in smaller sizes
(e.g., \citealt{wuDensityEccentricityKepler2013}).
\cite{owenKeplerPlanetsTale2013} found that the shape of the upper
envelope of the desert could be explained by the photoevaporation of
20\,$M_E$ planets with rocky cores of masses 10--15\,$M_E$. Simple
scaling of the upper envelope can also be crudely estimated here. In
the analytical model of \cite{owenEvaporationValleyKepler2017}, the
atmospheric lifetime scales with multiple factors, including the
orbital period, planetary radius and core radius. If the
planets below the radius gap are bare cores, the core size is
generally less than 2\,$R_E$. {P}lanets {in} the upper envelope of the
distribution are also generally larger than twice the core
size. In this
regime, assuming a constant evaporation timescale, we can rearrange
equation~(4) in \cite{owenEvaporationValleyKepler2017} and
obtain the dependence of the planetary radius on the orbital period:
\begin{equation}\label{pldesert}
\frac{R_p-R_c}{R_c}\propto P^{0.83},
\end{equation}
where $R_c$ is the core radius. An example of this scaling is
{depicted} in Figure~\ref{figrgap}. Note that many
factors could influence the upper envelope. Some of these planets in
the ``planet desert" are likely to be bare giant cores (e.g., the
recently discovered TOI-849 b,
\citealt{armstrongRemnantPlanetaryCore2020}).
{However,} some might hold a significant gaseous
envelope including several very-low-density planets (e.g.,
``Super-puffs") and hot Neptune-sized planets around
high-metallicity stars (``Hoptunes",
\citealt{dongLAMOSTTelescopeReveals2018}). Metallicity also plays an
important role in the atmospheric cooling process and the escaping
mass loss on short-period exoplanets.

The mechanism of {a} radius gap is less intuitive than
the planet desert. The black rectangular box in Figure~\ref{figcs}
{signifies} the region where small
exoplanets around the ``radius gap" lie in the ``cosmic shoreline"
plot. The planets with size between 1 and 6 Earth radii and
orbital period between 1 and 300 Earth days have escape
velocity between 10 and 30{\,}$\mathrm{km{\,}s^{-1}}$ and insolation flux
levels between 1$\times$ and 2000$\times$ the Earth's
value. Nevertheless, the ``cosmic shoreline" idea merely indicates a
boundary line instead of a gap.
\cite{owenEvaporationValleyKepler2017} and
\cite{owenAtmosphericEscapeEvolution2019b} elaborated that the
radius gap originates from the nonlinear dependence of the
mass loss timescale to the envelope mass (see fig.~4 in
\citealt{owenAtmosphericEscapeEvolution2019b}). The mass loss
timescale reaches a maximum (i.e., slowest erosion) when the
envelope doubles the core radius (hydrogen mass fraction is about
1\%) and creates a stable sub-Neptune group (1.7--4\,$R_E$) in the
radius-period diagram. Below this critical point, the mass loss
timescale drops very quickly below the Kelvin-Helmholtz timescale
(in a few Myrs) so that the envelope can be completely stripped, and
thus the ``evaporation gap" emerges. The bare-cores are thus left
behind to form the observed short-period super-Earths group
(1--1.7\,$R_E$) in Figure~\ref{figrgap}. Beyond the maximum point, as the
envelope mass further increases, the planetary size increases so
quickly---and so do the received XUV flux and the mass-loss
rate---that it overwhelms the increase {in} the envelope
mass. As a result, the mass loss timescale decreases towards a
minimum where the envelope mass matches the core mass. Beyond this
minimum point, compression of the atmosphere due to self-gravity
maintains a roughly fixed planetary size so that the mass loss
timescale increases again. Although the underlying physics of the
evaporation valley looks similar to the ``cosmic
shoreline{,}" the general cosmic shoreline theory in
\cite{zahnleCosmicShorelineEvidence2017} does not imply the
existence of a clear gap around the regime boundary due to a lack of
consideration of the nonlinear behavior of the mass loss
timescale. Also, there is a subtle difference. The ``cosmic
shoreline" separates the planets with atmospheres and airless
bodies, whereas the ``radius gap" separates the planets with and
without primordial hydrogen-helium envelopes. The planets
below the gap are not necessarily completely airless but could also
possess a significant amount of outgassed secondary
atmospheres such as water and carbon dioxide after the early
photoevaporation. {S}o-called ``bare cores" should be
more appropriately understood in the sense that their atmospheres
are too thin to have an important impact on their observed radii.

One can also analytically approximate the slope of the radius gap.
The bare-core boundary of the evaporation gap might just be another
version of the ``cosmic shoreline" with a fixed atmospheric
composition (like H$_2$-He) in the high XUV scenario in
\cite{zahnleCosmicShorelineEvidence2017}. Following the previous
scaling of the ``cosmic shoreline" in the XUV case, the
energy-limited escape formula (Eq.~(\ref{energylimit})) gives the
fractional mass loss rate that is proportional to
$L_{\mathrm{xuv}}R_p^3/M_p^2$. For a solid bare core, the
mass-radius relationship
(\citealt{lopezUNDERSTANDINGMASSRADIUSRELATION2014})
{yields} $M_p = \rho R_p^4$ (note that this is
different from what we have assumed in Sect.~\ref{sect:cos}
$M_p=\rho R_p^3$). From Kepler's third law,
$L_{\mathrm{xuv}} \propto L_{HE}/a^2 \propto P^{-4/3}$ where $P$ is
orbital period, one can obtain the size-period version of
the ``cosmic shoreline": $R_p=P^{-4/15}$. If we use the
``photon-limited" mass loss rate $\dot{M}\propto
L_{\mathrm{xuv}}R_p^2/M_p^2$ from
\cite{wangEvaporationLowmassPlanet2018}, the ``cosmic shoreline"
scaling becomes shallower{,} $R_p \propto P^{-2/9}$. Both
estimates are not very different from the scaling in
\cite{owenEvaporationValleyKepler2017} from a detailed treatment on
the physics of core and envelope mass evolution
\begin{equation}\label{plgap}
R_p\propto P^{-0.25}.
\end{equation}

Qualitatively, these analytical scalings indicate a decrease of the
transition radius as orbital radius increases, consistent with the
gap in the current Kepler-CKS sample (Fig.~\ref{figrgap}).
Quantitatively, the radius gap slope derived from a statistical
regression of the observational data
(\citealt{vaneylenAsteroseismicViewRadius2018}) follow{s}
$R\propto P^{-0.09}$.
\cite{martinezSpectroscopicAnalysisCaliforniaKepler2019} reported a
similar slope $R\propto P^{-0.11}$. Both are much shallower than the
analytical estimates above. Note that the current data showing the
radius gap only include short-orbit exoplanets, and hydrodynamic escape might not work well for planets with a
period larger than 30{\,}d
(\citealt{owenPlanetaryEvaporationUV2012}).

We also emphasize that all the systems are ``fossils" that record
XUV from an earlier time. Given that most XUV photons were emitted
when the star was very young, $L_{\mathrm{xuv}}$ is probably not a
constant and dependent on the stellar type. Thus the properties of
the radius gap are probably different around different types of
stars. Future statistics on cooler terrestrial exoplanets are needed
to unveil more details.

Besides photoevaporation that takes action after the
protoplanetary disk dissipation, alternative hypotheses have been
put forward to explain the radius gap. The first one is the
core-powered mass loss
(\citealt{ginzburgCorepoweredMasslossRadius2018};
\citealt{ginzburgSuperEarthAtmospheresSelfconsistent2016};
\citealt{guptaSculptingValleyRadius2019a,guptaSignaturesCorePoweredMassLoss2020}).
This mechanism argues that the core luminosity released from the
cooling of its primordial energy from planetary formation could
drive the atmospheric escape for Gyrs, even without
photoevaporation. The core-powered mass loss could also explain the
observed radius gap slope
(\citealt{ginzburgCorepoweredMasslossRadius2018,
guptaSculptingValleyRadius2019a}). The second hypothesis claims that
the radius gap is a natural result of planetary formation
pathways---planets above the gap are water-worlds
and the ones below are rocky
(\citealt{zengGrowthModelInterpretation2019}). The last one is the
impact erosion by planetesimals---planets below the gap
were bare cores with their primordial atmospheres stripped away, and
the ones above the gap grow enough volatiles to form secondary
atmospheres (\citealt{wyattSusceptibilityPlanetaryAtmospheres2019}).
The impact erosion could not only explain the cosmic shoreline
(\citealt{zahnleCosmicShorelineEvidence2017} but also reproduce the
radius gap, although the details need to be further investigated
(\citealt{wyattSusceptibilityPlanetaryAtmospheres2019}).

Is the radius gap a result of ``nurture" (i.e., photoevaporation,
impact erosion or core-powering mass loss) or ``nature"
(i.e., born to be, late formation in the gas-poor environment)? It
is not easy to distinguish these hypotheses. As mentioned above, the
observed slope appears shallower than the analytical scalings
(\citealt{vaneylenAsteroseismicViewRadius2018};
\citealt{martinezSpectroscopicAnalysisCaliforniaKepler2019}). In
theory, if the evaporation efficiency changes with orbital distance
and other factors, the predicted slopes could be different (e.g.,
\citealt{mordasiniPlanetaryEvolutionAtmospheric2020}). The slope is negative instead of positive
(\citealt{vaneylenAsteroseismicViewRadius2018}) {which }seems to
suggest that the stripped cores do not form in a gas-poor
environment after the disk dissipation. In the latter scenario,
\citet{lopezHowFormationTimescales2018} predicted a positive slope,
although no impact delivery or erosion was considered.
Investigations on the details of planetary accretion and evolution
processes in the disk environment will help improve
understanding of the planetary desert and radius gap such as effects
of the core mass and compositions (e.g.,
\citealt{owenEvaporationValleyKepler2017};
\citealt{jinCompositionalImprintsDensity2018};
\citealt{mordasiniPlanetaryEvolutionAtmospheric2020}) and stellar
and disk metallicity (e.g.,
\citealt{owenMetallicitydependentSignaturesKepler2018};
\citealt{guptaSignaturesCorePoweredMassLoss2020}). For example, in
the photoevaporation scenario, the radius gap should exhibit a trend
with early high-energy emission of stars. However, the current
analysis in \cite{loydCurrentPopulationStatistics2020} does not show
a correlation between the radius gap and stellar activity in near-UV emission. As mentioned before, it is the XUV flux in
the early stellar history rather than the current XUV flux
that matters for the escape rate. To date, uncertainty in
the XUV/X-ray history is large. One can eliminate this uncertainty
by analyzing multi-planetary systems. Recent work by
\cite{owenTestingExoplanetEvaporation2020} found that the current
dataset is consistent with their photoevaporation model, with a few
exceptions. Moreover, future observations on the atmospheric
compositions might also provide clues. For example, it was suggested
that planets close to the upper boundary of the radius
gap, i.e., the smallest ones in the sub-Neptune population, could
have helium-rich atmospheres due to preferential mass loss of
hydrogen over helium during photoevaporation (e.g.,
\citealt{huHeliumAtmospheresWarm2015a};
\citealt{malskyCoupledThermalCompositional2020}).

\section{Atmospheric Thermal Structure}
\label{sect:temp}
\subsection{Fundamentals}
\label{sect:tempfund}

The equilibrium temperature of a planet depends on the incoming
stellar flux, bond albedo and emissivity of the surface and the
atmosphere. The atmospheric albedo and emissivity are controlled by
the composition in the atmosphere, especially clouds (e.g.,
\citealt{marleyReflectedSpectraAlbedos1999}). The current dataset of
close-in gas giants does not suggest any correlation between the
inferred albedo and other planetary parameters
(\citealt{cowanStatisticsAlbedoHeat2011};
\citealt{hengUNDERSTANDINGTRENDSASSOCIATED2013};
\citealt{schwartzPhaseOffsetsEnergy2017};
\citealt{zhangPhaseCurvesWASP33b2018};
\citealt{keatingUniformlyHotNightside2019}) although it appears that
their bond albedos are systematically low as expected for hot,
cloud-free atmospheres (\citealt{cowanStatisticsAlbedoHeat2011}).
Temperature distribution in the atmosphere is controlled by energy
sources, sinks and transport processes. External energy
sources on exoplanets include various processes such as stellar
irradiation, high-energy particle precipitation and magnetic
Ohmic heating. The primary internal energy source on gaseous planets
and brown dwarfs is the heat release from gravitational contraction.
Geothermal heat from radioactive decay is usually negligible.
Deuterium burning is briefly important for young, less massive brown
dwarfs (e.g., \citealt{burrowsNongrayTheoryExtrasolar1997};
\citealt{spiegelDeuteriumburningMassLimit2011}). The atmosphere
mainly cools down through thermal emission to space. Atmospheric
loss processes, such as escape and condensation, can also change the
bulk energy of the atmosphere. Energy transport processes fall into
three primary types: dynamics, radiation, and conduction. Among all
dynamical processes, convection is more important in vertical energy
transport, while the horizontal energy transport is controlled by
other processes such as large-scale circulation, small-scale eddies
and waves, wave breaking and turbulent dissipation. Radiation
and conduction are usually important in vertical rather
than horizontal energy transport.

Convection represents a large overturning of bulk atmospheric mass
and the associated thermal energy and gravitational potential
energy. Vigorous convection can be considered as an adiabatic
process. As a result, convection tends to vertically smooth out the
entropy in the atmosphere, or potential temperature $\theta$, which
is defined as $\theta=(p/p_0)^{(\gamma-1)/\gamma}$ where $p_0$ is a
reference pressure. If the bulk vertical velocity scale is $w$, the
convective timescale is $\tau_{\mathrm{conv}}\sim H/w$, where $H=k_B
T/mg$ is the scale height of the convective atmosphere. It would be
useful to analyze the static stability of the atmosphere, which can
be measured by the buoyancy frequency, or
Brunt-V$\ddot{\mathrm{a}}$is$\ddot{\mathrm{a}}$l$\ddot{\mathrm{a}}$
frequency $N$, given by
\begin{equation}\label{bfreq}
N^2=g \frac{\partial\ln \theta}{\partial z}.
\end{equation}
{An a}tmosphere with low static stability ($N^2 < 0$)
tends to be convective. The temperature gradient follows the dry
adiabatic lapse rate $dT/dz=g/c_p$ in a dry atmosphere, and
follow{s} a shallower moist adiabat including the latent heat
release in condensation. If vertical compositional gradient exists
(i.e., lighter molecules on top of heavier molecules), tropospheric convection might not behave in a simple
Rayleigh-B\'enard type. Instead, double diffusive convection and
fingering might occur to result in less heat transport efficiency
and steeper vertical temperature profile in the deep atmosphere
(e.g., \citealt{stevensonSemiconvectionOccasionalBreaking1979};
\citealt{guillotCondensationMethaneAmmonia1995};
\citealt{leconteNewVisionGiant2012};
\citealt{tremblinFINGERINGCONVECTIONCLOUDLESS2015};
\citealt{leconteCondensationinhibitedConvectionHydrogenrich2017};
\citealt{tremblinThermocompositionalDiabaticConvection2019}).

Radiative energy is transferred via photon exchange among
atmospheric layers. Radiation will drive the vertical profile of
temperature to radiative equilibrium so that radiative heating and
cooling balance each other in each layer. The
radiative timescale depends on the temperature and opacities{ of}
gas and particle constituents in the atmosphere. In the optically
thick limit, usually applicable to the deep atmosphere, thermal
radiation can be approximated as a diffusion process. In the gray
limit---the atmospheric opacity does not depend on wavelength---the
radiative timescale $\tau_{\mathrm{rad},\infty}$ can be treated as
the diffusive timescale of temperature
 \begin{equation}
\tau_{\mathrm{rad},\infty}\sim\frac{p^2c_p\kappa_R}{g^2\sigma T^3},
\end{equation}
where $\kappa_R$ is the Rosseland{ }mean opacity.
$\sigma$ is the Stefan-Boltzmann constant. Here we omit the
prefactor close to unity. We can simplify the radiation as a
cooling-to-space process in the optically thin limit for the upper
atmosphere. The radiative timescale is
 \begin{equation}
\tau_{\mathrm{rad},0}\sim\frac{c_p}{\kappa_P\sigma T^3},
\end{equation}
where $\kappa_P$ is the Planck-mean opacity. If $\kappa$ and $T$ are
vertically constant, the radiative timescale is roughly constant
(independent of pressure) in the upper atmosphere
($\tau_{\mathrm{rad},0}$) but increases very rapidly with pressure
in the deep atmosphere ($\tau_{\mathrm{rad},\infty}$). The reason is
that, as the pressure increases towards the deep atmosphere, the
mean free path of the photon decreases so quickly that the transfer
efficiency decreases dramatically. The transition region between the
two regimes occurs at the thermal emission level $p_e$ where the
mean optical depth $p_e \bar{\kappa}/g\sim 1$. Here
$\bar{\kappa}=(\kappa_R \kappa_P)^{1/2}$. At this level,
$\tau_{\mathrm{rad},0}\sim\tau_{\mathrm{rad},\infty}\sim
p_ec_p/g\sigma T^3$, which can be considered as the mean radiative
timescale of the entire atmosphere (e.g.,
\citealt{showmanAtmosphericCirculationTides2002}).

Thermal conduction can smooth out{ the} vertical temperature
gradient by molecular collisions between adjacent atmospheric
layers. It can also be regarded as a thermal diffusion process. The
efficiency depends on the mean free path of the bulk gas components.
The conduction timescale $\tau_{\mathrm{cond}}$ is given by
 \begin{equation}
\tau_{\mathrm{cond}}\sim\frac{pc_ps_m}{g^2}(\frac{T}{k_B m})^{1/2},
\end{equation}
where $s_m$ and $m$ are the mean cross-section and mean mass of the
air molecule, respectively. The conduction timescale decreases
quickly towards the top of the atmosphere as long as the gas remains
collisional.

Crudely speaking, in {a} globally averaged sense, in
the deep atmosphere where the photon mean free path is short and the
radiative timescale is long, convection dominates the energy
transport process. Temperature follows the adiabat in this region,
and the potential temperature is homogenized vertically. Above the
convective region, radiation dominates the atmosphere, and the
temperature profile follows the radiative equilibrium. In the upper
atmosphere where the density is so low that thermal conduction
becomes efficient, energy is transported through molecular
collision, and the temperature (not potential temperature) gradient
tends to be smoothed out. Other dynamical processes could also be
important. For instance, wave energy deposition could also heat or
cool the upper atmospheric region where waves break, as has been
suggested for Jupiter's thermosphere
(\citealt{yelleJupiterThermosphereIonosphere2004}) and brown dwarf
WISE 0855 (\citealt{morleyBandSpectrumColdest2018}). Meridional
circulation and waves could also transport the energy from{ the}
polar auroral region to the equator on Saturn (e.g.,
\citealt{brownPoletopolePressureTemperature2020}). The vertical
temperature profile will be estimated analytically and discussed in
Section \ref{sect:verttemp}.

Vertical structures of temperature and compositions can be obtained
from the observed spectra, while the horizontal distribution could
be inferred from the light curves. For transmission spectra, transit
depth is determined by the line-of-sight optical depth, from which
one can invert the vertical optical depth, and density profile of
the species through inverse Abel transform (e.g.,
\citealt{phinneyRadioOccultationMethod1968}), and thus derive the
temperature from the density profile. However, it is challenging to
resolve degeneracies, such as that between the temperature and mean molecular weight
(\citealt{griffithDisentanglingDegenerateSolutions2014}). For
thermal emission spectra, temperature retrieval is a non-trivial
inversion problem. The basic principle is that the thermal emission
at different wavelengths is sensitive to different vertical layers
because the atmospheric optical depth is different. If the spectral
resolution is sufficiently high, a vertical profile of temperature
can be inverted from the thermal emission spectra. However, in
reality, this problem is often ill-defined mathematically due to a
finite number of data points, leading to non-unique solutions.

In the last decade{,} several successful inversion models have
been developed to retrieve information {o}n the
transmission and emission spectra of exoplanets and brown dwarfs.
The grid search method (e.g.
\citealt{madhusudhanTemperatureAbundanceRetrieval2009}) is usually
computationally expensive. Bayesian retrieval approaches are widely
used, including different techniques such as optimal estimation
gradient-descent (e.g.,
\citealt{lineInformationContentExoplanetary2012,
lineSystematicRetrievalAnalysis2013a};
\citealt{leeOptimalEstimationRetrievals2012};
\citealt{barstowConsistentRetrievalAnalysis2017}), nested sampling
(e.g., \citealt{bennekeHowDistinguishCloudy2013};
\citealt{bennekeStrictUpperLimits2015};
\citealt{waldmannTauRExNextGeneration2015};
\citealt{todorovWaterAbundanceDirectly2016};
\citealt{lavieHELIOSRETRIEVALOpensourceNested2017};
\citealt{kitzmannHeliosr2NewBayesian2020};
\citealt{oreshenkoRetrievalAnalysisEmission2017};
\citealt{macdonaldSignaturesNitrogenChemistry2017};
\citealt{gandhiRetrievalExoplanetEmission2018};
\citealt{fisherRetrievalAnalysis382018, fisherHowMuchInformation2019};
\citealt{seidelWindChangeRetrieving2020};
\citealt{zhangForwardModelingRetrievals2019};
\citealt{damianoExoReLBayesianInverse2020}), Markov chain Monte
Carlo (e.g.,
\citealt{waldmannTauRExNextGeneration2015,waldmannTRExIIRETRIEVAL2015};
\citealt{al-refaieTauRExIIIFast2019} and
\citealt{changeatAlfnoorRetrievalSimulation2020};
\citealt{madhusudhanHighRatioWeak2011,madhusudhanInferenceThermalInversions2010,
madhusudhanH2OAbundancesAtmospheres2014};
\citealt{lineSystematicRetrievalAnalysis2014,
lineUNIFORMATMOSPHERICRETRIEVAL2015,
lineUniformAtmosphericRetrieval2017};
\citealt{harringtonOpenSourceBayesianAtmospheric2015};
\citealt{cubillosCharacterizingExoplanetAtmospheres2016};
\citealt{cubillosCorrelatednoiseAnalysesApplied2017};
\citealt{blecicImplications3DThermal2017};
\citealt{wakefordHATP26bNeptunemassExoplanet2017};
\citealt{evansUltrahotGasgiantExoplanet2017};
\citealt{burninghamRetrievalAtmosphericProperties2017};
\citealt{mollierePetitRADTRANSPythonRadiative2019};
\citealt{zhangForwardModelingRetrievals2019}), and recently machine learning techniques (e.g.,
\citealt{waldmannDreamingAtmospheres2016};
\citealt{marquez-neilaSupervisedMachineLearning2018};
\citealt{soboczenskiBayesianDeepLearning2018};
\citealt{zingalesExoganRetrievingExoplanetary2018};
\citealt{cobbEnsembleBayesianNeural2019};
\citealt{hayesOptimizingExoplanetAtmosphere2020};
\citealt{himesAccurateMachineLearning2020};
\citealt{nixonAssessmentSupervisedMachine2020};
\citealt{johnsenMultilayerPerceptronObtaining2020}). More
description can be found in
\cite{madhusudhanAtmosphericRetrievalExoplanets2018}. It would be
desirable to do model intercomparison to cross validate all current
techniques to assess the advantage and disadvantage of each
retrieval method (e.g.,
\citealt{barstowComparisonExoplanetSpectroscopic2020}).

Regardless of the technical details, the uncertainties of the
retrieved temperature and composition primarily come from the
following uncertainty sources: {m}easurement uncertainty,
{r}elatively low spectral resolution in current data,
{f}inite wavelength coverage in the current instruments,
{l}aboratory data uncertainties of the optical properties
of gas and particles including opacities (e.g., absorption line
strength and spectral line shape parameters such as line-broadening
width), single scattering albedo and scattering phase functions,
{a}bundance uncertainties of the radiatively active gases
and particles that affect the observed spectra,{ and}
{t}he uncertainties in the forward model of radiative
transfer {applied} in retrieval. Among these factors,
(1) and (2) can be improved through more advanced observational
techniques and larger telescopes to enhance the {S/N}; (3) requires instruments with a wider range of
{IR} wavelengths; (4) needs to be significantly
improved in future laboratory experiments; (5) depends on the
chemistry and microphysics in the atmosphere but the gas and
particle abundances can also be jointly retrieved with the
temperature profile from observed spectra, and (6)
depends{ on} assumptions of the physics and chemistry
in the radiative transfer forward model, such as whether the
atmosphere can be assumed{ to be} horizontally homogenous (e.g.,
\citealt{lineInfluenceNonuniformCloud2016};
\citealt{fengImpactNonuniformThermal2016};
\citealt{plurielStrongBiasesRetrieved2020};
\citealt{taylorUnderstandingMitigatingBiases2020}), if the geometry
assumes the plane parallel{ case} or {a }spherical shell
(e.g., \citealt{caldasEffectsFully3D2019}), how to treat the gas and
particle scattering (\citealt{fisherRetrievalAnalysis382018}), and
if the non-LTE effect is important. Non-LTE is particularly
important for{ a} high-temperature and low-density medium,
although the claimed detection of non-LTE emission features on hot
Jupiters (e.g., \citealt{swainGroundbasedNearinfraredEmission2010};
\citealt{waldmannGROUNDBASEDNEARINFRAREDEMISSION2011})
{is} still controversial (e.g.,
\citealt{mandellNONDETECTIONOFLBANDLINE2011}). The inclusion of the
non-LTE effect also requires lab information o{n} the
collisional deactivation rates of the vibrational energy levels for
important radiatively active species.

Longitudinal information of substellar atmospheres is primarily
obtained from light curve observations. To date, the most useful
data to infer the temperature distribution in the photosphere are
emission light curves. The horizontal temperature distribution is
primarily controlled by inhomogeneously distributed external energy
sources, atmospheric thermal emission and horizontal heat
transport. The internal energy release through convection might be
higher in the polar region (in the direction of the rotational axis)
than the equatorial region on a fast-rotating body (e.g.,
\citealt{showmanAtmosphericDynamicsBrown2013}). However, the
resultant temperature difference is much smaller than that from
external sources, such as the equator-to-pole temperature contrast
due to incoming stellar irradiation. The influence of
planetary rotation, eccentricity and obliquity on the
external energy source distribution is also essential. Day-night
temperature contrast tends to be larger on slow{ly
}rotating planets than on fast{ }rotating planets.
That can be characterized by the ratio of the radiative timescale to
diurnal timescale and that to atmospheric dynamical timescale. In
the tidally locked configuration, planets will exhibit permanent
dayside and nightside. Seasonally varying temperature patterns are
expected on eccentric-tilted planets (more discussion in Sect.~\ref{sect:hjdyn}).

Atmospheric dynamics transport heat via including large-scale
circulation and small-scale eddies and waves. In the longitudinal
direction, zonal jets efficiently redistribute heat between the
dayside and nightside, leading to a small longitudinal
temperature contrast in the jet region. On the other hand,
substellar-to-anti-stellar circulation on slow{ly
}rotating planets or tidally locked planets will also
redistribute the energy from the dayside to the nightside. The upper
atmosphere on the nightside might also be warmed up by compressional
heating due to a strong downwelling flow. This dynamical effect has
been seen on the nightside of Venus
(\citealt{bertauxWarmLayerVenus2007}) and might also be crucial on
tidally locked exoplanets. Atmospheric waves such as Rossby and
Kelvin waves play essential roles in transporting energy in the
longitudinal direction. Meridional circulation cells and
eddies/waves redistribute heat among latitudes. For instance, on
terrestrial planets, Hadley-like circulation transports the excess
net heating from the equatorial region to the higher latitudes, but
the details could be tricky. Take the Earth as an example.
{A} Hadley cell transports the gravitational potential
energy poleward but sensible and latent heat equatorward (e.g.,
\citealt{shawTropicalSubtropicalMeridional2012}). At mid-latitudes
(e.g., in the Ferrell cell region), mid-latitude eddies from
transient baroclinic waves are responsible for poleward
heat transport (e.g., \citealt{vallisAtmosphericOceanicFluid2006}).

In Section \ref{sect:verttemp}, we will first discuss important
features in the vertical temperature profile, such as the
{RCB} and stratospheric
temperature inversion, as well as the influence of controlling
factors such as external and internal heat flux and opacity
distribution. Then we highlight the classification of close-in
exoplanets using temperature inversion and demonstrate spectral
statistical trends emerging from the {CMD}. We will discuss the thermal phase curve for
close-in exoplanets and brown dwarfs in Section \ref{sect:horitemp}
and rotational light curves for self-luminous bodies such as
directly imaged planets and free-floating brown dwarfs in Section
\ref{sect:rotlight}.

\begin{figure*}
   \centering
   \includegraphics[width=0.8\textwidth, angle=0]{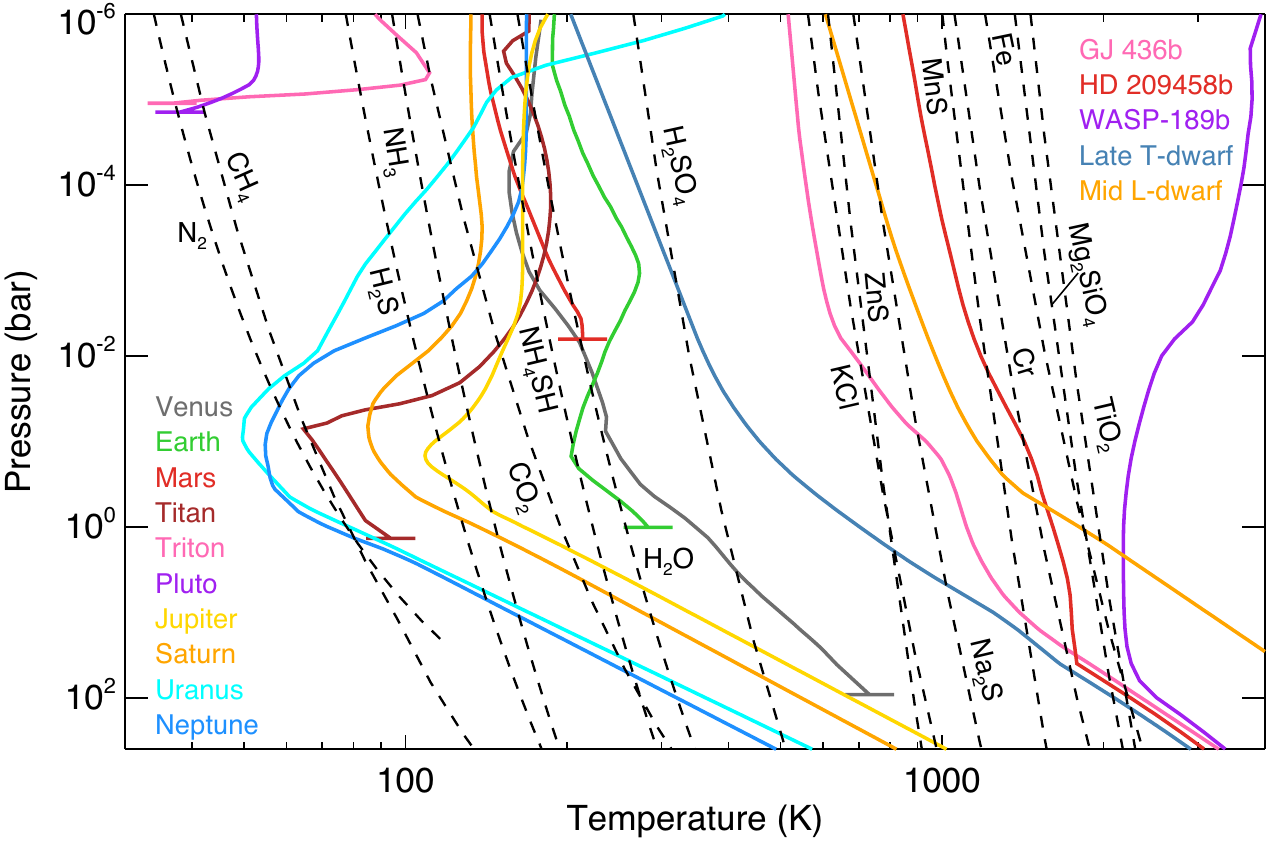}
   \caption{\baselineskip 3.8mm Typical temperature-pressure profiles ({\it solid}) on Solar System bodies, exoplanets and brown dwarfs. For exoplanets, we {display} typical profiles for a sub-Neptune (GJ 436 b), cooler hot Jupiter (HD 209458 b) and an ultra-hot Jupiter (WASP-189 b). For brown dwarfs, we show typical radiative-convective temperature profiles for a cooler late-T dwarf and a hotter mid-L dwarf. The {{\it dashed lines}} represent condensation curves of major condensable species. We assume solar metallicity for most condensates. We adopt volume mixing ratios of $10^{-5}, 2\times 10^{-2}, 1$ and $1$ for \ch{H2SO4}, \ch{CH4}, \ch{N2} and \ch{CO2}, respectively.}
   \label{figtprof}
   \end{figure*}

\subsection{Vertical Temperature Profile and Mid-IR Emission}
\label{sect:verttemp}

Typical vertical temperature profiles in the atmospheres on{
planets in} the Solar System and several exoplanets
and brown dwarfs show several important features (\ref{figtprof}).
First, most thick atmospheres are approximately in
radiative-convective equilibrium, characterized by a convective
region below and a radiative layer above. The two regions are
separated by the RCB.
Second, the thin atmospheres on Triton and Pluto are in
radiative-conductive equilibrium. Third, some planets develop
temperature inversion above their tropospheres, such as Earth,
Titan, giant planets and some ultra-hot Jupiters, but some
{do} not, including Venus, Mars, and brown dwarfs and
some hot Jupiters. Here we elaborate on the underlying processes
governing these behaviors.

To understand the RCB, let us first consider a dry, gray atmosphere
with external flux from the top $F_{\rm ext}$ and internal flux from
the bottom $F_{\rm int}$. The total luminosity of the planet is
equal to the sum of the external (incoming stellar irradiation) and
internal fluxes (self-luminosity). RCB should depend on the relative
strength of the external and internal fluxes. For simplicity, we
assume the temperature is roughly isothermal in the radiative zone
above the RCB. Below the RCB{,} the convective atmosphere is
characterized by an adiabat following $\nabla_{\rm ad}=\partial \ln
T/\partial \ln p=(\gamma-1)/\gamma$, where $\gamma$ is the adiabatic
index. For self-luminous exoplanets and brown
dwarfs where the external flux is negligible, the RCB pressure is
where the emission optical depth{ is} unity, $p_{\rm
rcb}=g/\kappa_R$. In the presence of external heat flux, the emission
temperature (skin temperature) of the atmosphere increases, and the
RCB progresses to the deeper atmosphere. In this case{,} the RCB
pressure level $p_{\rm rcb}$ can be scaled as
\begin{equation}\label{rcb}
p_{\rm rcb}\sim\frac{g}{\kappa_R}(1+\frac{F_{\mathrm{ext}}}{F_{\mathrm{int}}})^{\gamma/4(\gamma-1)}.
\end{equation}
For self-luminous bodies, $F_{\mathrm{ext}}=0$, we obtain $p_{\rm
rcb}=g/\kappa_R$. For highly irradiated planets, external flux is
much larger than internal flux. For an ideal diatomic gas
in a dry atmosphere, $\gamma=7/5$ and the RCB pressure is roughly
proportional to $F_{\mathrm{ext}}/F_{\mathrm{int}}$ (precisely, to
the 7/8 power). Consequently, a factor of two change
{in} the estimated internal heat will lead to a factor
of two change {in} the RCB pressure. Because many hot
Jupiters have inflated radii, their internal heat (entropy) might be
much higher than {a} non-inflated Jupiter. Thus{,}
their RCB should be located at a shallower level.
\cite{thorngrenIntrinsicTemperatureRadiative2019} pointed out that
the internal heat fluxes of inflated hot Jupiters could be much
larger than previous estimates, and thus the RCBs are located at
lower pressure levels. It is even possible that future
high-resolution observations can probe below the RCBs and detect the
properties of the convective region of some irradiated gas giants.

In reality, the temperature profile in the radiative zone is usually
not isothermal. For a gray, optically-thick atmosphere under radiative equilibrium, the temperature gradient
$\nabla_{\rm rad}$ can be calculated by radiation diffusion
\begin{equation}\label{reg}
\nabla_{\rm rad} = \frac{\partial \ln T}{\partial \ln p}=\frac{3\kappa_RpF_{\rm rad}}{16g\sigma T^4},
\end{equation}
where $F_{\rm rad}$ and $T$ are the radiative flux and temperature
at pressure $p$, respectively. If the radiative-equilibrium temperature
gradient $\nabla_{\rm rad}$ is steeper than the adiabat $\nabla_{\rm
ad}$, {the }atmospher{e} will be convectively
unstable. Therefore{,} RCB occurs at the pressure level where
the two temperature gradients are equal. Including this more
realistic consideration would further complicate
understanding the RCB.

More importantly, the atmospheric opacity has a significant
wavelength dependence. This ``non-gray" effect is the key to
understand{ing} many features in the vertical temperature
structure. First, because the radiative timescale and radiative
temperature gradient strongly depend on the emitting flux and
opacity (one can get some intuition from Eq.~(\ref{reg}) although
it is in a gray limit), the atmosphere might exhibit multiple RCBs
with alternating radiative and convective zones (e.g.,
\citealt{fortneyPlanetaryRadiiFive2007};
\citealt{marleyCoolSideModeling2015}). Imagine a convective region
below the first (top) RCB, and the atmosphere is optically thick. As
the temperature increases with pressure, the atmospheric emission
peak (e.g., the peak of the Planck function) shifts to shorter
wavelengths. It the emission peak happens to overlap with a
relatively transparent (low opacity) wavelength region, the
atmospheric energy would be carried outward by radiation instead of
convection. A second RCB emerges between an upper convective zone
(called the detached convective zone) and a deeper radiative zone.
The radiative zone ceases when the atmosphere becomes optically
thick again at a deeper level, where a third RCB forms. However, it
is difficult to detect the second and third RCBs because they
usually lie in very deep atmospheres.

The second important ``non-gray" effect is the temperature
inversion---increasing with decreasing pressure---in the radiative
zone. To elaborate this effect, we adopt the double-gray (or
semi-gray) atmosphere assumption, which assumes one gray opacity for
the visible and one gray opacity for the {IR}. The
globally averaged radiative equilibrium temperature $T$ can be
expressed as a function of pressure $p$ for a classical Milne
atmosphere (see derivation in the {a}ppendix in
\citealt{zhangRadiativeForcingStratosphere2013})
\begin{equation}\label{ret}
\begin{aligned}
T^4 (p)=&\,\frac{3F_{\mathrm{int}}}{4\sigma\pi} (\frac{2}{3}+\tau_{\mathrm{IR}})+\frac{3F_{\mathrm{ext}}}{4\sigma\pi} \Big{[}\frac{1+\alpha}{6\alpha}\\
&\quad+\frac{\alpha}{6} E_2 (\tau_{\mathrm{vis}})-\frac{1}{2\alpha} E_4 (\tau_{\mathrm{vis}})\Big{]}\,,
\end{aligned}
\end{equation}
where $E_n(x)=\int_1^{\infty}e^{-xt}/t^n dt$ is the exponential
integral (from the average over angles) and
$\alpha=\tau_{\mathrm{vis}}/\tau_{\mathrm{IR}}$ is a the ratio of
the visible opacity $\tau_{\mathrm{vis}}$  to the
{IR} opacity $\tau_{\mathrm{IR}}$ at pressure
level $p$. $F_{\mathrm{int}}$ and $F_{\mathrm{ext}}$ are the
internal heat flux and incoming stellar flux, respectively. See
similar expressions in other works for pure absorption (e.g.,
\citealt{hubenyPossibleBifurcationAtmospheres2003};
\citealt{hansenAbsorptionRedistributionEnergy2008};
\citealt{guillotRadiativeEquilibriumIrradiated2010};
\citealt{robinsonANALYTICRADIATIVECONVECTIVEMODEL2012};
\citealt{parmentierNongreyAnalyticalModel2014,parmentierNongreyAnalyticalModel2015})
and including scattering (e.g., \citealt{
hengEffectsCloudsHazes2012, hengANALYTICALMODELSEXOPLANETARY2014a}).

In this semi-gray framework, if the external flux is negligible,
e.g., on brown dwarfs or free-floating planets, the only heat source
is from the deep atmosphere. The optically thick lower atmosphere is
characterized by low static stability and vigorous vertical mixing
due to convection. Above the RCB, the radiative equilibrium
temperature profile in the absence of external heat source should
decrease with decreasing pressure (Eq.~(\ref{ret})). Also, see the
gray atmosphere results (Eq.~(\ref{reg})) and the brown dwarf
temperature profiles in Figure \ref{figtprof}. Note that this does
not mean temperature inversion is impossible in these
atmospheres. Other processes than radiative transfer might be
critical. For example, breaking of upward propagating waves from the
deep atmosphere might deposit the energy and heat the upper
atmosphere (e.g., \citealt{yelleJupiterThermosphereIonosphere2004};
\citealt{morleyBandSpectrumColdest2018}), although the detailed
mechanism is complicated because gravity waves might also cool the
upper atmosphere (\citealt{youngGravityWavesJupiter2005}).

If there is an external radiative forcing, the temperature profile
could develop an inversion profile more easily. Examples are the
thick atmospheres on Solar System planets (Fig.~\ref{figtprof}). In
the simple expression in Equation~(\ref{ret}), temperature inversion
could occur if the visible opacity $\tau_{\mathrm{vis}}$ exceeds the
{IR} opacity $\tau_{\mathrm{IR}}$, i.e., $\alpha >
1$. In this situation, local heating due to the absorption of
incoming stellar energy in the visible band is so large that the
atmosphere could not emit it away efficiently. As a result, an
Earth-like stratosphere forms. If temperature inversion
occurs, the static stability of the atmosphere significantly
increases with height, and diabatic mixing substantially
weakens. The level where the temperature inverts is the
``tropopause" because tropospheric dynamics such as
convective mixing {are} prohibited below that level.
Note that the tropopause and RCB in this context are different
because the former and the latter do not always coincide at the same
pressure level (e.g., \citealt{robinsonCommonBarTropopause2014}).

The nature of the tropopause is influenced by several physical
constraints from radiation, dynamics and thermodynamics. From
the radiation perspective (e.g.,
\citealt{manabeThermalEquilibriumAtmosphere1964};
\citealt{heldHeightTropopauseStatic1982};
\citealt{thuburnGCMTestsTheories1997,thuburnStratosphericInfluenceTropopause2000};
\citealt{robinsonANALYTICRADIATIVECONVECTIVEMODEL2012,robinsonCommonBarTropopause2014}),
the tropopause height---to first order---is a solution of
the temperature minimum that is consistent with the radiative
equilibrium upper atmosphere and the vertically mixed entropy flux
from the troposphere below. The lower boundary conditions, such as
surface temperature and surface opacity, play an important role.
{T}hermal inversion occurs above the RCB. Most thick
atmospheres in the Solar System {exhibit} temperature
inversion at approximately 0.1{\,}bar (Fig.~\ref{figtprof}), a result related to the atmospheric
{IR} opacity at{ the} surface lying between 1
and 10 in those atmospheres
(\citealt{robinsonCommonBarTropopause2014}). From the dynamical
constraint (e.g.,
\citealt{schneiderTropopauseThermalStratification2004}), large-scale
extratropical dynamics{,} such as horizontal transport of baroclinic eddies{,} play a dominant role in shaping the
temperature profile in the extraterrestrial region and thus the
tropopause height. This mechanism could be responsible for the
latitudinal distribution of the tropopause height in
Earth's atmosphere. In the moist atmosphere where
condensational species could saturate and form clouds, the
thermodynamic constraint is as important as other factors for
determining the tropopause height (e.g.,
\citealt{thompsonThermodynamicConstraintDepth2017}). On habitable
planets (e.g., \citealt{wordsworthWATERLOSSTERRESTRIAL2013}), the
saturation water vapor pressure in combination with the water vapor
radiative cooling greatly affects the temperature profile and thus
the tropopause height.

From the radiation constraint and the radiative equilibrium
temperature profile (Eq.~(\ref{ret})), UV or visible absorbers are
important to create the temperature inversion and stably stratified
upper atmosphere. On Solar System planets, the absorbers could be
ozone on Earth, methane on giant planets, and also haze particles on
Jupiter and Titan (e.g., \citealt{robinsonCommonBarTropopause2014};
\citealt{zhangAerosolInfluenceEnergy2015}). For exoplanets, titanium
oxides (TiO) and vanadium oxides (VO)---also major opacity sources
that dominate the visible spectra of M-dwarfs---have been proposed
to serve as stratospheric absorbers and might cause bifurcation of the temperature profile (e.g.,
\citealt{hubenyPossibleBifurcationAtmospheres2003};
\citealt{burrowsTheoreticalSpectralModels2007};
\citealt{fortneyUnifiedTheoryAtmospheres2008}). As mentioned in
Section~\ref{sect:char}, \cite{fortneyUnifiedTheoryAtmospheres2008}
systematically investigated the atmospheric temperature structures
of hot giant planets and suggested that these planets could be
classified into two categories: hotter ``pM" planets
and cooler ``pL" planets. The ``pM" planets show strong
thermal inversion caused by the TiO and VO opacity in the upper
atmosphere while the ``pL" class does not. Other
opacit{y} sources could also lead to thermal
inversions, such as sulfur-bearing haze particles (e.g.,
\citealt{zahnleAtmosphericSulfurPhotochemistry2009}). The other
option to create thermal inversion is to have the major
coolant vanishing quickly in the upper atmosphere, as in the case of
water on ultra-hot Jupiters (e.g.,
\citealt{arcangeliOpacityWaterDissociation2018};
\citealt{parmentierThermalDissociationCondensation2018};
\citealt{lothringerExtremelyIrradiatedHot2018}).

However, observational evidence of thermal inversion on
exoplanets has remained elusive for years. An isothermal atmosphere
will naturally produce blackbody-like spectra. A temperature profile
that decreases with height will generally {display}
absorption features. On the other hand, a strong emission feature in
the spectra is a possible signal of thermal inversion as
it implies that the upper layers are emitting more photons---and
thus might be hotter---than the underlying layers. Because of the
low-quality thermal emission spectra in a limited range of
wavelengths with contaminating star signals as well as
strong degeneracy between temperature and atmospheric composition,
searching and interpreting the specific emission features in
exoplanet spectra {have} not been very successful
(e.g., search for TiO by \citealt{singHSTHotJupiterTransmission2013}
and \citealt{hoeijmakersSearchTiOOptical2015}). The presumably
claimed thermal inversion on HD 209458 b (e.g.,
\citealt{burrowsTheoreticalSpectralModels2007};
\citealt{knutson68MmBroadband2008}) using Spitzer data
{was} later found to be not convincing (e.g.,
\citealt{diamond-loweNewAnalysisIndicates2014};
\citealt{schwarzEvidenceStrongThermal2015};
\citealt{lineNoThermalInversion2016}) when more constraints
{were} obtained from the HST Wide Field Camera 3 (WFC3) and ground-based photometry.
Other canonical hot Jupiters such as HD 189733 b and WASP-43 b
exhibit absorption instead of emission features in their thermal
spectra, implying no thermal inversion
(\citealt{grillmairStrongWaterAbsorption2008};
\citealt{kreidbergPRECISEWATERABUNDANCE2014};
\citealt{stevensonThermalStructureExoplanet2014};
\citealt{lineNoThermalInversion2016}).

Among the ultra-hot Jupiters with equilibrium temperatures higher
than 2200\,K (see Sect.~\ref{sect:char}), three of them have
recently been confirmed with temperature inversion. The observations
include TiO and \ch{H2O} dayside emissions on WASP-121 b
(\citealt{evansUltrahotGasgiantExoplanet2017}), CO emission on
WASP-18 b (\citealt{sheppardEvidenceDaysideThermal2017}) and TiO
emission on WASP-33 b
(\citealt{haynesSPECTROSCOPICEVIDENCETEMPERATURE2015};
\citealt{nugrohoHighresolutionSpectroscopicDetection2017}). Another
hot Jupiter HAT-P-7 b has also been suggested with atmospheric
thermal inversion (\citealt{mansfieldHSTWFC3Thermal2018}) but no
definitive emission feature has been confirmed yet. Some other
ultra-hot Jupiters like WASP-12 b and WASP-103 b
{display} absorption spectra that are consistent with
blackbodies, indicating possible isothermal atmospheres (e.g.,
\citealt{arcangeliOpacityWaterDissociation2018};
\citealt{parmentierThermalDissociationCondensation2018};
\citealt{kreidbergGlobalClimateAtmospheric2018}). The absorbers
responsible for thermal inversion on those ultra-hot
Jupiters were proposed as TiO/VO (e.g.,
\citealt{arcangeliOpacityWaterDissociation2018};
\citealt{parmentierThermalDissociationCondensation2018}) or haze and
soot particles (e.g., sulfur haze from
\citealt{zahnleAtmosphericSulfurPhotochemistry2009} as suggested for
WASP-18 b by \citealt{sheppardEvidenceDaysideThermal2017}), or
metals such as Na, Fe and Mg, SiO, metal hydrides and
continuous opacity like{ the} $\mathrm{H^-}$ ion
(\citealt{lothringerExtremelyIrradiatedHot2018},
\citealt{kitzmannPeculiarAtmosphericChemistry2018}). {T}hermal inversion is also partly attributed to
insufficient cooling of carbon monoxide in the upper atmosphere and
\ch{H2O} depletion due to thermal dissociation (\citealt{arcangeliOpacityWaterDissociation2018};
\citealt{parmentierThermalDissociationCondensation2018};
\citealt{lothringerExtremelyIrradiatedHot2018}). This suggests that
thermal inversion might be common on ultra-hot Jupiters
(\citealt{arcangeliOpacityWaterDissociation2018}). The lack of{
a} TiO/VO feature in the spectra of WASP-18 b (\citealt{arcangeliOpacityWaterDissociation2018}) could be due to the thermal dissociation of TiO/VO
(\citealt{lothringerExtremelyIrradiatedHot2018})), strong negative
ion opacities such as $\mathrm{H^-}$ or other metals, or an
oxygen-poor atmosphere
(\citealt{haynesSPECTROSCOPICEVIDENCETEMPERATURE2015}). Thermal
dissociation of hydrogen will also shape the day-night temperature
contrast on those ultra-hot Jupiters, which will be discussed in
Section~\ref{sect:horitemp}.

A systematic investigation of the vertical thermal structure on a
statistical sample is also possible. After the Spitzer telescope ran
out {of} cryogen in 2009, the
mid-{IR} bands centered at 3.6 and 4.5 microns
have provided the majority of thermal emission
observations on warm and hot exoplanets during their secondary
eclipses. In cloud-free atmospheres (i.e., no hazes or clouds) on
warm and hot \ch{H2}-dominated planets, the primary opacity sources
in the 3.6-micron channel are water and methane gases, while that at
4.5 microns {is} mainly carbon monoxide with some
contribution from water vapor. Therefore{,} the two channels
probe the thermal emission from two different vertical levels in the
atmosphere, although the weighting functions (i.e., the contribution
of each layer to the outgoing emission) from the two channels have
some overlap. We can estimate the temperature of the main emission
layer observed at each channel after translating the observed
thermal fluxes to the brightness temperatures ($T_B$) using the
Planck function. For warm hydrogen planets hotter than 600{\,}K, unless the C/O ratio is so large that the atmosphere is
oxygen-poor, atmospheric chemistry naturally favors CO over \ch{CH4}
(e.g., \citealt{mosesCompositionalDiversityAtmospheres2013}, see
Sect.~\ref{sect:gaschem}). A more considerable CO opacity than the
\ch{CH4} implies that the 4.5-micron observation probes at a higher
altitude than 3.6 microns. Thus, if $T_B$ at 4.5-micron
observation is higher than that at 3.6 microns, it might imply a
possible thermal inversion in the atmosphere.

\begin{figure*}
   \centering
   \includegraphics[width=0.85\textwidth, angle=0]{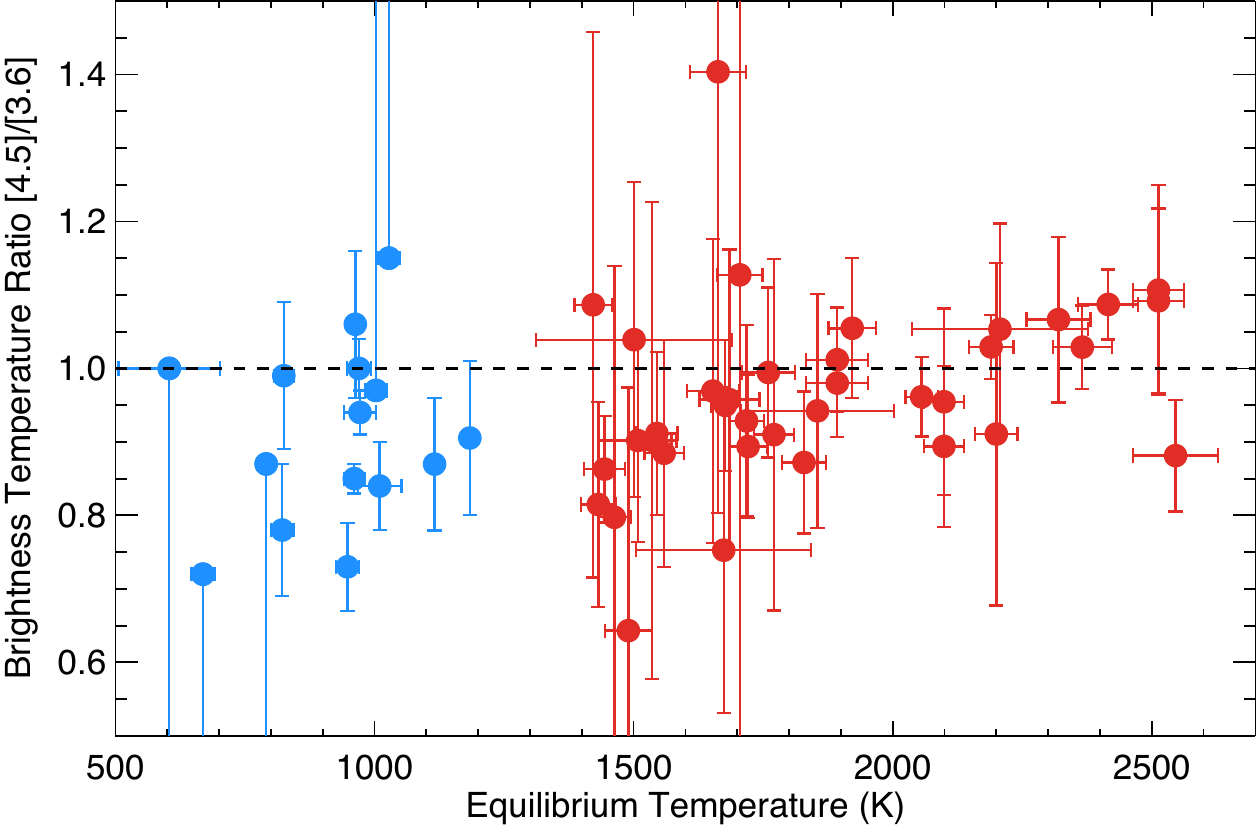}
   \caption{\baselineskip 3.8mm Ratio of the brightness temperatures at 3.6 and 4.5 micron Spitzer channels as a function of equilibrium temperature. The {{\it red dots}} are hot Jupiters from \cite{garhartStatisticalCharacterizationHot2020} and {{\it blue dots}} are cooler planets from \cite{kammerSPITZERSECONDARYECLIPSE2015} and \cite{wallackInvestigatingTrendsAtmospheric2019}. The horizontal {{\it dashed line}} indicates the ratio unity if the atmosphere behaves as a blackbody. A linear fit of the brightness temperature ratio as a function of equilibrium temperature across the entire sample {displays} a positive correlation, about 100$\pm$24 ppm change per Kelvin (\citealt{garhartStatisticalCharacterizationHot2020}).}
   \label{figteqratio}
   \end{figure*}

\cite{garhartStatisticalCharacterizationHot2020} compiled 78
secondary eclipse depths for a sample of 36 transiting hot Jupiters
in the warm {Spitzer} channels. Most of
the planets have smaller brightness temperatures at 4.5 microns than
at 3.6 microns. Exceptions include the ultra-hot Jupiters
discussed above that show ratio greater than unity, indicating
possible thermal inversion. This phenomenon is also consistent with
the emission features such as TiO detected in the spectra of those
planets. The Spitzer 3.6 and 4.5-micron data have also been
{ applied} in the thermal structure retrieval on
ultra-hot Jupiters (e.g., WASP-18 b, WASP-103 b and WASP-121
b) and provide important constraints on the determination of thermal inversion in their atmospheres. Several cooler planets
also {manifest} larger-than-unity ratios. HAT-P-26 b
is an extreme example that exhibits the brightness temperatures of
$\sim$2000{\,}K and $\sim$1400{\,}K (with
large uncertainties) at 4.5 and 3.6 microns, respectively. On the
other hand, a \ch{CH4}-rich and CO-poor atmosphere (i.e., large C/O
ratio) with the emission level higher at 3.6 than 4.5 microns could
also explain the larger-than-unity [4.5]/[3.6] ratio if the
atmospheric temperature is not vertically inverted.

Statistical analytics from
\cite{garhartStatisticalCharacterizationHot2020} also suggest
a weak trend {in} the ratio of the 4.5 micron $T_B$ to
the 3.6 micron $T_B$ (e.g., [4.5]/[3.6]) as a function of
equilibrium temperature (Fig.~\ref{figteqratio}). Previous studies
on a small sample of cooler bodies (e.g.,
\citealt{kammerSPITZERSECONDARYECLIPSE2015};
\citealt{wallackInvestigatingTrendsAtmospheric2019}) suggested no
evident trend of the [4.5]/[3.6] ratio with the equilibrium
temperature, planetary-mass or metallicity. When combining with
hotter 36 gas giants, the [4.5]/[3.6] ratio seems to increase with
equilibrium temperature by 100$\pm$24 ppm per Kelvin across the
range of 800--2500\,K (e.g.,
\citealt{garhartStatisticalCharacterizationHot2020}). Figure
\ref{figteqratio} {depicts} the overall trend that
seems deviated from blackbodies (horizontal dashed line). Despite
the uncertainty {in} the trend slope, it seems the
[4.5]/[3.6] ratio is smaller for colder planets and larger for
hotter planets. The interpretation of this trend is puzzling as is
not expected from current models (e.g.,
\citealt{burrowsTheorySecondaryEclipse2006};
\citealt{fortneyUnifiedTheoryAtmospheres2008}). This trend might
imply that the temperature structure becomes more and more
isothermal in the photosphere for hotter planets on which the
metallicity plays some unknown roles. {The }3D structure of the
temperature distribution or haze/cloud particles in the atmosphere
could also be possible reasons. More future observations need to
confirm this trend and understand the detailed mechanism behind it.

\subsection{Thermal Phase Curves on Tidally Locked Exoplanets}
\label{sect:horitemp}

Thermal phase curves have been observed on close-in exoplanets,
mostly on hot Jupiters. Even without fully resolved orbital
phase data, the averaged dayside and nightside flux difference could
provide important clues on the heat redistribution on these planets.
\cite{parmentierExoplanetPhaseCurves2018a} collected many data and
provided a thorough discussion of observational techniques and
potential problems. Most thermal phase curve observations
come from Spitzer 3.6 and 4.5-micron bands. Some light curves are
from the Kepler visible band, but the data have a significant
reflection stellar component. Recently observed ultra-hot Jupiters
show low albedos and their thermal phase curves can also be directly
obtained from the TESS band, such as WASP-18 b
(\citealt{shporerTESSFullOrbital2019}), WASP-19 b
(\citealt{wongTESSPhaseCurve2020}), WASP-121 b
(\citealt{daylanTESSObservationsWASP1212019};
\citealt{bourrierOpticalPhaseCurve2020}), WASP-100 b
(\citealt{jansenDetectionPhaseCurve2020}), KELT-9 b
(\citealt{wongExploringAtmosphericDynamics2019}) and WASP-33 b
(e.g., \citealt{vonessenTESSUnveilsPhase2020}). Here we just
highlight important trends on available data to date and try to
summarize the underlying mechanisms into a simple, self-consistent
framework.

To first order, thermal phase curves on
close-in exoplanets usually exhibit a sinusoidal shape, characterized by two
critical features: phase curve amplitude and phase offset from the
secondary eclipse. The thermal phase curve primarily probes the
horizontal temperature distribution with contributions from chemical
distributions in the photosphere. The temperature pattern is mainly
controlled by the day-night irradiation distribution and atmospheric
dynamics. The permanent day-night radiative forcing sets the
radiative equilibrium temperature distribution, while the
atmospheric dynamics such as waves and jets redistribute the heat
from the dayside to the night side and cause a deviation from equilibrium. Compared with the radiative equilibrium baseline,
the regulated day-night temperature contrast decreases, and the
longitude of the temperature maximum is shifted from the substellar
point. Thus the light curve amplitude and phase offset can be
{utilized} to diagnose the interplay between
radiation and dynamics in the atmosphere. Non-uniformly distributed
chemical species due to local chemistry, and dynamical transport
will further complicate the analysis. The detailed atmospheric
dynamics on close-in exoplanets will be discussed in Section~\ref{sect:hjdyn}. Here we present some simple scaling theories to
elaborate the underlying mechanisms governing the horizontal
temperature distribution to understand the thermal phase curves. We
will discuss the phase curve amplitude and day-night temperature
contrast in Section \ref{sect:phasecurve} and then
{address} the phase offset in Section
\ref{sect:offset}.

\subsubsection{Phase Curve Amplitude and Day-night Temperature Contrast}
\label{sect:phasecurve}

In principle, we can achieve scaling f{or} the
day-night temperature contrast by combining the horizontal momentum
equation, thermodynamic equation and continuity equation (see
{a}ppendix A in
\citealt{zhangEffectsBulkComposition2017}). In steady state, the
scaling equations can be represented as
\begin{subequations}\label{tscal}
\begin{align}
\frac{R\Delta T\ln (p_{s}/p)}{2 L}&\sim\frac{U^2}{L}+\Omega U+\frac{U}{\tau_{\mathrm{drag}}}
\\
\frac{\Delta T_{\rm eq}-\Delta T}{2\tau_{\mathrm{rad}}}&\sim\frac{wN^2H}{R}+\frac{qU}{c_pL}
\\
\frac{U}{L}&\sim\frac{w}{H}.
\end{align}
\end{subequations}
Here $R$ is the gas constant in units of
$\mathrm{J{\,}kg^{-1}{\,}K^{-1}}$. $\Delta
T_{\rm eq}$ is the day-night temperature contrast under radiative
equilibrium, and $\Delta T$ is the actual contrast. $\ln (p_{s}/p)$
is different in log-pressure from the deep pressure $p_s$ where the
temperature is horizontally homogeneous. This term approximates the
layer thickness in the log-pressure coordinate. $U$ and $w$ are the
typical horizontal and vertical wind scale{s}, respectively. $L$
and $H$ are the typical horizontal length scale and pressure scale
height, respectively. $N$ is the buoyancy frequency (Eq.~(\ref{bfreq})).

Equation (\ref{tscal}a) is a scaling of the horizontal momentum
equation in which we assumed a simple linear frictional term
characterized by $\tau_{\rm drag}$. On a close-in exoplanet with a broad
superrotating wind (i.e., eastward wind) pattern, the pressure
gradient force balances the nonlinear inertial term, the
Coriolis force and the drag force. This form essentially
tries to combine different momentum balance regimes discussed in
\cite{komacekATMOSPHERICCIRCULATIONHOT2016}. Equation~(\ref{tscal}c)
is the scaling of the continuity equation in which the vertical
divergence balances the horizontal divergence of the mass.

Equation (\ref{tscal}b) is the thermodynamic equation where the
horizontal advection of the temperature is assumed to be smaller
than the vertical entropy advection, so-called ``weak-temperature
gradient (WTG) approximation"
(\citealt{sobelWeakTemperatureGradient2001}). This assumption has
been {demonstrated to be} valid for typical close-in
hot Jupiters (\citealt{komacekATMOSPHERICCIRCULATIONHOT2016}) and
almost all cool terrestrial exoplanets (e.g.,
\citealt{pierrehumbertAtmosphericCirculationTideLocked2019}).
In{ a} traditional WTG framework, the radiative heating rate
balances the vertical entropy advection. But on the right-hand side
of Equation~(\ref{tscal}b), we include an additional heating source
$qU/c_p L$. This term collects several different possibilities other
than the traditional radiative heating, such as the thermal
dissociation of hydrogen on the dayside and recombination on the
nightside. In this case, $q=L_q \chi_H$ where $L_q$ is the bond
energy of hydrogen molecules, and $\chi_H$ is the mass mixing ratio
of the hydrogen atoms that recombine on the nightside and release
energy. This mechanism was suggested by
\cite{showmanAtmosphericCirculationTides2002} and has recently been
considered in the context of ultra-hot Jupiters (e.g.,
\citealt{bellIncreasedHeatTransport2018};
\citealt{komacekEffectsDissociationRecombination2018};
\citealt{tanAtmosphericCirculationUltrahot2019}). {O}ther possibilities have not been well investigated in the
context of exoplanetary atmospheres, for another example, the
photodissociation of species (mainly hydrogen) and
recombination. This case is similar to the previous one except that
$\chi_H$ depends on UV intensity from the central star.
Another possibility is downwelling compressional heating on
slow{ly }rotating planets in analogy with the upper
atmosphere of Venus (e.g., \citealt{bertauxWarmLayerVenus2007}).
In{ the} pressure coordinate, this term should be regarded as
the non-hydrostatic effect of adiabatic cooling and
compressional heating. Also, dissociation of hydrogen changes the
atmosphere's molecular weight and leads to expansion cooling on the
dayside. On the nightside, the recombination results in molecular
weight increase and subsequent compression heating. This mechanism
has been included in recent simulations of
\cite{tanAtmosphericCirculationUltrahot2019} but was not quantified
separately. Another heating mechanism could be shock heating with
the dissipating {KE} associated with mean
flows (e.g., \citealt{hengEXISTENCESHOCKSIRRADIATED2012}). The final
possibility is heat transport via interactions with the magnetic
field, but which might be too complicated to be represented by a
simple term like $qU/c_p L$.

For a scaling theory, one can assume $\Delta T_{\rm eq}\sim T_{\rm
eq}$ for tidally locked planets and horizontal length scale $L$ is
the planetary radius $R_p$. One can also use isothermal sound speed
$(RT_{\rm eq})^{1/2}$, which differs from the adiabatic sound
speed by a factor of $\gamma$, to approximate the fastest gravity
wave speed $NH$ in the isothermal limit and the cyclostrophic wind
speed induced by the day-night temperature difference in radiative
equilibrium $U_{\rm eq} = (R\Delta T_{\rm eq}\ln
(p_{s}/p)/2)^{1/2}$. If we simply assume the depth of the
temperature variation $\ln (p_{s}/p)\sim 2$, the dynamical
timescales of both wave propagation and wind advection across the
planet are comparable and can be approximated by
$\tau_{\mathrm{dyn}} = R_p(RT_{\rm eq})^{-1/2}$. The solution of the
scaling equation set (\ref{tscal}) is
\begin{subequations}\label{tscalsol}
\begin{align}
\frac{\Delta T}{\Delta T_{\rm eq}}&\sim1-2\alpha_1^{-1}(\sqrt{1+\alpha_2^2}-\alpha_2),
\\
\frac{U}{U_{\rm eq}}&\sim \sqrt{1+\alpha_2^2}-\alpha_2.
\end{align}
\end{subequations}

The non-dimensional parameters $\alpha_1$ and $\alpha_2$ are defined
as
\begin{subequations}\label{alphadef}
\begin{align}
\alpha_1&=\frac{\tau_{\mathrm{dyn}}}{\tau_{\mathrm{rad}}}(1+\frac{q}{c_pT_{\rm eq}})^{-1},
\\
\alpha_2&=\alpha_1^{-1}+(\Omega+\tau^{-1}_{\mathrm{drag}})\tau_{\mathrm{dyn}}/2.
\end{align}
\end{subequations}
If $q$ is zero (i.e., no extra heating mechanism), the solution is
consistent with that in \cite{zhangEffectsBulkComposition2017}.
Although the detailed dynamical and thermodynamical mechanisms are
not elucidated in this simple scaling theory, the solution implies
that the bulk atmospheric behavior such as the temperature and wind
is governed by dimensionless numbers: $\Omega\tau_{\mathrm{dyn}},
\tau_{\mathrm{dyn}}/\tau_{\mathrm{drag}},
\tau_{\mathrm{dyn}}/\tau_{\mathrm{rad}}$ and $q/c_pT_{\rm eq}$. An
additional number $\tau_{\mathrm{vis}}/\tau_{\mathrm{IR}}$ is
important for the vertical temperature profile and radiative
transfer (the term $\alpha$ in Eq.~(\ref{ret})). These five
parameters highlight important processes of planetary rotation, wave
dynamics, frictional drag, radiative transfer and (hydrogen) latent
heat. The first three numbers $\tau_{\mathrm{dyn}}\Omega,
\tau_{\mathrm{dyn}}/\tau_{\mathrm{drag}},
\tau_{\mathrm{dyn}}/\tau_{\mathrm{rad}}$ come from comparing the
dynamical timescale (flow advection timescale) $\tau_{\mathrm{dyn}}$
with that in other processes such as the rotational timescale
$\Omega^{-1}$, drag timescale $\tau_{\mathrm{dyn}}$ and radiative
timescale $\tau_{\mathrm{rad}}$, respectively. Based on these
numbers we can demarcate the atmospheric dynamics on tidally locked
exoplanets into several regimes in Section \ref{sect:hjdyn}.

The first number $\Omega\tau_{\mathrm{dyn}}$ is also equivalent to
the inverse of the ``WTG parameter" $\Lambda=c_0/\Omega R_p$
introduced in
\cite{pierrehumbertAtmosphericCirculationTideLocked2019} for
terrestrial planets. Here $c_0$ is the isothermal sound speed based
on the potential temperature of the adiabatic region above the
surface. $\Lambda$ can be considered as the ratio of the Rossby
deformation radius to the planetary radius. For synchronously
rotating terrestrial planets, scaling laws are different because the
surface flux needs to be taken into account to estimate the heat
redistribution between the dayside and nightside. Scalings in the
rocky planet regime have been derived in previous studies (e.g.,
\citealt{wordsworthAtmosphericHeatRedistribution2015};
\citealt{kollTemperatureStructureAtmospheric2016};
\citealt{kollScalingTheoryAtmospheric2019}).

The third dimensionless number
$\tau_{\mathrm{dyn}}/\tau_{\mathrm{rad}}$ is particularly important
in understanding the thermal phase curve on tidally locked
exoplanets. A strong radiative relaxation tends to maintain the
day-night thermal contrast towards the radiative equilibrium state,
while atmospheric winds and waves redistribute the heat and reduce
the thermal contrast. With other factors unchanged, both the
radiative timescale $\tau_{\mathrm{rad}}\propto T_{\rm eq}^{-3}$ and
the dynamical timescale $\tau_{\mathrm{dyn}}\propto T_{\rm
eq}^{-1/2}$ decrease with increasing temperature, but with a
different dependence---the former decreases faster than the latter.
Thus a hotter exoplanet tends to be more radiatively controlled,
leading to a larger day-night temperature contrast. This trend has
been confirmed by {two-dimensional (}2D{)} and 3D numerical
simulations of hot Jupiters (e.g.,
\citealt{perez-beckerAtmosphericHeatRedistribution2013};
\citealt{komacekATMOSPHERICCIRCULATIONHOT2016};
\citealt{komacekAtmosphericCirculationHot2017}). The theory can also be applied to tidally locked planets in
the habitable zone. Because of their relatively low
temperatures, they should have small day-night temperature contrast
in the free atmospheres, and thus are located
in the WTG regime. 3D terrestrial climate simulations on those
planets have also confirmed this behavior (e.g.,
\citealt{wordsworthAtmosphericHeatRedistribution2015};
\citealt{kollTemperatureStructureAtmospheric2016};
\citealt{haqq-misraDemarcatingCirculationRegimes2018};
\citealt{pierrehumbertAtmosphericCirculationTideLocked2019}).

\begin{figure*}
   \centering
   \includegraphics[width=0.9\textwidth, angle=0]{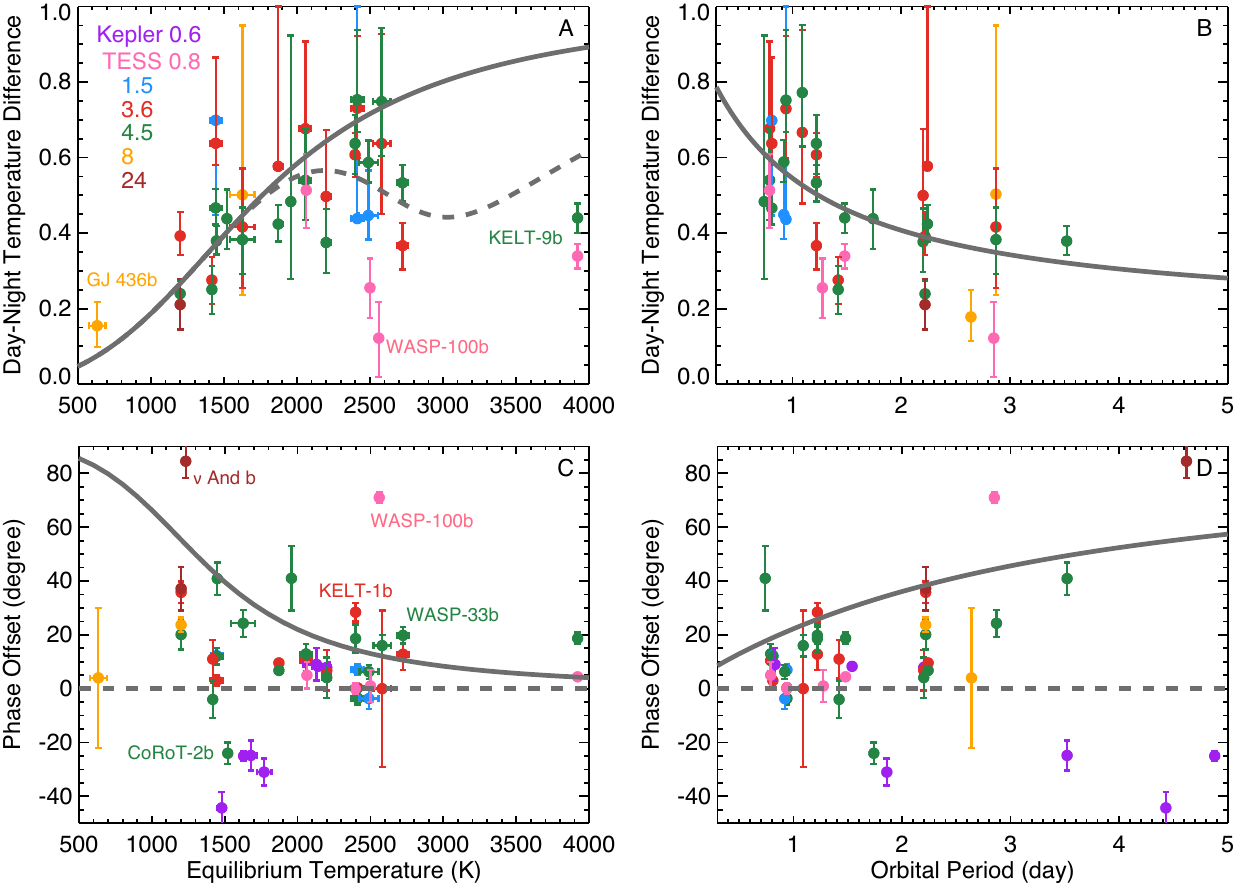}
   \caption{\baselineskip 3.8mm Day-night temperature contrast and phase offset versus equilibrium temperature and orbital period from observed phase curves at different wavelengths (in $\mathrm{\mu}m$, colored points). The simple analytical scaling predictions (grey curves) are mainly for illustrative purposes. The theories assume Jupiter size, no drag and radiative timescale $\tau_{\mathrm{rad}}=3\times10^4 (1400/T_{\rm eq} )^3$\,s (\citealt{iroTimedependentRadiativeModel2005}). The {{\it solid}} and {{\it dashed}} curves in (A) assume 3-day orbit planets with and without hydrogen dissociation and recombination respectively. The hydrogen dissociation energy source $q$ is calculated at the pressure 50{\,}mbar and the dayside temperature using Eq.~(\ref{tdn}) and the Saha equation (e.g., \citealt{bellIncreasedHeatTransport2018} and \citealt{tanAtmosphericCirculationUltrahot2019}). For (B) and (D), we adopted a relationship between the orbital period $P$ in days and equilibrium temperature around a {S}un-like star $T_{\rm eq}=1380(P/3)^{-1/3}${\,}K. Most data are collected from Table~1 in \cite{parmentierExoplanetPhaseCurves2018a} (see reference therein) and we converted their flux contrasts to temperature contrasts. In addition, we included new Spitzer observations for HD{ }149026 b (\citealt{zhangPhaseCurvesWASP33b2018}), WASP-33 b (\citealt{zhangPhaseCurvesWASP33b2018}), KELT-1 b (\citealt{beattySpitzerPhaseCurves2019}), KELT-9 b (\citealt{mansfieldEvidenceDissociationRecombination2020}), CoRoT-2 b (\citealt{dangDetectionWestwardHotspot2018}) and Qatar-1 b (\citealt{keatingSmallerExpectedBrightspot2020}), as well as recent TESS phase curve observations for WASP-18 b (\citealt{shporerTESSFullOrbital2019}), WASP-19 b (\citealt{wongTESSPhaseCurve2020}), WASP-121 b (\citealt{daylanTESSObservationsWASP1212019}), KELT-9 b (\citealt{wongExploringAtmosphericDynamics2019}) and WASP-100 b (\citealt{jansenDetectionPhaseCurve2020}). The WASP-43 b data are from the reanalysis by \cite{mendoncaRevisitingPhaseCurves2018}. We used the averaged day-night temperature as the approximated $T_{\rm eq}$ for WASP-110\,{\,}b because the calculated $T_{\rm eq}$ with zero albedo is still smaller than both observed day and night temperatures (\citealt{jansenDetectionPhaseCurve2020}).}
   \label{figteqtrend}
   \end{figure*}

If we take $q$ as the latent heat, the fourth dimensionless number
$q/c_pT_{\rm eq}$ can also be considered as the inverse Bowen ratio
(\citealt{bowenRatioHeatLosses1926}). The Bowen ratio is the ratio
of sensible heat flux (heat transfer flux between the
surface and the atmosphere) to the latent heat flux and depends on
details such as the temperature gradient and
condensational process. Here we generalize this concept to compare
the latent heat with the atmosphere's thermal energy
{utilizing} $q/c_pT_{\rm eq}$. This number is useful
for diagnosing the importance of latent heat release from
condensable species such as water and silicate (see
Sect.~\ref{sect:bddyn}).

Figure \ref{figteqtrend}(A) summarizes the observed day-night
temperature contrasts on tidally locked gas giants that were
inferred from the amplitude of thermal phase curves (see summary in
\citealt{komacekAtmosphericCirculationHot2017} and more data from
\citealt{parmentierExoplanetPhaseCurves2018a}).  It looks like there
might be an increasing trend with equilibrium temperature in the
current sample (Fig.~\ref{figteqtrend}(A), especially the 4.5-micron
data). However, data uncertainties are too large to confirm the
trend, which has also been pointed out in the day-night flux
contrast in \cite{parmentierExoplanetPhaseCurves2018a}. The curves
from the simple scaling theory are {displayed} mainly
for illustrative purposes. In reality, these planets likely have
different detailed properties such as temperature profile, opacity
(especially clouds) and frictional drag. The realistic
mechanisms should be more complicated than the discussion here (see
\citealt{parmentierTransitionsCloudComposition2016}), and more
observations with better data quality are needed for further
analysis.

In order for the non-traditional heat source $q$ to take effect, $q$
needs to be comparable{ to} or larger than the thermal energy
$c_p T_{\rm eq}$ (Eq.~(\ref{alphadef}a)), which is about 0.6\,eV
($c_p \sim3.5\,R$) for a 2000\,K hot hydrogen atmosphere. The latent
heat of most condensable species released from the intramolecular
bonds is at{ the} 0.1--10\,eV level and is generally unimportant
in hot, solar-metallicity atmospheres because the species is not
abundant (see Sect.~\ref{sect:bddyn}). On the other hand, hydrogen
bond energy is $\sim4.5$\,eV. If a large fraction of hydrogen
molecules are dissociated on the dayside and recombine on the night
side, the heat release could exceed the thermal energy $c_p T$ by a
factor of 10 or more. \cite{bellIncreasedHeatTransport2018}
quantified this effect by calculating the hydrogen atom fraction due
to thermal dissociation and found that this mechanism is important
for planets hotter than $\sim2500$\,K. Some recently characterized
ultra-hot Jupiter such as WASP-33 b ($T_{\rm eq} \sim 2723 $\,K,
\citealt{zhangPhaseCurvesWASP33b2018}) and KELT-9 b (dayside $T_{\rm
eq} \sim 4600 $\,K, \citealt{gaudiGiantPlanetUndergoing2017};
\citealt{wongExploringAtmosphericDynamics2019};
\citealt{mansfieldEvidenceDissociationRecombination2020}) might fall
into this regime as they show smaller day-night temperature contrast
than expected. \cite{komacekEffectsDissociationRecombination2018}
included this term in a scaling theory. They found that the
ultra-hot Jupiters could have a lower day-night temperature contrast
than the cooler ones. Figure~\ref{figteqtrend}(A)
illustrate{s} an example (dashed curve) if one includes
the thermal dissociation and recombination in the scaling theory and
the day-night temperature contrasts on planets in the ultra-hot
Jupiter regime ($T_{\rm eq}
>$2200\,K) decrease with equilibrium temperature and seem to
explain the day-night contrast of WASP-33 b and KELT-9 b
qualitatively.

The simple scaling theory also implies a nonlinear dependence
of the day-night temperature contrast on the equilibrium temperature
in the ultra-hot Jupiter regime. After about 2200\,K, the day-night
temperature contrast first decreases with $T_{\rm eq}$ but increases
again after $T_{\rm eq} >$ 3000\,K. This is because the thermal
dissociation has reached the limit beyond about 3000\,K and atomic hydrogen dominates the entire dayside atmosphere in our
simple scaling. In other words, as the atomic hydrogen fraction
$\chi_H$ increases with $T_{\rm eq}$ (Saha equation,
\citealt{bellIncreasedHeatTransport2018}), the hydrogen latent heat
term $q=L_q \chi_H$ first increases but is saturated at about
3000{\,}K when $\chi_H\sim 1$. Beyond this temperature, a
hotter planet is more radiative dominated (see the previous
discussion on $\tau_{\mathrm{dyn}}/\tau_{\mathrm{rad}}$) and the
day-night temperature contrast increases with $T_{\rm eq}$ again.
This nonlinear behavior is {signified as}
the dashed curve in Figure \ref{figteqtrend}(A), which needs to be
confirmed by more realistic dynamical simulations and observational
data for planets with $T_{\rm eq} >$ 3000\,K. 3D numerical models
have been {applied} to investigate the effects of
hydrogen thermal dissociation and recombination
(\citealt{tanAtmosphericCirculationUltrahot2019}). The models
{indicate} that hydrogen atoms produced by the thermal
dissociation on the dayside mostly recombine at the terminators
before{ being} transported to the nightside. Although the
nightside atmosphere also increases due to this mechanism, the
terminators are heated significantly in a 3D model. Consequently,
the decrease of day-night temperature contrast is mainly due to the
dayside cooling rather than the nightside warming. The
photodissociation of hydrogen (not included in current
models) due to high-energy UV stellar flux might also be important
but is probably limited {to} pressure level less
than $10^{-5}$ bar (see an example atomic hydrogen profiles in
\citealt{mosesDisequilibriumCarbonOxygen2011}). It looks unlikely to
impact the photospheric temperature distribution. However, a strong
magnetic effect might also occur on the zonal flow in this
high-temperature regime, as the atmospheres should be partially
ionized, the influence of which on the day-night temperature
contrast for ultra-hot Jupiters has yet to be explored.

\begin{figure*}
   \centering
   \includegraphics[width=0.8\textwidth, angle=0]{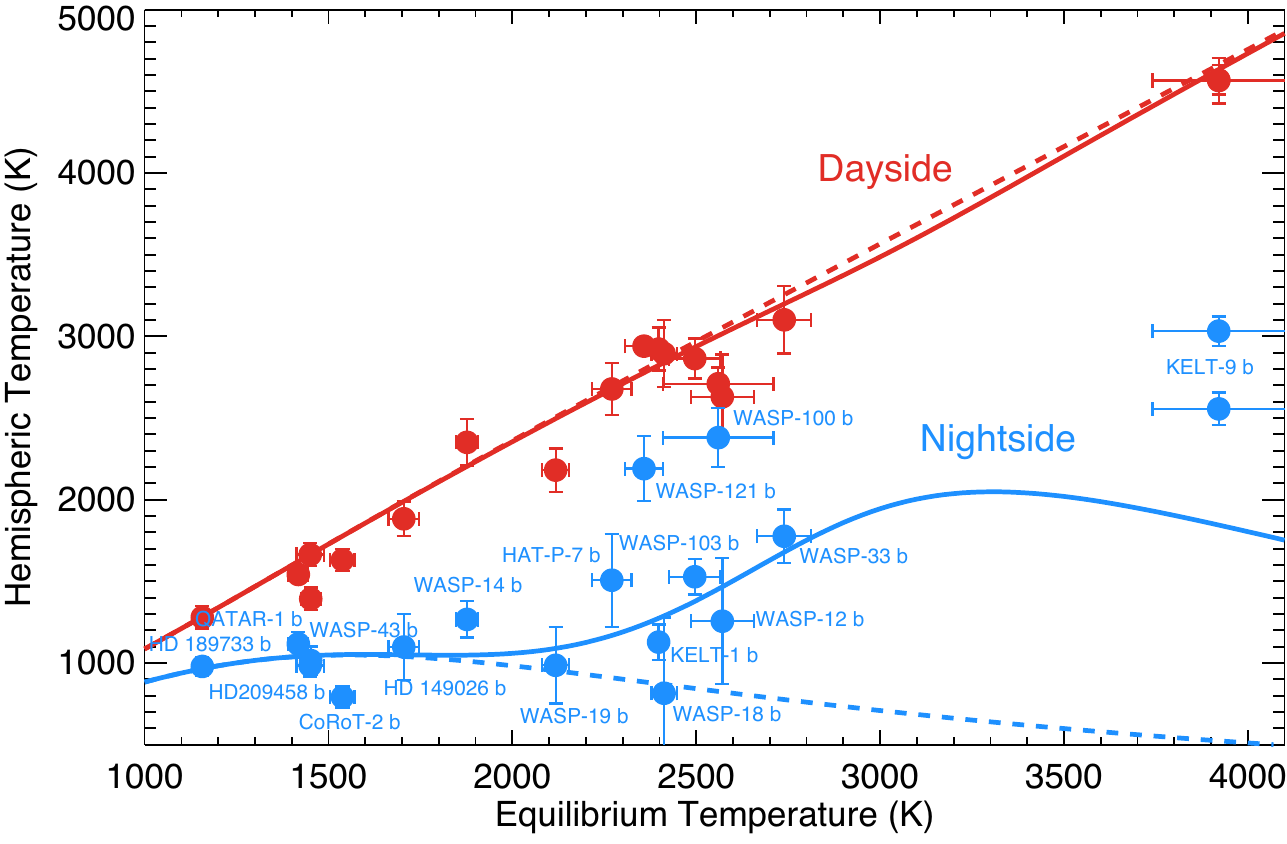}
   \caption{\baselineskip 3.8mm Brightness temperatures on the  dayside and nightside from an ensemble of hot Jupiters. Among all 17 samples, 12 hot Jupiters (CoRoT-2 b, HAT-P-7 b, HD 149026 b, HD 189733 b, HD 209458 b, WASP-12 b, WASP-14 b, WASP-18 b, WASP-19 b, WASP-33 b, WASP-43 b and WASP-103 b and a brown dwarf KELT-1 b are from \cite{keatingUniformlyHotNightside2019}. KELT-9 b is from \cite{mansfieldEvidenceDissociationRecombination2020} and \cite{wongExploringAtmosphericDynamics2019}. WASP-121 b is from \cite{daylanTESSObservationsWASP1212019}. Qatar-1 b is from\cite{keatingSmallerExpectedBrightspot2020}. WASP-100 b is from \cite{jansenDetectionPhaseCurve2020}. The analytical curves are plotted using the scaling theory in Eq.~(\ref{tdn}) with ({\it solid}) and without ({\it dashed}) the thermal dissociation and recombination of hydrogen on ultra-hot Jupiters. The input parameters are the same as{ in} Fig.~\ref{figteqtrend}(A). We have assumed a Jupiter size{d} planet in a 3-day orbit, no atmospheric drag and an analytical radiative timescale{ of} $\tau_{\mathrm{rad}}=3\times10^4 (1400/T_{\rm eq} )^3${\,}s from \cite{iroTimedependentRadiativeModel2005}. As in Fig.~\ref{figteqtrend}, we used the averaged day-night temperature as the approximated $T_{\rm eq}$ for WASP-110 b. One should not focus on the goodness of fit of the analytical models for the data because these curves are mainly for illustrative purposes. The real behaviors of the atmospheres should be much more complicated.}
   \label{figdnt}
   \end{figure*}

The dayside and nightside brightness temperatures can also be
separately derived from well-characterized phase curves.
Figure~\ref{figdnt} {displays} 16 hot Jupiters and an
irradiated (close-in) brown dwarf across a large range of $T_{\rm
eq}$ at 3.6 and 4.5 microns from Spitzer
(\citealt{keatingUniformlyHotNightside2019};
\citealt{keatingSmallerExpectedBrightspot2020};
\citealt{jansenDetectionPhaseCurve2020}). Their dayside temperatures
roughly scale linearly with the equilibrium temperature (assuming
zero albedo) from 1000--3000\,K, but the nightside temperatures
remain the same ($\sim$1100\,K) when $T_{\rm eq}$ changes from 1100
to 2500\,K, and then increase after $T_{\rm eq} >$ 2500\,K
(\citealt{keatingUniformlyHotNightside2019}). The hottest planet
KELT-9 b (TESS channel data from
\citealt{wongExploringAtmosphericDynamics2019} and the 4.5 micron
data from \citealt{mansfieldEvidenceDissociationRecombination2020})
also seem to follow this trend. The different temperature
trends in the two hemispheres can be roughly understood as a result
of increasing heat redistribution efficiency between the day and
nightsides as the planet gets hotter
(\citealt{keatingUniformlyHotNightside2019}). Based on the scaling
theory in Equation~(\ref{tscal}), we can further separate the
dayside and nightside temperatures assuming $2T^4_{\rm
eq}=T^4_{\mathrm{day}}+T^4_{\mathrm{night}}$ and
$(T_{\mathrm{day}}-T_{\mathrm{night}})/T_{\mathrm{day}}\sim\Delta
T/\Delta T_{\rm eq}$
\begin{subequations}\label{tdn}
\begin{align}
T_{\mathrm{day}}\sim(\frac{2}{1+\epsilon^4})^{1/4}T_{\rm eq},
\\
T_{\mathrm{night}}\sim\epsilon T_{\mathrm{day}},
\end{align}
\end{subequations}
where $\epsilon$ is the ratio of the nightside to dayside
temperatures and defined as
$\epsilon=2\alpha_1^{-1}(\sqrt{1+\alpha_2^2}-\alpha_2)$, and
$\alpha_1$ and $\alpha_2$ are from Equation~(\ref{tscalsol}a). The
scaling results {demonstrate} a decent explanation of
the increasing trend of the dayside temperature with $T_{\rm eq}$
(Fig.~\ref{figdnt}). In fact, given that the dayside temperature is
generally much larger than the nightside ($\epsilon< 1$), the
dayside temperature in Equation (\ref{tdn}a) can be well
approximated by $T_{\mathrm{day}}\sim2^{1/4}T_{\rm eq}$ in the limit
of $\epsilon^4\sim0$. The predicted nightside temperature also stays
roughly constant below $T_{\rm eq}=2500$\,K, implying that the
deposited stellar energy on the dayside is not efficiently
transported to the nightside as the stellar flux increases. This is
because the increasing dominance of radiation over wave
dynamics as $T_{\rm eq}$ increases
($\tau_{\mathrm{rad}}/\tau_{\mathrm{dyn}}\propto T^{-5/2}_{\rm
eq}$). The observed increasing nightside temperature after $T_{\rm
eq}>2500$\,K implies that the redistribution efficiency increases
again.

Including the hydrogen dissociation and recombination mechanism in
the ultra-hot Jupiter regime can explain the nightside
temperature trend better. Nevertheless, after about 3200\,K, the
nightside temperature decreases again in our simple theory because
the heating efficiency has reached {its} maximum, which
needs to be confirmed in future 3D models and observations. Note the
explanation of WASP 121 b and KELT 9 b is not very good by our
simple theory. Using the 3D GCM from
\cite{tanAtmosphericCirculationUltrahot2019},
\cite{mansfieldEvidenceDissociationRecombination2020} could explain
the day-night temperature difference of KELT 9~b although the phase
offset was not explained. Again, we emphasize the analytical
theories here are mainly for illustrative purposes because the
theory is oversimplified without including details of the radiative
transfer, tracer transport and opacity sources, in
particular, clouds. For example, as stated before, 3D simulations
from \cite{tanAtmosphericCirculationUltrahot2019} show that the
hydrogen dissociation and recombination could not significantly heat
the nightside atmosphere of ultra-hot Jupiters. Alternately, some
studies (\citealt{keatingUniformlyHotNightside2019},
\citealt{beattySpitzerPhaseCurves2019}) also proposed that nightside
thick clouds could mitigate the emission temperature variation
across the equilibrium temperature range and help maintain the
uniform brightness temperature on the nightside of cooler hot
Jupiters (see Sect.~\ref{sect:cloud} for discussion). Realistic 3D
dynamical simulations with cloud formation and future observations
at different wavelengths could elucidate the underlying mechanism
through the analysis of temperature and cloud distributions as well
as the wavelength dependence of the thermal emission.

It is also interesting to show a possible trend of
day-night temperature contrast with the planetary orbital period
(e.g., \citealt{stevensonSPITZERPHASECURVE2017}). Two factors are
influencing this trend. For a fixed star type, the orbital period
correlates with equilibrium temperature. As these close-in
exoplanets are expected to be tidally locked and synchronously
rotating, the orbital period correlates with the rotation period.
For a shorter-period planet, the equilibrium temperature is higher,
and the rotation is also faster. From the scaling theory, one should
also expect that a faster rotation and a stronger drag maintain a
larger spatial temperature gradient. Also, a hotter planet tends to
have a smaller heat redistribution efficiency. That implies that the
day-night temperature contrast for a fixed stellar type is higher
for a planet with a shorter orbital period. For a
{S}un-like star, one can approximate the equilibrium
temperature as $T_{\rm eq}=1380(P/3)^{-1/3}$\,K. Put in the
analytical theory and Figure~\ref{figteqtrend}(B) illustrates a
decreasing trend of the day-night temperature contrast with increasing orbital period. This conclusion is also consistent
with recent simulations on tidally locked planets with very rapid
rotation rates (\citealt{tanAtmosphericCirculationTidally2020}) that
show larger day-night contrast (see more discussion in
Sect.~\ref{sect:hjdyn}). However, one should also be cautious
because the planets in Figure~\ref{figteqtrend} are orbiting around
different types of stars, and there is no clear correlation between
the equilibrium temperature and orbital period in this sample. So if
there is any trend, it might be more related to the rotation rate
($\Omega$) dependence instead of the temperature ($T_{\rm eq}$)
dependence. Future observations need to separate the two factors
(i.e., $\Omega$ and $T_{\rm eq}$) {by employing}
more statistically significant data for planets around each
stellar type.

\subsubsection{Phase Offset}
\label{sect:offset}

Thermal phase curves on tidally locked exoplanets usually exhibit
phase offset. In the absence of dynamics, the hottest spot at the
same pressure level is located at the substellar point, and the peak
of the thermal phase curve occurs right at the secondary eclipse.
Heat redistribution by atmospheric jets and waves shift{s} the
hot spot away from the substellar point and lead{s} to a phase
offset of the light curve peak before the secondary eclipse. This
behavior was first predicted in a 3D atmospheric model
(\citealt{showmanAtmosphericCirculationTides2002}) and later was
detected in the observation of a hot Jupiter
(\citealt{knutsonMapDaynightContrast2007}). In a kinematic picture
(e.g., \citealt{cowanStatisticsAlbedoHeat2011};
\citealt{zhangEffectsBulkComposition2017}), the phase offset is
controlled by the horizontal heat transport and radiative
relaxation. Strong radiative damping maintains the horizontal
temperature distribution in the equilibrium
substellar-to-anti-stellar pattern, leading to a small phase offset
in the thermal phase curve. A strong longitudinal heat transport
would likely advect the hot spot away from the substellar point,
thus increasing the phase offset. A more realistic analysis using a
dynamical model in \cite{hammondWavemeanFlowInteractions2018} found
that the hot spot phase shift is a result of zonal flow Doppler
shifting the stationary wave response. Strong damping reduces the
forced wave response and brings the response in phase with the
forcing while in a weak damping case, the Doppler shift by the zonal
jet leads to a large phase offset. Quantitatively, the phase offset
$\delta$ can be estimated based on the relative magnitude of the
radiative timescale and dynamical timescale
\begin{equation}\label{off}
\delta\sim\tan^{-1}(\tau_{\mathrm{rad}}/\tau_{\mathrm{dyn}}).
\end{equation}

\cite{zhangEffectsBulkComposition2017} proposed a more complicated
formula, but the idea is similar. The theory can explain the
idealized 3D GCM results in \cite{zhangEffectsBulkComposition2017}.
Because the ratio $\tau_{\mathrm{rad}}/\tau_{\mathrm{dyn}}\propto
T^{-5/2}_{\rm eq}$ decreases with the equilibrium temperature, the
phase offset becomes smaller as the planets get hotter and thus more
radiatively controlled. This scaling predicts a trend that generally
agrees with the {IR} observational data (Fig.~\ref{figteqtrend}(C)). Exceptions will be discussed later.

For a fixed stellar type, the equilibrium temperature decreases with
an increasing orbital period, and the phase offset increases with
the orbital period. See an analytical curve assuming a
{S}un-like star in Figure~\ref{figteqtrend}(D) for
illustration. However, there is no clear dependence of the phase
offset on the orbital period for current characterized planets
orbiting around different types of stars (also see fig.~3 in
\citealt{parmentierExoplanetPhaseCurves2018a}). A larger size of the
planet sample is needed to analyze the statistical properties for
planets around each stellar type. In general, the analytical scaling
{i}n both Figures 6(C) and 6(D) predict{s} a larger phase
offset than the observations because we have estimated the dynamical
timescale based on the isothermal sound wave speed
$\tau_{\mathrm{dyn}} = R_p(RT_{\rm eq})^{-1/2}$ and the heat
transport might be overestimated, leading to a larger phase offset
than in the real atmospheres.

Despite a large scattering in the current data, this simple,
first-principle scaling seems to do a decent job to explain the
first-order, systematical behavior of the thermal phase curves on
tidally locked planets. However, the caveats of this scaling are
also evident. First, it does not include the feedbacks between
radiation and dynamics. It only considers the horizontal heat
transport and neglects the vertical entropy advection that seems
essential for many gas giants. 3D GCM simulations with{ a}
realistic radiation scheme overestimate the phase offset
(\citealt{parmentierExoplanetPhaseCurves2018a}), implying more
complicated dynamics therein. For rapid{ly }rotating
tidally locked planets, the hot spot could also be shifted westward
by the off-equatorial Rossby waves in addition to the eastward
offset by the eastward propagating Kelvin waves and mean flow at the
equator (e.g., \citealt{leeSimplified3DGCM2020};
\citealt{tanAtmosphericCirculationTidally2020}). These effects were
not considered in the simple scaling Equation~(\ref{off}).

Second, this theory predicts the amplitude of the light curve should
correlate with the phase offset---a flatter phase curve is
associated with a larger phase offset. However, we do not observe a
clear correlation between the day-night contrast and phase offset in
the current dataset (no{t} shown here). More precise
observations with smaller errors are needed to unveil any potential
correlation here. When including the reanalysis of Spitzer data of
WASP-43 b from \cite{mendoncaRevisitingPhaseCurves2018},
\cite{beattySpitzerPhaseCurves2019} further suggest that
there is no clear trend {in} the phase offset as a
function of $T_{\rm eq}$. Instead, the atmospheric clouds might have
significantly altered the thermal phase curves.

Third, there are some outliers. Some hot Jupiters show westward
phase offsets (i.e., the peak of the light curves occurs after the
secondary eclipse) at the thermal wavelengths. For
instance, CoRoT-2 b exhibits a clear westward hot spot
phase shift at 4.5 microns (Fig.~\ref{figteqtrend}(C),
\citealt{dangDetectionWestwardHotspot2018}). The possible westward phase offsets on
HD 149026 b at Spitzer 3.6 microns is a bit suspicious
(\citealt{zhangPhaseCurvesWASP33b2018}). Qatar-1 b might also show
{display} westward offset at Spitzer 4.5 microns but
the data {are} also consistent with zero
(\citealt{keatingSmallerExpectedBrightspot2020}. Moreover, HAT-P-7 b exhibits strong variability between the
eastward and westward offset in the phase curve
(\citealt{armstrongVariabilityAtmosphereHot2016}), {as do} Kepler 76 b
(\citealt{jacksonVariabilityAtmosphereHot2019}), WASP-12 b (e.g.,
\citealt{bellMassLossExoplanet2019}) and WASP-33 b
(\citealt{zhangPhaseCurvesWASP33b2018};
\citealt{vonessenTESSUnveilsPhase2020}). These first-order data-model
discrepancies suggest several missing physical processes in
the current understanding of the phase offset on those planets, such
as non-synchronous rotation dynamics
(\citealt{rauscherAtmosphericCirculationObservable2014};
\citealt{showman3DAtmosphericCirculation2015}), magnetic effects
(\citealt{batyginNonaxisymmetricFlowsHot2014};
\citealt{rogersConstraintsMagneticField2017};
\citealt{hindleShallowwaterMagnetohydrodynamicsWestward2019}),
partial cloud coverage in the east hemisphere of CoRoT-2 b or
planetary obliquity (e.g., \citealt{rauscherModelsWarmJupiter2017};
\citealt{ohnoAtmospheresNonsynchronizedEccentrictilted2019,ohnoAtmospheresNonsynchronizedEccentrictilted2019a};
\citealt{adamsSignaturesObliquityThermal2019}).

Other than the considerable variation between eastward and
westward offset, some ultra-hot Jupiters {manifest} a
much larger eastward phase offset than expected. For example,
WASP-33 b shows large phase shifts in both warm Spitzer channels,
implying an arguably increasing trend of the phase offset with
equilibrium temperature beyond some critical value
(\citealt{zhangPhaseCurvesWASP33b2018}; \citealt{vonessenTESSUnveilsPhase2020}). Again WASP-33 b is strange
because the recent TESS observations from
\citealt{vonessenTESSUnveilsPhase2020} also
{indicated} a large westward offset (28.7$\pm$7.1
degrees, see Fig.~\ref{figteqtrend}(C)), implying a sign of time
variability. Recently, \cite{jansenDetectionPhaseCurve2020} analyzed
the TESS phase curves of WASP-100 b and reported an eastward hotspot
offset of $71^{+2}_{-4}$ degrees. It suggests that some new physics
might affect the flow pattern in the ultra-hot regime, albeit a
slightly cooler, irradiated brown dwarf KELT-1 b
(\citealt{beattySpitzerPhaseCurves2019}) is also showing a
puzzling{ly} large phase offset (Fig.~\ref{figteqtrend}(C)). The
recently characterized KELT-9 b is also puzzling. While the large
phase curve amplitude{ is} expected in the high-temperature
regime, the large phase shift (18.7 degrees) at 4.5
microns is not expected
(\citealt{mansfieldEvidenceDissociationRecombination2020}). The
recent 3D model, including hydrogen dissociation and recombination
(\citealt{tanAtmosphericCirculationUltrahot2019}), could explain the
day-night contrast of this planet, but the model could not explain
the large phase shift. The magnetic effect may play an essential
role in the heat redistribution on this hot and ionized planet. More
ultra-hot Jupiter observations are needed to fully reveal any
possible statistical correlation of the phase offset with
equilibrium temperature in the high-temperature regime.

Amplitude and phase offset on the observed thermal phase curves also
depend on wavelengths, which probe different vertical
layers in the atmospheres. Theoretically, both the radiative
timescale and dynamical timescale change vertically, leading to
different $\tau_{\mathrm{rad}}/\tau_{\mathrm{dyn}}$ at different
layers and the resulting thermal phase curves at different
wavelengths. Phase-resolved emission spectroscopy is promising to
probe the phase shift of the phase curves among different
wavelengths (\citealt{stevensonThermalStructureExoplanet2014}).
Furthermore, in short wavelengths such as HST
near-{IR} band 1.5 micron (Fig.~\ref{figteqtrend}(C)), the phase curves are influenced by the
reflection of the stellar light (e.g.,
\citealt{shporerStudyingAtmosphereDominatedHot2015};
\citealt{parmentierTransitionsCloudComposition2016}). For the light
curves observed in the Kepler band at 0.6 micron, significant
westward phase offsets are detected on cooler planets ($<$ 1800\,K),
indicating brightest spots on the western hemispheres due to cloud
reflection of the stellar light at short wavelengths (see purple
dots in Fig.~\ref{figteqtrend}(C),
\citealt{parmentierTransitionsCloudComposition2016}). The nightside and western hemispheres are the coldest regions on the tidally
locked exoplanets, facilitating mineral and metal cloud formation
there. For hotter planets (e.g., $>$2000\,K), the thermal emission
components dominate the thermal phase curves, and thus the peak
phases are shifted before the secondary eclipse, even in the optical
wavelengths. This has also been seen in recently observed phase
curves of ultra-hot Jupiters using TESS (pink dots in Fig.~\ref{figteqtrend}(C), \citealt{shporerTESSFullOrbital2019};
\citealt{wongTESSPhaseCurve2020};
\citealt{daylanTESSObservationsWASP1212019};
\citealt{bourrierOpticalPhaseCurve2020};
\citealt{wongExploringAtmosphericDynamics2019};
\citealt{jansenDetectionPhaseCurve2020}). For example, the TESS
phase curve of KELT-9 b
(\citealt{wongExploringAtmosphericDynamics2019}) shows a smaller
eastward phase shift (4.4 degrees) than the Spitzer band data (18.7
degrees, \citealt{mansfieldEvidenceDissociationRecombination2020}),
implying that the two wavelengths probe different vertical levels on
that planet.

Non-synchronized rotation, orbital eccentricity and planetary
obliquity could further complicate the thermal structure evolution
and thermal phase curve behaviors. If the planet is orbiting far
away from the star where the gravitational tidal effect is weak and
the tidal circularization timescale is long, the planet is not
expected to be tidally locked. Eccentricity is not easy to be damped
by the gravitational tides as the planet migrates inward. The
planetary rotation axis might also be misaligned with the orbital
normal, resulting in non-zero planetary obliquities, as evidenced by
many planets in the Solar System. In {a} multi-planet
system, even the close-in planets might have non-zero obliquities in
highly compact systems
(\citealt{millhollandObliquitydrivenSculptingExoplanetary2019}).
Non-synchronized rotation induces a movement of the substellar point
along the longitude and alters the diurnal cycle of the stellar
forcing (e.g.,
\citealt{showmanThreedimensionalAtmosphericCirculation2015};
\citealt{pennThermalPhaseCurve2017, pennAtmosphericCirculationThermal2018}). A fast rotation could also homogenize the longitudinal
temperature distribution. The details of the dynamics will be
discussed in Section \ref{sect:hjdyn}.

Both the eccentricity and obliquity have large impacts on the
observed thermal phase curve. Orbital eccentricity causes the
``eccentricity season" in which the star-planet distance, and thus
the atmospheric temperature, changes with the orbit phase. Planetary
obliquity could also lead to a strong seasonal cycle due to the tilt
of the rotation axis. Planets orbiting in a highly eccentric orbit
sweep very fast near the periapse due to Kepler's second law,
resulting in a highly skewed thermal phase curve and some possible
oscillation pattern due to the pseudo-synchronous rotation (i.e.,
\citealt{langtonObservationalConsequencesHydrodynamic2007};
\citealt{lewisAtmosphericCirculationEccentric2010};
\citealt{katariaAtmosphericCirculationSuper2014}). An extreme
example of this case could be HD 80606 b with an eccentricity of
$\sim$0.93
(\citealt{katariaThreedimensionalAtmosphericCirculation2013}). On an
oblique planet, the substellar point migrates back and forth between
the northern and southern hemispheres in one orbit. As a result, the
peak of the thermal emission in the phase curve varies from case to
case, depending on the obliquity and viewing geometry (e.g.,
\citealt{rauscherModelsWarmJupiter2017};
\citealt{ohnoAtmospheresNonsynchronizedEccentrictilted2019,ohnoAtmospheresNonsynchronizedEccentrictilted2019a};
\citealt{adamsReassessingExoplanetLight2018}). In some cases, the
phase offset could occur after the secondary eclipse, i.e., westward
phase shift.
\cite{ohnoAtmospheresNonsynchronizedEccentrictilted2019a} provided
an intuitive understanding that the summed fluxes control the light
curve shape from the shifted hot spot projected onto the orbital
plane and the pole heated at the summer solstice.
\cite{adamsReassessingExoplanetLight2018} developed a thermal
radiative model to explore the full phase light curves and suggested
high-obliquity signatures might be linked to the recently detected
abnormal phase offset signals on some exoplanets such as WASP-12 b,
CoRoT-2 b and (possibly) HD 149026 b. Future observations on those
abnormal signals might be useful to constrain the planetary
obliquity and eccentricity.

\subsection{Rotational Light Curves on Brown Dwarfs and Directly Imaged Planets}
\label{sect:rotlight}

For directly imaged exoplanets and brown dwarfs, disk-integrated
photometric modulation {has} been observed and studied
for two decades, starting from
\cite{tinneySearchingWeatherBrown1999} shortly after the first
detected brown dwarfs (see \citealt{billerTimeDomainBrown2017} and
\citealt{artigauVariabilityBrownDwarfs2018} for a more detailed
review). Rotational light curves are not only useful for
constraining the rotational rates, but also the weather on those
worlds. Regarding the light curve behaviors, there are two aspects.
The first one is photometric light curves in emission caused by
self-rotation and spatial heterogeneity. The amplitude of the light
curves ranges from sub-percent to tens of percent on brown dwarfs
(e.g., \citealt{artigauPhotometricVariabilityT22009};
\citealt{metchevWeatherOtherWorlds2011};
\citealt{radiganLargeamplitudeVariationsTransition2012};
\citealt{apaiHSTSPECTRALMAPPINGTRANSITION2013};
\citealt{heinzeWeatherOtherWorlds2013};
\citealt{yangHSTRotationalSpectral2014};
\citealt{radiganINDEPENDENTANALYSISBROWN2014};
\citealt{metchevWeatherOtherWorlds2015};
\citealt{heinzeWeatherOtherWorlds2015};
\citealt{buenzliCloudStructureNearest2015};
\citealt{lewCloudAtlasDiscovery2016};
\citealt{yangExtrasolarStormsPressuredependent2016};
\citealt{miles-paezWeatherOtherWorlds2017};
\citealt{apaiZonesSpotsPlanetaryscale2017};
\citealt{manjavacasCloudAtlasDiscovery2017};
\citealt{zhouCloudAtlasRotational2018};
\citealt{lewCloudAtlasWeak2020}) and on directly imaged exoplanets
(e.g., \citealt{billerVariabilityYoungTransition2015};
\citealt{zhouDiscoveryRotationalModulations2016};
\citealt{lewCloudAtlasWeak2020, manjavacasCloudAtlasHubble2019,
manjavacasCloudAtlasRotational2019,
miles-paezCloudAtlasVariability2019, zhouCloudAtlasHighcontrast2019,
zhouCloudAtlasHighprecision2020}). The shapes of the light curves
are sometimes irregular rather than a simple sinusoidal shape, with
more than one peak in the curves within one rotation. The second one
is the temporal variability of the rotational light curves in both
amplitude and shape (e.g.,
\citealt{apaiHSTSPECTRALMAPPINGTRANSITION2013};
\citealt{yangExtrasolarStormsPressuredependent2016};
\citealt{apaiZonesSpotsPlanetaryscale2017}). There are both
short-term and long-term variabilities, associated with the weather
change in the photospheres. \cite{apaiZonesSpotsPlanetaryscale2017}
summarized three important puzzling behaviors in the temporal
variability: (1) single-peaked light curves splitting into
double-peaked (e.g.,
\citealt{radiganLargeamplitudeVariationsTransition2012};
\citealt{yangExtrasolarStormsPressuredependent2016}), (2) rapid
transitions from very low-amplitude ($<$0.5\%) to high-amplitude
($\sim$5\%) (\citealt{yangExtrasolarStormsPressuredependent2016}),
and (3) recurring features embedded in the irregularly evolving
light curves (e.g., \citealt{karalidiAeolusMarkovChain2015}).

Observed light curves on brown dwarfs depend on the spectral type
and the observed wavelength. It looks{ like} almost all L and T
dwarfs are variable with amplitudes larger than 0.2\%. Early surveys
with limited data sample could not conclude whether the fraction of
objects showing rotational variability is uniformly distributed
across the spectral type or not
(e.g.,\citealt{wilsonBrownDwarfAtmosphere2014};
\citealt{radiganSTRONGBRIGHTNESSVARIATIONS2014}). Later statistical
studies seem to support that brown dwarfs in the L/T transition
region tend to exhibit stronger variability and higher amplitude
than the objects outside the transition (e.g.,
\citealt{radiganINDEPENDENTANALYSISBROWN2014};
\citealt{erikssonDetectionNewStrongly2019}). L dwarfs with IR
variability larger than 2\% are generally limited within the red,
low-gravity objects (\citealt{metchevWeatherOtherWorlds2015}). The
wavelength dependence implies a pressure-dependent behavior in the
photosphere because different wavelengths probe at different layers.
The most noticeable phenomenon is the so-called ``phase offset"
between 0 and 180 degrees (e.g.,
\citealt{buenzliVerticalAtmosphericStructure2012a};
\citealt{radiganLargeamplitudeVariationsTransition2012};
\citealt{billerWeatherNearestBrown2013};
\citealt{yangExtrasolarStormsPressuredependent2016}). Here the
``phase" is loosely defined because the shapes of the light curves
are not always sinusoidal, especially in high-resolution
observations. The ``phase offset" phenomenon just states that the
peaks at different wavelengths occur at a different observational
time{s} in the same rotational period. Moreover, on some objects
the phase lags correlate with the probed pressure level at different
wavelength (\citealt{buenzliVerticalAtmosphericStructure2012a};
\citealt{yangExtrasolarStormsPressuredependent2016}). The dependence
of the rotational variability on the gravity or rotational period is
still an open question.

The mechanisms behind the observed behaviors of the light curves are
not completely understood. The rotational light curve itself is a
result of the inhomogeneous distributions of temperature and opacity
sources, i.e., the weather patterns in the photosphere rotating in
and out of the view to the observer (e.g.,
\citealt{apaiHSTSPECTRALMAPPINGTRANSITION2013};
\citealt{zhangAtmosphericCirculationBrown2014};
\citealt{crossfieldObservationsExoplanetAtmospheres2015a}). The
spatial distribution of the clouds is the leading hypothesis (e.g.,
\citealt{radiganLargeamplitudeVariationsTransition2012};
\citealt{apaiHSTSPECTRALMAPPINGTRANSITION2013,apaiZonesSpotsPlanetaryscale2017})
such as the ``patchy clouds scenario" with completely depleted cloud
holes (e.g., \citealt{ackermanPrecipitatingCondensationClouds2001};
\citealt{burgasserSpectraDwarfsNearinfrared2002};
\citealt{marleyPatchyCloudModel2010};
\citealt{morleySpectralVariabilityPatchy2014}) or the
``thin-and-thick cloud scenario" (e.g.,
\citealt{apaiHSTSPECTRALMAPPINGTRANSITION2013}). Dust-cloud break-up
was also proposed to cause the L/T transition (see
Sect.~\ref{sect:bdcloud} for discussion about clouds on brown
dwarfs). Alternatively, temperature could also vary with longitude
due to moist convection  (e.g.,
\citealt{zhangAtmosphericCirculationBrown2014};
\citealt{tanEffectsLatentHeating2017}), cloud radiative feedback
(\citealt{tanAtmosphericVariabilityDriven2019};
\citealt{tanAtmosphericCirculationBrown2020}), thermal wave
propagation (\citealt{robinsonTemperatureFluctuationsSource2014}),
trapped waves in the bands
(\citealt{apaiZonesSpotsPlanetaryscale2017}) and dynamical
modulations due to upward propagating waves
(\citealt{showmanAtmosphericCirculationBrown2019}). It is also
important to keep in mind that clouds and temperature are tightly
coupled together due to condensational processes, radiative
feedback and atmospheric circulation. The
{IR} opacity of the clouds could trap the
radiative flux from the bottom so that the top of the clouds
continues cooling off. As a result, more condensable species are
transported upward to form more clouds, leading to a positive
feedback, or so-called ``cloud radiative instability" (e.g.,
\citealt{gieraschRadiativeInstabilityCloudy1973};
\citealt{tanAtmosphericVariabilityDriven2019}). The intrinsic
timescales of the atmospheric processes such as convection and wave
propagation might control the temporal evolution timescales of the
light curves, but the dominant causes are still elusive. Dynamics on
these self-luminous bodies will be discussed in Section
\ref{sect:bddyn}.

The ``phase offset" in the multi-wavelength observations implies a
strong vertical variation in the horizontal distributions of
temperature and clouds. Spectroscopic measurements on brown dwarfs
should be able to provide more clues on the underlying mechanisms of
the rotational modulation and its time variability (e.g.,
\citealt{apaiHSTSPECTRALMAPPINGTRANSITION2013};
\citealt{morleySpectralVariabilityPatchy2014}). On the other hand,
cloud-free mechanisms have also been suggested to cause variability
at the L/T transition. \cite{tremblinCLOUDLESSATMOSPHERESDWARFS2016}
claimed that the brown dwarf variability could be a result of
surface heterogeneity of carbon monoxide or temperature due to
atmospheric waves although the details were not elucidated. If this
were true, gas (e.g., \ch{CH4}) absorption bands should exhibit
abnormal amplitudes in L/T transition objects. This seems not
consistent with recent observations (e.g.,
\citealt{buenzliCloudStructureNearest2015};
\citealt{billerTimeDomainBrown2017}).

How to test possible mechanisms underlying the rotational light
curves? It is hard to spatially resolve the weather patterns on
these distant objects except the very close ones, such as Luhman
16B, which has been mapped using the Doppler imaging technique
(\citealt{crossfieldGlobalCloudMap2014}), although the data are
still much noisier than the bright stellar counterparts. To date,
continuous monitoring of objects over multiple rotations is a
successful method to break degeneracies in surface brightness
distribution and time-evolution (e.g.,
\citealt{apaiMappingUltracoolAtmospheres2019}).
\cite{apaiZonesSpotsPlanetaryscale2017} compiled light curves of 32
rotations for six brown dwarfs from Spitzer, along with simultaneous
HST time-evolving spectra for some of the rotations. The analysis
showed that beating patterns of the planetary-scale waves---rather
than large bright spots---modulate the cloud thickness in the zonal
bands on L/T brown dwarfs and produce the rotational modulation
and light curve variability. Polarimetric observations
(e.g., \citealt{goldmanPolarisationVerylowmassStars2009};
\citealt{miles-paezOpticalNearinfraredLinear2017}) often provide new
insights on the oblateness of the body, cloud grain
properties and atmospheric banding, but it was difficult to
achieve sufficient sensitivity for brown dwarfs.  Recently,
\cite{millar-blanchaerDetectionPolarizationDue2020} successfully
detected polarized signals from Luhman 16AB. The degree of linear
polarization is about 300{\,}ppm for Luhman 16A and about
100{\,}ppm for 16B. The data imply cloud patchiness and
banded structures on 16A, but the interpretation for 16B is still
ambiguous.

On the other hand, giant planets in the Solar System might provide
clues because their rotational light curves can be understood
together with corresponding global maps. For example, Jupiter and
Neptune exhibit strong rotational modulations (e.g.,
\citealt{gelinoVariabilityUnresolvedJupiter2000};
\citealt{karalidiAeolusMarkovChain2015};
\citealt{simonNeptuneDynamicAtmosphere2016};
\citealt{staufferSpitzerSpaceTelescope2016};
\citealt{geRotationalLightCurves2019}). The photometric variability
also depends on the wavelength, with the amplitude ranging from less
than one percent to tens of percent. With sufficient data, the
atmospheric patterns and jet speed can be derived from rotational
light curves (e.g., \citealt{karalidiAeolusMarkovChain2015};
\citealt{simonNeptuneDynamicAtmosphere2016}). Most of these studies
focused on reflective lights. Based on high-resolution spatial maps
of Jupiter from UV to mid-IR (e.g.,
\citealt{simonFirstResultsHubble2015};
\citealt{fletcherMidinfraredMappingJupiter2016}),
\cite{geRotationalLightCurves2019} conducted a comprehensive study
on Jupiter including both reflective light curves (UV and visible)
and emission light curves (mid-IR). The peak-to-peak amplitudes of
Jupiter's light curves range from less than 1\% up to 4\% at most
wavelengths, but the amplitude exceeds 20\% at 5 microns. The
rotational modulations originate mainly in the cloudless belts
instead of the cloudy zones. Important discrete patterns responsible
for the rotational modulation include the Great Red Spot (GRS),
expansions of the North Equatorial Belt (NEB), patchy clouds in the
North Temperate Belt (NTB) and a train of hot spots in the
NEB. The temporal variation of the light curves is caused by
periodic events in the belts and longitudinal drift of the GRS and
patchy clouds in the NTB.

The thermal emission light curves on Jupiter shed light on brown
dwarfs and directly imaged planets. There are two mechanisms found
for modulating Jupiter's light curves. For small rotational
variability (i.e., 1\% level in the mid-IR), the surface
inhomogeneity is induced by the spatial distribution of temperature
and opacities of gas and aerosols. On the other hand, the vertical
distribution of clouds is important for the 20\% variation at 5
microns. At this wavelength, the large photometric modulation is
induced by holes in the upper clouds at wavelengths of atmospheric
windows where the gas has little opacity. Note that all giant
planets in the Solar System are zonally banded. Whether this is true
for brown dwarfs and directly imaged exoplanets is still uncertain
although some recent observations have shed light on it (e.g.,
\citealt{apaiZonesSpotsPlanetaryscale2017};
\citealt{millar-blanchaerDetectionPolarizationDue2020}). See Section
\ref{sect:bddyn} for discussion.

Most information obtained from the rotational light curves is the
surface inhomogeneity across longitude. However, the latitudinal
properties of brown dwarfs and directly imaged planets can also be
inferred, in the statistical sense, if we know their inclination
angles to the observer. The inclination angles of the brown dwarfs
with known rotation periods can be derived from the line-of-sight
rotational velocities (e.g.,
\citealt{radiganSTRONGBRIGHTNESSVARIATIONS2014};
\citealt{vosViewingGeometryBrown2017}). Both the light curve
amplitude and {$J-K$} color anomaly (i.e., after
subtraction of the mean color in the same spectral type) correlate
with the inclination. The viewing geometry might mostly explain the
former because a higher inclined (pole-on) object should exhibit
smaller rotational modulation since fewer features can rotate in and
out of view. Also, the analysis of Jupiter implies that Jupiter has
larger rotational modulation at lower latitudes (both at NEB and
SEB) than the higher latitudes. If the brown dwarfs are also banded
like Jupiter, this will contribute to the amplitude-inclination
trend observed by \cite{vosViewingGeometryBrown2017}.

The latter behavior ({$J-K$} anomaly vs.
inclination)---an equator-on object tends to be redder---is
interesting. Why does an object tend to be redder at lower
latitudes? If the temperature is not systematically lower in the
equatorial region, it is perhaps due to more clouds or larger cloud
particle sizes forming at lower latitudes (e.g.,
\citealt{kirkpatrickDiscoveriesNearinfraredProper2010};
\citealt{vosViewingGeometryBrown2017}). However, the physical
mechanism for this cloud distribution has yet to be explored in
detail. One possibility is surface gravity at the equator
is smaller than the polar region on a fast-rotating, oblate object
with a larger equatorial radius than the poles. Thus the cloud might
extend to lower pressure levels (also depends on the mixing).
Alternatively, if we again use Jupiter as an analog, Jupiter shows
more ammonia clouds at lower latitudes (e.g., the equatorial zone
and the south tropical zone).  The reason appears to be related to
the global circulation pattern in the deep atmosphere of Jupiter
recently revealed by the Juno mission (e.g.,
\citealt{boltonJupiterInteriorDeep2017}). The microwave observations
(\citealt{liDistributionAmmoniaJupiter2017}), as well as the VLT
radio observations (\citealt{depaterJupiterAmmoniaDistribution2019},
also see \citealt{showmanDynamicalImplicationsJupiter2005}), found
that ammonia gas is enriched in the equatorial region where thick
ammonia clouds form, while it is largely depleted in the
off-equatorial region{,} {t}he mechanism of
which is still unknown. If this mechanism (for a different kind of
cloud) also occurs on those early-L and early-T brown dwarfs
in \cite{vosViewingGeometryBrown2017}, it could also explain the
observed reddening at lower latitudes. We will discuss more global
dynamics on brown dwarfs and directly imaged exoplanets in Section
\ref{sect:bddyn}, where we will see simulations on rapid{ly
}rotating brown dwarfs that indeed show lower emission at the
equator (Fig.~\ref{figbd}).

\section{Atmospheric Composition}
\label{sect:chem}
\subsection{Fundamentals}
\label{sect:chemfund}

{A} planetary atmosphere is mostly composed of gas
molecules and atoms. Suspended solids and liquid particles,
so-called aerosols, are also ubiquitous. In the hot upper atmosphere
where ionization processes play a role, the plasma phase
is a significant fraction. In principle, atmospheric compositions
are determined by the accreting volatile materials during planetary
formation and by subsequently atmospheric evolution. The primordial
atmosphere (proto-atmosphere) is formed through the accretion of gas
and dust from the forming disk environment and should be mostly
composed of hydrogen and helium with minor constituents depending on
the location and composition of the formation feeding zone.
Significant amount{s} of ices {in} various
forms can be accreted onto planets residing outside the snow lines.
The subsequent evolution track is very different between
small and large planets---how to define ``large" and ``small" is not
very clear, perhaps related to the radius gap discussed in
Section~\ref{sect:radgap}. In general, large planets (such as gas
giants, ice giants and possibly sub-Neptunes) retain
the{ir} primordial hydrogen and helium envelopes, whereas
planets with the size of Earth or Mars lost their proto-atmospheres.
The secondary atmospheres on small planets might be formed
and evolve through either various ingassing and
degassing processes after the loss of primordial atmospheres
(\citealt{catlingAtmosphericEvolutionInhabited2017}). Examples of
degassing processes are outgassing from the magma ocean, volcanic
eruptions on terrestrial planets and core erosion on giant
planets. Examples of ingassing processes include surface weathering,
subduction during plate tectonics on terrestrial planets, and helium
rains on giant planets (e.g.,
\citealt{stevensonDynamicsHeliumDistribution1977}). Exchange of
chemical compositions between atmospheres and planetary interiors
implies that the secondary atmospheric compositions are closely
related to the redox state of planetary interiors. There are also tertiary processes to exchange the atmospheric
composition with the space environment, such as atmospheric escape,
stellar wind injection, asteroid{ and} comet impacts,
and late disk accretion in long-lived disks (e.g.,
\citealt{kralFormationSecondaryAtmospheres2020}).

\begin{figure*}
   \centering
   \includegraphics[width=0.85\textwidth, angle=0]{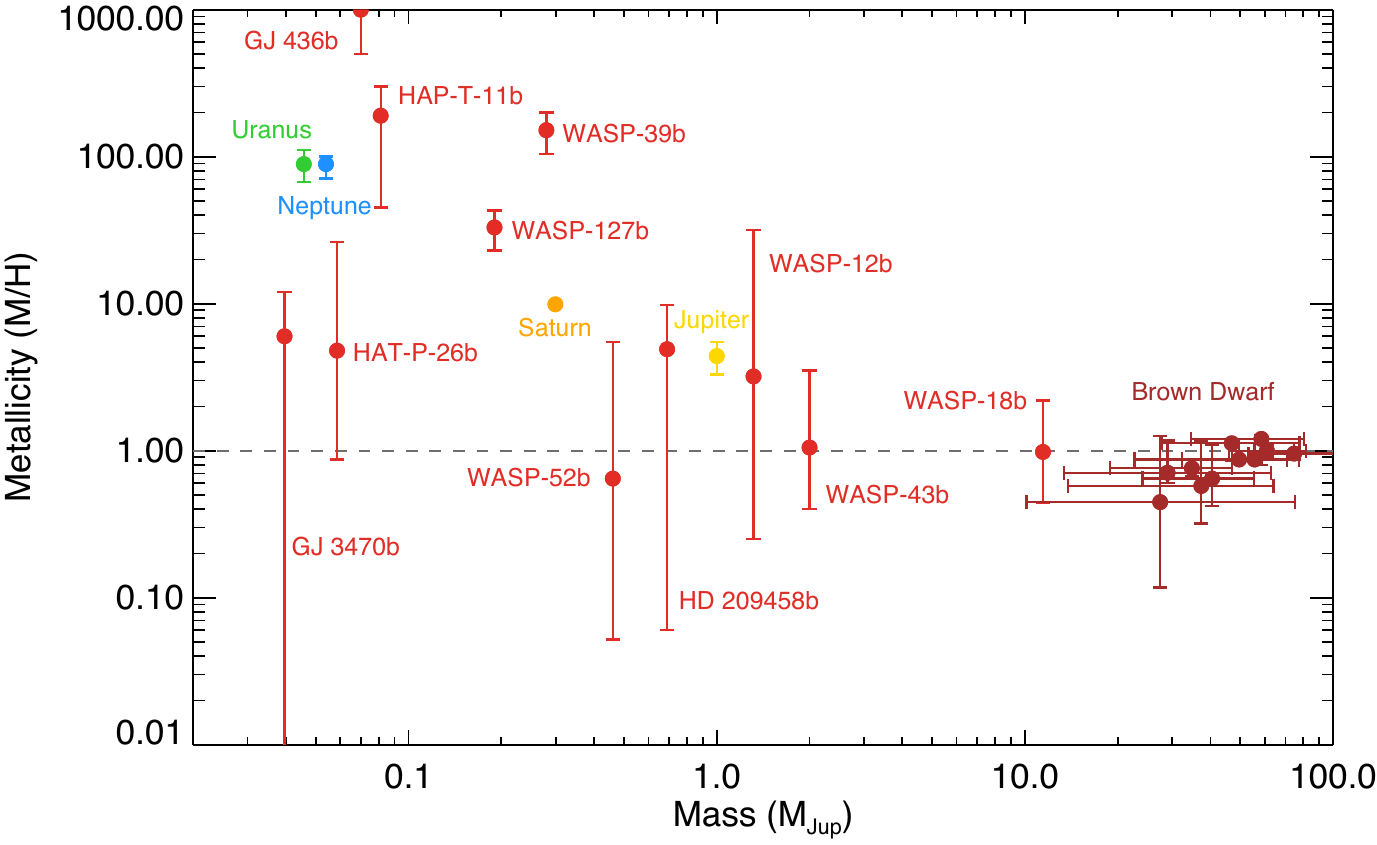}
   \caption{\baselineskip 3.8mm Derived metallicities of four giant planets in the Solar System, exoplanets ({\it red}) and some brown dwarfs ({\it black}). Exoplanet data are obtained from compilation in \cite{wakefordHATP26bNeptunemassExoplanet2017} and updates from \cite{morleyFORWARDINVERSEMODELING2017},  \cite{brunoWASP52bEffectStarspot2019},  \cite{bennekeSubNeptuneExoplanetLowmetallicity2019}, \cite{chachanHubblePanCETStudy2019} and \cite{spakeSupersolarMetallicityAtmosphere2019}. Brown dwarf data are from \cite{lineUniformAtmosphericRetrieval2017}.}
   \label{figmet}
   \end{figure*}

Atmospheric compositions provide clues on planetary formation and
evolution processes. Three categories seem within observational
reach. The first one is metallicity---the relative ratios of heavy
elements to hydrogen. The second one is the ratio of carbon and
oxygen. The third one is the ratio between the refractory materials
(rocks and real metals) and the volatiles (e.g., ices). This ratio
determines the planet type. The ratio of carbon to oxygen is
particularly important for atmospheric chemistry as it affects the
redox of the planets. In general, the elemental ratios change with
distance in the protoplanetary disk, suggesting the bulk composition
of the planets should also change if they were born at different
locations in the disk. The C/O ratio might be inferred from atmospheric measurements, but that on gas giants in the Solar
System is highly uncertain because the main oxygen-bearing species,
\ch{H2O}, condenses as clouds, and \ch{CH4} condenses on Uranus and
Neptune (\citealt{atreyaDeepAtmosphereComposition2020}).

On the other hand, for high-temperature exoplanets, carbon
and oxygen abundances in the photosphere can be retrieved directly
from the spectra. However, to get the bulk composition in the entire
planet, one still needs to assume that the elements are well mixed
in the planetary interiors and atmospheres. This assumption could be
violated due to interactions among transport, chemistry and
phase change processes. Noble gases are typically important among
all trace species because of their chemically inert and
non-condensable nature. However, directly detecting the abundance
of{ a} noble gas except helium is very difficult without an
entry probe. Helium, on the other hand, has been detected from the
Helium 10830 \r{A} triplet lines in possibly evaporating atmospheres
on several exoplanets (see Sect.~\ref{sect:stabilityfund}). There is
also an issue with possible depletion of helium due to helium rain
that could also dissolve neon
(\citealt{niemannGalileoProbeMass1996}).

From transmission and emission spectroscopy on both transit and
directly imaged planets, many species have been detected in
substellar atmospheres, including several atomic species in the UV
such as H, He, C, O, K, Na, Si, Mg, Ti, Fe, Ca, Li, and metallic
oxides TiO, VO and AlO, as well as \ch{H2O}, CO, \ch{CH4},
\ch{NH3} and HCN detected in the {IR}.
Check Table 1 in \cite{madhusudhanExoplanetaryAtmospheresKey2019}
for the latest summary and references therein. On high
{S/N} spectra of brown dwarfs, more
metallic species have been detected, such as Rb, Cs, as well as the
hydrides MgH, CaH, CrH and FeH
(\citealt{kirkpatrickNewSpectralTypes2005}). The currently inferred
metallicities of substellar atmospheres from the observed abundances
of photospheric \ch{H2O}, CO and \ch{CH4} seem to show a
decreasing trend with increasing planetary mass (Fig.~\ref{figmet}).
Also see \cite{welbanksMassMetallicityTrends2019} for the individual
mass-metallicity relations derived from each species such as
\ch{H2O}, \ch{CH4}, \ch{Na} and \ch{K}. Despite large uncertainties
in the data, the trend in Figure~\ref{figmet} seems consistent with
four giant planets in the Solar System. The retrieved metallicities
of several brown dwarfs are roughly consistent with the solar value
(\citealt{lineUniformAtmosphericRetrieval2017}). This trend probably
implies that smaller planets in general accreted less hydrogen and
helium fraction during their formation across different disk
environment{s}.

Isotopic compositions are particularly useful in understanding the
evolution of the atmosphere. Enhancement of the deuterium to
hydrogen (D/H) ratio on Uranus and Neptune relative to Jupiter and
Saturn by a factor of 2-3 indicates the icy giants accreted more
deuterium-rich icy blocks in the protoplanetary disk
(\citealt{hartoghOceanlikeWaterJupiterfamily2011};
\citealt{atreyaDeepAtmosphereComposition2020}).    The exceedingly
large D/H ratio in Venus{'} atmosphere is evidence of past
atmospheric escape
(\citealt{donahueVenusWasWet1982,mcelroyEscapeHydrogenVenus1982}).
Exchange processes between the interior/surface and the atmosphere
can be inferred from the isotopic signatures of helium, argon,
carbon, oxygen, sulfur and so on. Detecting isotopes on
exoplanet atmospheres is still difficult using current facilities,
but it will be possible to infer the D/H ratio from \ch{CH3D} or HDO
in the mid-{IR} thermal emission spectra
(\citealt{morleyMeasuringRatiosExoplanets2019}). To compare isotopic
abundances between the planets and their formation environment, one
must also understand the atmospheric isotopic compositions on stars
(e.g., \citealt{crossfieldUnusualIsotopicAbundances2019}).

\begin{figure*}
   \centering
   \includegraphics[width=0.8\textwidth, angle=0]{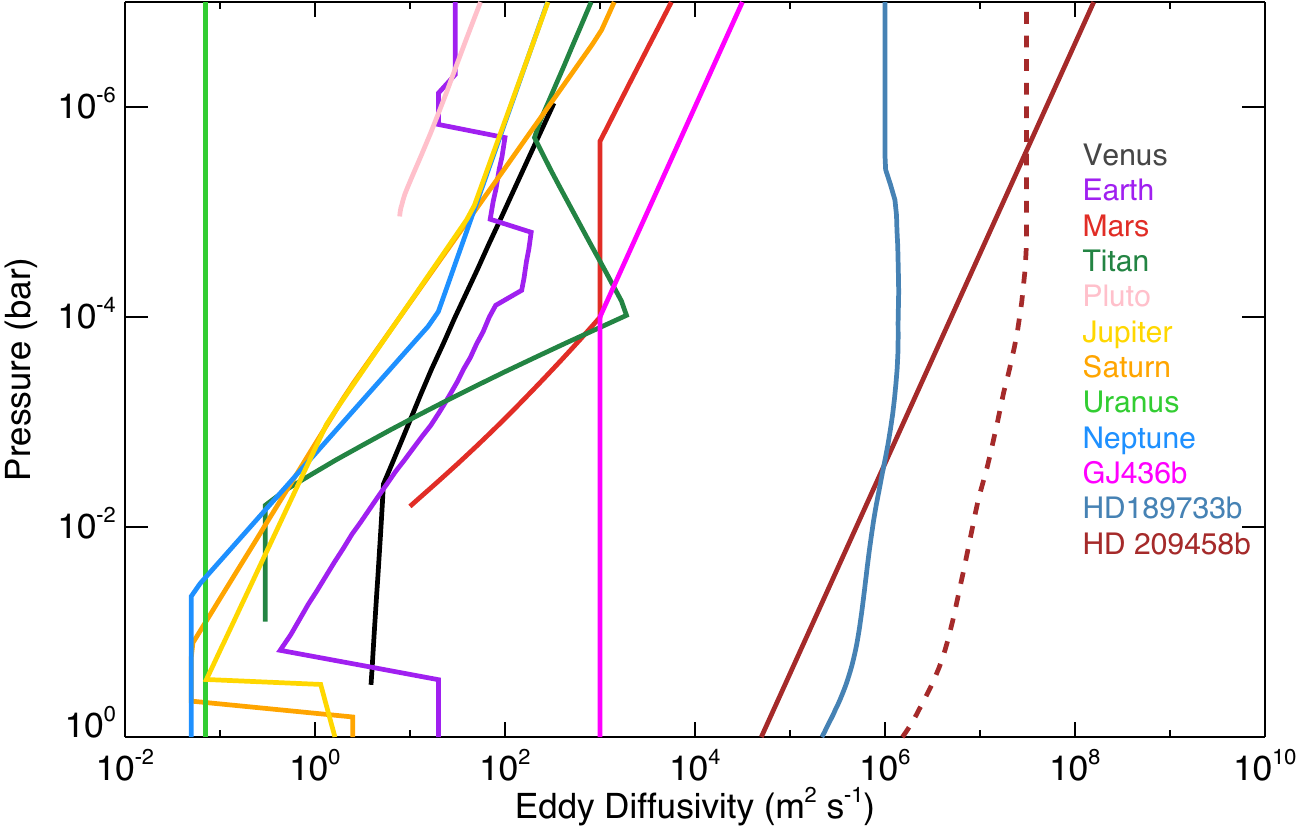}
   \caption{\baselineskip 3.8mm Vertical profiles of eddy diffusivity ($K_{zz}$) in typical 1D chemical models of planets in and out of the Solar System from \cite{zhangGlobalmeanVerticalTracer2018a}. Data sources: \cite{zhangSulfurChemistryMiddle2012} for Venus, \cite{allenVerticalTransportPhotochemistry1981} for Earth, \cite{nairPhotochemicalModelMartian1994} for Mars, \cite{liNonmonotonicEddyDiffusivity2014, liVerticalDistributionC3hydrocarbons2015} and  for Titan, \cite{wongPhotochemistryPlutoAtmosphere2017} for Pluto, \cite{mosesPhotochemistryDiffusionJupiter2005} for Jupiter, Saturn, Uranus and Neptune, \cite{mosesCompositionalDiversityAtmospheres2013} for GJ 436 b and \cite{mosesDisequilibriumCarbonOxygen2011} for HD 189733 b and HD 209458 b. For HD 209458 b, we {display} eddy diffusivity profiles assumed in a gas chemistry model ({\it dashed}, \citealt{mosesDisequilibriumCarbonOxygen2011}) and that derived from a 3D particulate tracer transport model ({\it solid}, \citealt{parmentier3DMixingHot2013}). For some brown dwarfs cooler than 750\,K, \cite{milesObservationsDisequilibriumCO2020} derived the $K_{zz}$ ranging from $1-10^4{\,}\mathrm{m^2{\,}s^{-1}}${ }in the deep atmosphere (below{ the} 1 bar level, not shown here).}
   \label{figkzz}
   \end{figure*}

Given the elemental abundances, atmospheric abundances are
controlled by temperature, chemistry and transport processes.
{E}quilibrium chemistry drives the atmosphere towards
thermodynamical equilibrium, given a sufficiently long time.
Disequilibrium processes---including photochemistry, ion chemistry,
biochemistry (for life-bearing planets) and phase
change---force the atmosphere out of thermodynamic
equilibrium. Atmospheric tracer transport by winds, waves and
turbulences also results in chemical disequilibrium. In
{one-dimensional (}1D{)} chemical models, vertical
transport is conventionally approximately by{ a} diffusion
process (\citealt{andrewsMiddleAtmosphereDynamics1987};
\citealt{yungPhotochemistryPlanetaryAtmospheres1998}), the strength
of which is characterized by vertical eddy diffusivity
($K_{zz}$). The chemical transport timescale is $\tau_{\rm
trans}=H^2/K_{zz}$, where $H$ is usually taken as the pressure scale
height. Figure \ref{figkzz} illustrates several $K_{zz}$ profiles
that were empirically determined for Solar System planets,
as well as the theoretical predictions from models of some typical
exoplanets. There seems no obvious trend for the $K_{zz}$ within
Solar System planets, but hotter exoplanets may have
larger diffusivities than colder Solar System planets.

The diffusive approximation generally works well for 1D models but
has some caveats because the physical underpinning of
$K_{zz}$ is elusive. There are approximately three regimes from the
bottom of the atmosphere to the top. In a bottom convective
atmosphere, vertical transport probably is well-approximated by eddy
diffusion according to the traditional Prandtl mixing length theory
(e.g., \citealt{prandtlBerichtUberUntersuchungen1925};
\citealt{gieraschStudyThermalDynamical1968};
\citealt{gieraschEnergyConversionProcesses1985};
\citealt{smithEstimationLengthScale1998};
\citealt{ackermanPrecipitatingCondensationClouds2001};
\citealt{bordwellConvectiveDynamicsDisequilibrium2018}). $K_{zz}$ is approximate{ly} a product of a
convective velocity and a typical vertical length scale. This scaling applies to
convective atmospheres on directly imaged planets and brown dwarfs,
as well as the deep convective part on close-in irradiated
exoplanets. The convective velocity scaling will be discussed in
Section \ref{sect:bddyn}. Basically, in the slow rotation regime
(e.g., \citealt{claytonPrinciplesStellarEvolution1968};
\citealt{stevensonTurbulentThermalConvection1979};
\citealt{showmanAtmosphericDynamicsBrown2013}), flows tend to be
radially isotropic and mixing length theory predicts:

\begin{equation}\label{ckzz}
K_{zz, {\rm slow}}\sim(\frac{\alpha g F l^4}{\rho c_p})^{\frac{1}{3}},
\end{equation}
where $\alpha$ is the thermal expansivity and equals $1/T$
for isobaric expansion, $g$ is gravity, $F$ is the convective heat
flux (internal heat flux), $l$ is the mixing length, $\rho$ is the
air density and $c_p$ is the specific heat at constant pressure. The
vertical length scale $l$ is usually assumed to be the pressure
scale height $H$ but it also changes with the chemical
timescale---the shorter-lived species have smaller length scales
(e.g., \citealt{smithEstimationLengthScale1998};
\citealt{bordwellConvectiveDynamicsDisequilibrium2018}). The $K_{zz,
slow}$ can also be applied to the equatorial region on rapidly
rotating planets because the rotational effect (Coriolis effect) is
not important at the equator (\citealt{wangNewInsightsJupiter2015}).

In the rapid rotation regime (e.g.,
\citealt{golitsynConvectionStructureFast1981,golitsynGeostrophicConvection1980};
\citealt{boubnovExperimentalStudyConvective1986,boubnovTemperatureVelocityField1990};
\citealt{fernandoEffectsRotationConvective1991};
\citealt{showmanScalingLawsConvection2011};
\citealt{showmanAtmosphericDynamicsBrown2013}), large-scale flows
tend to align along columns parallel to the rotation axis
(\citealt{houghIXApplicationHarmonic1897};
\citealt{proudmanMotionSolidsLiquid1916};
\citealt{taylorMotionSolidsFluids1917}). Both the velocity scaling
and vertical length scale are different from
{those} in the slow{ly} rotating regime. If we
take a characteristic length scale as $l=w/\Omega$ where $w$ is the
vertical velocity, the $K_{zz}$ scaling is (e.g.,
\citealt{wangNewInsightsJupiter2015})
\begin{equation}
K_{zz, {\rm rapid}}\sim\frac{\alpha g F}{\rho c_p \Omega^2}
\end{equation}
where $\Omega$ is the rotational rate.

In the low-density upper atmosphere (e.g., Earth's mesosphere),
waves such as gravity waves generated from the lower atmosphere
propagate vertically and break, leading to strong vertical mixing of
the chemical tracers. \cite{lindzenTurbulenceStressOwing1981} first
parameterized $K_{zz}$  from the turbulence and stress
from those breaking gravity and tidal waves (also see
\citealt{schoeberlNonzonalGravityWave1984,strobelParameterizationLinearWave1981,strobelVerticalConstituentTransport1987}).
For energy-conserved waves, wave amplitude increases as density
drops, suggesting that $K_{zz}$ decreases with pressure in a fashion
of $\rho^{-1/2}$. For an isothermal atmosphere, approximately
$K_{zz}\propto p^{-1/2}$.

The behavior in the middle part---the stably stratified
atmosphere---is also complicated. For example, in Earth's
stratosphere, tracer transport is controlled by both large-scale
overturning circulation and vertical wave mixing
(\citealt{huntenVerticalTransportAtmospheres1975};
\citealt{holtonDynamicallyBasedTransport1986}). The eddy diffusivity
should be considered to be an effective parameter for global-mean
tracer transport. The magnitude of $K_{zz}$ depends on
many other factors and may differ from planet to planet. Just like
the eddy diffusivity in the convective medium depends on the
chemical tracer itself (e.g.,
\citealt{smithEstimationLengthScale1998};
\citealt{bordwellConvectiveDynamicsDisequilibrium2018}), that in
{a} stratified atmosphere has similar behavior. Several
studies (e.g., \citealt{holtonDynamicallyBasedTransport1986};
\citealt{parmentier3DMixingHot2013};
\citealt{zhangGlobalmeanVerticalTracer2018a,zhangGlobalmeanVerticalTracer2018};
\citealt{komacekVerticalTracerMixing2019}) found that the
parameterized $K_{zz}$ depends not only on atmospheric dynamics but
also the tracer itself such as the tracer chemistry and trace
distributions. Also, in stratified atmospheres on hot
planets, turbulent vertical transport driven by vertical shear
instabilities could contribute to the vertical mixing (e.g.,
\citealt{fromangSheardrivenInstabilitiesShocks2016};
\citealt{menouTurbulentVerticalMixing2019}).
\cite{menouTurbulentVerticalMixing2019} derived the eddy mixing
coefficients in double-diffusive shear instabilities using the
secular Richardson number and turbulent viscosity. The derived
$K_{zz}$ is inversely proportional to pressure square{d} (i.e.,
$K_{zz}\propto p^{-2}$).

Although{ the} 1D model is still useful for a first-order
global-mean situation (e.g.,
\citealt{yungPhotochemistryPlanetaryAtmospheres1998}), several
efforts have been put forward {a}t simulating the tracer
distributions in 3D dynamical models with simplified chemical
schemes (e.g.,
\citealt{drummondEffectMetallicityAtmospheres2018,drummondEffectsConsistentChemical2016,drummondImplicationsThreedimensionalChemical2020};
\citealt{steinrueckEffect3DTransportinduced2019}). Based on the
tracer distributions in 3D models, one could also derive 1D
equivalent eddy diffusion coefficients (e.g.,
\citealt{parmentier3DMixingHot2013};
\citealt{charnay3DModelingGJ1214b2015a}).
\cite{zhangGlobalmeanVerticalTracer2018a,zhangGlobalmeanVerticalTracer2018}
specifically investigated the regimes of global-mean vertical tracer
mixing in stratified planetary atmospheres. They found that $K_{zz}$
strongly depends on the large-scale circulation strength, horizontal
mixing due to eddies and waves, and local tracer sources and sinks
due to chemistry and microphysics. The first regime is for a
short-lived tracer with chemical equilibrium abundance uniformly
distributed across the globe, and global-mean vertical tracer mixing
behaves diffusively. Unlike the traditional assumption, different
chemical species in a single atmosphere should, in principle, have
different eddy diffusion profiles. The second regime is for a
short-lived tracer with a non-uniform distribution of the chemical
equilibrium abundance. A significant non-diffusive component in this
regime might lead to a negative $K_{zz}$ under the diffusive
assumption. In the third regime where the tracer is long-lived with
the tracer material surface significantly controlled by dynamics,
global-mean vertical tracer transport is also largely influenced by
non-diffusive effects.

\cite{zhangGlobalmeanVerticalTracer2018a,zhangGlobalmeanVerticalTracer2018}
derived an analytical solution of $K_{zz}$ and validate that
against 2D and 3D global-mean vertical mixing properties over a wide
parameter space. For stably stratified atmospheres on tidally locked
exoplanets, if chemical equilibrium abundance{ is} uniformly
distributed, the analytical solution of $K_{zz}$ can be approximated
using the continuity Equation~(\ref{tscal}c) and
Equation~(\ref{tscalsol}b)
\begin{equation}\label{kzz}
K_{zz, {\rm strat}}\sim\frac{(RT_{\rm eq})^{5/2}}{g^2R_p}(\sqrt{1+\alpha_2^2}-\alpha_2)(1+\zeta)^{-1},
\end{equation}
where $\alpha_2$ is given in Equation (\ref{alphadef}b) and $\zeta$
is the ratio of the vertical transport timescale $H/w$ to the
chemical timescale $\mathrm{\tau_{chem}}$
\begin{equation}
\zeta=\frac{H}{w\mathrm{\tau_{chem}}}\sim\frac{R_p}{(RT_{\rm eq})^{1/2}(\sqrt{1+\alpha^2}-\alpha)\mathrm{\tau_{chem}}}.
\end{equation}
Also see another derivation in
\cite{komacekVerticalTracerMixing2019}. It can be shown that the
effective 1D eddy diffusivity given by Equation~(\ref{kzz}) is
smaller for a shorter-lived species and increases with the chemical
timescale. The asymptotic value in the very long-lived limit is
$K_{zz}\sim Hw$, which is the traditionally adopted value. For a
chemically inert tracer, this scaling predicts about
$10^3{\,}\mathrm{m^2{\,}s^{-1}}$ for a tidally
locked planet with $T_{\rm eq}\sim$300\,K and about
$10^6{\,}\mathrm{m^2{\,}s^{-1}}$ for $T_{\rm
eq}\sim$1000\,K. This is also more or less consistent with the
values in Figure~\ref{figkzz}.

The interplay among transport and chemical processes leads to three
chemical regimes in the atmosphere from the bottom to the top. The
reaction rate in equilibrium chemistry highly depends on
temperature because thermal energy is needed to overcome
the activation barrier of both forward and backward reactions. In
the deep atmosphere where the temperature is high, the reactions are
generally so fast that the atmosphere is typically assumed to be in
thermochemical equilibrium. In it, the Gibbs free energy (including
chemical potential) reaches {its} minimum at a given
temperature. The reaction rates drop as temperature decreases with
altitude. If the chemical reaction is not as efficient as the
transport, the tracer distribution is dynamically
``quenched{,}'' meaning that the atmospheric dynamics
homogenize the molar fractions of the species above the quenching
point. The quench point (e.g.,
\citealt{smithEstimationLengthScale1998}) is approximately where the
transport timescale (e.g., $H^2/K_{zz}$) equals the
timescale of the rate-limiting step in the thermochemical pathways.
The middle atmosphere is in the photochemistry-dominate{d}
regime due to incoming {UV} photons from
central stars or{ the} interstellar medium. Photochemistry is
efficient because high-energy photons break the chemical
bonds and produce meta-stable radicals that provide sufficient
energy to overcome the activation barrier and speed up the neural
chemical reactions. In the top layers of the atmosphere, such as the
thermosphere and ionosphere, electrons and chemical ions play
dominant roles in the chemistry. Tracer transport due to the
electromagnetic field in the plasma environment also operates
differently from the underlying neural atmosphere. Phase change,
such as cloud formation or photochemical haze formation, would
further complicate the chemical process.

Lastly, for tidally locked planets, the large dayside and nightside
temperature difference would imply very different chemistry and
cloud compositions (e.g.,
\citealt{parmentierTransitionsCloudComposition2016};
\citealt{venotGlobalChemistryThermal2020};
\citealt{powellFormationSilicateTitanium2018,powellTransitSignaturesInhomogeneous2019}),
but the horizontal transport would try to homogenize, or even
quench{,} the tracer distributions in the horizontal direction.
Chemical-transport models in 2D (e.g.,
\citealt{agundezPseudo2DChemical2014};
\citealt{venotGlobalChemistryThermal2020}) and 3D (e.g.,
\citealt{cooperDynamicsDisequilibriumCarbon2006};
\citealt{parmentier3DMixingHot2013};
\citealt{linesSimulatingCloudyAtmospheres2018};
\citealt{drummondEffectMetallicityAtmospheres2018,drummondEffectsConsistentChemical2016,drummondImplicationsThreedimensionalChemical2020};
\citealt{steinrueckEffect3DTransportinduced2019}) have shed light on
those behaviors but remain to be confirmed by observations in{
the} future. Non-uniformly distributed chemical tracers, if they are
radiatively active, would impact the transmission and emission
spectra, transit light curves,{ and} thermal phase curves on
close-in exoplanets (e.g.,
\citealt{venotGlobalChemistryThermal2020}) but we did not
discuss{ it} in detail in this review. In the
following{,} we will first talk about the gas chemical species
and the atmospheric compositional diversity in Section
\ref{sect:gaschem} and focus on clouds and hazes in Section
\ref{sect:cloud}.

\subsection{Gaseous Compositional Diversity}
\label{sect:gaschem}

\begin{figure*}
   \centering
   \includegraphics[width=0.8\textwidth, angle=0]{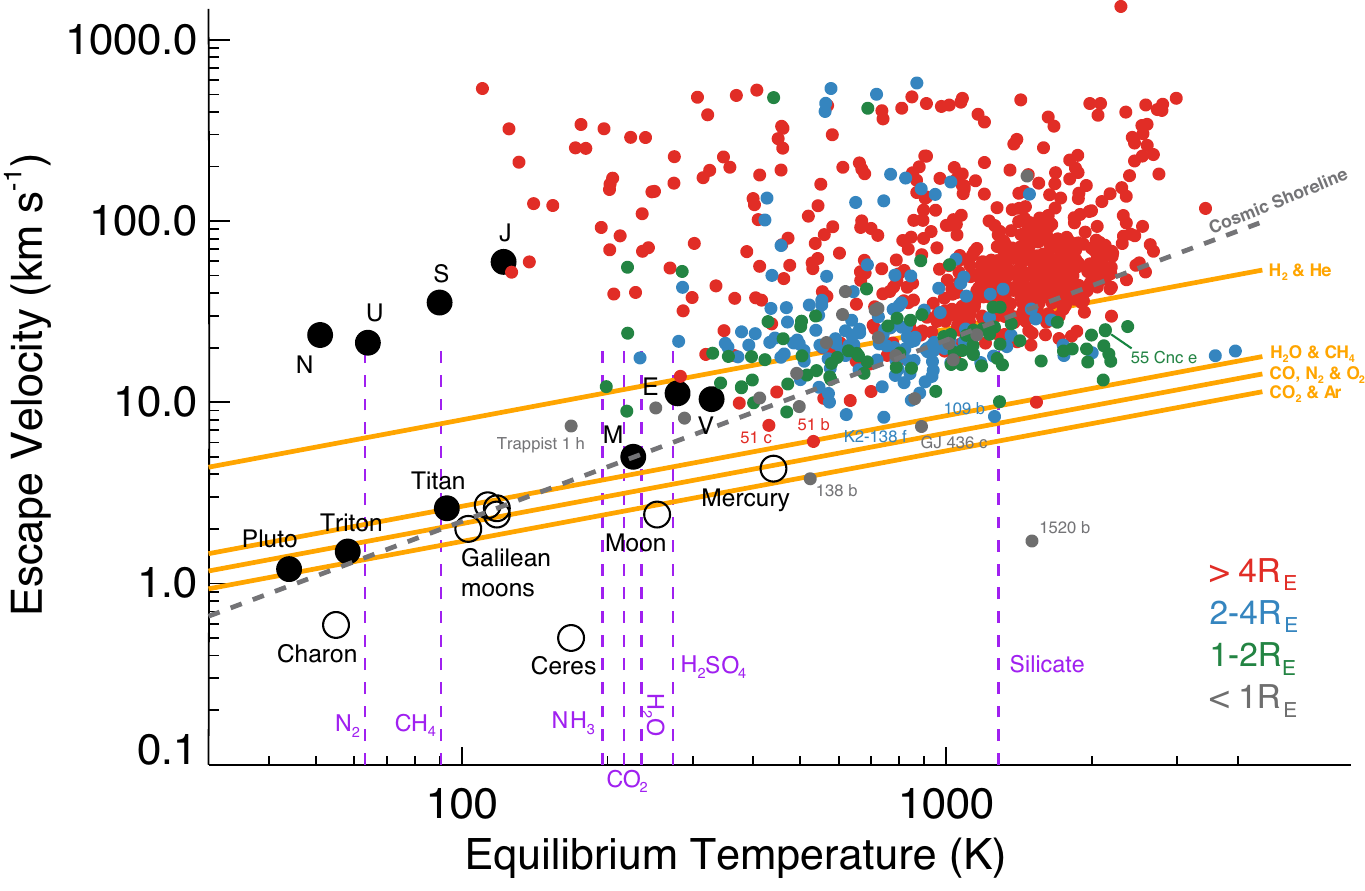}
   \caption{\baselineskip 3.8mm Compositional diversity map of substellar atmospheres as a function of equilibrium temperature and escape velocity. Hypothetic{al} regime boundaries ({\it orange}) for different compositions {are }shifted based on their approximate molecular weight differences. The {{\it purple}} lines {signify} rough condensational temperatures for{ a} variety of species. Cosmic shoreline ({\it black dashed}{ line}) is from \cite{zahnleCosmicShorelineEvidence2017}. The presence or absence of an atmosphere on Solar System objects is indicated by filled or open symbols, respectively. The extrasolar planets are color-coded in terms of their size range. Some typical planets with small escape velocities are highlighted, such as the ``Super-puffs" Kepler 51 b and 51 c{,} and Trappist 1 h.}
   \label{figgchem}
   \end{figure*}

In this section, we first discuss the bulk compositions and then
talk about the minor species in the atmospheres, as well as the
important controlling factors. Even though recent studies show that
the overall ratios of C/O and
magnesium-to-silicon (Mg/Si) in solar-metallicity stars are not very
compositionally diverse
(\citealt{bedellChemicalHomogeneitySunlike2018}), the ratios in the
protoplanetary disks significantly change with the radial distance
due to the ice lines of condensable species such as water, carbon
monoxide and carbon dioxide
(\citealt{madhusudhanChemicalConstraintsHot2014}). The formation
environment of the planets and their subsequent migration, as well
as the associated atmospheric formation and evolution processes such
as accretion, outgassing, impact, condensation and escape, could
lead to a wide range of elemental ratios and metallicities in the
atmospheres (e.g., \citealt{elkins-tantonRangesAtmosphericMass2008};
\citealt{schaeferChemistryAtmospheresFormed2010,schaeferVaporizationEarthApplication2012};
\citealt{lupuAtmospheresEarthlikePlanets2014}). To first
order, we can simply categorize planetary atmospheres into several
regimes in terms of their bulk compositions across the entire
parameter space of planetary mass, temperature,
metallicity and elemental ratios. The currently confirmed
exoplanets with estimated mass{es}, radii and equilibrium
temperature{s}, as well as large Solar System bodies, are
{displayed} in Figure \ref{figgchem}. Planets within
different size ranges are color-coded. Here we crudely
summarize them in terms of escape velocity, equilibrium
temperature, and ratios of hydrogen/carbon/oxygen (H/C/O) and
highlighted several important aspects related to the bulk
compositional diversity.

We first consider the condensation and evaporation processes in
which temperature plays a key role. The bulk compositions are
normally simple chemical compounds made of hydrogen, carbon, oxygen,
nitrogen and sulfur. The condensational temperatures for
those compounds are usually low due to their weak intermolecular
bonds or hydrogen bonds. In Figure~\ref{figgchem}, we roughly
mark their condensational temperatures (triple point
temperatures) {with} vertical dashed lines. \ch{N2} and
\ch{CH4} condense below 100\,K. \ch{NH3}, \ch{CO2},
\ch{H2O} and \ch{H2SO4} condens{e} at around
200--300\,K.
If planets are colder than their condensational temperatures,
corresponding compounds will be primarily locked in the ice or
liquid phase and thus their abundances in the atmosphere will be
limited by the saturation vapor pressure. Take Earth as an example.
Earth's atmosphere will be a steam atmosphere dominated by several
hundred bars of water vapor if the ocean completely evaporates. On
the other hand, in the high-temperature regime beyond 1000{\,}K, elements are not tied up in condensates. It is possible
to evaporate the rocks and metals at the surface and form a silicate
atmosphere, as proposed for 55 Cancri e (e.g.,
\citealt{demoryVariabilitySuperEarth552016}) and the proposed
``super comets" such as Kepler 1520 b
(\citealt{rappaportPossibleDisintegratingShortperiod2012};
\citealt{perez-beckerCatastrophicEvaporationRocky2013}) and K2-22 b
(\citealt{sanchis-ojedaK2ESPRINTProjectDiscovery2015}). In gas-melt
equilibrium with the magma ocean, the atmospheric composition could
be dominated by Na, K, Fe, Si, \ch{SiO}, O and \ch{O2} as the
major atmospheric species
(\citealt{itoTheoreticalEmissionSpectra2015}).

A thermal escape of species is likely to have a large impact on
atmospheric composition. The escape rate of an individual species
depends on its molecular weight. The ad-hoc orange lines in Figure
\ref{figgchem} are by no means quantitative boundaries. The
\ch{H2}\&He line also seems to divide the larger planets (radius
larger than four Earth radii, red dots) and the smaller ones into
two groups. Larger planets are reasonably represented by hydrogen or
hydrogen-helium atmospheres. For planets smaller than one
Earth radius, the compositional candidates are mostly restricted to
higher molecular weight in the atmosphere, if there is any. From the
perspective of atmospheric thermal escape, atmospheres could be
dominated by many possible molecules such as water, \ch{N2},
\ch{O2}, CO, \ch{CO2} and \ch{SO2} (and even argon?). Other
escape mechanisms such as solar-wind stripping might further
constrain the atmospheric composition in this small terrestrial
planet regime.

The mid-size planets between one and four Earth radii, namely
sub-Neptunes, mostly reside between the \ch{H2}\&He line and the
\ch{H2O}\&\ch{CH4} line. Note that the \ch{H2}\&He line also goes
through the sub-Neptunes, {as} does the ``Cosmic
Shoreline{.}" Therefore \ch{H2}\&He atmospheres are still
possible on these bodies. Although we do not
distinguish the mini-Neptunes (if we define them as
hydrogen-dominated) and super-Earths (non-hydrogen-dominated), it
looks {to be more challenging for} smaller
and hotter sub-Neptunes to retain a
low-molecular-weight atmosphere than the bigger and colder
ones. We
expect the atmospheric composition in the sub-Neptune regime might
be highly diverse since almost all kinds of compositions are
possible on those planets.

\begin{figure*}
   \centering
   \includegraphics[width=0.6\textwidth, angle=0]{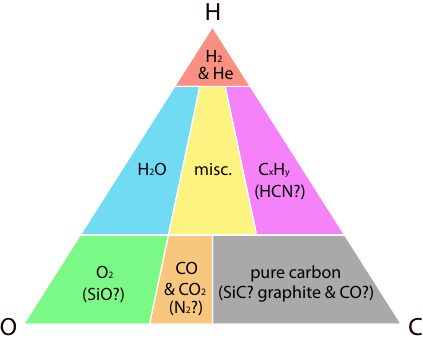}
   \caption{\baselineskip 3.8mm Compositional diversity map of substellar atmospheres as a function of H/C/O ratio. This ternary plot is modified from Fig.~4 in \cite{huPhotochemistryTerrestrialExoplanet2014} for sub-giant planets. $\mathrm{C_xH_y}$ represents hydrocarbons. This picture could be very different if we add other elements such as nitrogen, sulfur and silicon (e.g., \citealt{mosesCompositionalDiversityAtmospheres2013}; \citealt{zahnleAtmosphericSulfurPhotochemistry2009}).}
   \label{figtri}
   \end{figure*}

To further classify these atmospheres, we introduce
thermoequilibrium chemistry, which assumes the atmosphere
composition is solely dependent on temperature and
elemental abundances. This has been investigated by a number of
works (e.g., \citealt{loddersAtmosphericChemistryGiant2002};
\citealt{visscherAtmosphericChemistryGiant2006,visscherAtmosphericChemistryGiant2010};
\citealt{kemptonAtmosphericChemistryGJ2012};
\citealt{mosesCompositionalDiversityAtmospheres2013,mosesDisequilibriumCarbonOxygen2011};
\citealt{lineThermochemicalPhotochemicalKinetics2011};
\citealt{venotChemicalModelAtmosphere2012};
\citealt{huPhotochemistryTerrestrialExoplanet2014}; \citealt{mbarekCloudsSuperEarthAtmospheres2016};
\citealt{tsaiVULCANOpensourceValidated2017}). The most important
three elements are hydrogen, carbon and oxygen. The
dependence of the composition on the H/C/O ratios is summarized in a
ternary plot in Figure \ref{figtri} based on simulation results in
\cite{huPhotochemistryTerrestrialExoplanet2014} for a typical
sub-Neptune temperature range (500--1200\,K). The low-metallicity
atmospheres are probably still hydrogen-dominated. As the
metallicity increases from top to bottom, the composition starts to
diversify. The atmosphere would be more oxygen-rich to the left and
more carbon-rich to the right. If oxygen dominates over the carbon
but not hydrogen, {a} water world (steam atmosphere) is
a possibility; the other end member is {a}
hydrocarbon-dominated atmosphere if carbon dominates over oxygen.
Higher-order (more than two carbons in a molecule) hydrocarbon
atmospheres ($\mathrm{C_xH_y}$) are thermochemically favorable and
{there is }no need to invoke photochemistry to break
the chemical bonds in methane
(\citealt{huPhotochemistryTerrestrialExoplanet2014}).

In the high metallicity regime, hydrogen compounds are no longer
important. If the C/O ratio is low, the atmosphere might be
dominated by molecular oxygen without biochemistry. Further
photolysis could produce ozone. As the C/O ratio increases, the bulk
composition shifts to CO and \ch{CO2}, similar to the atmospheres on
Venus and Mars. If the C/O ratio is high, the extra carbon atoms
will not be able to combine with other elements so that the bulk
composition could be dominated by carbon (graphite?). In this
regime, graphite is actually stable for a large range of temperature
and metallicity conditions (e.g.,
\citealt{mosesCompositionalDiversityAtmospheres2013}). If graphite
is abundant, a large fraction of carbon would be requested in the
condensed graphite form, reducing the C/O ratio to near unity and resulting in a
\ch{CO}- or \ch{CO2}-rich atmosphere. Condensed graphite might also be a source
of the haze particles (Sect.~\ref{sect:cloud}). Nevertheless,
details o{n} the graphite chemistry have yet to
be explored.

The above scenarios are just end-members. If the bulk metallicity
has a relatively balanced H/C/O ratio, there will be a range of
possible atmospheric compositions. Temperature plays a crucial role
in determining what the atmosphere is made of (the ``misc." regime
in Fig.~\ref{figtri}). Including less abundant elements such as
nitrogen (e.g.,
\citealt{mosesCompositionalDiversityAtmospheres2013};
\citealt{mosesCHEMICALCONSEQUENCESRATIO2013}), sulfur (e.g.,
\citealt{zahnleAtmosphericSulfurPhotochemistry2009}) and silicon
would further complicate the classification. For example,
$\mathrm{C_xH_y}$-dominated atmospheres might not exist because
species like \ch{HCN} can dominate instead. {A}
\ch{N2}-rich atmosphere could be common. If silicon is present, SiC
or SiO might also dominate the high metallicity atmospheres under
certain conditions (Fig.~\ref{figtri}). Moreover, disequilibrium
chemistry such as photochemistry, ion-chemistry and vertical
mixing will change the abundances of trace species in the
atmosphere, but whether these processes could alter the bulk
composition is an open question.

In addition to the bulk composition discussed above, minor
constituents are also important and have a notable impact on the
spectra, light curves and radiative energy balance of the
substellar atmospheres. Key gaseous species providing very important
opacities include \ch{H2O}, \ch{CH4} and other hydrocarbons, CO,
\ch{CO2}, \ch{NH3}, HCN, \ch{O3} and some sulfur-bearing
species such as \ch{H2S} and \ch{SO2}. Again, the minor species also
depends on many factors. Other than temperature and elemental
ratios, the abundances and distributions of minor chemicals also
crucially depend on disequilibrium processes induced by
photochemistry and transport.

In general, for low-metallicity hydrogen-dominated
atmospheres, the minor species are usually hydrogen compounds like
\ch{H2O}, \ch{CH4}, and \ch{NH3} with photochemically generated
hydrocarbons and nitriles. For high-metallicity
atmospheres composed of \ch{H2O}, \ch{CO}, \ch{CO2} or
\ch{N2}, molecules with more than one heavy atoms per molecule and
oxidized photochemical products such as \ch{O2}, \ch{O3} and \ch{NO}
are abundant. Here we mainly focus on the hydrogen-helium
atmosphere. The C/O ratio in the hydrogen atmosphere is an important
factor. \cite{madhusudhanRatioDimensionCharacterizing2012} proposed
a classification scheme based on irradiation (essentially the
temperature) and the C/O ratio in hydrogen atmospheres. The boundary
between {a} C-rich atmosphere and {an}
O-rich atmosphere is C/O$\sim$1. It was claimed that C-rich
atmospheres are not likely to have the thermal inversion because TiO
and VO are not abundant. The O-rich, haze-free
atmospheres could only develop thermal inversion in the
high-temperature regime, but the low-temperature regime does not,
similar to the pM and pL class{es} in
\cite{fortneyUnifiedTheoryAtmospheres2008}. Note that the calculations for C-rich atmospheres in
\cite{madhusudhanRatioDimensionCharacterizing2012} have neglected
the possible contribution of carbon-based aerosols, which
could easily produce atmospheric inversion.

The detection of the C/O ratio in {an}
exoplanetary atmosphere is important. From the formation point of
view, it is expected{ that} the C/O ratios for most stars should
be less than unity because oxygen is more cosmically abundant than
carbon (e.g., \citealt{fortneyCarbontooxygenRatioMeasurement2012};
\citealt{brewerSpectralPropertiesCool2016}), but the formation and
evolution of the planetary atmospheres will diversify the ratios.
To date, there is no firm evidence in any
extrasolar gas giants with C/O larger than unity. WASP-12~b was
claimed to be a ``carbon-rich giant planet"
(\citealt{madhusudhanCarbonrichGiantPlanets2011}) but refuted by a
subsequent work by
\cite{kreidbergDETECTIONWATERTRANSMISSION2015}. For other planets{,} the upper
limits of C/O have also been reported to be smaller than unity
(e.g., \citealt{lineSystematicRetrievalAnalysis2014};
\citealt{bennekeStrictUpperLimits2015};
\citealt{barstowConsistentRetrievalAnalysis2017}). Also,
\cite{wallackInvestigatingTrendsAtmospheric2019} analyzed several
colder planets under 1000{\,}K and suggested a possible
correlation between the derived \ch{CH4}/(CO+\ch{CO2}) ratio and
stellar metallicity.

Depending on the temperature, thermochemistry predicts two important
regimes. High-temperature atmospheres tend to have O-bearing species
(e.g.,  \ch{H2O}, \ch{CO}) for a small C/O ratio and C-bearing
species (e.g., \ch{HCN}, \ch{C2H2}, and \ch{CO}) for a large C/O
ratio (\citealt{mosesCHEMICALCONSEQUENCESRATIO2013}).
\ch{N2} is the dominant nitrogen species. If the temperature
is sufficiently high, atomic neutrals and ions of refractory
elements such as Mg, Mg+, Fe, Fe+, Ca, Ca+, Na, Na+, K, K+, Al, as
well as their molecular forms, stay in the gas phases that have been
detected (see Sect.~\ref{sect:stabilityfund}). As
the atmospheric{ temperature} decreases, \ch{CH4} and \ch{NH3}
emerge. In the cool regime ($<$ 1000{\,}K),
\ch{CH4}, \ch{H2O} and \ch{NH3} become the main reservoirs of
carbon, oxygen and nitrogen, respectively.

Nevertheless, the photospheric constituents are not likely to be in
thermochemical equilibrium because of the transport-induced
quenching and photochemistry. To further investigate the quenching
mechanisms, it is necessary to understand the important pathways in
the interconversion of \ch{N2} $\leftrightarrow$ \ch{NH3} and CO
$\leftrightarrow$ \ch{CH4} and identify necessary rate-limiting
steps. The chemical timescales of those steps can thus be compared
with the vertical transport timescale to determine the quenching
points in the deep atmosphere. The \ch{N2}/\ch{NH3} quench point
usually occurs deeper than the CO-\ch{CH4}-\ch{H2O} quench point.
Many efforts have been made but the chemical mechanisms are still elusive (e.g.,
\citealt{mosesCompositionalDiversityAtmospheres2013,mosesDisequilibriumCarbonOxygen2011};
\citealt{lineThermochemicalPhotochemicalKinetics2011};
\citealt{huPhotochemistryTerrestrialExoplanet2014};
\citealt{hengAnalyticalModelsExoplanetary2016,tsaiVULCANOpensourceValidated2017,tsaiConsistentModelingAtmospheric2018};
\citealt{venotChemicalModelAtmosphere2012,venotNewChemicalScheme2015,venotNewChemicalScheme2020}).
Nevertheless, uncertainties associated with the laboratory-measured
rate coefficients of those quenching reactions, especially those
time-limiting steps, hinder the predictive power of the abundances
of important species and the subsequent interpretation of the
observed spectra. {For a} review of
detailed chemical cycles{,} refer to
\cite{mosesChemicalKineticsExtrasolar2014} and
\cite{madhusudhanExoplanetaryAtmospheresChemistryFormation2016}.

The thermochemical carbon cycle can be summarized as \ch{CH4} +
\ch{H2O} $\leftrightarrow$ CO + 3\ch{H2}. The rates of \ch{CH4}
$\rightarrow$ CO and CO $\rightarrow$ \ch{CH4} conversion depend on
the efficiency to form and break the strong C-O bond, respectively.
It was proposed{ that} the rate limiting step of \ch{CH4}
$\rightarrow$ CO is the reaction of the OH and \ch{CH3} radicals,
e.g., \ch{CH3} + OH $\rightarrow$ \ch{CH2OH} + H or \ch{CH3} + OH +
M $\rightarrow$ \ch{CH2OH} + H + M
(\citealt{mosesDisequilibriumCarbonOxygen2011,
tsaiConsistentModelingAtmospheric2018}), where M is the ambient bulk
gas molecule. {Also,} that of CO $\rightarrow$
\ch{CH4} is perhaps \ch{CH3OH} + M $\rightarrow$ \ch{CH3} + OH + M
(\citealt{mosesDisequilibriumCarbonOxygen2011}) or \ch{CH3OH} + H
$\rightarrow$ \ch{CH3} + \ch{H2O} (e.g.,
\citealt{venotAtmosphericChemistryWarm2014,venotNewChemicalScheme2015,venotNewChemicalScheme2020};
\citealt{zahnleMethaneCarbonMonoxide2014}). The carbon
interconversion cycle \ch{CH4} + \ch{H2O} $\leftrightarrow$ CO +
3\ch{H2} is also considered as the main pathway controlling water abundances.

For \ch{N2} $\leftrightarrow$ \ch{NH3} interconversion, the net
cycle can be written as \ch{N2} + 3\ch{H2} $\leftrightarrow$
2\ch{NH3}{,} {b}ut the rate limiting steps are
highly uncertain (see discussion in
\citealt{mosesChemicalKineticsExtrasolar2014}). For \ch{N2}
$\rightarrow$ \ch{NH3}, the rate-limiting step is speculated as NH +
\ch{NH2} $\rightarrow$ \ch{N2H2} + H, 2\ch{NH2} $\rightarrow$
\ch{N2H2} + \ch{H2} or \ch{N2H3} + M $\rightarrow$ \ch{N2H2} + H +
M, depending on the temperature and pressure conditions (e.g.,
\citealt{mosesDisequilibriumCarbonOxygen2011}). For
\ch{NH3}$\rightarrow$N2, the rate-limiting step could be just the
reverse reactions of the above, such as \ch{N2H2} + H $\rightarrow$
NH + \ch{NH2}. Mechanisms become more complicated if we further
include carbon-bearing species HCN in the pathways. For example, the interconversion pathway between \ch{NH3} and HCN
is \ch{NH3} + CO $\leftrightarrow$ HCN + \ch{H2O} in the warm
atmosphere where CO is domina{n}t over \ch{CH4}. In {a}
relatively cold atmosphere where \ch{CH4} is more abundant, the
pathway becomes \ch{NH3} + \ch{CH4} $\leftrightarrow$ HCN +
3\ch{H2}.

For the dominant species, such as CO in {a} deep and
warm atmosphere, transport-induced quenching does not affect their
abundances too much because they are the primary elemental carrier
already. Efficient transport quenching occurs for the species that
are less abundant at and below the quenching point
(\citealt{mosesChemicalKineticsExtrasolar2014}). For example,
in warm or hot Jupiter atmospheres, \ch{CH4} is not
predicted to be abundant in thermochemical equilibrium in the
observable regions of the atmosphere. However, there is a greater
\ch{CH4} mixing ratio at the quenching point, so the disequilibrium
quenching ends with more \ch{CH4} than expected in the photosphere.
For colder planets where \ch{CH4} dominates the observations in
thermochemical equilibrium, dynamical quenching transports CO
upward, leading to a greater-than-expected CO abundance. On the
other hand, some species are also less affected by quenching because
of fast chemistry. For example, \ch{CO2} is mostly controlled
by fast interconversion in the {H$_2$O-CO-CO$_2$} chemical
network.

One interesting case is the young, directly imaged planets and brown
dwarfs. Their temperature-pressure gradient is large, and the
temperature-pressure profile crosses the CO-\ch{CH4} equal-abundance
boundary somewhere above the quenching point but below the
observable atmosphere. As a result, dynamical quenching is
significant for these objects. The expected relative abundances of
CO and \ch{CH4} can completely switch places compared to what is
expected in thermochemical equilibrium.
\cite{mosesCOMPOSITIONYOUNGDIRECTLY2016} found that dynamical
quenching on young Jupiters leads to CO/\ch{CH4} and
\ch{N2}/\ch{NH3} ratios much larger than chemical-equilibrium
predictions, while the mixing ratio of \ch{H2O} is a factor of a few
less than its chemical-equilibrium value.

In the mode-data comparison, the lack of detection of spectral
features of \ch{CH4} on some low-mass sub-Neptunes such as GJ 436 b
(e.g., \citealt{stevensonPossibleThermochemicalDisequilibrium2010};
\citealt{knutsonFeaturelessTransmissionSpectrum2014}), GJ 1214 b
(e.g., \citealt{beanOpticalNearinfraredTransmission2011};
\citealt{kreidbergCloudsAtmosphereSuperEarth2014}) and GJ 3470 b
(\citealt{bennekeSubNeptuneExoplanetLowmetallicity2019}) is not
consistent with the equilibrium methane abundances predicted by
cloudless \ch{H2}-rich chemical models, revealing our incomplete
understanding of the mechanisms. The current hypotheses include
high-metallicity atmosphere
(\citealt{mosesCompositionalDiversityAtmospheres2013};
\citealt{venotAtmosphericChemistryWarm2014}), Helium-rich atmosphere
(\citealt{huHeliumAtmospheresWarm2015a};
\citealt{malskyCoupledThermalCompositional2020}),
hotter-than-expected interiors so that \ch{CH4} is quenched in low
abundances (\citealt{agundezPUZZLINGCHEMICALCOMPOSITION2014};
\citealt{morleyFORWARDINVERSEMODELING2017}) and \ch{CH4}
photodissociation by a high-energy stellar flux such as
Lyman-$\alpha$ penetrating into the stratosphere
(\citealt{miguelEffectLymanRadiation2015}). The last possibility is
debatable because the hot thermosphere on top of the stratosphere
might absorb most of the incoming high-energy flux. The hot interior
hypothesis is particularly interesting because it implies some
unknown heat source that might be related to the tidal
heating due to the non-zero eccentric orbit (\citealt{agundezPUZZLINGCHEMICALCOMPOSITION2014})
or obliquity-induced tides (\citealt{millhollandTidallyInducedRadius2019}). The
interconversion between CO and \ch{CH4} has also been suggested to
play a very important role in substellar atmosphere evolution on
brown dwarfs (e.g.,
\citealt{tremblinCloudlessAtmospheresYoung2017,tremblinThermocompositionalDiabaticConvection2019}),
which will be discussed in Section \ref{sect:bddyn}.

Last but not the least, photochemical and ion-chemical processing of
quenched species in the upper atmosphere will further
complicate the chemical pathways and observations. For close-in
planets, the stellar high-energy flux is strong enough to make
notable impact o{n} the vertical profiles of the chemical
compositions and the observed spectra (e.g.,
\citealt{liangSourceAtomicHydrogen2003,
liangInsignificancePhotochemicalHydrocarbon2004};
\citealt{yelleAeronomyExtrasolarGiant2004};
\citealt{kemptonAtmosphericChemistryGJ2012};
\citealt{koskinenThermosphericCirculationModel2007};
\citealt{kopparapuPhotochemicalModelCarbonrich2011};
\citealt{mosesDisequilibriumCarbonOxygen2011,
mosesCompositionalDiversityAtmospheres2013};
\citealt{kemptonAtmosphericChemistryGJ2012}; \citealt{
huPhotochemistryTerrestrialExoplanet2014}). {However,}
the photolysis can also be important f{or} directly
imaged planets (\citealt{mosesCOMPOSITIONYOUNGDIRECTLY2016}).
Photolysis of CO, \ch{CH4} and \ch{NH3} could produce{ a}
significant amount of \ch{CO2}, HCN and hydrocarbons like \ch{C2H2}
and even photochemical hazes (e.g.,
\citealt{lavvasAerosolPropertiesAtmospheres2017};
\citealt{gaoSulfurHazesGiant2017};
\citealt{horstHazeProductionRates2018};
\citealt{heLaboratorySimulationsHaze2018,hePhotochemicalHazeFormation2018,heSulfurdrivenHazeFormation2020};
\citealt{moranChemistryTemperateSuperEarth2020}; \citealt{
kawashimaTheoreticalTransmissionSpectra2018,
kawashimaDetectableMolecularFeatures2019,
kawashimaTheoreticalTransmissionSpectra2019};
\citealt{lavvasPhotochemicalHazesSubNeptunian2019};
\citealt{fleuryPhotochemistryHotH2dominated2019};
\citealt{ohnoCloudsFluffyAggregates2020};
\citealt{adamsAggregateHazesExoplanet2019,
gaoDeflatingSuperpuffsImpact2020}), which will be discussed in the
following Section \ref{sect:cloud}. {For a} detailed review of the photochemistry{,}
refer to \cite{mosesChemicalKineticsExtrasolar2014} and
reference therein. Photochemistry has been shown to significantly alter the
chemical compositions of terrestrial exoplanetary atmospheres and
the interpretation of their spectra (e.g.,
\citealt{selsisSignatureLifeExoplanets2002};
\citealt{seguraBiosignaturesEarthlikePlanets2005};
\citealt{seguraAbioticProductionO22007};
\citealt{huPhotochemistryTerrestrialExoplanet2012,huPhotochemistryTerrestrialExoplanet2013};
\citealt{rugheimerEffectUVRadiation2015};
\citealt{arneyPaleOrangeDot2016, arneyPaleOrangeDots2017};
\citealt{meadowsHabitabilityProximaCentauri2018};
\citealt{lincowskiEvolvedClimatesObservational2018};
\citealt{chenHabitabilitySpectroscopicObservability2019}). One
interesting fact is that photochemistry could produce false
positives of gaseous biosignatures. For example, there
{is} a variety of ways to produce abiotic molecular
oxygen from the photolysis of water and \ch{CO2} by strong UV flux
(\citealt{gaoStabilityCO2Atmospheres2013};
\citealt{tianHighStellarFUV2014};
\citealt{wordsworthABIOTICOXYGENDOMINATEDATMOSPHERES2014};
\citealt{lugerExtremeWaterLoss2015};
\citealt{harmanAbiotiCO2Levels2015,harmanAbiotiCO2Levels2018}) on
terrestrial exoplanets, especially on planets around M-dwarfs with
high-energy fluxes. On the other hand, ion-chemistry is particularly
important to understand the composition and energy balance in the
upper thermosphere and the detailed mechanisms of
atmospheric escape (\citealt{yelleAeronomyExtrasolarGiant2004};
\citealt{garciamunozPhysicalChemicalAeronomy2007};
\citealt{scheucherProximaCentauriStrong2020a}), but many of the
chemical reaction coefficients have large uncertainties at this
moment.

\subsection{Clouds and Hazes}
\label{sect:cloud}

A growing body of evidence suggests that spectra of exoplanets and
brown dwarfs are significantly affected by the presence of
aerosols---condensational clouds and chemical hazes---that are also
ubiquitous in all substantial atmospheres of{ planets in} the
Solar System. The most prominent evidence is from the
muted spectral features. For example, if an exoplanetary atmosphere
is cloud-free, its transmission spectra at optical wavelengths would
exhibit a Rayleigh scattering slope with sharp spectral features
from alkali metals like sodium and potassium if hot enough. In the
presence of high-altitude aerosols, however, the spectral slope and
metal absorption peaks are significantly reduced and may even
disappear. Similarly, in the {IR}, the predominant
gas (e.g., water, methane) rotational-vibrational features seen in a
clear atmosphere could also be blocked by the presence of aerosols.

As noted in Section \ref{sect:char} and reference therein, such
flattened transmission spectra have been seen for many hot Jupiters
and cooler and smaller planets. The mean particle size and cloud top
pressures have been retrieved from some of their spectra (e.g.,
\citealt{kreidbergCloudsAtmosphereSuperEarth2014};
\citealt{knutsonFeaturelessTransmissionSpectrum2014};
\citealt{morleyTHERMALEMISSIONREFLECTED2015};
\citealt{bennekeStrictUpperLimits2015};
\citealt{bennekeSubNeptuneExoplanetLowmetallicity2019}). For
example, the cloud tops on GJ 1214 b
(\citealt{kreidbergCloudsAtmosphereSuperEarth2014}), GJ 436 b
(\citealt{knutsonFeaturelessTransmissionSpectrum2014}) and HD 97658
b (\citealt{knutsonHubbleSpaceTelescope2014}) are as high as the 0.1
mbar pressure level. For GJ 3470 b, the cloud top is at a lower
altitude (\citealt{bennekeSubNeptuneExoplanetLowmetallicity2019}).
These high-altitude aerosols cause trouble in retrieving atmospheric
compositions on sub-Neptunes. For example, GJ 1214 b could be made
of water, hydrogen or other heavier elements (e.g.,
\citealt{miller-ricciNatureAtmosphereTransiting2010};
\citealt{rogersThreePossibleOrigins2010}), but the flattened spectra
are not useful {for} distinguishing among these
candidates due to the lack of detected molecular features
(\citealt{kreidbergCloudsAtmosphereSuperEarth2014}). Atmospheric
windows with lower cloud opacity at longer wavelengths are needed to
solve this problem.

In addition to the spectral evidence, spatial information
o{n} the substellar atmospheres also indicates the
existence of aerosols. The rotational light curves of brown dwarfs
are a good example of the influence{ of clouds}
on thermal emission (see Sect.~\ref{sect:rotlight}). The
reflection light curves in the Kepler bands also demonstrate the
importance of aerosols. For example, significant westward
phase offsets in the visible wavelengths in
Figure~\ref{figteqtrend}(C) probably come from cloud
reflection (\citealt{parmentierTransitionsCloudComposition2016}).
The depletion of condensable vapors could also be a result
of cloud condensation. A recent observation on an ultra-hot Jupiter
WASP-76 b ($T_{\rm eff}\sim2190$\,K) using high-dispersion transit
spectroscopy found asymmetry {in} the atomic iron
signature in the atmosphere
(\citealt{ehrenreichNightsideCondensationIron2020}). The iron
absorption is absent on the nightside close to the morning
terminator in contrast to the other limb, indicating that the iron
is possibly condensing on the nightside.

Aerosols on exoplanets and brown dwarfs have both direct and
indirect sources. The direct sources include dust emission from the
surface or dust infall from space. The surface sources are
common on terrestrial planets, such as volcanic ash and sea salt on
Earth, and dust storms on Mars. For example, it was proposed that
radiatively active mineral dust emitted from the surface could
postpone planetary water loss and impact the habitability of
Earth-like exoplanets (\citealt{boutleMineralDustIncreases2020}).
The atmospheres of ``Super-puffs" Kepler 51 b and 51 d might also be
dusty because of the outflow of tiny grains from the surface
(\citealt{wangDustyOutflowsPlanetary2019}). The infalling dust could
come from meteoric dust sources (see
\citealt{gaoBimodalDistributionSulfuric2014} for the Venus case) or
directly from the protoplanetary disk (e.g., PDS 70 b and c,
\citealt{wangKeckNIRC2Band2020}).

The indirect sources refer to atmospheric condensation and
chemical processes. For example, hazes and clouds on exoplanets and
brown dwarfs have been predicted to form from either condensation of
salt, silicate and metal vapors  (e.g., \citealt{ackermanPrecipitatingCondensationClouds2001}; \citealt{morleyNEGLECTEDCLOUDSDWARF2012};
\citealt{ohnoMicrophysicalModelingMineral2018};
\citealt{gaoMicrophysicsKClZnS2018};
\citealt{gaoSedimentationEfficiencyCondensation2018};
\citealt{ormelARCiSFrameworkExoplanet2019};
\citealt{ohnoCloudsFluffyAggregates2020}), or
coagulation of particles generated by atmospheric chemistry (e.g.,
\citealt{lavvasAerosolPropertiesAtmospheres2017};
\citealt{horstHazeProductionRates2018};
\citealt{fleuryPhotochemistryHotH2dominated2019};
\citealt{heSulfurdrivenHazeFormation2020};
\citealt{moranChemistryTemperateSuperEarth2020}). In the condensate
cloud scenario, clouds form when the condensable species become
supersaturated, ranging from KCl and ZnS in the cooler regime to
\ch{Mg2SiO4}, \ch{TiO2}, MnS, Cr, Fe, corundum (\ch{Al2O3}),
calcium-aluminates and calcium-titanates (e.g., perovskite
\ch{CaTiO3}) in hotter atmospheres (e.g.
\citealt{visscherAtmosphericChemistryGiant2006,visscherAtmosphericChemistryGiant2010},
\citealt{loddersExoplanetChemistry2010}). Some L-dwarf spectra
{manifest} possible broad absorption features at
around 9{\,}${\upmu}$m that could result
from the SiO vibrational band
(\citealt{cushingSpitzerInfraredSpectrograph2006}).

Photochemical hazes on exoplanets have been hypothesized to form via
atmospheric photochemistry and ion chemistry of methane, nitrogen
and sulfur (e.g., \citealt{kemptonAtmosphericChemistryGJ2012};
\citealt{morleyQUANTITATIVELYASSESSINGROLE2013};
\citealt{zahnlePHOTOLYTICHAZESATMOSPHERE2016};
\citealt{lavvasAerosolPropertiesAtmospheres2017};
\citealt{gaoSulfurHazesGiant2017};
\citealt{horstHazeProductionRates2018};
\citealt{heLaboratorySimulationsHaze2018,hePhotochemicalHazeFormation2018,heSulfurdrivenHazeFormation2020};
\citealt{kawashimaTheoreticalTransmissionSpectra2018, kawashimaTheoreticalTransmissionSpectra2019}; \citealt{kawashimaDetectableMolecularFeatures2019};
\citealt{lavvasPhotochemicalHazesSubNeptunian2019};
\citealt{fleuryPhotochemistryHotH2dominated2019};
\citealt{moranChemistryTemperateSuperEarth2020};
\citealt{adamsAggregateHazesExoplanet2019};
\citealt{gaoDeflatingSuperpuffsImpact2020};
\citealt{ohnoSuperRayleighSlopesTransmission2020}),
as analogues of hazes in Solar
System atmospheres. These chemical haze particles may be highly
porous like those on Titan and Pluto, where chemically produced
``macromolecules" or ``monomers" coagulate into large fluffy
aggregates (e.g.,
\citealt{lavvasAerosolGrowthTitan2013,lavvasEnergyDepositionPrimary2011};
\citealt{gaoConstraintsMicrophysicsPluto2017}). This production of
photochemical hazes in warm atmospheres has been confirmed by
laboratory experiments (e.g.,
\citealt{horstHazeProductionRates2018};
\citealt{heLaboratorySimulationsHaze2018,hePhotochemicalHazeFormation2018,
heSulfurdrivenHazeFormation2020};
\citealt{fleuryPhotochemistryHotH2dominated2019};
\citealt{moranChemistryTemperateSuperEarth2020}), in the relevant
temperature range from 300--1500\,K. Yet, particle formation
mechanisms and their impacts on substellar atmospheres and
observations remain poorly understood.

Detailed simulations have also been conducted to understand aerosol
formation in the warm and hot regime. Pioneering work from
\cite{ackermanPrecipitatingCondensationClouds2001} and subsequent
works (e.g., \citealt{saumonEvolutionDwarfsColorMagnitude2008};
\citealt{marleyPatchyCloudModel2010};
\citealt{morleyNEGLECTEDCLOUDSDWARF2012,morleySpectralVariabilityPatchy2014,morleyTHERMALEMISSIONREFLECTED2015})
simulated 1D cloud profiles in substellar atmospheres based on{
an} idealized, homogeneous chemical equilibrium framework. They assumed that the
species immediately condenses to particles once supersaturated
but did not simulate the particle growth such as coagulation
in \cite{ormelARCiSFrameworkExoplanet2019} and \cite{ohnoCloudsFluffyAggregates2020}.
A sophisticated, kinetic, brown dwarf and exoplanet grain chemistry
model is described in a series of papers by Helling and
collaborators (e.g.,
\citealt{hellingConsistentSimulationsSubstellar2008,
witteDustBrownDwarfs2009, witteDustBrownDwarfs2011,
woitkeDustBrownDwarfsII2003, woitkeDustBrownDwarfsIII2004,
woitkeDustBrownDwarfs2020};
\citealt{hellingDetectabilityDirtyDust2006,
hellingDustBrownDwarfs2008,hellingDustBrownDwarfsI2001,
hellingDustBrownDwarfsV2006, hellingMineralCloudsHD2016,
hellingUnderstandingAtmosphericProperties2019,
samraMineralSnowflakesExoplanets2020,
hellingMineralCloudHydrocarbon2020}, see reviews in
\citealt{hellingAtmospheresBrownDwarfs2014} and
\citealt{hellingExoplanetClouds2019}). The model considered more
complicated mixtures of dust grains, but it does not include certain
important grain growth processes, such as the vital effect of grain
surface energy on the condensation process (``Kelvin effect", e.g.,
\citealt{rossowCloudMicrophysicsAnalysis1978};
\citealt{pruppacherMicrophysicsCloudsPrecipitation1980};
\citealt{seinfeldAtmosphericChemistryPhysics2016}). More recently,
microphysical models originating from Earth science have been
successfully applied to hot Jupiters
(\citealt{powellFormationSilicateTitanium2018,powellTransitSignaturesInhomogeneous2019};
\citealt{gaoAerosolCompositionHot2020}) and smaller planets
(\citealt{gaoMicrophysicsKClZnS2018}) to simulate multiple cloud
layers via processes such as nucleation, coagulation, condensation,
sedimentation, evaporation and transport. These models can predict
the particle size distributions that have shown to be important for
spectral simulations, but they do not consider the mixture of
condensable species.

Photochemical haze models including coagulations microphysics (e.g.,
\citealt{lavvasAerosolPropertiesAtmospheres2017};
\citealt{gaoSulfurHazesGiant2017};
\citealt{kawashimaTheoreticalTransmissionSpectra2018,
kawashimaDetectableMolecularFeatures2019,
kawashimaTheoreticalTransmissionSpectra2019};
\citealt{lavvasPhotochemicalHazesSubNeptunian2019};
\citealt{adamsAggregateHazesExoplanet2019,
gaoDeflatingSuperpuffsImpact2020}) have been applied to both
Jupiter-sized planets and smaller planets. Spectral slopes in
optical transmission spectra on some exoplanet{s} are
observed{ to be} steeper than the Rayleigh slopes (so-called
``super-Rayleigh slopes",
\citealt{sedaghatiDetectionTitaniumOxide2017};
\citealt{pinhasH2OAbundancesCloud2019}
\citealt{welbanksMassMetallicityTrends2019};
\citealt{mayMOPSSIIExtreme2019}). It was suggested
that photochemical haze produced in the upper atmosphere could
result in an increasing trend of atmospheric opacity with altitude,
which might explain the super-Rayleigh slopes of transit
depth toward blue in optical wavelength{s} (e.g.,
\citealt{lavvasPhotochemicalHazesSubNeptunian2019};
\citealt{ohnoSuperRayleighSlopesTransmission2020}). The sub-Neptune
GJ 1214 b has been {regarded} as the test bed for
those haze models owing to the surprisingly flat spectra observed in
the near-IR (e.g.,
\citealt{kreidbergCloudsAtmosphereSuperEarth2014}). Models reached a consensus
that a very high metallicity is required to explain the spectral
flatness of this planet. The detected radio emission from a close-in
planet HAT-11 b (\citealt{desetangsHint150MHz2013}) indicates the
existence of lightning inside the clouds, suggesting that the
particle charge could be important (e.g.,
\citealt{hellingDustCloudLightning2013,hellingLightningChargeProcesses2019,hodosanExolightningRadioEmission2017}).
{However,} the effect of charging on particle
microphysical growth has not been investigated in detail for
exoplanets and brown dwarfs.

Several computationally expensive 3D simulations have also been
performed with particles for exoplanets (e.g.,
\citealt{parmentier3DMixingHot2013};
\citealt{charnay3DModelingGJ1214bs2015,charnay3DModelingGJ1214b2015a};
\citealt{oreshenkoOpticalPhaseCurves2016};
\citealt{leeModellingLocalGlobal2015,leeDynamicMineralClouds2016,leeDynamicMineralClouds2017};
\citealt{linesExonephologyTransmissionSpectra2018,linesSimulatingCloudyAtmospheres2018};
\citealt{romanModeledTemperaturedependentClouds2019}) and for brown
dwarfs (\citealt{tanEffectsLatentHeating2017}; \citealt
{tanAtmosphericCirculationBrown2020}){,} {b}ut
the microphysics is usually simplified to increase the computational
efficiency. A few models with fully coupled cloud microphysics,
radiative and transfer and 3D dynamics can only perform very
short-term integrations ($\sim$40-60{\,}d,
\citealt{leeDynamicMineralClouds2016};
\citealt{linesExonephologyTransmissionSpectra2018}). In 3D models
where the large-scale dynamics can be resolved, particles are
advected by atmospheric circulation and their sizes are found
distributed inhomogeneously across the globe (e.g.,
\citealt{linesExonephologyTransmissionSpectra2018,linesSimulatingCloudyAtmospheres2018}).
However, the 3D model results typically produce
flat spectra in the UV and visible, in contrast with
observations which {display} slopes across the
wavelengths. Detailed diagnosis is still needed to {identify} the underlying mechanisms. See{ a} brief
discussion in Section~\ref{sect:hjdyn} and {for }detailed
discussions refer to a recent review by
\cite{hellingExoplanetClouds2019}.

\begin{figure*}
   \centering
   \includegraphics[width=0.85\textwidth, angle=0]{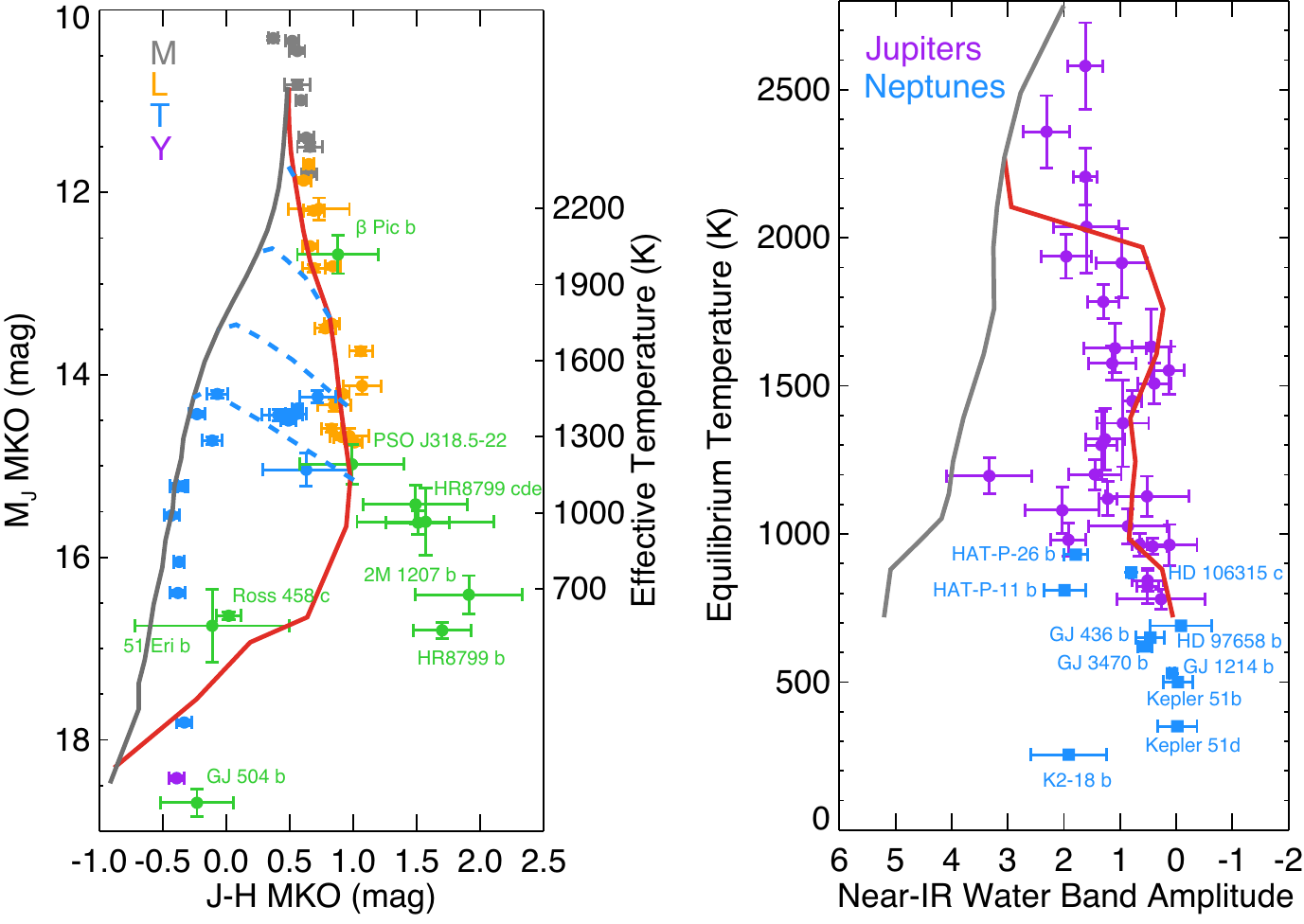}
   \caption{\baselineskip 3.8mm Trends of near-IR cloudiness proxies on brown dwarfs (colored spectral types) and exoplanets with theoretical model curves ({\it grey}: clear-sky models; {\it red}: cloudy models). {\it Left}: near-IR {CMD} for brown dwarfs (see legends for spectral types) and directly imaged planets ({\it green}). The effective temperatures are estimated based on the {$J$ }band magnitude. The brown dwarf data are from the MKO weighted averages in a large compilation in \href{http://www.as.utexas.edu/~tdupuy/plx/Database_of_Ultracool_Parallaxes.html}{Database of Ultracool Parallaxes} (\citealt{dupuyHAWAIIINFRAREDPARALLAX2012}; \citealt{dupuyDistancesLuminositiesTemperatures2013}; \citealt{liuHAWAIIINFRAREDPARALLAX2016}). The data sources for directly imaged planets are the same as in Fig.~1. The theoretical models are from \cite{marleyPatchyCloudModel2010}. The {{\it dashed}} lines are approximate ``patchy cloud" scenarios from the simulations in \cite{marleyPatchyCloudModel2010}. {\it Right}: the near-IR water band amplitude (i.e., $A_H$) as a function of equilibrium temperature from transmission spectra on tidally locked exoplanets. Hot Jupiter ({\it purple}) data are from \cite{fuStatisticalAnalysisHubble2017} and updated in \cite{gaoAerosolCompositionHot2020}. Warm Neptune ({\it blue}) data are compiled by \cite{crossfieldTrendsAtmosphericProperties2017} with additional planets:  Kepler 51 b and d (\citealt{libby-robertsFeaturelessTransmissionSpectra2020}), K2-18 b (\citealt{bennekeWaterVaporClouds2019}) and HD 106315 c (\citealt{kreidbergTentativeEvidenceWater2020}). The water band information for K2-25 b is not available. The theoretical models for hot Jupiters are from \cite{gaoAerosolCompositionHot2020}.}
   \label{figcloud}
   \end{figure*}

Although individual objects deserve investigation in detail,
statistical trends from observations could put strong constraints on
the complicated aerosol formation processes on brown dwarfs and
exoplanets in a single framework. Clouds have been proposed to
significantly influence not only the spectral sequence of the
emission light from brown dwarfs and directly imaged planets, but
also the near-{IR} water signals in transmission
spectra of close-in giant planets and sub-Neptunes
(Sect.~\ref{sect:char}). We first discuss the brown dwarfs and
directly imaged planes in Section~\ref{sect:bdcloud} and close-in
exoplanets in Section~\ref{sect:planetcloud}.

\subsubsection{Spectral Trends on Brown Dwarfs and Directly Imaged Planets}
\label{sect:bdcloud}

As mentioned in Section~\ref{sect:rotlight}, clouds are
important {f}o{r} understand{ing} the rotational
light curves and their variability of an individual body. The impact
of clouds also shows up in well-characterized near-IR
{CMD} (Fig.~\ref{figcloud}). For
example, TiO conversion to \ch{TiO2} and TiO and VO condensation
(such as perovskite \ch{CaTiO3}) from thermochemical models (e.g.,
\citealt{loddersAlkaliElementChemistry1999};
\citealt{burrowsChemicalEquilibriumAbundances1999};
\citealt{allardLimitingEffectsDust2001};
\citealt{loddersAtmosphericChemistryGiant2002}) have been suggested
to cause the M/L transition. On the other hand, the cloudless model
could not explain the reddening of the M-L trend
(Fig.~\ref{figcloud}), due to the onset of \ch{H2}
collision-induced absorption and \ch{CH4} bands as the atmospheres
cool down (e.g., \citealt{saumonEvolutionDwarfsColorMagnitude2008}).
As mentioned before, observations on L dwarfs show possible evidence
for silicate grain absorption in Spitzer IRS data (e.g.,
\citealt{cushingSpitzerInfraredSpectrograph2006}), indicating the
importance of clouds in controlling the spectral sequence. On the
other hand, unlike warm and cloudy L dwarfs, T
dwarfs are generally thought to be cold and clear in their
photospheres. In the late-T to Y, the color reversal (blue to red)
is probably due to both the disappearance of
{$J$ }band alkali metal opacity because the metals
are bound into molecules
(\citealt{liuDiscoveryHighlyUnequalmass2010};
\citealt{kirkpatrickFurtherDefiningSpectral2012}) and the emergence
of \ch{NH3} absorption in the {$H$} band (e.g.,
\citealt{loddersAtmosphericChemistryGiant2002};
\citealt{burrowsDwarfsTheoreticalSpectra2003}). Chloride and sulfide
clouds (e.g., \citealt{morleyNEGLECTEDCLOUDSDWARF2012};
\citealt{beichmanWISEDwarfsProbes2014}) might also contribute to the
T and Y sequences.

1D numerical models (e.g.,
\citealt{tsujiDustPhotosphericEnvironment2002,tsujiDustPhotosphericEnvironment2004,tsujiTransitionDwarfsColorMagnitude2003,tsujiUnifiedModelPhotospheres2001};
\citealt{allardLimitingEffectsDust2001};
\citealt{ackermanPrecipitatingCondensationClouds2001};
\citealt{saumonEvolutionDwarfsColorMagnitude2008};
\citealt{marleyPatchyCloudModel2010}) with silicate clouds could
explain the redward sequence of L dwarfs. Cloud-free models agree
with the blueward sequence of T dwarfs (also plotted in
Fig.~\ref{figcloud}). The very red colors of very
low-gravity objects (VL-G, see Fig.~1) are still difficult to
explain (e.g., \citealt{charnaySelfconsistentCloudModel2018}). In
{a} low-gravity environment, how could the cloud
particles be lofted in the very high photosphere resulting in red
color? One possible explanation is that those cloud
particles are tiny so that sedimentation is not{ as}
efficient {as} vertical mixing, but they are
still big enough to affect the near-IR emission. However, a detailed microphysical
model with a more realistic treatment of the particle size
distribution and cloud radiative feedback is needed to investigate
this possibility in the VL-G regime.

A big unsolved puzzle is the L/T transition where brown
dwarfs {exhibit} almost the same effective temperature
($\sim1400\pm200$\,K), but the{ir} color changes abruptly from
red to blue (\citealt{kirkpatrickNewSpectralTypes2005}). While for
directly imaged planets, it is not obvious if a similar sharp
transition exists or not (Fig.~\ref{figcloud}, also see Fig.~\ref{figcmd}). Traditional 1D models
predict much more gradual color change than the observed sharp L/T
transition on brown dwarfs. The hypotheses with clouds to explain
the sharp L/T transition can be grouped into two main categories.
The first one can be called the ``rain out" or ``downpour" scenario
(e.g., \citealt{knappNearInfraredPhotometrySpectroscopy2004};
\citealt{tsujiTransitionDwarfsColorMagnitude2003}
\citealt{tsujiDustPhotosphericEnvironment2004};
\citealt{burrowsDwarfModelsTransition2006};
\citealt{cushingAtmosphericParametersField2008};
\citealt{saumonEvolutionDwarfsColorMagnitude2008};
\citealt{stephens814UpmuSPECTRA2009}). During the L/T transition,
the cloud particle size changes and might cause a sudden cloud drop
and clear the atmosphere.
\cite{saumonEvolutionDwarfsColorMagnitude2008} combined a cloud
model from \cite{ackermanPrecipitatingCondensationClouds2001} with
planetary evolutionary models to simulate the near-IR
{CMD}. In the
\cite{ackermanPrecipitatingCondensationClouds2001} model, a
parameter called $f_{\rm sed}$ is prescribed as the efficiency of
sedimentation compared with vertical mixing. If
$f_{\rm sed}$ is kept constant for all objects, the color change is
too gradual to explain the sharp L/T transition.
\cite{saumonEvolutionDwarfsColorMagnitude2008} changed the $f_{\rm
sed}$ rapidly during the L/T transition to simulate the change of
particle size. They could successfully reproduce the near-IR
{CMD}. However, the detailed
microphysics of this cloud particle size evolution has not been
elucidated.

The second hypothesis can be called ``patchy cloud" scenario
(e.g.,\citealt{ackermanPrecipitatingCondensationClouds2001};
\citealt{burgasserEvidenceCloudDisruption2002};
\citealt{marleyPatchyCloudModel2010}), which might be analogous to
the belt-zone structure and 5-micron hot spots in the NEB on
Jupiter. If the cloud fraction in the atmosphere changes rapidly
during the L/T transition, such as dust breakup forming cloud holes,
hot air emits from the deep atmosphere and the disk-averaged
color could suddenly transition blueward. This patchy cloud scenario
could also produce large rotational light curves of brown dwarfs,
consistent with observations. Moreover, light curve
variability is often observed in the L/T transition objects,
suggesting a pretty dynamic weather pattern regime. Recent
observations (e.g.,
\citealt{apaiHSTSPECTRALMAPPINGTRANSITION2013,apaiZonesSpotsPlanetaryscale2017})
suggested other mechanisms than cloud holes to cause the light curve
variability, such as thin-thick cloud distribution, spots and
trapped waves (see discussion in Sect.~\ref{sect:rotlight}). Whether
these mechanisms could lead to a rapid color change during the L/T
transition has yet to be investigated. A thorough understanding of
the patchy cloud scenario requires a 3D convective model with cloud
radiative feedback (e.g.,
\citealt{tanAtmosphericCirculationBrown2020}).

Alternatively, cloud-free models have been proposed to explain brown dwarf spectra. Atmospheric retrieval work found that a
nearly isothermal photosphere could explain the muted near-IR features
in L dwarfs (e.g.,
\citealt{burninghamRetrievalAtmosphericProperties2017}). The problem
is that an isothermal atmosphere might strongly violate
convective-radiative equilibrium in brown dwarf atmospheres such as
in \cite{ackermanPrecipitatingCondensationClouds2001}. With strongly
pressure-dependent opacities like H$_2$-H$_2$ {collision-induced
absorption (}CIA{)} and broadening of various molecular and atomic
(like K and Na) lines, the atmospheric lapse rate is usually large.
It was recently proposed that fingering convection
(\citealt{tremblinFINGERINGCONVECTIONCLOUDLESS2015,
tremblinCloudlessAtmospheresYoung2017}) or thermo-chemical
instability (\citealt{tremblinCLOUDLESSATMOSPHERESDWARFS2016}) might
cause the shallower temperature gradient.
\cite{tremblinCLOUDLESSATMOSPHERESDWARFS2016} also claimed that
cloud-free models could explain the spectral sequence as the result
of thermochemical instabilities in the CO/\ch{CH4} transition in the
case of the L/T boundary and the \ch{N2}/\ch{NH3} transition in the
case of the T/Y boundary. The details of this mechanism have not
been completely worked out. As noted in Section~\ref{sect:rotlight},
it looks{ like} this mechanism could not be responsible for the
light curve variability seen in the L/T transition because the
observed light curves and their variability do not show very
different behaviors between the gas absorption cores and the outside
continuum. To date, photospheric clouds and cloud variability remain
the most probable mechanism for the observed rapid L/T color change
and the light curve variability. This proposed cloud-free mechanism
is still under debate. See more discussion on atmospheric dynamics
and convection in Section~\ref{sect:bddyn}.

\subsubsection{Spectral Trends on Close-in Exoplanets}
\label{sect:planetcloud}

Statistical clear-to-cloudy trends for transiting exoplanets
beg{a}n to emerge thanks to the increasing number of
observed transmission spectra. \cite{singContinuumClearCloudy2016}
found that the near-IR water feature amplitude is correlated with
two spectral indices. The first one is the relative strength between
optical scattering and near-IR absorption. The
second one is the relative absorption strength between the near-IR
and mid-IR. These observations suggest that clouds play an
important, systematic role in shaping the transmission spectra. They
also pointed out that hot Jupiters do not exhibit a strong
relationship between temperature and cloud signatures,
whereas brown dwarfs have a very obvious spectral sequence.
Figure~\ref{figcmd} shows the hot Jupiters on top of the brown dwarf
near-IR {CMD}. The color of hot
Jupiters has a much larger scatter, indicating their cloud
formation is more complicated and diverse than brown dwarfs.
\cite{singContinuumClearCloudy2016} attributed the reason to the
difference {in} vertical temperature structures between
hot-Jupiter atmospheres and {those} on brown dwarfs.
Because of intense stellar irradiation, hot Jupiters possess much
steeper pressure-temperature profiles than {do} field
brown dwarfs. However, because cloud condensation curves are also
steep (i.e., stronger dependence of pressure than temperature, see
Fig.~\ref{figtprof}), a small temperature change will significantly
change the cloud base pressure to a much larger extent on hot
Jupiters than on brown dwarfs. Also, due to the diversity of
planetary metallicity, gravity and radiative properties, the
cloud materials could be cold trapped at depth (at $\sim$1--100 bar)
on some hot-Jupiters but not the others (e.g.,
\citealt{parmentier3DMixingHot2013};
\citealt{powellFormationSilicateTitanium2018}), which would also
increase the cloud variability in exoplanetary photospheres.

Planetary clouds are diverse and complicated,
but current hot Jupiter data do suggest some possible
cloudiness trend as a function of temperature (e.g.,
\citealt{stevensonQuantifyingPredictingPresence2016};
\citealt{fuStatisticalAnalysisHubble2017};
\citealt{gaoAerosolCompositionHot2020}). Using a larger size of
samples, \citealt{fuStatisticalAnalysisHubble2017} found that the
near-IR water spectral strength $A_H$---defined as the transit depth
difference between the near-IR water band and the underlying continuum
in units of the scale height---increases with the equilibrium
temperature $T_{\rm eq}$ from 500--2500\,K. A further analysis in
\cite{gaoAerosolCompositionHot2020} {indicates} a
seemingly non-monotonic trend among hot Jupiters. $A_H$ increases
with $T_{\rm eq}$  when $T_{\rm eq} < 1300 $\,K and $T_{\rm eq}  >
1600 $\,K while the opposite trend seems to exist for planets
located within 1300\,K $ < T_{\rm eq}  < 1600 $\,K
(Fig.~\ref{figcloud}).

Using a 1D aerosol microphysics model,
\cite{gaoAerosolCompositionHot2020} proposed a mechanism for the
non-monotonic $A_H-T_{\rm eq}$ trend on hot Jupiters
(Fig.~\ref{figcloud}). They showed that aerosol opacity in
the HST WFC3 channel is dominated by silicates when $T_{\rm eq} >
950 $\,K, while iron and sulfur clouds do not form efficiently due
to their higher nucleation energy barriers. The kinetic model
results are different from the prediction from thermochemical
models. The atmospheres are relatively clear when $T_{\rm
eq}>2200$\,K, which is too hot for global-scale silicate clouds to
form, although clouds might still{ be} present on the nightside
(see the last paragraph of this section). As $T_{\rm eq}$ decreases,
the formation of high-altitude silicate clouds increases the
cloudiness. Meanwhile, as the planets get cold, the cloud layers
also progressively move to the deeper atmosphere, resulting in
relatively clear atmospheres. Below 950\,K, due to rising methane
abundances and photodissociation rates, high-altitude photochemical
hazes form and damp the near-IR water features. The future search of possible
spectral features of the aerosols at longer wavelengths, such as the silicate feature
at 10 microns, is the key to test this hypothesis
(e.g., \citealt{ormelARCiSFrameworkExoplanet2019};
\citealt{powellTransitSignaturesInhomogeneous2019};
\citealt{gaoAerosolCompositionHot2020}).

The dominant role of silicate clouds in the high-temperature range
echoes the earlier work {o}n brown dwarfs (e.g.,
\citealt{saumonEvolutionDwarfsColorMagnitude2008};
\citealt{marleyPatchyCloudModel2010}) that {utilizes}
the silicate clouds to explain the spectral sequence evolution and
L/T transition. The above hot Jupiter cloudiness trend (the right
panel {i}n Fig.~\ref{figcloud}) seems to share
similarities with the near-IR {CMD}
of brown dwarfs (the left panel on Fig.~\ref{figcloud}), despite
that the former diagnoses the transmission properties of the
atmospheres and the latter probes the emission. First, both data
sets are not consistent with the clear-sky models in the
high-temperature regime. Second, the 39 hot Jupiter samples seem to
also exhibit a turn in the cloudiness index $A_H$ at around $T_{\rm
eq}\sim1400$\,K, which is reminiscent of the brown dwarf L/T
transition although the turn of the exoplanet curve looks weaker.
Note that 1400\,K is also roughly the effective temperature when the
brown dwarfs change their self-emission color (also related to the
cloud opacity). {Such} similarity might not be a
coincidence. Instead, it looks like a smoking gun that both the
brown dwarf L/T transition and hot Jupiter cloudiness trend share
some common behaviors of high-temperature clouds, although the
underlying mechanism has yet to be investigated in detail. On
the other hand, it would also be interesting to analyze if there is
any statistical trend in the reflective spectra on those hot
Jupiters because silicate clouds could significantly increase planetary albedo (e.g.,
\citealt{marleyReflectedSpectraAlbedos1999}).

Cooler and smaller planets might also show some statistical
clear-to-cloudy trends with equilibrium temperature
(Fig.~\ref{figcloud}). From hot to cold, these planets include
HAT-P-26 b, HD 106315 c, HAT-P-11 b, HD 97658 b, GJ 436 b, GJ 3470
b, GJ 1214 b, Kepler 51 b, Kepler 51 d, K2-25 b, and K2-18 b. Among
those, HAT-P-26 b shows a relatively clear atmosphere
(\citealt{wakefordHATP26bNeptunemassExoplanet2017}). HD 106315 c
{displays} some weak water feature in the near IR
band(\cite{kreidbergTentativeEvidenceWater2020}). HAT-P-11 b
{manifests} water features
(\citealt{fraineWaterVapourAbsorption2014}) but might be partially
cloudy{, as} indicated by a nearly flat optical transmission
spectrum from HST STIS recently
(\citealt{chachanHubblePanCETStudy2019}). HD 97658 b
(\citealt{knutsonHubbleSpaceTelescope2014}), GJ 436 b
(\citealt{knutsonFeaturelessTransmissionSpectrum2014}), GJ 3470 b
(\citealt{ehrenreichNearinfraredTransmissionSpectrum2014}),
and GJ 1214 b (\citealt{kreidbergCloudsAtmosphereSuperEarth2014})
are cloudy. Super-puffs Kepler 51 b and d show very flat
transmission spectra in the near IR
(\citealt{libby-robertsFeaturelessTransmissionSpectra2020}), perhaps
due to aloft tiny dust particles
(\citealt{wangDustyOutflowsPlanetary2019}) or high-altitude
photochemical hazes (\citealt{gaoDeflatingSuperpuffsImpact2020}).
The current data o{n} K2-25 b are consistent with a flat
spectrum (\citealt{thaoZodiacalExoplanetsTime2020}), implying a
cloudy atmosphere or a high-molecular-weight atmosphere. The coolest
one, K2-18 b ($T_{\rm eq}\sim$255{\,}K),
{exhibits} water features in the near-IR but could{
also} be partially cloudy
(\citealt{bennekeWaterVaporClouds2019};
\citealt{tsiarasWaterVapourAtmosphere2019}). It looks that the
atmosphere might become clear again when the temperature drops below
about 400{\,}K. Is there another non-monotonic trend from
1000{\,}K to 300{\,}K? Using the six planets
in this list, \cite{crossfieldTrendsAtmosphericProperties2017} first
hypothesized that the water band amplitude changes either with the
hydrogen and helium mass fraction or the equilibrium temperature.
\cite{fuStatisticalAnalysisHubble2017} analyzed both Jupiter- and
Neptune-size{d} planets together. The entire sample
{exhibits} the $A_H$ dependence on the equilibrium
temperature. With the new {s}uper-puff data,
\cite{libby-robertsFeaturelessTransmissionSpectra2020} revisited
this statistical trend and concluded that the clear-to-cloudy trend
is more consistent with the equilibrium temperature dependence
(Fig.~\ref{figcloud}) instead of the metallicity dependence (see
their fig.~16).

The underlying mechanisms behind this seemingly clear-to-cloudy
trend o{f} equilibrium temperature on cooler and smaller
exoplanets have not been investigated. The mechanism is not likely
the same as the high-temperature ``condensation clouds" scenario
proposed by \cite{gaoAerosolCompositionHot2020} for giant planets as
the clouds (e.g., ZnS or KCl) tend to condense at a deeper
atmosphere as temperature decreases
(\citealt{crossfieldTrendsAtmosphericProperties2017}). It might be
more consistent with the photochemical hazes as methane becomes more
important in the low-temperature regime (e.g.,
\citealt{morleyTHERMALEMISSIONREFLECTED2015};
\citealt{gaoAerosolCompositionHot2020}).
\cite{morleyTHERMALEMISSIONREFLECTED2015} pointed out that
atmospheres on sub-Neptunes change from haze-free to hazy
atmospheres at around 800--1100{\,}K due to the onset of
\ch{CH4} (see their fig.~\ref{figkzz}). In principle, the
photochemical haze formation depends on UV intensities and plasma
environment. Both factors depend on the star-planet distance and are
likely to be positively correlated with the planetary equilibrium
temperature. Therefore the total haze precursors (and presumably the
total haze abundances) decrease as the irradiation level decreases,
{a}s does the temperature if other factors are fixed.
However, the observed data show that colder atmospheres between
400--800\,K appear to be hazier (Fig.~\ref{figcloud}), except for
the coldest one, K2-18 b, where the planetary atmosphere is located
in the habitable zone and appears relatively clear. This planet thus
might be explained by its low UV irradiation level.
\cite{crossfieldTrendsAtmosphericProperties2017} conjectured that
variations in haze formation altitude (cloud top) could play a role.
Systematic, microphysical modeling of photochemical haze formation
from 300--1000\,K that takes into account variations in the rates
and locations of haze production, haze transport and the
impact of condensate clouds is needed to understand any possible
trends better and explore the role of parameters such as
metallicity, temperature and stellar UV fluxes.

Inhomogeneous aerosol distributions on exoplanets also impact
observations on tidally locked exoplanets. The detailed
microphysical simulations in
\cite{powellFormationSilicateTitanium2018,powellTransitSignaturesInhomogeneous2019}
imply that hot Jupiters might have very distinct transmission
spectra between the eastern and western limbs: the eastern one has
sloped spectra, and the western has flatter spectra. It remains to
be confirmed because the current techniques can only observe limb-averaged spectra. Inhomogeneous aerosol coverage would
cause a distorted transit light curve due to the different
absorption radius on the eastern and western limbs (e.g.,
\citealt{lineInfluenceNonuniformCloud2016};
\citealt{kemptonObservationalDiagnosticDistinguishing2017};
\citealt{powellTransitSignaturesInhomogeneous2019}).

As noted in Section \ref{sect:rotlight}, orbital phase curves and
rotational light curves are also heavily modulated by clouds on
brown dwarfs and exoplanets. Clouds emit {IR}
light as well as scatter and reflect incoming starlight, such that
light curves, including both thermal emission and optical
reflection, can be greatly affected by the spatial distribution of
clouds. Light curve observations (e.g.,
\citealt{parmentierTransitionsCloudComposition2016};
\citealt{parmentierExoplanetPhaseCurves2018a}) suggest that clouds
on tidally locked exoplanets are inhomogeneously distributed and
cause phase offset with respect to the secondary eclipse. Therefore,
the simple explanation in Section \ref{sect:rotlight} should be much
more complicated because of the existence of clouds. Besides,
thermal phase curves of some hot Jupiters are difficult to interpret
without invoking clouds on the nightsides (e.g., WASP-43 b,
\citealt{katariaAtmosphericCirculationHot2015};
\citealt{stevensonSPITZERPHASECURVE2017}). On the other hand,
asymmetries in the optical light curve are useful for constraining
particle distributions and properties. The optical light curves of
Kepler-7 b constrain the spatial distribution of aerosols and their
composition and particle size (e.g.,
\citealt{garciamunozProbingExoplanetClouds2015}; also see
\citealt{demoryInferenceInhomogeneousClouds2013};
\citealt{hengUNDERSTANDINGTRENDSASSOCIATED2013};
\citealt{webberEFFECTLONGITUDEDEPENDENTCLOUD2015};
\citealt{estevesCHANGINGPHASESALIEN2015};
\citealt{estevesCHANGINGPHASESALIEN2015};
\citealt{parmentierTransitionsCloudComposition2016}). Several other
exoplanets also exhibit a westward offset bright spot, indicating
cloud reflection (e.g.,
\citealt{shporerStudyingAtmosphereDominatedHot2015}; Kepler-7 b, 8
b, 12 b{ and} 41 b). A recently observed westward offset
bright spot on CoRoT-2 b
(\citealt{dangDetectionWestwardHotspot2018}) at 4.5 microns might
also be related to cloud distributions. {U}nderstand{ing} the inhomogeneous aerosol distribution
on exoplanets requires simulating microphysics coupled with the
dynamical transport of aerosols under various conditions. Recent 3D
model{ing} efforts have made some progresses but could not
explain the data (e.g.,
\citealt{linesExonephologyTransmissionSpectra2018,linesSimulatingCloudyAtmospheres2018}).
However, the lack of laboratory data o{n} the model input
parameters further hinders quantitative conclusions
{about} the mechanisms. Also, although clouds could be
bright and increase the planetary albedo, to date there is no
apparent correlation between the geometric albedo and the incident
stellar flux (e.g., \citealt{cowanStatisticsAlbedoHeat2011};
\citealt{hengUNDERSTANDINGTRENDSASSOCIATED2013};
\citealt{schwartzPhaseOffsetsEnergy2017};
\citealt{zhangPhaseCurvesWASP33b2018};
\citealt{keatingUniformlyHotNightside2019}).

\cite{keatingUniformlyHotNightside2019} derived the dayside and
nightside temperature of hot Jupiters from the Spitzer phase curves
(Fig.~\ref{figdnt}). The nightside brightness temperatures across a
broad range of $T_{\rm eq}$ ($<$2500\,K) are roughly the same
($\sim$1100\,K). Although atmospheric theory predicts that the
nightside temperature could behave more uniform{ly} than the
dayside (see discussion in Sect.~\ref{sect:horitemp}), this trend
might also be explained by the ubiquitous existence of clouds on the
nightside (\citealt{keatingUniformlyHotNightside2019}). If
the clouds---a strong opacity source that affects the emission
temperature---form at roughly the same temperature across the
parameter space, the outgoing thermal flux might just be controlled
by, but not necessarily equal to, the cloud base temperature. This
theory also seems consistent with the above hypothesis from
\cite{gaoAerosolCompositionHot2020} that thick clouds on close-in
gas giants are dominated by a single component, such as silicates.
The silicate condensation temperature is about 1400{\,}K
near the cloud base, depending on the actual pressure-temperature
profile. That also implies the Spitzer channels can only probe the
emission near the thick cloud top, or wherever the cloud opacity
reaches unity, rather than the cloud base emission. Again, future
observations at other thermal wavelengths are needed to disentangle
the contributions of the uniformity of brightness temperature on hot
Jupiters from the dynamical heat transport and that from the
nightside clouds.

\section{Atmospheric Dynamics}
\label{sect:dyn}
\subsection{Fundamentals}
\label{sect:dynfund}

The atmospheric flow pattern is primarily controlled by differential
heating, drag and planetary rotation. The external energy
source comes from the top (i.e., the stellar irradiation) or the
bottom (i.e., convection or surface fluxes outside the domain).
Depending on the energy flux distribution and atmospheric energy
transport processes (e.g., radiation, conduction and
convection), spatially inhomogeneous heating causes temperature
anomalies and pressure gradient and drives the atmospheric movement.
Thus the chemistry of the opacity sources from radiatively active
gas and particles greatly influences the atmospheric dynamics. Drag
exerts the momentum (and energy) exchange with the atmospheric flow
via surface friction (on terrestrial planets), magnetic effect (for
deep and hot ionized flow) or small-scale dissipative viscous
processes.

In a rotating frame, the Coriolis effect plays an important role in
shaping the fluid motion. Consider a deep, convective atmosphere on
a fast-rotating giant planet. Rotation and convection tend to
homogenize the entropy, leading to a barotropic fluid regime---small
density variation on an isobar (constant pressure surface). {The
}Taylor-Proudman theorem (\citealt{houghIXApplicationHarmonic1897};
\citealt{proudmanMotionSolidsLiquid1916};
\citealt{taylorMotionSolidsFluids1917}) predicts that the wind flow
behaves constant as vertical columns paralleled with the rotation
axis and net exchange across the columns is not permitted.
Atmospheric flows move freely with the Taylor columns in the
east-west directions around the rotation axis. On the other hand, in
a shallow, stratified, rotating atmosphere, the horizontal motion is
usually much larger than the vertical{ case} because the
vertical velocity is suppressed due to the large aspect ratio,
vertical stratification and rapid rotation (e.g.,
\citealt{showmanAtmosphericCirculationExoplanets2010}). The
atmosphere is approximately in hydrostatic equilibrium
{o}n a large scale. With appropriate
approximations---hydrostatic, shallow-fluid and traditional
approximations (see
\citealt{holtonDynamicMeteorologyStratosphere2016};
\citealt{vallisAtmosphericOceanicFluid2006}), the equation set
(\ref{basiceq}) introduced in Section~\ref{sect:process} can be
reduced into the so-called ``primitive equations{.}" In
this simplified system, hydrostatic equilibrium implies that the
fluid parcel is incompressible in the pressure coordinate, and
gravity disappears in the equations (e.g.,
\citealt{vallisAtmosphericOceanicFluid2006}). The effect of gravity
is thus only limited in the radiation via determining the column
density and opacity between two pressure levels but not on the fluid
dynamics directly (see numerical examples in
\citealt{katariaAtmosphericCirculationNineHotJupiter2016} with
different gravities).

From the force balance point of view, one can characterize planetary
atmospheric dynamics in several regimes using a dimensionless
number: the Rossby number $\mathrm{Ro}=U/\Omega L$, where $U$ is the
typical wind speed, $L$ is the characteristic length scale of the
atmospheric flow and $\Omega$ is the rotational rate. In a
slow{ly} rotating atmosphere such as on Venus, $\mathrm{Ro}\gg$
1, the horizontal motion is controlled by the balance between the
inertial force (centrifugal force for Venus) and the pressure
gradient, residing in the cyclostrophic regime. As the rotation rate
increases, the Coriolis force becomes as important as the inertial
terms. In the intermediate regime where $\mathrm{Ro}\sim$ 1,
multi-way force balance applies among pressure gradient, Coriolis
force, nonlinear advection and atmospheric drag. For example,
large hurricanes are balanced by the pressure gradient, Coriolis
force and centrifugal force (``gradient wind balance"). In
a fast-rotating atmosphere, $\mathrm{Ro}\ll$ 1, the pressure
gradient tends to be balanced with the Coriolis force, leading to
the geostrophic regime. In this regime, the latitudinal temperature
gradient from equator to pole is associated with a positive zonal
wind shear leading to a faster east (or slower west) zonal wind at
higher altitude. This is called the thermal wind balance.

The characteristic length scale $L$ is important. The typical length
scale is the Rossby deformation radius $L_R\sim NH/\Omega$ where $N$
is the buoyancy frequency (Eq.~(\ref{bfreq})), and $H$ is the
pressure scale height. Flow with length scale larger than the Rossby
deformation radius is influenced by planetary rotation,
whereas small-scale flow is typically affected more by local
processes (such as buoyancy). If one takes the wind scale $U$ as
$NH$, $L_R$ corresponds to a length scale where the $\mathrm{Ro}$ is
equal to one. Rossby deformation radius is a natural length scale of
many atmospheric phenomena such as geostrophic adjustment,
baroclinic instabilities and the interaction of convection
with the environment (\citealt{vallisAtmosphericOceanicFluid2006}).

In a shallow atmosphere, planetary sphericity also plays a role
because the vertical component of the Coriolis force is changing
with latitude $\phi$, characterized by the Coriolis parameter
$f=2\Omega\sin\phi$. The local Rossby number can be written as
$U/fL$. The local Rossby number is larger at lower latitudes and
smaller at higher latitudes. The dynamical regime consequently could
be different from latitude to latitude. Not only does the Rossby
deformation radius change with latitude, but also the horizontal
fluctuations of pressure, density or potential
temperature---they are proportional to each other---might be
controlled by different mechanisms at different latitudes. At low
latitudes, the local $\mathrm{Ro}$ is large; the pressure gradient
is balanced by the inertial term
(\citealt{charneyNoteLargeScaleMotions1963}). The horizontal
potential temperature fluctuation $\theta_h$ is estimated as
\begin{equation}\label{lRo}
\frac{\Delta\theta_h}{\theta}\sim\frac{U^2}{gD}\sim \mathrm{Fr}~~~~~~~~~\mathrm{Ro}>1.
\end{equation}
Here $\mathrm{Fr}$ is the Froude number $ \mathrm{Fr}=U^2/gD$ for a
flow depth of $D$. $\mathrm{Fr}$ can be described as the square of
the ratio of wind speed to the gravity wave speed. It characterizes
the relative strength of the inertia of a fluid particle to the
gravity. In the extratropics or middle-latitude where the local
$\mathrm{Ro}$ is small, geostrophy leads to potential
temperature fluctuation
(\citealt{charneyNoteLargeScaleMotions1963})
\begin{equation}\label{sRo}
\frac{\Delta\theta_h}{\theta}\sim\frac{fUL}{gD}\sim \frac{\mathrm{Fr}}{\mathrm{Ro}}~~~~~~~~~\mathrm{Ro}<1.
\end{equation}
Because the local Rossby number is smaller than one in this regime,
the density perturbation in the middle latitudes is expected to be
larger than{ in} the tropics
(\citealt{charneyGeostrophicTurbulence1971}). For the tropical
regime, one can invoke the WTG approximation as described in
Section~\ref{sect:phasecurve} (e.g.,
\citealt{sobelWeakTemperatureGradient2001}).

From the view of vorticity---the curl of the velocity field,
planetary rotation is a fundamental vorticity in the system. The
vertical component of the planetary vorticity changes with latitude
as the Coriolis parameter $f$ changes. This is called the
$\beta$-effect where $\beta = \partial f/a\partial\phi$ is the
meridional gradient of $f$. Because the fluid parcel tries to
conserve its total vorticity (more precisely, potential vorticity,
see \citealt{holtonDynamicMeteorologyStratosphere2016}), the
vorticity gradient provides a restoring force f{or} the
meridional disturbance, producing Rossby waves. {A }Rossby wave
plays a significant role in the formation of zonal jets
via interaction with the mean flow. This ``eddy-driven" jet
formation mechanism causes multiple jet streams in the middle
latitudes on giant planets and terrestrial planets. The
characteristic length scale of the jet width is naturally related to
$\beta$. \cite{rhinesWavesTurbulenceBetaplane1975} pointed out the
jet width should be scaled as $L_{\rm
jet}\sim\pi(2U_e/\beta)^{1/2}$, where $U_e$ is the eddy velocity
scale. This Rhines scale, although it is primarily from a 2D
turbulent flow argument, could be related to the jet width on the
multiple jets on 3D giant planets. There is another jet width scale
that is more associated with the zonal jet velocity ($U_{\rm jet}$)
and potential vorticity gradients $L_{\rm jet}\sim\pi(2U_{\rm
jet}/\beta)^{1/2}$ (e.g.,
\citealt{williamsPlanetaryCirculationsBarotropic1978};
\citealt{lianDeepJetsGasgiant2008};
\citealt{scottStructureZonalJets2012}). The low latitudes have a
larger $\beta$, leading to larger anisotropy and large waves,
whereas at high latitudes, the inertial advection dominates over the
$\beta$ effect resulting in a more turbulent atmosphere. This has
been demonstrated in 2D shallow-water simulations (e.g.,
\citealt{showmanNumericalSimulationsForced2007};
\citealt{scottEquatorialSuperrotationShallow2008}).

From {an} energetics point of view, the local energy
imbalance from the external or internal sources leads to
fluctuations of temperature and density on isobars that create the
available potential energy (APE)---only a small fraction of the
total potential energy that is then converted into KE. Qualitatively, in the framework of the classical
Lorenz energy cycle
(\citealt{lorenzAvailablePotentialEnergy1955,lorenzNatureTheoryGeneral1967}),
both the APE and KE are partitioned into zonal mean and
eddy (deviation from the zonal mean) components. The
energy cycle starts from the production of the mean APE and eddy
APE, the conversion among the four energy components, and the{
eventual} loss of KE through frictional dissipation. The energy cycle could be complicated and requires a
detailed analysis of the entire system, in particular, radiative energy flow in the atmosphere. One can see the discussion
in \cite{peixotoPhysicsClimate1992} for Earth and
\cite{schubertPlanetaryAtmospheresHeat2013} for other Solar System
planets. For exoplanets, this cycle has not been analyzed in detail
yet. The mean APE can be converted into the mean KE through the
formation of thermally direct, overturning meridional circulation as
well (e.g., \citealt{liLorenzEnergyCycle2007}). For zonal
jet formation, another source of mean KE is from the mean
APE to the eddy APE, then to the eddy KE and eventually the
mean KE. Conversion from the mean to the eddy APE is done by
generation and growth of eddies such as non-axisymmetric waves and
other disturbances through many processes, such as convection, shear
instabilities and baroclinic instability. The conversion from
eddy KE to mean KE is through the eddy momentum
convergence into the mean flow. Those eddy energies can be
cascade{d} into a larger scale through the ``inverse cascade"
process in a quasi-2D regime for a large-aspect-ratio fluid like a
shallow atmosphere, in contrast to the 3D turbulence where
{KE} is cascade{d} into the smaller
scale and eventually lost via viscous dissipation.
{For} detailed discussion{,} refer to
textbooks such as \cite{vallisAtmosphericOceanicFluid2006}.

The atmosphere can be considered to be a heat engine or a
refrigerator. A classic heat engine extracts energy from a hot
region and transfers it to a cold region. In this process, it
converts part of the energy into work. The heat engine efficiency is
the ratio of work it has done to the input heat. In a
convective atmosphere on terrestrial planets (or even a local
weather system such as a hurricane), the flux is carried upward from
the hot boundary layer near the surface and emitted in the top, cold
atmosphere. In this process{,} the atmosphere is doing work to
produce {KE}, which is eventually lost in
frictional or viscous dissipation (e.g.,
\citealt{emanuelAirseaInteractionTheory1986};
\citealt{peixotoPhysicsClimate1992};
\citealt{rennoNaturalConvectionHeat1996};
\citealt{emanuelMoistConvectiveVelocity1996};
\citealt{schubertPlanetaryAtmospheresHeat2013}). The adiabatic
processes in those systems can be analogous to the classic Carnot
engine. The Carnot efficiencies for the terrestrial atmospheres in
the Solar System are estimated to be less than 27.5\%, 13.2\%,
4.4\% and 4.1\% for Venus, Earth, Mars and Titan,
respectively (\citealt{schubertPlanetaryAtmospheresHeat2013}).
On tidally locked terrestrial planets, the
day-night temperature difference in the boundary layer
{i}s large, but the temperature in the free atmosphere is
roughly homogenous because of small wave-to-radiative timescales
(see Sect.~\ref{sect:phasecurve}). As a result, the heat engine mainly works by the
day-night overturning circulation.
\cite{kollTemperatureStructureAtmospheric2016} analyzed this system
and found out the heat engine efficiency could be estimated as
$(T_d-T_{\rm eq})/T_d$ where $T_d$ is the dayside temperature and
$T_{\rm eq}$ is the equilibrium temperature.

Hot Jupiter atmospheres {can} also be regarded as a
heat engine (e.g.,
\citealt{goodmanThermodynamicsAtmosphericCirculation2009};
\citealt{kollAtmosphericCirculationsHot2018}). However, the
atmospheres are highly irradiated from the top and thus highly
stratified. A Carnot cycle, which assumes adiabatic expansion and
compression and isothermal heat addition and removal, is not a good
analogy. Instead, \cite{kollAtmosphericCirculationsHot2018} proposed
that one could approximate hot Jupiter atmospheres using
the Ericsson cycle, which assumes isothermal expansion and
compression and isobaric heat addition and removal. The heat engine efficiency of the Ericsson cycle is always smaller
than{ that of} the Carnot cycle. However, note that the heat engine concept
is just a crude analogy. Circulation in some parts of the atmosphere
could behave as a refrigerator, a reverse model of a heat engine. It
might occur in those thermally indirect circulations (forced
motions), for example, the wave-forced circulation in the lower
stratosphere (e.g.,
\citealt{newellStratosphericEnergeticsMass1964}). The
anti-Hadley-like behavior in the equatorial region seen in some
dynamical models on tidally locked planets (e.g.,
\citealt{charnay3DModelingGJ1214bs2015}) might also act more like a
refrigerator, rather than a classic heat engine.

\begin{figure*}
   \centering
   \includegraphics[width=0.85\textwidth, angle=0]{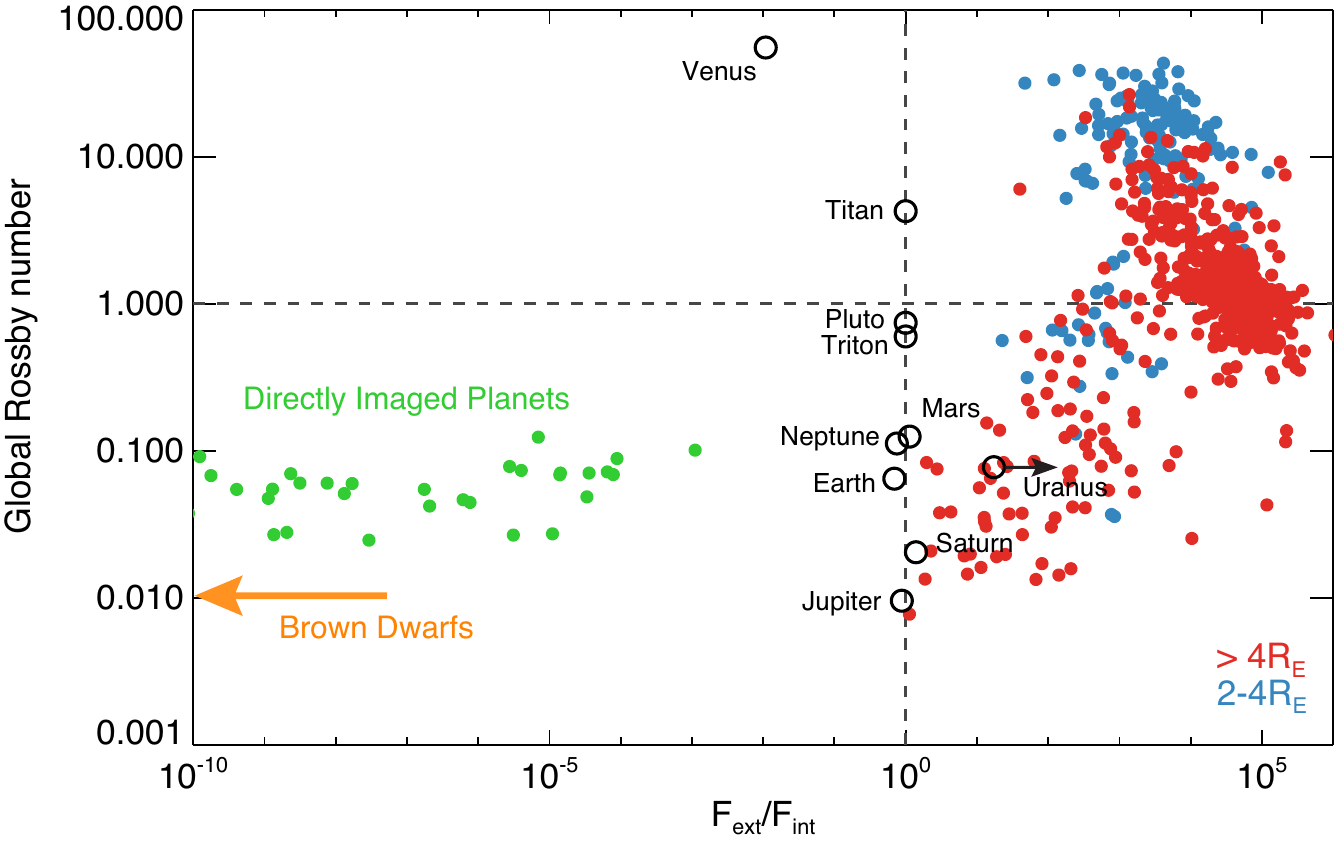}
   \caption{\baselineskip 3.8mm Classification of substellar atmospheric dynamics using global Rossby number $\mathbbm{Ro}=U/\Omega R_p$ and the ratio of the external to internal fluxes ($F_{\rm ext}/F_{\rm int}$, see Eq.~(\ref{rcb})). For terrestrial atmospheres in the Solar System, the internal fluxes (surface fluxes) are calculated {utilizing} blackbody emission based on the surface temperature. The Uranus value is the upper limit from Voyager (\citealt{pearlAlbedoEffectiveTemperature1990}). For exoplanets, only planets with size larger than two Earth radii are considered. All sub-Neptunes are assumed to have hydrogen atmospheres. For self-luminous, directly imaged exoplanets, we calculated their internal fluxes by subtracting incoming stellar fluxes from their observed emission fluxes. For close-in exoplanets, the internal fluxes are difficult to estimate and might correlate with their ages. Here we made an assumption (probably an oversimplification) {by applying} the current internal flux of Jupiter (7.485{\,}$\mathrm{W{\,}m^{-2}}$) from \cite{liLessAbsorbedSolar2018}. For planets less than 0.2{\,}AU from their host stars, we assume they are tidally locked; for those located more than 0.2{\,}AU, we estimated the rotational period using the mass scaling $v_e\sim v_0(M/M_J)^{1/2}$, where $v_0=10{\,}\mathrm{km{\,}s^{-1}}$ and $M_J$ is the mass of Jupiter. Field brown dwarfs and rogue planets without host stars have no external fluxes. Global Rossby numbers of Solar System bodies are calculated based on realistic winds, but that of the exoplanets are estimated from the isothermal sound speed using equilibrium temperature for tidally locked planets and effective temperature (based on their observed luminosity) for non-tidally locked planets (see text).}
   \label{figrof}
   \end{figure*}

Because both forcing and rotation play key roles in atmospheric
motion, in Figure~\ref{figrof} we classify the atmospheres in terms
of the two parameters. First, {utilizing} the ratio
of external stellar flux to internal flux or
surface flux from below, we can classify the atmospheres into three
regimes (\citealt{showmanIlluminatingBrownDwarfs2016}): externally
forced, internally forced and forced by both external and internal
sources. Most close-in planets such as tidally locked hot Jupiters
and sub-Neptunes are mainly forced by the external source from the
central star. Most brown dwarfs and directly imaged planets are
mainly forced from their internal fluxes. For
planets and brown dwarfs located {at an}
intermediate distance from the star, both the stellar forcing and
internal flux are important. All Solar System atmospheres
seem to fall in this regime (Fig.~\ref{figrof}) but with different
reasons. We consider the surface flux as the internal flux for
terrestrial atmospheres, comparable to the external solar flux. For
giant planets and brown dwarfs, low-mass, self-luminous, substellar
evolution models show that their internal luminosity highly depends
on their mass and age (e.g., \citealt{burrowsTheoryBrownDwarfs2001};
\citealt{phillipsNewSetAtmosphere2020}). As these objects become
older, their radii decrease very slowly after about 1{\,}Gyr due to Coulomb and electron degeneracy, but their
internal luminosity continues decreasing via radiative cooling to
space. The giant planets in the Solar System happen to have a
roughly similar magnitude of external and internal fluxes (maybe
except Uranus) at their current ages. {Y}oung hot
Jupiters and highly irradiated brown dwarfs could also lie in this
regime, for example, the recently discovered close-in brown dwarf
rotating around a white dwarf (J1433,
\citealt{santistebanIrradiatedBrowndwarfCompanion2016}).

To first order, if an atmosphere is mostly irradiated by
external flux from the top, the photosphere is stably
stratified ($N^2>0$). {This} is generally the case for
close-in exoplanets such as hot Jupiters. However, as noted in
Section \ref{sect:verttemp}, inflated hot Jupiters could
have much higher internal heat than non-inflated
Jupiter{s} (e.g.,
\citealt{thorngrenIntrinsicTemperatureRadiative2019}). As a result,
their {RCBs} could lie in
the photospheres (Eq.~(\ref{rcb})). In this case, convection must
also be taken into account to understand the dynamics of inflated
hot Jupiters. On the other hand, if an atmosphere is mostly forced
by internal heat, convection could dominate the
atmospheric behavior, at least in the deep atmosphere. However, in the upper atmosphere where it is optically thin,
the atmosphere could still{ be} stably stratified, but could also be significantly perturbed by
upward propagating waves from below (e.g.,
\citealt{showmanAtmosphericCirculationBrown2019}). For an atmosphere with
both important external and internal fluxes such as Jupiter, the
dynamical nature might be more complicated. For example, it was
hypothesized that equatorial superrotation on Jupiter is
produced by upward propagating Rossby waves (e.g.,
\citealt{schneiderFormationJetsEquatorial2009}) generated by the
internal heat flux where moist convection associated with
water condensation could play a vital role (e.g.,
\citealt{lianGenerationEquatorialJets2010}). However, the
off-equatorial jets might be produced from baroclinicity induced
by differential heating with latitude (e.g.,
\citealt{liuMechanismsJetFormation2010}).

To characterize the rotational effect of the entire planet, we
define a ``global Rossby number" $\mathbbm{Ro}=U/\Omega R_p$ using
the planetary radius $R_p$ as the length scale $L$. Adopting typical
wind speeds on Solar System bodies, in Figure \ref{figrof}
we show that slowly rotating planets such as Venus (may also include
Titan) are in the ``tropical regime" ($\mathbbm{Ro} \gg$ 1). Earth,
Mars and giant planets are in the ``geostrophic regime"
($\mathbbm{Ro} \ll$ 1). Triton and Pluto fall in the intermediate
regime ($\mathbbm{Ro} \sim$ 1). To estimate $\mathbbm{Ro}$ of
exoplanets and brown dwarfs, here we approximate $U$ using a typical
isothermal sound speed $(RT_{\rm eq})^{1/2}$. Thus the global Rossby
number is a ratio of the isothermal sound speed to the equatorial
velocity $\Omega R_p$ of the planet. $\mathbbm{Ro}$ is also the
inverse of the dimensionless number $\Omega \tau_{\mathrm{dyn}}$ we
introduced in Section \ref{sect:phasecurve}. When we
{utilize} the isothermal sound speed to approximate the
wind velocity, $\mathbbm{Ro}$ is also similar to the ``WTG
parameter" ($\Lambda=c_0/\Omega R_p$, see
Sect.~\ref{sect:phasecurve}) introduced in
\cite{pierrehumbertAtmosphericCirculationTideLocked2019} for
terrestrial planets. If $\Lambda \gg$ 1, we expect a global WTG
behavior, i.e., weak horizontal temperature gradients on the entire
planet. On the other hand, temperature gradients are strong in the
regime of $\Lambda\ll$ 1. If $\Lambda$ is order unity, one
expects WTG behavior near the equator, but strong
temperature gradients in the extratropics. Note that some studies
also {apply} the thermal wind expression $U\sim
R\Delta\theta/\Omega R_p$ to estimate the wind speed so that the
Rossby number is redefined as a thermal Rossby number (e.g.,
\citealt{mitchellTransitionSuperrotationTerrestrial2010};
\citealt{wangComparativeTerrestrialAtmospheric2018}).

To estimate the rotation rate $\Omega$ of the planets, we first
consider planets that are not greatly slowed down by the tidal
effect. The equatorial velocity $v_e$ of giant planets and brown
dwarfs seem{s} to follow an empirical scaling law with the
planetary mass (e.g., \citealt{snellenFastSpinYoung2014};
\citealt{allersRADIALROTATIONALVELOCITIES2016};
\citealt{bryanConstraintsSpinEvolution2018})
\begin{equation}\label{vscal}
v_e\sim v_0(\frac{M}{M_J})^{1/2},
\end{equation}
where $v_0=10{\,}\mathrm{km{\,}s^{-1}}$ and
$M_J$ is the mass of Jupiter. Thus the global Rossby number can be
empirically expressed as
\begin{equation}
\mathbbm{Ro}\sim 0.08(\frac{H}{H_J})^{1/2}(\frac{R_J}{R_p}),
\end{equation}
where $H$ is the pressure scale height and $H_J =
25{\,}\mathrm{km}$ is roughly the pressure scale height
of Jupiter's upper troposphere. $R_J$ is the radius of Jupiter. As
{visible} in Figure~\ref{figrof}, self-luminous brown
dwarfs almost certainly lie in the geostrophic regime. For young,
hot giant planets, even though the scale height could be ten times
larger than Jupiter's, $\mathbbm{Ro}$ could be smaller than unity
for a large range of temperatures. For terrestrial planets in the
habitable zone, $\mathbbm{Ro}$ is generally smaller than unity, even
for a low-mass small planet like Mars with a nitrogen atmosphere.

For a synchronously rotating planet, the rotation period is the same
as the orbital period and related to the equilibrium temperature via
Kepler's third law. In this regime, the global Rossby
number is
\begin{equation}
\mathbbm{Ro}\sim(\frac{a}{0.03~\mathrm{AU}})^{5/4}(\frac{R_J}{R_p})(\frac{m_a}{m_H})^{1/2},
\end{equation}
where $a$ is the semi-major axis and $m_a/m_H$ is the ratio
of mean molecular mass of the atmosphere to
hydrogen. The global Rossby number does not depend on the stellar
mass if we use the stellar mass-luminosity relationship $L\propto
M^4$. The reason is that, given the same semi-major axis, as the
stellar mass increases, the planetary orbital period
decreases and the stellar luminosity increases. Both the
planetary rotation rate and equilibrium temperature
increase; their effects almost cancel out in the global Rossby
number.

As Figure \ref{figrof} {demonstrates}, $\mathbbm{Ro}$
is around unity for a hot Jupiter in a typical 3-day orbit. For
smaller planets such as hot sub-Neptunes, $\mathbbm{Ro}$ could be
larger, and many of their atmospheres are in the tropical regime.
{However,} those with fast rotation (e.g., very
close-in planets) lie in the geostrophic regime (Fig.~\ref{figrof}).
Also, for Earth-like planets with an atmosphere of heavier
molecules, the global Rossby number (and thus the WTG parameter
$\Lambda$) could exceed unity if the small planet is relatively far
from the star and is still tidally-locked. For example, GJ 1132 b
and LHS 1140 and the Trappist I planets in the habitable zone might
have global Rossby numbers larger than unity and thus are in the
tropical regime (the WTG regime in
\citealt{pierrehumbertAtmosphericCirculationTideLocked2019}). This
regime is different from the mid-latitude climate on Earth in which
geostrophy is important but could resemble the
tropics{ on Earth} where the WTG approximation is applicable
(e.g., \citealt{sobelWeakTemperatureGradient2001}).

In the following discussion, we will summarize our understanding of
two specific populations of exoplanets and brown dwarfs in terms of
their forcing pattern. The first category is the close-in, highly
irradiated gaseous planets such as hot Jupiters and sub-Neptunes,
warm Jupiters and warm Neptunes. The observable atmospheres
of this type are mostly stably stratified. The second population is
weakly irradiated planets that{ are} locate{d} far away from
their host stars, such as directly imaged planets and brown dwarfs,
on which the internal flux plays an important role. We will focus
more on the convective nature of this category. For highly
irradiated brown dwarfs or young hot Jupiters with comparable
external and internal fluxes, we only briefly discuss here due to
the lack of sufficient constraints from observations yet. Again,
note that most Solar System planets are in this regime
(Fig.~\ref{figrof}). Lastly, we will briefly discuss the terrestrial
planets in the habitable zone and highlight the uniqueness of this
climate regime in the presence of liquid water.

\subsection{Highly Irradiated Planets}
\label{sect:hjdyn}

The most famous examples in the highly irradiated exoplanet
population are the synchronously rotating planets locked by stellar tides. The observational characterization of this type
has been discussed in Section \ref{sect:horitemp}. For the flow pattern, one can infer the jet speed (dynamical
timescale) by analyzing the temperature distribution, as revealed by
the hot spot phase shift in the thermal phase curve (e.g.,
\citealt{showmanAtmosphericCirculationTides2002};
\citealt{knutsonMapDaynightContrast2007}), but the
presence of clouds greatly complicates the thermal emission. Eclipse mapping
techniques (e.g., \citealt{rauscherEclipseMappingHot2007};
\citealt{dewitConsistentMappingDistant2012}) could also be useful to
map the spatial inhomogeneity on those distant objects. On the other
hand, directly probing the wind speed on those planets is possible
using Doppler techniques (e.g.,
\citealt{snellenOrbitalMotionAbsolute2010};
\citealt{showmanDopplerSignaturesAtmospheric2013}). An ultra-high
resolution cross-correlation method has been applied for a specific
atom or a molecule (such as CO, Mg and Fe) to measure the
planet radial velocity and even wind-induced redshift/blueshift
for both transiting (e.g., \citealt{snellenOrbitalMotionAbsolute2010};
\citealt{dekokDetectionCarbonMonoxide2013};
\citealt{birkbyDetectionWaterAbsorption2013};
\citealt{brogiDETECTIONMOLECULARABSORPTION2013};
\citealt{wyttenbachSpectrallyResolvedDetection2015};
\citealt{loudenSpatiallyResolvedEastward2015};
\citealt{brogiROTATIONWINDSEXOPLANET2016};
\citealt{birkbyDiscoveryWaterHigh2017};
\citealt{salzDetectionHeLambda2018};
\citealt{flowersHighresolutionTransmissionSpectrum2019};
\citealt{ehrenreichNightsideCondensationIron2020})
and non-transiting planets (e.g., \citealt{brogiSignatureOrbitalMotion2012};
\citealt{rodlerWEIGHINGNONTRANSITINGHOT2012};
\citealt{brogiCarbonMonoxideWater2014}). For example, a blueshift of
$2 \pm 1{\,}\mathrm{km{\,}s^{-1}}${ }was
reported on HD 209458{ }b {utilizing} the CO
lines by \cite{snellenOrbitalMotionAbsolute2010} and
{considering} both CO and \ch{H2O} by
\cite{brogiROTATIONWINDSEXOPLANET2016}. A blueshift of several
$\mathrm{km{\,}s^{-1}}$ was detected on HD 189733 b
utlizing the atomic sodium doublet (\citealt{wyttenbachSpectrallyResolvedDetection2015}),
the \ch{H2O} and \ch{CO} infrared lines (\citealt{brogiROTATIONWINDSEXOPLANET2016}, \citealt{flowersHighresolutionTransmissionSpectrum2019}),
and the Helium I triplet (e.g., \citealt{salzDetectionHeLambda2018}).

{Employing} time-resolved ultra-high
resolution spectra, one can even derive the wind speed at separate
limbs on tidally locked planets. For example, on HD 189733 b,
\cite{loudenSpatiallyResolvedEastward2015} resolved a redshift of
$2.3^{+1.5}_{-1.3}{\,}\mathrm{km{\,}s}^{-1}$
on the leading limb and a blueshift of
$5.3^{+1.4}_{-1.0}{\,}\mathrm{km{\,}s}^{-1}$
on the trailing limb, suggesting an equatorial super rotating jet.
The entire 3D wind structure on this planet could be complicated.
A recent reanalysis showed that the sodium doublet data on HD 189733 b
are consistent with a super-rotating wind, a day-to-night flow,
or a very strong vertical wind (close to the escape velocity)
in the upper atmosphere (\citealt{seidelWindChangeRetrieving2020}).
The vertical flow structure could be better constrained by future
observations in a broader range of wavelengths. {However,}
cloud condensation of those metals could
limit the application of this technique using metals as the tracers.
A recent effort on an
ultra-hot Jupiter WASP-76 b only detected the wind speed on the
trailing limb from iron lines. {I}ron vapor was not
detected due to a significant depletion {i}n the
nightside and around the morning terminator, probably a result of
cloud condensation in the lower atmosphere
(\citealt{ehrenreichNightsideCondensationIron2020}).

The day-night temperature difference and wind speed can be estimated
using the scaling equation set (\ref{tscal}) in
Section~\ref{sect:phasecurve}. The thermal phase offset for a clear
atmosphere was also estimated {utilizing} kinematic
wind transport. Nevertheless, these scaling theories do not provide
insights on the detailed mechanisms of the dynamics, such as the
origin of the equatorial superrotating jet, development of the
day-night flow pattern, wave-adjustment dynamics, eddy-eddy
interaction, eddy-mean flow interaction and turbulent energy
transfer. The weather of tidally locked planets is further
complicated owing to the interaction between dynamics and
radiation, chemistry, cloud microphysics and electromagnetic
field. {For a} detailed review of the
dynamics on tidally locked giant planets{,} refer to
\cite{showmanAtmosphericCirculationExoplanets2010},
\cite{hengAtmosphericDynamicsHot2015} and
\cite{showmanATMOSPHERICDYNAMICSHOT2020} and that for terrestrial
planets refer to \cite{showmanAtmosphericCirculationTerrestrial2013}
and \cite{pierrehumbertAtmosphericCirculationTideLocked2019}.

\begin{figure*}
   \centering
   \includegraphics[width=0.9\textwidth, angle=0]{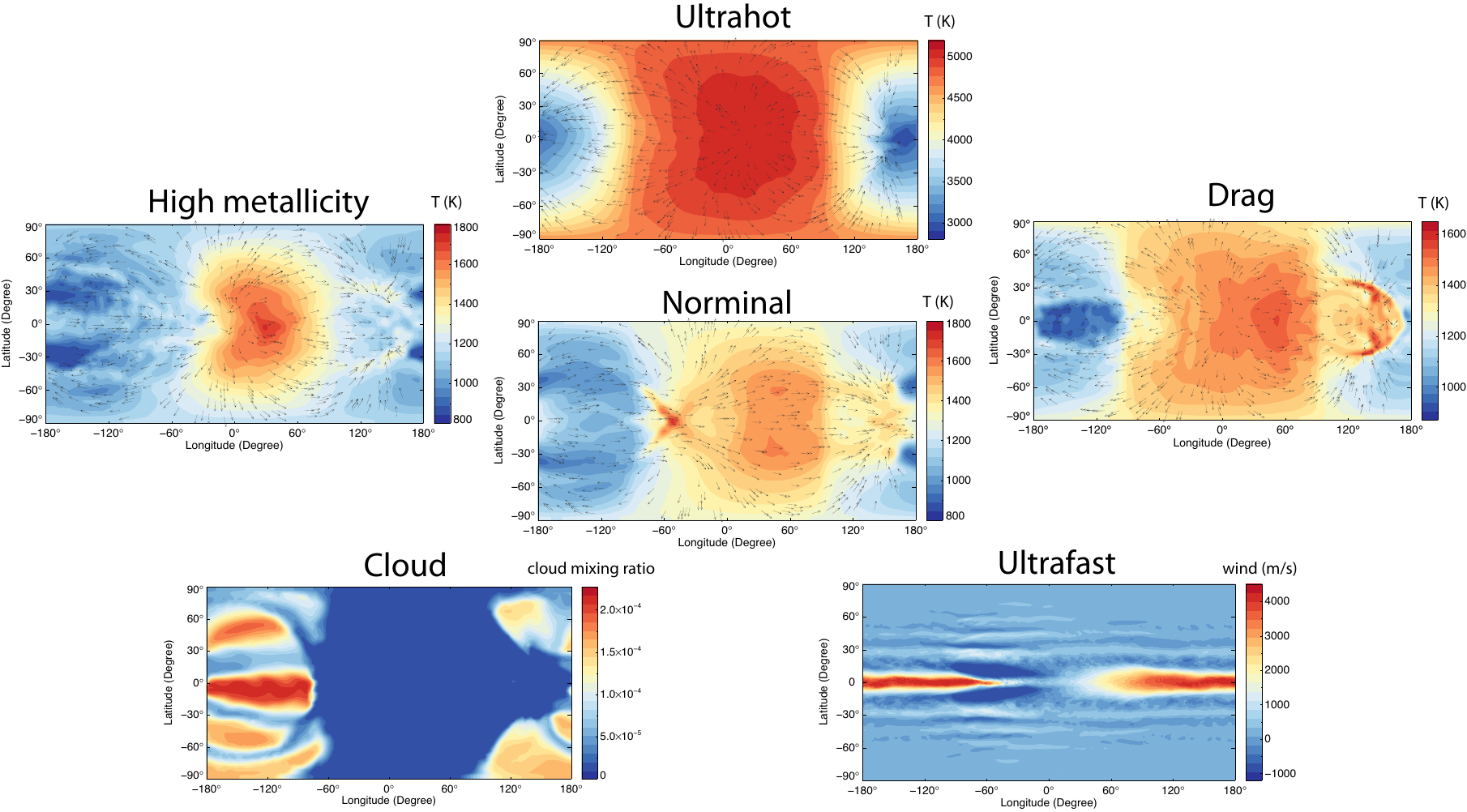}
   \caption{\baselineskip 3.8mm Typical patterns of tidally locked giant planets for six dynamical regimes: ``nominal{,}" ``drag{,}" ``high metallicity{,}" ``ultrahot{,}" ``cloud" and ``ultrafast{.}" We plotted the horizontal maps at 100{\,}Pa of the temperature for the first four cases, the cloud mass mixing ratio for the ``cloud" case, and the wind pattern for the ``ultrafast" case. Simulations were performed using the 3D global MITgcm with a gray radiative transfer scheme (from Xianyu Tan) except the ``ultrafast" case that {implemented} a Newtonian cooling scheme (\citealt{tanAtmosphericCirculationTidally2020}).  The basic planetary parameters, such as size and gravity, are similar to {those} of HD 209458 b. The ``normal" case is assumed solar metallicity and drag-free. Based on the ``nominal" case, a linear frictional drag is applied in the ``drag" case. The atmospheric metallicity is increased to 10$\times$ solar in the ``high metallicity" case. The equilibrium temperature is 3000{\,}K in the ``ultrahot" case. The ``cloud" case assumes{ a} magnesium silicate cloud ($\mathrm{Mg_2SiO_4}$) with cloud radiative feedback. The rotational period of the ``ultrafast" case is 2.5{\,}h.}
   \label{fighj}
      \end{figure*}

Here we just briefly summarize our current understanding of the
mechanisms under different regimes. In particular, in light of the
scaling in Section~\ref{sect:phasecurve}, the bulk atmospheric flow
of a tidally locked exoplanet is governed by dimensionless numbers:
$\Omega\tau_{\mathrm{dyn}},
\tau_{\mathrm{dyn}}/\tau_{\mathrm{drag}},
\tau_{\mathrm{dyn}}/\tau_{\mathrm{rad}}$ and $q/c_pT_{\rm eq}$. We
can roughly characterize the atmospheric dynamics on tidally locked
gas giants into four regimes with an emphasis of each dimensionless
number: canonical tropical regime
($\tau_{\mathrm{dyn}}/\tau_{\mathrm{rad}}$, ``nominal"),
fast-rotating geostrophic regime ($\Omega\tau_{\mathrm{dyn}}$,
``ultrafast"), strong drag regime
($\tau_{\mathrm{dyn}}/\tau_{\mathrm{drag}}$, ``drag"){
and} ultra-hot regime ($q/c_pT_{\rm eq}$, ``ultrahot"). In addition,
to highlight the importance of opacit{y} sources, we
have two more regimes: high metallicity regime (``high metallicity")
and cloud regime (``cloud"). For cool and small planets,
compositional diversity could also greatly impact the dynamics.
Figure~\ref{fighj} summarizes {all the} six
regimes with a representative dynamical pattern (temperature, flow
or cloud tracer) for each regime from 3D GCM simulations.

Canonical drag-free hot Jupiter simulations from the 3D general
circulation models {exhibit} a strong, broad eastward
(superrotating) jet at the equator and westward wave patterns
off the equator (Fig.~\ref{fighj}, ``nominal" regime). The
temperature pattern is shifted to the east compared with the
stationary day-night radiative forcing pattern centered at the
substellar point. These temperature and flow patterns showed up in
the first hot Jupiter GCM results from
\cite{showmanAtmosphericCirculationTides2002} and the temperature
offset was later confirmed by observations in
\cite{knutsonMapDaynightContrast2007}. The subsequent 3D hot Jupiter
models qualitatively agree with the Showman and Guillot results
(e.g., \citealt{cooperDynamicsDisequilibriumCarbon2006};
\citealt{dobbs-dixonWavelengthDoesNot2017};
\citealt{showmanAtmosphericCirculationHot2009};
\citealt{rauscherThreedimensionalModelingHot2010};
\citealt{hengAtmosphericCirculationTidally2011a};
\citealt{pernaEffectsIrradiationHot2012};
\citealt{mayneUnifiedModelFullycompressible2014};
\citealt{mendoncaThorNewFlexible2016};
\citealt{caroneEquatorialAntirotatingDay2019};
\citealt{mayneLimitsPrimitiveEquations2019};
\citealt{deitrickTHORMajorImprovements2020}; \citealt{geRotationalLightCurves2020}). The
underlying mechanism of the equatorial superrotating wind is,
however, not easy to understand. According to Hide's
theorem (\citealt{hideDynamicsAtmospheresMajor1969};
\citealt{schneiderAxiallySymmetricSteadystate1977a}), in a steady
axisymmetric atmosphere with diffusion, a local maximum in absolute
angular momentum cannot be maintained away from boundaries by the
mean flow. Thus{,} a local maximum in angular momentum such as
equatorial superrotation must imply upgradient eddy
momentum fluxes that balance the diffusion of angular momentum.

Where do the upgradient eddy momentum fluxes come from in hot
Jupiter atmospheres? The key to understanding the mechanism dates
back to the ``Matsuno-Gill model" in the Earth{'s} tropics.
\cite{matsunoQuasigeostrophicMotionsEquatorial1966} considered
freely propagating, linear wave modes on a $\beta$ plane.
\cite{gillSimpleSolutionsHeatinduced1980} analyzed the atmospheric
modes in response to the stationary, longitudinal forcing in the
tropics. The excited large-scale wave modes,
including Kelvin wave{s}, Rossby waves, mixed Rossby-gravity
waves and gravity waves, can be trapped in an equatorial
width characterized by the equatorial Rossby deformation radius
($NH/\beta)^{1/2}$. The canonical hot Jupiters with the global Rossby number
around unity exhibit those modes under stationary day-night stellar
forcing. In particular, the standing, eastward Kelvin modes
and westward Rossby modes form a chevron-shaped feature, with
northwest-southeast tilts in the northern hemisphere and
southwest-northeast tilts in the southern hemisphere. These
horizontal eddy patterns have been shown to feed angular
momentum to the equatorial superrotating jet and maintain it on
tidally locked planets using a simpler one and a half layer shallow
water model (\citealt{showmanMatsunoGillModelEquatorial2010};
\citealt{showmanEquatorialSuperrotationTidally2011}).

Later on,
\cite{tsaiTHREEDIMENSIONALSTRUCTURESEQUATORIAL2014} performed a 3D
analysis of the resonance of the Rossby waves and vertical wavefront
tilt that provides vertical eddy-momentum flux to influence the jet
acceleration and deceleration.
\cite{debrasAccelerationSuperrotationSimulated2020} further extended
the analysis to arbitrary drag and radiative timescales and
highlighted nonlinear feedbacks in the system on the onset
of the prograde jet. \cite{mayneResultsSetThreedimensional2017}
argued the angular momentum transferred by mean meridional
circulation, aside from the eddies, is also important for the jet
maintenance. Motivated by the roles of thermal tides {i}n
generation of the superrotating jets on slowly rotating
planets in the Solar System such as Venus and Titan, recently
\cite{mendoncaAngularMomentumHeat2020} performed a detailed wave
analysis and pointed out that the semi-diurnal tides excited by the
stellar forcing play an important role in jet generation.
Semi-diurnal eddy features have been seen in previous 3D models. For
example, \cite{showmanEquatorialSuperrotationTidally2011} emphasized
that an important difference between their 2D shallow-water cases
and full 3D cases is that the 3D models develop pronounced
mid-latitude Rossby-wave anticyclonic gyres on the dayside and
cyclonic gyres on the nightside in both hemispheres, resulting
from feedback from the mean flow on the eddies (also see
discussion in
\citealt{pierrehumbertAtmosphericCirculationTideLocked2019}).

The day-night temperature difference and hot spot phase offset on
tidally locked planets are caused by eastward group propagation of
Kelvin waves and the equatorial superrotating flow (e.g.,
\citealt{showmanEquatorialSuperrotationTidally2011};
\citealt{perez-beckerAtmosphericHeatRedistribution2013};
\citealt{komacekATMOSPHERICCIRCULATIONHOT2016}). More precisely, the
hot spot offset is caused by the zonal flow Doppler shifting the
stationary wave response because the mean zonal wind on hot Jupiters
has a horizontal velocity close to those waves.
\cite{hammondWavemeanFlowInteractions2018} demonstrated this
mechanism in a 2D system with a horizontally shearing flow. The
kinematic scaling of day-night temperature difference and hot spot
phase shift using the dimensionless number
$\tau_{\mathrm{dyn}}/\tau_{\mathrm{rad}}$ in Section
\ref{sect:offset} still holds true (e.g.,
\citealt{cowanStatisticsAlbedoHeat2011};
\citealt{zhangEffectsBulkComposition2017}){,}
{b}ut there is some difference in the effect of radiative
damping. In the kinematic theories, the radiative damping directly
controls the temperature distribution, relaxing it to the day-night
equilibrium temperature pattern. In the theory of
\cite{hammondWavemeanFlowInteractions2018}, the damping weakens the
forced wave response, relaxing the phase of the response to the
forcing phase. Thus{,} the dynamical models from
\cite{hammondWavemeanFlowInteractions2018} highlight the importance
of the wave response to the mean{ }flow. Note that the
mean flow generation is also closely related to the waves. To date,
the complete 3D picture of nonlinear wave-mean-flow interaction and
the influence on heat redistribution has not been
thoroughly analyzed. Note that the magnetic field, clouds and
other complicated factors play roles in real hot Jupiter
atmospheres, and may{ }be more important roles than the simple
dynamics argument above.

Some fast-rotating tidally locked planets (Fig.~\ref{fighj},
``ultrafast") lie in the geostrophic regime with small Rossby
numbers. This emerging regime includes both planets with an orbital
period around one Earth day or smaller (e.g., WASP-12 b, WASP-103 b,
WASP-18 b, WASP-19 b, NGTS-7A b and TOI 263.01) and several
super-fast rotating brown dwarfs with rotation period within two
hours (e.g., NLTT 5306, WD\,0137--349, EPIC 21223532 and WD\,1202--024). Not surprisingly, some of these planets are very close
to their host stars, so they are also ultra-hot Jupiters, but the
temperature on rapid rotators could also be mild if the host stars
are cool. \cite{leeSimplified3DGCM2020} presented the first 3D
simulation in this regime on the atmosphere of WD\,0137--349B around a
white dwarf. They found a large day-night temperature contrast and
multiple, alternating east-west jet patterns.
\cite{tanAtmosphericCirculationTidally2020} performed 3D simulations
to explore the atmospheric dynamics in this geostrophic regime
systematically. As expected, geostrophic adjustment is important.
Because the equatorial Rossby deformation radius is small, the
meridional extent of the temperature pattern is confined within a
very narrow region around the equator. A big difference from the
nominal case is in the zonal mean zonal wind pattern. Instead of a
broad equatorial jet in the tropical regime, multiple off-equatorial
jets emerge on a fast-rotating hot Jupiter (Fig.~\ref{fighj}), the
formation mechanism of which is associated with the baroclinic waves
induced by the equator-to-pole stellar forcing. The day-night
temperature difference is larger than that on the slower-rotating
planets because a stronger rotation can support a more significant
isobaric temperature difference in the geostrophic regime. As the
scaling prediction in Equations~(\ref{tscalsol}) and
Figure~\ref{figteqtrend}(B) {demonstrate}, the day-night
contrast decreases with rotation period for tidally locked planets.
Also, the hot spot phase shift is not necessarily eastward in this
regime because the substellar temperature is shifted by far-extended
westward Rossby waves in the subtropics to compensate the eastward
Kelvin mode at the equator.
\cite{tanAtmosphericCirculationTidally2020}
{reported} that, as the rotation rate changes, the
equatorial jet width scales well with the equatorial Rossby
deformation radius and the off-equatorial jet width scales well with
the Rhines length.

Ultra-hot Jupiters are also a recent emerging population. They are
not necessarily fast rotators but just receive large stellar flux so
that their temperature exceeds $\sim$ 2200{\,}K. This is
the ``ultrahot" regime in Figure \ref{fighj}. As summarized in
Section \ref{sect:verttemp}, high-temperature chemistry will have
significant impact on the vertical temperature structure (e.g.,
\citealt{evansUltrahotGasgiantExoplanet2017};
\citealt{sheppardEvidenceDaysideThermal2017};
\citealt{haynesSPECTROSCOPICEVIDENCETEMPERATURE2015};
\citealt{nugrohoHighresolutionSpectroscopicDetection2017}). In
particular, thermal dissociation of hydrogen on the dayside and
recombination at the terminator and on the nightside could influence
the horizontal distribution of temperature (e.g.,
\citealt{bellIncreasedHeatTransport2018};
\citealt{komacekEffectsDissociationRecombination2018};
\citealt{tanAtmosphericCirculationUltrahot2019}) and thus the
$q/c_pT_{\rm eq}$ is important.
\cite{parmentierThermalDissociationCondensation2018} investigated
the local thermal chemistry (without tracer transport) and their
radiative feedback on observational signatures such as spectra and
thermal phase curves. \cite{tanAtmosphericCirculationUltrahot2019}
studied the effects of hydrogen dissociation and recombination on
the dynamics with tracer transport, but their gray
radiative transfer scheme did not take into account the influence of
detailed thermochemistry of other species on the temperature
distribution. With hydrogen dissociation and recombination, the
eastward equatorial jets become weaker as temperature increases,
suggesting less horizontal eddy forcing due to suppressed
horizontal---both day-night and equator-pole---temperature contrast
(\citealt{tanAtmosphericCirculationUltrahot2019}). Interestingly,
westward equatorial winds (in the zonal-mean sense) emerge at the
lower pressure level above the superrotating wind when hydrogen
dissociation and recombination are included. The westward winds
become more pronounced when the temperature exceeds 2400{\,}K. In the simulations, westward winds are accelerated by
vertical eddies that overcome the eastward forcing by horizontal
eddies, but the detailed mechanism has yet to be explored. Also, for
the same stellar type, an ultra-hot Jupiter is usually rotating
faster than a cooler Jupiter, and thus the rotational effect also
needs to be taken into account. When the rotational effect is
included, some of the previously seen westward jets at the low
pressure disappear
(\citealt{tanAtmosphericCirculationUltrahot2019}).

If frictional drag is strong, the flow pattern of a tidally locked
planet can be significantly altered (``drag" regime in Fig.~\ref{fighj}). The drag force could come from multiple sources. For
terrestrial planets, frictional drag from the surface sets the lower
boundary condition of the flow. For hot Jupiter atmospheres that
could be partially ionized, Lorentz force due to magnetic field
should play a role (e.g., \citealt{pernaMagneticDragHot2010}).
Small-scale vertical turbulent mixing (e.g.,
\citealt{liCirculationDissipationHot2010};
\citealt{ryuTurbulencedrivenThermalKinetic2018}) and breaking
gravity waves (e.g., \citealt{lindzenTurbulenceStressOwing1981})
could also be considered as drag forces exerted on the large-scale
flow. If we simplify the drag effect as linear friction, one can
understand the flow pattern in terms of force balance (e.g.,
\citealt{showmanDopplerSignaturesAtmospheric2013}). A three-way
balance of the frictional drag, Coriolis force and the
pressure gradient causes the horizontal eddy wind to rotate
clockwise in the northern hemisphere and counterclockwise in the
southern hemisphere, leading to equatorward-eastward and
poleward-westward velocity tilts and thus driving equatorial
superrotation, in addition to its direct damping effect
f{rom} the wind itself. If the frictional drag is
stronger than both the Coriolis force (i.e., large
$\Omega\tau_{\mathrm{drag}}$) and nonlinear inertial force
(i.e., large $\tau_{\mathrm{dyn}}/\tau_{\mathrm{drag}}$), the
horizontal wind is directly controlled by the balance of the drag
force and the pressure gradient and exhibits a day-night divergent
flow pattern instead of an east-west jet pattern. This strong drag
effect has been investigated in all current dynamical models (e.g.,
see a large grid of idealized 3D simulations in
\citealt{komacekATMOSPHERICCIRCULATIONHOT2016}). Therefore, by
observing the horizontal wind pattern on tidally locked giant
planets (e.g., through the cross-correlation technique), one might
infer the strength of the atmospheric drag. For very rapidly
rotating tidally locked planets, if drag is strong, the thermal
phase curves could actually show a near alignment of peak flux to
secondary eclipse (\citealt{tanAtmosphericCirculationTidally2020}),
as observed in close-in brown dwarfs orbiting white dwarfs, for
example, NLTT 5306 (\citealt{steeleNLTT5306Shortest2013}),
WD\,0137--349
(\citealt{casewellMultiwavebandPhotometryIrradiated2015};
\citealt{longstaffEmissionLinesAtmosphere2017}), EPIC 21223532
(\citealt{casewellFirstSub70Min2018}) and WD 1202--024
(\citealt{rappaportWD1202024Shortestperiod2017}).

The magnetic drag effect could be particularly important for hot
planets. Strong magnetic fields ($\sim$20--120{\,}G) have been detected on four hot Jupiters from
the energy released in the Ca II K line during star-planet
interactions (HD 179949 b, HD 189733 b, $\tau$ Boo b and $\nu$ And
b, by \citealt{cauleyMagneticFieldStrengths2019}). Crudely, one can
estimate the importance of magnetic effect in an ionized medium
using a non{-}dimensional number called ``plasma-$\beta$"---the
ratio of the plasma pressure ($p$) to the magnetic pressure
($B^2/8\pi$) where $B$ is the background magnetic field.  The
magnetic effect dominates when $\beta\ll 1$, such as in the solar
corona. The plasma pressure dominates when $\beta\gg 1$, such as in
the solar interior. We can further use the Alfv\'{e}n Mach number
$M_A$ for subsonic flow. $M_A^2=U^2/U_A^2=4\pi pM_a^2/B^2$ is the
ratio of the flow speed $U$ to the Alfv\'{e}n speed $U_A=(B^2/4\pi
\rho)^{1/2}$ where $\rho$ is the plasma density.
$M_a=U/(p/\rho)^{1/2}$ is the Mach number to the isothermal sound
speed. If $M_A$ is small, the magnetic field controls the flow. For
example, for a strong magnetic field strength of 100{\,}G, if the wind velocity is subsonic with
$M_a\sim0.1$, $M_A$ reaches unity at 0.1{\,}bar where the
magnetic force (Lorentz force) could be as important as the pressure
force in a fully ionized atmosphere.

The realistic MHD effect could
only be more complicated than a simple drag effect because of
partial ionization and feedbacks of the flow pattern to the magnetic
field. In particular, the non-ideal MHD effects from Ohmic
resistivity, Hall effect and ambipolar diffusion could play a role
in {a} partially ionized medium like hot Jupiter
atmospheres, leading to incomplete coupling between the atmosphere
and magnetic field. Most published studies to date have
neglected feedbacks (\citealt{pernaMagneticDragHot2010,
pernaOhmicDissipationAtmospheres2010,
pernaEffectsIrradiationHot2012};
\citealt{menouMagneticScalingLaws2012};
\citealt{rauscherThreedimensionalAtmosphericCirculation2013};
\citealt{hindleShallowwaterMagnetohydrodynamicsWestward2019}) but
there are some recent efforts considering more realistic MHD
situation (e.g., \citealt{batyginNonaxisymmetricFlowsHot2014};
\citealt{rogersMagneticEffectsHot2014};
\citealt{rogersMagnetohydrodynamicSimulationsAtmosphere2014};
\citealt{rogersConstraintsMagneticField2017}).

Exoplanets have a large range of metallicities (Fig.~\ref{figmet})
that could influence the atmospheric dynamics. For example, the
simulated temperature pattern on a hot Jupiter using ten times solar
metallicity looks different from the nominal case (Fig.~\ref{fighj}, ``high metallicity" case). The metallicity effect could
be more influential for small planets as their bulk composition
might not be hydrogen. As {depicted} in Figure
\ref{figmet}, for sub-Neptunes and smaller planets, compositional
diversity of the bulk atmosphere greatly increases, ranging from low
molecular mass atmospheres of \ch{H2} to higher molecular
atmospheres of water, \ch{CO2}, \ch{N2} or other species (see
Sect.~\ref{sect:gaschem}). Compared with the hydrogen case, three
important effects need to be taken into account in these
higher-metallicity atmospheres: molecular weight, heat
capacity and radiative opacity
(\citealt{zhangEffectsBulkComposition2017}). Take GJ 1214 b as an
example. The simulated atmospheric flow pattern greatly changes
{with} different assumptions o{n} the bulk
composition (\ch{H2}, \ch{H2O} or \ch{CO2}) or metallicity in
hydrogen atmospheres, or the presence of cloud or haze particles in
the atmosphere (e.g.,
\citealt{katariaAtmosphericCirculationSuper2014};
\citealt{charnay3DModelingGJ1214bs2015,
charnay3DModelingGJ1214b2015a}). A detailed characterization of the
dynamics on those planets requires further observations in
longer wavelengths to probe deep below the high-altitude particle
layers to determine the atmospheric composition. From
{an} atmospheric dynamics point of view, it would be
good to keep in mind that the global Rossby numbers for smaller and
hot planets (e.g., warm Neptunes) are likely to be higher than
for{ their} Jovian-sized counterpart (e.g., warm Jupiters) and
thus their climate states lie in the large-Rossby-number regime
(Fig.~\ref{figrof}), which means their day-night contrast is
generally smaller than their gas giant counterpart.

Clouds on tidally locked exoplanets might need a separate
discussion, given their importance {i}n observations. As
mentioned in Section~\ref{sect:offset}, cloud particles in the
atmospheres could significantly distort the phase curves and might
cause {a} westward offset of the bright spot in the
Kepler band (Fig.~\ref{figteqtrend}) and regulate the emission
temperature on the nightside (Fig.~\ref{figdnt}). Moreover, as a
large opacity source, clouds could impact the radiative flux
exchange in the atmospheric layers and influence the dynamics.
\cite{romanModeledTemperaturedependentClouds2019} investigated the
cloud radiative feedback to the atmospheric temperature patterns in
a 3D GCM using a parameterized cloud scheme but without tracer
transport. To fully understand their effects, 3D distribution of the
cloud tracers needs to be resolved in fully coupled radiative
hydrodynamical simulations with cloud physics and tracer transport.
To date, only two planets have been simulated in the fully coupled
fashion: HD 189733 b (\citealt{leeDynamicMineralClouds2016}) and HD
209458 b (\citealt{linesSimulatingCloudyAtmospheres2018}), but the
simulations were so computationally expensive that only short-term
integrations were performed. We demonstrate a simple, fully coupled
case in Figure~\ref{fighj}, the ``cloud" case,
{implementing} a gray radiative transfer scheme and
assuming constant particle size, but with tracer transport
and cloud radiative feedback. It can be seen that cloud mass
distribution is highly non-uniform across the globe and seems to
follow the temperature distribution well in this case. In
more realistic simulations in \cite{leeDynamicMineralClouds2016} and
\cite{linesSimulatingCloudyAtmospheres2018}, particle size seems to
anti-correlate with temperature, with smaller particles in
the equatorial region on the dayside, and larger particles in high latitudes and the nightside. The cloud distributions are
also largely shaped by the circulation pattern. For example, the
cloud simulations for HD 209458 b show three distinct zonal bands
with one at the equator and two off-equatorial bands (e.g.,
\citealt{linesSimulatingCloudyAtmospheres2018}), roughly
correlat{ing} with the zonal-mean zonal wind pattern.
Clouds are also highly variable, which might provide temporal
evolutions of detected spectral features. See
\cite{hellingExoplanetClouds2019} for more discussions.

Unlike the classic tidally locked planets in circular orbits, the
climates on other irradiated planets located further from their host
stars are significantly influenced by orbital eccentricity,
self-rotation and planetary obliquity. Due to stellar tidal
effect, the timescale of the orbital circularization scales as
$a^{13/2}$ where $a$ is the semi-major axis
(\citealt{goldreichSolarSystem1966}) and that of the spin
synchronization depends on $a^6$
(\citealt{bodenheimerTidalInflationShortPeriod2001}). As a result,
the orbit of a close-in planet is not necessarily circular. In fact,
observations {indicate} that some short-period
exoplanets have high eccentricities, for example, hot Jupiter
HAT-P-2 b ($e\sim 0.51$), HD 80606 b ($e\sim 0.93$) and sub Neptune
GJ 436 b ($e\sim 0.15$). The large difference of this regime
compared with the planets in circular orbits originates from the
large temporal variation of the stellar flux. If the radiative
timescale is short, the atmosphere will experience a significant
``eccentricity season{.}" Several studies (e.g.,
\citealt{langtonHydrodynamicSimulationsUnevenly2008};
\citealt{lewisAtmosphericCirculationEccentric2010};
\citealt{katariaThreedimensionalAtmosphericCirculation2013};
\citealt{lewisAtmosphericCirculationEccentric2014};
\citealt{ohnoAtmospheresNonsynchronizedEccentrictilted2019,ohnoAtmospheresNonsynchronizedEccentrictilted2019a})
have investigated the atmospheric dynamics of eccentric exoplanets.
They found that spatial patterns of the atmospheric
temperature and circulation are qualitatively similar to that of
planets in circular orbits, although the magnitudes of the
temperature fluctuation and wind velocity could change with time.
The eccentric orbit significantly influences the shape of the
thermal light curve because of intense stellar heating
during perihelion and non-uniform orbital velocity of the
planet passage.

Outside the synchronization zone, planets have much faster
self-rotation rates and are likely to lie in the geostrophic regime
(Fig.~\ref{figrof}). Warm Jupiters are good examples. Note that
these planets are still highly irradiated by the central star. The
stellar flux is still stronger than the expected interior flux by
several orders of magnitude.
\cite{showmanThreedimensionalAtmosphericCirculation2015}
investigated the influences of planetary rotation on 3D atmospheric
dynamics on non-synchronized giant planets. A non-synchronized
planet with a slow rotation rate and a high incoming stellar flux is
dominated by an equatorial superrotating jet like a canonical hot
Jupiter, whereas a planet with a fast rotation rate and a low
stellar flux develops mid-latitude jets, like on our Jupiter and
Saturn. Nevertheless, these fast rotators {manifest} a
westward flow at{ the} equator instead of equatorial
superrotation {like }on our Jupiter, perhaps due to a lack of
moist processes in the high-temperature regime such as
water condensation or insignificance of the internal heat from below, which were proposed to be important to drive the
equatorial eastward flow on Jupiter and Saturn (e.g.,
\citealt{schneiderFormationJetsEquatorial2009};
\citealt{lianGenerationEquatorialJets2010}).
\cite{pennThermalPhaseCurve2017,pennAtmosphericCirculationThermal2018}
investigated non-synchronized terrestrial exoplanets. They
{demonstrated} that the substellar point moves
westward due to rapid rotation, and the hot spot is shifted eastward
from the substellar point. But if the gravity waves are faster than
the substellar point movement, the hot spot could be shifted
westward, resulting in different thermal phase curves.

Planetary obliquity is much more easily damped by stellar
tides than eccentricity
(\citealt{pealeOriginEvolutionNatural1999}). The non-synchronized
planets could have non-zero eccentricities and obliquities.
\cite{rauscherModelsWarmJupiter2017} investigated the 3D dynamics on
planets in a circular orbit and demonstrated that the atmospheric
flow pattern significantly varies with obliquity. Using a 2D
shallow-water model,
\cite{ohnoAtmospheresNonsynchronizedEccentrictilted2019,ohnoAtmospheresNonsynchronizedEccentrictilted2019a}
unified previous studies of non-synchronized planets with different
orbital eccentricities, rotation rates and planetary obliquities.
They classified the atmospheric dynamics into five regimes
{considering} the radiative timescale and obliquity
(see Fig.~1 in
\citealt{ohnoAtmospheresNonsynchronizedEccentrictilted2019}). If the
radiative timescale is shorter than the rotation period, the
atmosphere {displays} a time-varying day-night
contrast and a day-to-night flow pattern (regime I). When the
radiative timescale is longer than the rotation period but shorter
than the orbital period, the temperature pattern is controlled by
diurnal mean insolation. For obliquity smaller than
$\sim18^{\circ}$ (regime II), the temperature distribution is
longitudinally homogeneous with an equator-to-pole gradient. An
eastward flow is dominant in this regime. For obliquity larger than
$18^{\circ}$(regime III), the atmosphere is heated in the polar
region, resulting in a westward wind on the heated hemisphere but an
eastward flow on the other hemisphere. If the radiative timescale is
longer than the orbital period, the temperature field is dominated
by the annual mean insolation. For obliquity smaller than
$54^{\circ}$ (regime IV), the atmosphere exhibits an equator-to-pole
temperature gradient and an eastward flow on the entire planet. For
obliquity larger than $54^{\circ}$ (regime V), the temperature
gradient is from the pole to the equator, and a westward flow
dominates. Compared with the complicated dynamical behavior, the
behaviors of thermal phase curves in the entire parameter space can
only be more complex because of the wide range of view geometry
(e.g., \citealt{rauscherModelsWarmJupiter2017};
\citealt{ohnoAtmospheresNonsynchronizedEccentrictilted2019a};
\citealt{adamsSignaturesObliquityThermal2019}), as discussed in
Section \ref{sect:horitemp}.

Observations {indicate that} atmospheres of highly
irradiated planets appear to be dynamically variable. Dramatic
short-term variability of the peak brightness offset in the Kepler
light curve has been observed on hot Jupiter HAT-P-7 b
(\citealt{armstrongVariabilityAtmosphereHot2016}). Recently{,}
another hot Jupiter Kepler 76 b has been observed to exhibit large
variability in reflection and emission on a timescale of tens of
days (\citealt{jacksonVariabilityAtmosphereHot2019}). Hot Jupiter
atmospheric flows should be generally rotationally stable (e.g.,
\citealt{liCirculationDissipationHot2010};
\citealt{menouTurbulentVerticalMixing2019}) but the equatorial jets
could also be potentially unstable due to barotropic
Kelvin-Helmholtz instability and vertical shear instabilities (e.g.,
\citealt{fromangSheardrivenInstabilitiesShocks2016}). In general,
transients in the atmosphere could also come from several mechanisms
such as barotropic and baroclinic instabilities
(\citealt{pierrehumbertAtmosphericCirculationTideLocked2019}),
large-scale atmospheric waves
(\citealt{komacekTemporalVariabilityHot2019}), large-scale
oscillations due to wave-mean-flow interactions
(\citealt{showmanAtmosphericCirculationBrown2019}), as well as mean-flow interaction with the magnetic field
(\citealt{rogersConstraintsMagneticField2017}). Hot Jupiter
simulations found that the globally averaged temperature can be
time-variable at the 0.1\%-1\% level and the variation of globally
averaged wind speeds is at the 1\%-10\% level
(\citealt{komacekTemporalVariabilityHot2019}). The abundances of
atmospheric chemical tracers, either gas or clouds, could also vary
significantly with time (e.g., \citealt{parmentier3DMixingHot2013}).
Relatively long-term variability could result from the eccentricity
and obliquity seasons, as seen in the thermal phase curves on
eccentric planets. The long-term variation of the climate is related
to orbital dynamics such as the Milankovitch cycles, including periodic changes of obliquity, axial precession, apsidal
precession and orbital inclination (e.g.,
\citealt{spiegelGENERALIZEDMILANKOVITCHCYCLES2010};
\citealt{deitrickExoMilankovitchCyclesOrbits2018};
\citealt{deitrickExoMilankovitchCyclesII2018}){,}
{b}ut these timescales might be too long for
observations. Some close-in planets might be experiencing rapid
orbital decay (e.g., WASP 12 b,
\citealt{maciejewskiDepartureConstantperiodEphemeris2016};
\citealt{patraApparentlyDecayingOrbit2017}), which could also induce
interesting time variability in a decadal timescale.

\subsection{Weakly irradiated planets and brown dwarfs}
\label{sect:bddyn}

For a distant planet located far from its host star, the internal
heat flux (i.e., self-luminosity) plays a dominant role in the
atmospheric dynamics. Extreme cases in this regime are free-floating
planets and field brown dwarfs. To date{,}
direct imag{ing} is the best observational
technique to characterize these atmospheres, inherited from traditional stellar astronomy. Compared with the close-in planets,
observational data on directly imaged planets have a much better
quality because of much less stellar contamination. High-resolution
spectra provide clues {about} the vertical distributions
of temperature and opacities from chemical tracers, while
time-domain photometry, such as rotational light curve and
Doppler imaging{,} can be {utilized} to unveil
their horizontal distributions. The steady patterns in the
rotational light curves suggest the mean-state of the surface
inhomogeneity; the temporal variability of those curves indicates
the short-term and long-term weather patterns---both are closely
related to atmospheric dynamics. High-resolution
spectroscopy has been {applied} to measure the
rotational line broadening and infer the self-rotation rate of
planets and brown dwarfs (e.g., \citealt{snellenFastSpinYoung2014};
\citealt{allersRADIALROTATIONALVELOCITIES2016};
\citealt{bryanConstraintsSpinEvolution2018}) but with this technique
alone we are not at the stage to separate the surface wind from the
internal solid-body rotation. Recently, combining with
radio wave observations to infer the rotational rate of the internal
magnetic field,  \cite{allersMeasurementWindSpeed2020} successfully
detected differential rotation between the photosphere
(from the IR rotational light curves) and the deep interior (from
the periodic radio burst, e.g.,
\citealt{williamsROTATIONPERIODMAGNETIC2015}) of 2MASS
J10475385+2124234, a T6.5 brown dwarf 10.6\,pc away. The
atmosphere is rotating faster than the interior, suggesting a strong
superrotating (eastward) wind in the photosphere with a speed of
$650\pm 310{\,}\mathrm{m{\,}s^{-1}}$. This
behavior is similar to Jupiter and Saturn, where the global-mean
zonal wind is also superrotating, mostly from the broad eastward jet
at the equator. The same analysis {implies} that the
Jupiter wind speed at{ its} equator is about 106{\,}s$\mathrm{m{\,}s^{-1}}$
(\citealt{allersMeasurementWindSpeed2020}), close to the cloud
tracking results from Cassini (e.g.,
\citealt{porcoCassiniImagingJupiter2003}). The global-mean zonal
wind on Saturn is probably also superrotating based on its strong
eastward equatorial wind, but the value is not well constrained due
to the uncertainty of Saturn's solid-body rotation rate.

The atmospheres of planets and brown dwarfs in this regime are
fast-rotating and strongly convective. A na\"{i}ve picture of these
atmospheres is an adiabatic temperature profile in the deep
atmosphere, rotationally symmetric weather pattern, and
homogeneously distributed chemical tracers due to dynamical
quenching. The realistic picture is much more complicated and
significantly deviate{s} from the above description. The
existence of rotational light curves on these bodies implies strong
spatial inhomogeneity in their photospheres. Ammonia in the deep troposphere of Jupiter retrieved from the Juno spacecraft data also shows substantial
variation across latitude (e.g.,
\citealt{boltonJupiterInteriorDeep2017};
\citealt{liDistributionAmmoniaJupiter2017}) and also hints that the
traditional quenching framework might not be sufficient to
understand the tracer transport behavior in the convective
atmosphere.

When studying the dynamics of distant directly imaged
planets, the first problem is the bottom boundary. These planets do
not have surfaces at the bottom, raising the question of whether the
flow on these atmospheres is ``shallow" or ``deep{.}"
This question is twofold. First, is the dominant atmospheric motion
horizontal, vertical or intrinsically
{3D}? Second, is the weather pattern in
the upper atmosphere connect{ed} to the deep
atmosphere?

We first discuss the first question related to the preferential
direction of heat transport in the atmosphere. One can analyze the
potential temperature gradient in the horizontal direction versus
the vertical{ one}. For the large-scale dynamics in the
photosphere, the horizontal gradients of potential temperature in
the regimes of low and high Rossby numbers were discussed using
Equations~(\ref{lRo}) and (\ref{sRo})
(\citealt{charneyNoteLargeScaleMotions1963}) respectively. By
scaling the entire momentum equation{,} we can combine the two
regimes {applying} one unified scaling of the
horizontal (latitudinal) potential temperature gradient
$\Delta\theta_h/\theta\sim \mathrm{Fr}(1+\mathrm{Ro}^{-1})$. The
vertical potential temperature gradient is related to the static
stability $N^2\sim g\Delta\theta_v/\theta D$, where $D$ is the flow
depth. Thus the ratio of the horizontal to vertical potential
temperature contrasts, or sometimes called the ``baroclinic
criticality" $\xi$, can be scaled as
\begin{equation}
\xi\sim \mathrm{Ri}^{-1}(1+\mathrm{Ro}^{-1}),
\end{equation}
where the Richardson number $\mathrm{Ri}\sim N^2D^2/U^2$
characterizes the atmospheric stratification versus the vertical
wind shear. The atmosphere is subject to free convection if
$\mathrm{Ri}$ is smaller than 0.25. For reference, the baroclinic
criticality $\xi$ for Earth's atmosphere is about unity. This
scaling is consistent with an alternative derivation in
\cite{allisonRichardsonNumberConstraints1995}. It implies that in
the tropical regime, the horizontal to vertical potential
temperature slope ratio is not dependent on the rotation rate,
$\xi\sim \mathrm{Ri}^{-1}$. For rapidly rotating planets
(geostrophic regime), $\xi\sim \mathrm{Ri}^{-1}
\mathrm{Ro}^{-1}\sim\Omega R_pU/gH$.
\cite{komacekScalingRelationsTerrestrial2019} used the turbulent
cascade scaling (e.g., \citealt{heldScalingTheoryHorizontally1996})
and achieved a more detailed scaling of baroclinic criticality to
the planetary parameters for rapidly rotating planets, (see their
eq.~(6)). It shows a strong dependence of $\xi$ on the rotation
rate, scale height and planetary size, qualitatively
consistent with our simple scaling{ described} above.

\cite{allisonRichardsonNumberConstraints1995} classified the
dynamics of planetary atmospheres in the Solar System in a
$\mathrm{Ri}-\mathrm{Ro}$ diagram. They found three regimes.
Slow-rotating planetary atmospheres like Venus and Titan have both
large Richardson ($\sim$10) and Rossby ($\sim$10--100) numbers and
small $\xi$ (note that
\citealt{allisonRichardsonNumberConstraints1995} used the ratio
$1/\xi$). In this regime, a large vertical potential temperature
gradient with a small horizontal temperature contrast is developed in a relatively stratified
atmosphere due to large-scale Hadley-like
circulation.
{The atmospheres of }Earth and Mars lie in the
geostrophic regime with a smaller $\mathrm{Ro}$ ($\sim$0.1--1) but a
similar $\mathrm{Ri}$ with Venus and Titan. In this regime, eddies
transport heat effectively in both upward and poleward directions.
As a result, the baroclinic criticality $\xi$ is around unity. The
third regime is the giant planet regime with both small Richardson
($\sim$1) and Rossby ($\sim$0.01) numbers. Strong vertical
convection transports heat efficiently, leading to almost vertical
isentropes and a large baroclinic criticality $\xi$. The fluid
motion in those atmospheres behaves more like{ the}
{2D case}.

Without knowing the exact temperature and wind structures, it is
difficult to estimate the exact Richardson number for an atmosphere
outside the Solar System. The tidally locked giant planets likely
lie in a different regime from the directly imaged planets and brown
dwarfs. {T}idally locked giant planets generally have
higher Rossby numbers (Fig.~\ref{figrof}), and their photospheres
are more stably stratified. These planets should mostly occupy a
similar corner to Venus and Titan in the $\mathrm{Ri}-\mathrm{Ro}$
diagram. On the other hand, for directly imaged planets and brown
dwarfs, the global Rossby numbers appear to be comparable to that of
Jupiter (Fig.~\ref{figrof}). Their Richardson number could be as
large as the gas giants in the Solar System as well. Thus directly
imaged planets and brown dwarfs have both lower $\mathrm{Ri}$ and
$\mathrm{Ro}$ because of their convective nature and fast rotation.
They are likely to overlap with Jupiter and Saturn in the
diagram. Their horizontal and vertical potential temperature
contrast $\xi$ should be small even with a strong horizontal motion
because strong vertical convection could efficiently homogenize the
entropy in the vertical direction, or along the direction of the
rotational axis if the flow is deep.

How deep is the atmospheric flow? This is not an easy question to
answer. Insights come from recent
observations from{ the} Juno spacecraft and Cassini {``Grand
}Finale{''} mission. A deep zonal flow could perturb
the gravitational fields of giant planets. The perturbation signals
could be measured by precise spacecraft tracking during the orbit
(e.g., \citealt{kaspiGravitationalSignatureJupiter2010,
kaspiInferringDepthZonal2013};
\citealt{liuPredictionsThermalGravitational2013};
\citealt{caoGravityZonalFlows2017};
\citealt{kongEffectEquatoriallySymmetric2018,
kongOriginJupiterCloudlevel2018, kongSaturnGravitationalField2018}).
The latest data suggest that the surface zonal jets could extend to
the interiors, a depth of about 3000{\,}km on Jupiter
(e.g., \citealt{kaspiJupiterAtmosphericJet2018};
\citealt{guillotSuppressionDifferentialRotation2018}) and about
9000{\,}km on Saturn (e.g.,
\citealt{iessMeasurementImplicationsSaturn2019};
\citealt{galantiSaturnDeepAtmospheric2019}). The winds are coupled
with the magnetic field and might be damped in the region where Lorentz drag becomes important (e.g.,
\citealt{caoZonalFlowMagnetic2017};
\citealt{kaspiJupiterAtmosphericJet2018}). The same deep winds have
also been inferred on ice giants Uranus and Neptune from
Voyager data, to a depth of about 1000{\,}km
(\citealt{kaspiAtmosphericConfinementJet2013}). If this is a generic
behavior on giant planets, the atmospheric flows on
tidally locked planets, directly imaged planets and brown
dwarfs are potentially deep. Th{is} deep circulation
might be the key for explanation f{or} inflated radii of hot Jupiters (e.g.,
\citealt{showmanAtmosphericCirculationTides2002};
\citealt{youdinMechanicalGreenhouseBurial2010};
\citealt{tremblinAdvectionPotentialTemperature2017};
\citealt{sainsbury-martinezIdealisedSimulationsDeep2019}). On the
other hand, because those atmospheres are much hotter than the cold
gas giants in the Solar System, the thermal ionization rate should
be much higher at the same pressure level, implying that the
influence of the magnetic field could be much more important, as we
briefly discussed in Section \ref{sect:hjdyn}. Therefore, the deep
flows on hot planets might cease at a shallower level than that on
the cold gas giants. For example, if an electric conductivity of
1{\,}$\mathrm{S{\,}m^{-1}}$ is sufficient to
influence the zonal jets in the interior of Jupiter at about
$5\times 10^4${\,}bar
(\citealt{guillotSuppressionDifferentialRotation2018}), the magnetic
breaking might effectively impact the zonal jets at a pretty
shallow level on a hot gas giant (could be as shallow as
$\sim$100{\,}bar, see fig.~4 in
\citealt{wuOHMICHEATINGSUSPENDS2013}). {A }3D MHD simulation
coupling the realistic radiative photosphere and the convective
interior is required to investigate the details but is challenging
with current computational facilities.

Neglecting the effect of MHD, purely hydrodynamic simulations of the
deep atmospheres have been performed on directly imaged planets and
brown dwarfs by \cite{showmanAtmosphericDynamicsBrown2013}. In these
models with convective heat flux from below, rotation plays an
important role in organizing the large-scale flow pattern. In the
slowly rotating regime ($\mathrm{Ro}>$1), rotation is not important.
The rising convective plumes originate from the bottom and rise
upward quasi-radially. Convection also appears globally isotropic.
In this regime, the traditional mixing length theory (e.g.,
\citealt{claytonPrinciplesStellarEvolution1968};
\citealt{stevensonTurbulentThermalConvection1979}, also see Section
\ref{sect:chemfund}) predicts that the vertical velocity scales as
$w\sim(\alpha g F l/\rho c_p)^{1/3}$ (see the notations under
eq.~(\ref{ckzz})). The temperature fluctuation in this regime is
$\Delta T\sim(F^2/\rho^2 c_p^2\alpha gl)^{1/3}$. On the other hand,
if the body is rapidly rotating ($\mathrm{Ro}<$1), planetary
rotation organizes the large-scale flow to align along with columns
parallel to the rotation axis, i.e., the Taylor columns. Assuming
the buoyancy force balance{s} vertical Coriolis forces in{
an} isotropic flow, one can deduce the velocity scale $w\sim(\alpha
g F/\rho c_p \Omega)^{1/2}$ (e.g., \citealt{
golitsynGeostrophicConvection1980,
golitsynConvectionStructureFast1981};
\citealt{boubnovExperimentalStudyConvective1986,boubnovTemperatureVelocityField1990};
\citealt{fernandoEffectsRotationConvective1991}). The horizontal
temperature perturbation is $\Delta T\sim(F\Omega/\rho c_p\alpha
g)^{1/2}$. For instance, for a rapidly rotating brown dwarf with a
heat flux of $10^7{\,}\mathrm{W{\,}m^{-2}}$,
the temperature fluctuation at 1{\,}bar is about  2{\,}K (\citealt{showmanAtmosphericDynamicsBrown2013}). The
simulations predict that the polar temperature is larger by about
one Kelvin than the equatorial temperature because convection occurs
more efficiently at high latitude.

In addition to the Rossby number $\mathrm{Ro}$, several
dimensionless numbers are useful to characterize the behaviors of
convection and deep flow structure on weakly irradiated giant
planets and brown dwarfs. The Rayleigh number $\mathrm{Ra}$, Ekman
number $\mathrm{E}$ and Prandtl number $\mathrm{Pr}$ are
defined as follows
 \begin{subequations}
 \begin{align}
\mathrm{Ra}=\frac{\alpha g FD^4}{\rho c_p \nu \kappa^2},
\\
\mathrm{E}=\frac{\nu}{\Omega D^2},
\\
\mathrm{Pr}=\frac{\nu}{\kappa},
 \end{align}
\end{subequations}
where $D$ is the thickness of the convective layer, $\kappa$ is the
thermal conductivity and $\nu$ is the kinematic viscosity.
The Rayleigh number measures the strength of the thermal convection.
The Ekman number measures the significance of rotation. The Prandtl number evaluates the relative
importance between thermal conduction and momentum
transport. Thermal convection occurs if the Rayleigh number is
larger than the critical Rayleigh number $\mathrm{Ra_{cr}}\sim
O(E^{4/3})$ (e.g.,
\citealt{robertsThermalInstabilityRotatingfluid1968}, but note that
the Roberts' paper used the Taylor number
$\mathrm{Ta}=\mathrm{E}^{-2}$). The Prandtl number for gas is on the
order{ of} unity. {V}iscosity of the gas giants
is very low, leading to a high Rayleigh number and low Ekman number.
The viscosity is also very uncertain. For example, the viscosities
from molecular diffusion and that from turbulent diffusion could
differ by several orders of magnitude. The extremely low viscosity
also imposes a great computational challenge in realistic 3D models
on gas giants. In fact{,} the dynamical regime in current
simulations is far from realistic situations (e.g.,
\citealt{showmanScalingLawsConvection2011}).

Nevertheless, if the underlying physics governing the thermal
convection in a rapidly rotating atmosphere is universal,
dimensionless numbers provide useful insights into the dynamical
regimes of exoplanets and brown dwarfs. For small Ekman number and
Prandtl number of order unity,
\cite{schubertDynamicsGiantPlanet2000} classified four important
regimes in terms of the ratio of $\mathrm{Ra}$ to
$\mathrm{Ra_{cr}}$: (1) if $\mathrm{Ra}/\mathrm{Ra_{cr}}<1$,
convection is inhibited and geostrophic flows along the azimuthal
direction are possible. (2) For
$1<\mathrm{Ra}/\mathrm{Ra_{cr}}<O(1)$, convection occurs in the form
of azimuthally propagating waves in a columnized configuration
parallel to the rotation axis. (3) For
$O(1)<\mathrm{Ra}/\mathrm{Ra_{cr}}<\mathrm{Ra^*}/\mathrm{Ra_{cr}}$
where $\mathrm{Ra^*}$ is another critical number. In this regime,
small-scale convection disturbs the columns chaotically. (4)
$\mathrm{Ra}/\mathrm{Ra_{cr}}>\mathrm{Ra^*}/\mathrm{Ra_{cr}}$, the
strong nonlinear advection significantly dominate{s} over the
Coriolis effect. \citet{showmanScalingLawsConvection2011} derived
the scaling of mean jet speed with heat flux and viscosity in two
regimes. In the regime where convection is weakly
nonlinear, the jet speed scales approximately with $F/\nu$. On the
other hand, if the convection is strongly nonlinear, the jet speed
has a weaker dependence on the heat flux in the form of
$(F/\nu)^{1/2}$.

We can scale these numbers from Jupiter's values to the exoplanet
and brown dwarf regime to roughly estimate their behaviors compared
to the highly nonlinear, vigorous and chaotic Jovian
atmosphere. Here we assume the viscosity on these bodies is the same
as that of Jupiter, and the depth of the convection zone is the
planetary size that is roughly the Jupiter radius (e.g., $D\sim
R_J$). Similar to the previous work by
\cite{schubertDynamicsGiantPlanet2000}, we obtain the dependence of
$\mathrm{Ra}$ and $\mathrm{E}$ on the mass and temperature:
$\mathrm{Ra}\propto gF\propto MT_{eff}^4$ and $\mathrm{E}\propto
\Omega^{-1}\propto M^{-0.5}$. For the rotation rate, we have
{applied} the velocity scaling (Eq.~(\ref{vscal})).
Thus the $\mathrm{Ra}${ values} on hot, massive exoplanets and
brown dwarfs are orders of magnitude higher than Jupiter's value,
while the $\mathrm{E}$ does not change too much with the mass. In
other words, the effect of rotation on these bodies is not
much stronger than that on Jupiter, while the thermal convection
could be very different. \cite{schubertDynamicsGiantPlanet2000}
claimed that for very massive, rapidly rotating bodies, the
$\mathrm{Ra}$ is large, and the convection should be fully 3D and
chaotic. On these bodies, bands of alternating zonal winds like on
Jupiter may not be expected.

The horizontal wind speed in the convective region may not be large,
but \cite{showmanAtmosphericDynamicsBrown2013} argued that the
upward propagating waves could drive the mean flow in the overlying
stratified layers, leading to large-scale circulation and fast
horizontal flows. This wave-derive{d} flow impacts the
temperature, wind and tracer distributions, and thus the
observational signatures in the photosphere. The shallow weather
layer in the photosphere, in the simplest picture, can be understood
as a forced-dissipative system. The stratified layer is forced
mechanically by the convection below the {RCB}, and the momentum and energy are dissipated by
viscous friction and radiation. Unlike the unforced
free{ly}-evolving turbulent fluid (e.g.,
\citealt{choEmergenceJetsVortices1996}), the behavior of the
forced-dissipative system strongly depends on the relative strength
of the forcing and dissipation. 2D model simulations in
\cite{zhangAtmosphericCirculationBrown2014} show that banded zonal
flow patterns spontaneously emerge from the interaction between turbulence and planetary rotation if the bottom heat flux is
strong or radiative dissipation is weak. On the other hand, if the
internal forcing is weak or radiative dissipation is strong,
atmospheric turbulence damps quickly before self-organizing into
large-scale jets. Transient eddies and isotropic turbulences
dominate the weather pattern.  Jupiter appears to lie in the first
regime (strong-forcing and weak-damping), but some hot brown dwarfs
or directly imaged exoplanets might lie in the latter (weak-forcing
and strong-damping). In a more detailed picture, a variety of waves
generated in a 3D convective, rotating atmosphere such as gravity
waves and Rossby waves propagate upward and dump the momentum in the
stratified layer. Those waves could force a large-scale circulation
pattern (\citealt{showmanAtmosphericDynamicsBrown2013}) as well as
multiple zonal jets in the photosphere
(\citealt{showmanAtmosphericCirculationBrown2019}).  In particular,
the equatorial region could exhibit vertically stacked eastward and
westward jets that emerge at the top of the atmosphere and migrate
downward over time in a periodic fashion. This behavior resembles
the oscillations that were observed in the equatorial regions on
Earth (quasi-biennial oscillation, QBO, with a period of
$\sim$2{\,}yr,
\citealt{baldwinQuasibiennialOscillation2001}), Jupiter
(quasi-quadrennial oscillation, QQO, with a period of $\sim$4{\,}yr,
\citealt{leovyQuasiquadrennialOscillationJupiter1991}) and
Saturn (the semiannual oscillation, SAO, with a period of
$\sim$15{\,}yr,
\citealt{ortonSemiannualOscillationsSaturn2008};
\citealt{fouchetEquatorialOscillationSaturn2008}). These peculiar
wave-mean-flow interaction behaviors on brown dwarfs and directly
imaged exoplanets are potentially detectable in future observations.

Although the formation of zonal jets and banded structure on
directly imaged exoplanets and brown dwarfs is theoretically
possible, searching for banded structures is challenging.
Nevertheless, several lines of observational evidence suggest that
brown dwarfs might exhibit zonally banded patterns in the
photospheres. The first one is the recent detection of differential
rotation using {IR} photometry and radio
observations (\citealt{allersMeasurementWindSpeed2020}). The second
one is the light curve variability observed by
\cite{apaiZonesSpotsPlanetaryscale2017} that suggests the beating of
trapped waves in the atmospheric bands on brown dwarfs. The third
one is the recently detected polarimetric signals from Luhman 16AB
(\citealt{millar-blanchaerDetectionPolarizationDue2020}). The linear
polarization signal of 300{\,}ppm on Luhman 16A prefers
banded structures rather than oblateness. It would be interesting to
apply the above techniques to other brown dwarfs as well as the
directly imaged exoplanets and reveal possible banded weather
patterns.

If we neglect the photochemistry for weakly irradiated objects
(which is not necessarily true, see
\citealt{mosesCOMPOSITIONYOUNGDIRECTLY2016}), dynamical mixing is
the primary mechanism that drives the chemical tracers out of thermochemical equilibrium. In general, convection, large-scale
circulation, and associated waves and eddies all contribute to the
tracer mixing. Empirically, chemical models can be
{applied} to constrain the vertical mixing under
diffusive approximation (e.g.,
\citealt{mosesCOMPOSITIONYOUNGDIRECTLY2016};
\citealt{milesObservationsDisequilibriumCO2020}. Theoretically, as
discussed in Section \ref{sect:chemfund}, {the }conventional
prescription of eddy mixing in a convective atmosphere has
been parameterized using the Prandtl mixing length theory (e.g.,
\citealt{prandtlBerichtUberUntersuchungen1925};
\citealt{smithEstimationLengthScale1998}) without taking into
account the effect of local chemistry.
\cite{freytagRoleConvectionOvershoot2010} simulated dust grains by
solving the fully compressible equations of radiative hydrodynamics
in a 2D local model. They found that convectively excited gravity
waves are important for vertical mixing in the atmospheres of M
dwarfs and brown dwarfs. They also {identified} a
discrepancy between the derived eddy diffusivity and that from the
mixing length theory.
\cite{bordwellConvectiveDynamicsDisequilibrium2018} performed local
2D and 3D hydrodynamic simulations with tracer transport in a
non-rotating convective atmosphere.  They modified the traditional
mixing length theory by introducing a new length scale---the scale
height of the reacting species under chemical equilibrium---and
achieved a better scaling of the averaged 1D eddy mixing strength.
This conclusion is in line with that from 2D and 3D global
simulations for stably stratified atmospheres in
\cite{zhangGlobalmeanVerticalTracer2018a,zhangGlobalmeanVerticalTracer2018}
and \cite{komacekVerticalTracerMixing2019} that found that tracer
chemistry needs to be taken into account in the estimate of eddy
diffusivity.

Chemical species could also have important feedbacks on dynamics.
Radiatively active species modulate the {IR}
opacity distribution and change the radiative energy distribution in
the atmosphere. Besides the radiative effect, the atmospheric
dynamics and temperature structure could be affected by the change
of mean molecular weight during the chemical processes on brown
dwarfs and young giant planets
(\citealt{tremblinFINGERINGCONVECTIONCLOUDLESS2015,tremblinCLOUDLESSATMOSPHERESDWARFS2016,tremblinCloudlessAtmospheresYoung2017,tremblinThermocompositionalDiabaticConvection2019}).
Chemical transitions of \ch{CO} $\leftrightarrow$ \ch{CH4} and
\ch{N2} $\leftrightarrow$ \ch{NH3} {were} proposed to
change the vertical gradient of the mean molecular weight in the
atmosphere and trigger thermo-chemical instability, similar to the
fingering convection (compositional convection) in Earth oceans. The
resulting turbulent transport could lead to a reduction of the
thermal gradient near the photosphere and might explain the NIR-band
reddening of very low-gravity bodies and the onset of L-T transition
in a cloud-free atmosphere
(\citealt{tremblinFINGERINGCONVECTIONCLOUDLESS2015,
tremblinCLOUDLESSATMOSPHERESDWARFS2016,tremblinCloudlessAtmospheresYoung2017}).
However, \cite{leconteWhyCompositionalConvection2018} subsequently
pointed out that turbulent transport should homogenize the potential
temperature (entropy) instead of the temperature and
increase---instead of decrease---the temperature gradient. A recent
paper by \cite{tremblinThermocompositionalDiabaticConvection2019}
argued that the radiative source (or other energy sources) was not
negligible in the photosphere, which is therefore diabatic instead
of adiabatic. In this radiative-convective regime, compositional
convection could indeed lead to a reduction of the temperature
gradient, but their simulations are limited to quite a small domain.
A large-scale simulation including necessary physics is further{ needed} to explore this idea of ``diabatic
convection{,}" and quantify whether or not a cloudless
model can indeed explain the redness of very low-G bodies and the
L-T transition. On the other hand, it is elusive how this cloudless
theory could explain the rotational light curves of brown dwarfs and
directly imaged planets. To date, clouds remain a better candidate
responsible for the observed rotational modulations.

\begin{figure*}
   \centering
   \includegraphics[width=0.9\textwidth, angle=0]{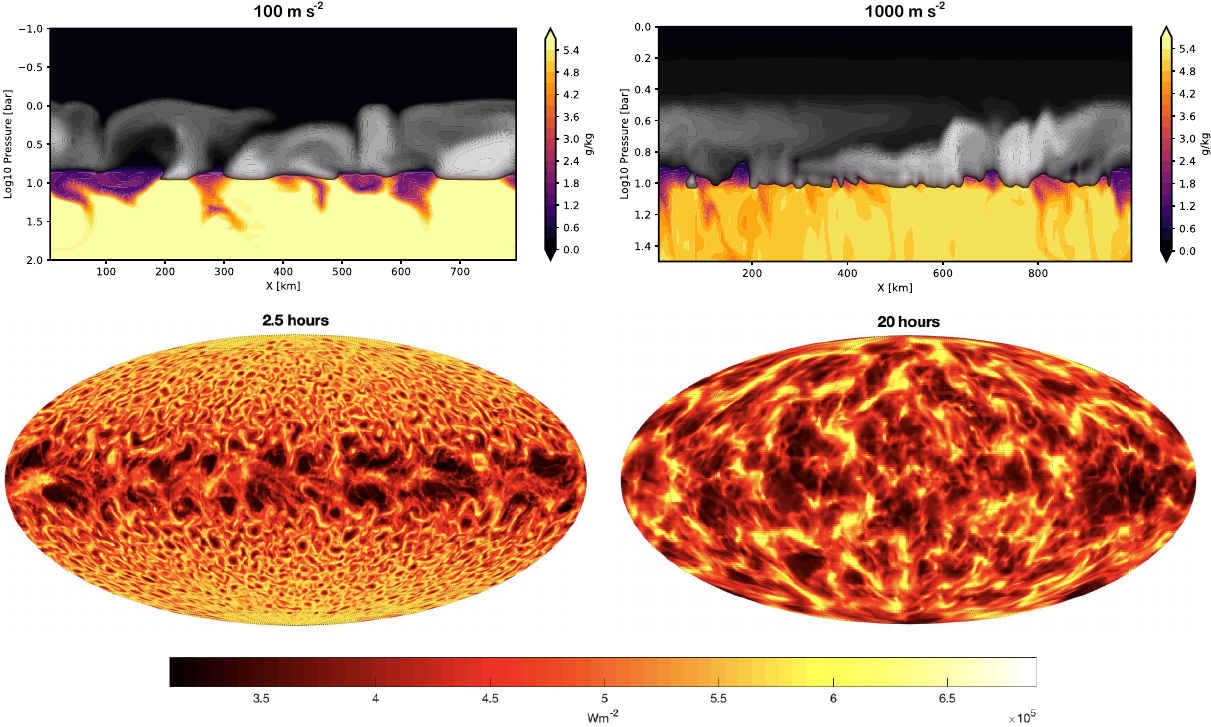}
   \caption{\baselineskip 3.8mm Simulations of atmospheric dynamics and \ch{MgSiO3} cloud formation in a convective, self-luminous atmosphere on a Jupiter size{d} body. {{\it Top}}: 2D local non-hydrostatic simulations of a low-gravity ({{\it left}}, $g=10^2{\,} \mathrm{m{\,}s^{-2}}$) and a high-gravity ({{\it right}}, $g=10^3{\,} \mathrm{m{\,}s^{-2}}$) object. The plots are from \cite{zhangSaltCloudsMetal2019} {employing} the SNAP model (\citealt{liSimulatingNonhydrostaticAtmospheres2019}; \citealt{geRotationalLightCurves2020}). The white-grey patches are clouds and color represents the mass fraction of the silicate vapor. Given other parameters{ being} fixed, the vertical extent of the clouds is more compact as gravity increases. {{\it Bottom}} (from Xianyu Tan, also see his local 3D simulations in \citealt{tanAtmosphericCirculationBrown2020}): 3D global GCM simulations of a rapid{ly }rotating ({\it left}, period 2.5 hours) and a slow{ly }rotating ({\it right}, period 20 hr) atmosphere. The color represents the outgoing thermal radiation flux at the top of the atmosphere. The patchy structure size increases as the rotation period increases.}
   \label{figbd}
   \end{figure*}

It looks{ like} clouds are inevitable in
interpreting many kinds of observations of exoplanets and brown
dwarfs from previous sections. Also, of all chemical tracers, cloud species might
play the most complicated role in atmospheric dynamics. Condensable
cloud species {have} three main effects on the
atmospheric dynamics: virtual effect, latent heat effect and
radiative effect. Some theoretical models have tried to explore
these effects in the context of the directly imaged planets and
brown dwarfs (Fig.~\ref{figbd}).

In the following, we discuss these effects sequentially. The virtual
effect, or the mass-loading effect, originated from the
fact that condensable cloud species often have different
molecular weights from the background air. The mean molecular weight
of the mixed air is different due to the mixing ratio change of condensable vapor during the condensation and evaporation
processes. The change {in} local density would also
change the static stability of the atmosphere and affect the
dynamics. The virtual effect is particularly significant in hydrogen atmospheres because the background hydrogen is much
lighter than all condensable species ranging from water to silicate.
For Solar System planets, water cloud condensation and evaporation
near the cloud layers could change the vertical density
gradient---which is characterized by a quantity called ``virtual
temperature"---and stabilize the atmosphere against convection
(e.g., \citealt{guillotCondensationMethaneAmmonia1995};
\citealt{liMoistConvectionHydrogen2015}).

The virtual effect critically depends on metallicity in
the atmosphere. If condensable vapor is not abundant, the
stabilization effect is weak. Imagine an air parcel
at the top of the atmosphere {that} is cooled down by
radiation. If
there are not many condensable species, the air density will just
increase due to cooling, and the parcel will sink. If the
condensable species is abundant---quantitatively, its mixing ratio
is larger than the critical value given by equation~(17) in
\cite{leconteCondensationinhibitedConvectionHydrogenrich2017}---as
the parcel cools, it will first unload the heavier condensable via
precipitation, and the air parcel could actually become lighter than
the surroundings and stay aloft. Stable stratification is developed
to suppress the convection. The convective {APE} is accumulated below the stratification until the
top-layer temperature further drops so that the air parcel is denser
than the environment. Then convection starts, and a large amount of
the stored energy is suddenly released, resulting in a sizeable
erupting storm. This mechanism has been proposed to explain the
quasi-periodic giant storms in Saturn's troposphere occurring about
every 30{\,}yr
(\citealt{liMoistConvectionHydrogen2015}). Whereas{,}
water in the deep atmosphere of Jupiter is probably not abundant
enough to trigger this periodic behavior, explaining the lack of
observed giant water storm eruptions on Jupiter. For hot giant
planets and brown dwarfs where the silicate could be the major
condensates, it is also possible to have similar periodic
storms---with a period related to the radiative cooling
timescale---if the metallicity is sufficiently high. Detailed
numerical simulations have yet to be performed to explore this
possibility.

The virtual effect on the convection suppression via cloud formation
can also significantly impact the temperature structure. Normally,
if a dry convection is suppressed with molecular weight gradient but
heat is still allowed to transport, the atmosphere could go into the
double-diffusive convection regime
(\citealt{sternSaltfountainThermohalineConvection1960};
\citealt{stevensonSemiconvectionOccasionalBreaking1979};
\citealt{rosenblumTurbulentMixingLayer2011};
\citealt{leconteNewVisionGiant2012};
\citealt{garaudDoublediffusiveConvectionLow2018}) in which the
temperature gradient can be greatly reduced. However, in the
presence of cloud formation,
\cite{leconteCondensationinhibitedConvectionHydrogenrich2017}
show that if the condensation occurs much faster than the
vapor diffusion, local condensable vapor abundance is almost
instantaneously controlled by the temperature change as if the heat
and vapor diffuse at the same efficiency. Condensation thus
suppresses the double-diffusive instability. As a result, the heat
is transported through the slower radiation process near the cloud
formation level in the stable layer. The temperature in the deep
atmosphere below the clouds could be much hotter than the
conventional estimate.

The second effect of the clouds on atmospheric dynamics is the
latent heat effect. Latent heat release during cloud
formation facilitates moist convection in the atmosphere. As
mentioned in Section \ref{sect:phasecurve}, the significance of the
latent heat effect of different condensable species can be evaluated
using the inverse Bowen ratio of $q/c_p T$
(\citealt{bowenRatioHeatLosses1926}). Here we adopt the temperature
as the condensational temperature $T_c$ and let $q=\chi L$ where
$\chi$ is the mass mixing ratio of the condensable species that is
primarily determined by metallicity (or surface
condition). $L$ and $c_p T_c$ are the latent and thermal heat energy
at $T_c$. The inverse Bowen ratio $\chi L/c_p T_c$ for water is
about 0.02 if we take $\chi\sim 1\%$. But for silicates (e.g.,
\ch{MgSiO3}) on hotter hydrogen atmospheres, the ratio is about
$10^{-6}$ if we take $\chi\sim 0.05\%$. Therefore{,} water has a
much larger energetic effect on the moist convection in a cold
atmosphere than silicate in a hot atmosphere.
\cite{tanEffectsLatentHeating2017} investigated the importance
of silicate latent heating on the atmospheric dynamics on
brown dwarfs. They found the latent heat from silicate condensation
is small, but the produced eddies in the moist convection could
still form zonal jets and storms. The storms are patchy with a
temporal evolution on a timescale of hours to days. But the
temperature perturbation due to the silicate condensation is
localized and only on the order of 1 Kelvin. When averaged over the
observed disk, the moist convective storms seem difficult to
reproduce the observed large amplitude of the rotational light
curves from one to tens of percent on variable brown dwarfs.

Instead, the cloud radiative effect might be the key to
understand{ing} the atmospheric dynamics and cloud variability
on hotter giant planets and brown dwarfs. Unlike the close-in
planets where clouds both reflect the stellar light and interact
with the atmospheric {IR} emission, clouds affect
the directly imaged planets and brown dwarfs, mostly via
{IR} opacity. Spatially inhomogeneously
distributed clouds such as cloud patchiness and vertical extent
could strongly impact the radiative budget and modify the horizontal
and vertical temperature distributions, which will substantially
influence the atmospheric dynamics.

Two types of cloud radiative feedbacks could exist. The first one is
1D, local, spontaneous variability. Consider a local, optically
thick cloud column that is perturbed to a higher altitude, resulting
in a lower emission to space at the cloud top and a larger heating
rate trapped inside the clouds. Vertical motion is enhanced to
balance excess heating. Consequently, cloud condensate is mixed
upward to increase the cloud top height further. It is positive
feedback. 1D simulations in
\cite{tanAtmosphericVariabilityDriven2019} coupled the radiative
transfer with cloud formation and mixing demonstrated that cloud
radiative instability could produce temperature variability up to
hundreds of Kelvins on a timescale of one to tens of hours in brown
dwarf atmospheres. This 1D, spontaneous variability looks{ to
be} a promising mechanism to explain the light curves and their
variability. This type of radiative feedback might occur only on
{a} scale smaller than the Rossby deformation radius
and in the convective portion of the atmosphere. On a larger scale,
the geostrophic adjustment takes the role due to rapid rotation. The
second type of cloud radiative feedback occurs on a larger scale and
for a relatively stratified atmosphere, for example, on some L
dwarfs where clouds only condense in the upper stratified part of
the atmosphere (e.g.,
\citealt{tsujiDustPhotosphericEnvironment2002};
\citealt{morleyNEGLECTEDCLOUDSDWARF2012}). In the absence of
convection, 1D spontaneous variability would not occur. In
this scenario, large-scale cloud radiative instability occurs in the
form of 2D or 3D flows with a range of unstable modes
(\citealt{gieraschRadiativeInstabilityCloudy1973}). Imbalanced
radiative heating and cooling lead to strong temperature contrast,
which could drive an overturning circulation. This circulation would
transport the clouds horizontally and vertically as feedback. It
might be a mechanism to maintain the patchy clouds on directly
imaged planets and brown dwarfs
(\citealt{tanAtmosphericCirculationBrown2020}).

Several theoretical steps are required to fully understand the cloud
radiative effect in 3D convective atmospheres on directly imaged
planets and brown dwarfs. First, local, non-hydrostatic models with
a simple cloud formation scheme and radiative feedback can shed
light on the system's details. For example, how is the horizontally
inhomogeneous heating in the {IR} opaque clouds
produced that drives the turbulen{ce} and circulation?
How do the circulation feedback to both the tracer
transport and the cloud formation? Is there cloud
self-aggregation occurring in this high-temperature regime? Does the
radiative instability play a role in episodic storms in 3D?
Preliminary 2D results from \cite{zhangSaltCloudsMetal2019} show
richness of the physics in these local tests (Fig.~\ref{figbd}),
including highly variable cloud fraction change, severelly depleted
vapor in the downwelling region and strong dependence of the gravity
with many compact clouds in the high-gravity regime.

Second, 3D large-scale simulations with cloud formation in rotating
atmospheres are also important in understanding how rotation impacts
the clouds and storms. Local simulations by
\cite{tanAtmosphericCirculationBrown2020} on an $f$-plane (constant
Coriolis parameter) {demonstrate} that vigorous
circulation can be driven and self-sustained by cloud radiative
feedback. The local wind speeds can reach
$10^3{\,}\mathrm{m{\,}s^{-1}}$, and horizontal
temperature contrast could be up to a few hundred Kelvin. Strong
rotation suppresses the vertical extent of the clouds. The 3D global
simulations (Fig.~\ref{figbd}) found that storm size is
generally larger at low latitudes and smaller at high latitudes.
Cloud thickness also reaches {its} maximum at the
equator and decreases toward high latitudes.  At mid and high
latitudes, the storm size scales inversely proportional{ly} to
the Coriolis parameter $f$, i.e., storm size is smaller in
higher latitude{s} and on more rapid rotators (Fig.~\ref{figbd}). Equatorial waves greatly modulate thick clouds and
clouds holes in the low latitudes. As a result, brightness
variability originates from the inhomogeneously distributed thick
and thin clouds and cloud holes, as well as the propagation of equatorial storms. This finding is consistent with the
mechanisms proposed for the observed light curve change on Jupiter
(\citealt{geRotationalLightCurves2019}). The outgoing thermal
radiation could vary locally by a factor of two due to variations in
cloud opacity and temperature structure. In an equator-on geometry,
the disk-integrated variation could be large enough to explain the
observed light curve amplitudes on brown dwarfs (Fig.~\ref{figbd}). If these distant bodies are observed from the pole-on
geometry, the rotation modulation is mainly contributed by the
evolution of turbulent eddies and storm themselves.
Detailed discussions refer to a recent review by
\cite{showmanATMOSPHERICDYNAMICSHOT2020}.

Lastly, because the cloud radiative properties critically depend on
the shape and size distribution of the cloud particles (see
Sect.~\ref{sect:cloud}), microphysics in the cloud formation is
essential. As mentioned before, integrating the microphysical
calculation in 3D dynamical simulations is very computationally
expensive and, to date, can only be integrated at a short timescale
(e.g., \citealt{leeDynamicMineralClouds2016} for simulations on a
tidally locked planet). Future work is needed to achieve an
efficient parameterization of the cloud microphysics for
3D dynamical models.

\subsection{Terrestrial Climates in the Habitable Zone}
\label{sect:hpdyn}

Last but certainly not the least, we briefly talk about the climate
regimes on habitable terrestrial planets. We only keep it brief due
to a lack of sufficient observational data for exoplanets at this
moment. We will also focus more on the dynamics and climate patterns
rather than detailed radiation or chemistry. The
{h}abitable {z}one is defined as the region
where the planetary temperature or planetary climate allows liquid water on the surface. Sunlit liquid water is the key to
habitability as we understand it today. For example, water is an
excellent solvent to allow many chemical and biological reactions
due to its polar arrangement of oxygen and hydrogen atoms in the
molecular structure. From the atmospheric perspective, sunlit water
is also the key to understanding the climates of habitable planets.
Water makes the climate on habitable planets
fundamentally different from the other terrestrial planets outside
the habitable zone, such as that on the hot terrestrial planet 55
Cancri e,{ and} planets like Venus or{ the}
{present-day} Mars. Currently, the big challenges
in the field are not only to observe and characterize these mild
terrestrial atmospheres but also to understand the complex behaviors
of {a} moist climate in theory.

\begin{figure*}
   \centering
   \includegraphics[width=0.85\textwidth, angle=0]{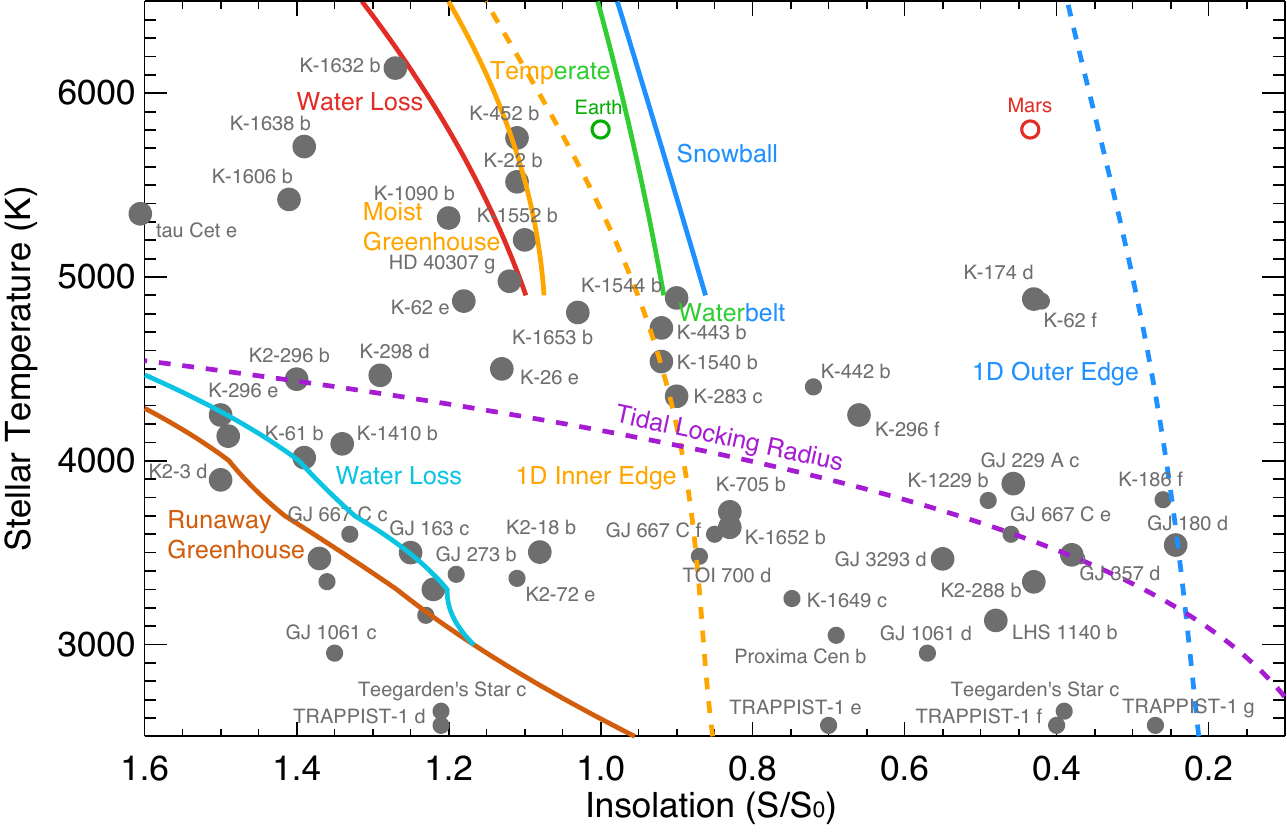}
   \caption{\baselineskip 3.8mm Circumstellar climate zones as functions of relative stellar flux in units of the Earth insolation $S_0$ and stellar temperature with Earth, Mars and currently known 59 potentially habitable exoplanets. Among them{,} 55 planets are compiled by the \href{http://phl.upr.edu/projects/habitable-exoplanets-catalog}{Habitable Exoplanets Catalog}. We also included very recently discovered planets TOI-700 d (\citealt{gilbertFirstHabitableZone2020};  \citealt{rodriguezFirstHabitableZone2020}), GJ 180 d (\citealt{fengSearchNearbyEarth2020}), GJ 229 A c (\citealt{fengSearchNearbyEarth2020}) and Kepler 1649 c (\citealt{vanderburgHabitablezoneEarthsizedPlanet2020}). Small and large dots represent planets with radii smaller and larger than 1.5{\,}$R_E$, respectively. The {{\it dashed}} curves {signify} the inner ({\it orange}) and outer ({\it blue}) edges of habitable zones from 1D radiative-convective models in \cite{kopparapuHABITABLEZONESMAINSEQUENCE2013}, respectively. The {{\it purple dashed}} curve indicates the tidal locking distance. Above the tidal locking radius, the climate simulations assuming the planets have rotation period of 24{\,}hr are from the ``warm start" 3D GCMs in \cite{wolfConstraintsClimateHabitability2017}. The colored solid lines {mark} boundaries between possible climate zones, from left to right: ``Moist Greenhouse" ({\it left of the orange curve}), ``Temperate" ({\it between orange and green{ curves}}), ``Waterbelt" ({\it between green and blue{ curves}}) and ``Snowball" ({\it right of the blue{ curve}}). Below the tidal locking radius, planets are assumed{ to be} synchronously rotating around their host stars with three climate zones near the inner edge of the habitable zone from 3D moist simulations (\citealt{kopparapuHabitableMoistAtmospheres2017}), from left to right: ``thermal runaway" ({\it left of the brown curve}), ``moist stratosphere" ({{\it between the brown and cyan curves}}) and ``mild climate" ({\it right of the cyan{ curve}}).}
   \label{fighz}
   \end{figure*}

The complexity primarily comes from the fact that water has three
important phases: vapor, liquid and ice. Each phase has
significant but different roles in the terrestrial climate system.
In the vapor phase, water vapor is a strong greenhouse gas with
strong {IR} opacity across most wavelengths. In
the liquid phase, the ocean regulates the climate on a timescale of
decades and longer. Also, liquid water is the catalyst in the carbon
cycle, including \ch{CO2} dissolution, surface erosion and
probably plate tectonics, and thus plays a part in regulating the
climate on a geological timescale. In the ice phase, water ice
floats on top of liquid water, contributing to the
planetary surface albedo. Besides, both liquid and ice clouds
significantly impact the energy budget of the system through cloud
albedo and also through cloud opacity. The phase transition between
water vapor and liquid/ice is associated with substantial latent
heat exchange. As mentioned in Section~\ref{sect:bddyn}, water ranks
among the top of all species in terms of the inverse Bowen ratio.

Consequently, the climate on a habitable planet with a large amount
of liquid water is naturally in the moist regime, in which the water
latent heat change{s} and hydrological cycle dominate{s}
many climate behaviors. One must understand water before we
understand the moist climate on habitable planets. The fact is, even
though Earth is a well-observed planet, we have not fully
understood the moist climate dynamics yet. For example, the
fundamental mechanism of the Madden-Julian oscillation in the
Earth{'s} tropics has not been fully explained
(\citealt{maddenDetection40501971}), and cloud feedbacks remain
among the largest sources of uncertainties contributing to the
current climate model diversity
(\citealt{webbOriginsDifferencesClimate2013}). Also, one should keep
in mind that desert planets with little water on their surfaces
could also be abundant in the habitable zone around M dwarfs (e.g.,
\citealt{tianWaterContentsEarthmass2015}). Whether the role of water
on the climates and habitability on these arid planets is important
or not needs further analysis.

The classical habitable zone boundaries for inner and outer edges
are estimated from 1D radiative-convective models (e.g.,
\citealt{kastingHabitableZonesMain1993};
\citealt{selsisHabitablePlanetsStar2007};
\citealt{kitzmannCloudsAtmospheresExtrasolar2010};
\citealt{wordsworthGliese581dHabitable2010};
\citealt{kalteneggerMODELSPECTRAFIRST2011};
\citealt{zsom1DMicrophysicalCloud2012};
\citealt{kopparapuHABITABLEZONESMAINSEQUENCE2013};
\citealt{rugheimerSpectralFingerprintsEarthlike2013};
\citealt{grenfellSensitivityBiosignaturesEarthlike2014};
\citealt{kopparapuHABITABLEZONESMAINSEQUENCE2014};
\citealt{rugheimerEffectUVRadiation2015};
\citealt{turbetHabitabilityProximaCentauri2016};
\citealt{yangDifferencesWaterVapor2016};
\citealt{meadowsHabitabilityProximaCentauri2018}). For example, the
estimates in Figure \ref{fighz} (from
\citealt{kopparapuHABITABLEZONESMAINSEQUENCE2013}) show the dependence
of insolation required for the habitable zone on stellar
temperature. The inner and outer boundaries are not merely
controlled by the insolation itself. The
second-order dependence comes from different stars
emitting different spectral energy distribution{s} to which the
atmosphere responds differently. In the atmosphere, water vapor absorption is
strong in the near{-IR}, while ice/snow/cloud
albedo and atmospheric scattering are strong in the short
wavelengths. As a consequence, given the same insolation, a planet
orbiting a hotter star will absorb less stellar light in the
atmosphere than a planet orbiting a colder star, pushing the
habitable zone closer to the hotter star (Fig.~\ref{fighz}).

3D theoretical dynamical models have been put forward to understand
the moist climate in a larger parameter space for Earth-like planets
(non-tidally-locked planets) in the habitable zone (e.g.,
\citealt{abeHabitableZoneLimits2011};
\citealt{boschiBistabilityClimateHabitable2013};
\citealt{shieldsEffectHostStar2013};
\citealt{yangStabilizingCloudFeedback2013};
\citealt{leconteIncreasedInsolationThreshold2013};
\citealt{yangStabilizingCloudFeedback2013};
\citealt{shieldsEffectHostStar2013};
\citealt{wolfDelayedOnsetRunaway2014};
\citealt{shieldsSpectrumdrivenPlanetaryDeglaciation2014};
\citealt{kaspiATMOSPHERICDYNAMICSTERRESTRIAL2015};
\citealt{yangStrongDependenceInner2014};
\citealt{shieldsSpectrumdrivenPlanetaryDeglaciation2014};
\citealt{wolfEvolutionHabitableClimates2015a};
\citealt{pierrehumbertDynamicsAtmospheresNondilute2016};
\citealt{shieldsEffectOrbitalConfiguration2016};
\citealt{poppTransitionMoistGreenhouse2016};
\citealt{godoltAssessingHabitabilityPlanets2016};
\citealt{wangEFFECTSOBLIQUITYHABITABILITY2016};
\citealt{dingConvectionCondensiblerichAtmospheres2016};
\citealt{wolfConstraintsClimateHabitability2017};
\citealt{kilicMultipleClimateStates2017};
\citealt{kopparapuHabitableMoistAtmospheres2017};
\citealt{wayResolvingOrbitalClimate2017};
\citealt{wayClimatesWarmEarthlike2018};
\citealt{adamsAquaplanetModelsEccentric2019};
\citealt{yangSimulationsWaterVapor2019}, also see a recent review in
\citealt{shieldsHabitabilityPlanetsOrbiting2016} and
\cite{kopparapuCharacterizingExoplanetHabitability2019} and the
white paper by \citealt{wolfImportance3DGeneral2019}). When
complicated 3D climate dynamics {are} considered, the
boundaries could be drastically different from the 1D model
predictions (Fig.~\ref{fighz}). The 1D and 3D models show that the
outer edge is dominated by many factors including albedo, greenhouse
gas inventory, \ch{CO2} collapse, clouds, carbonate-weather
feedback, surface pressure and so on (e.g.,
\citealt{kastingHabitableZonesMain1993};
\citealt{kopparapuHABITABLEZONESMAINSEQUENCE2013};
\citealt{turbetCO2CondensationSerious2017}). The inner edge could
also be strongly affected by the planetary rotation rate that
largely modulates the cloud distribution and planetary albedo (e.g.,
\citealt{yangStrongDependenceInner2014};
\citealt{wayWasVenusFirst2016}).

Climate on fast-rotating terrestrial planets can be classified into
several regimes (e.g.,
\citealt{goldblattHabitabilityWaterworldsRunaway2015};
\citealt{wolfConstraintsClimateHabitability2017}). Using the global
mean surface temperature ($T_s$) on 3D simulations as a proxy,
\citet{wolfConstraintsClimateHabitability2017} defined four
potentially stable climate states that are separated by abrupt
climatic transitions (Fig.~\ref{fighz}): snowball ($T_s<$235\,K),
waterbelt (235\,K$<T_s<$250\,K), temperate (275\,K$<T_s<$315\,K) and
moist greenhouse ($T_s<$330\,K). Those states are in stable
equilibrium where the incoming stellar flux balances the outgoing
thermal radiation, and the states are resilient against any small
perturbation. In the snowball state that might have occurred in Earth's
Neoproterozoic glaciations (0.75 to 0.54 billion years
ago), the ocean surface is globally covered by ice (e.g.,
\citealt{kirschvinkLateProterozoicLowLatitude1992};
\citealt{hoffmanNeoproterozoicSnowballEarth1998};
\citealt{liuLargeEquatorialSeasonal2020}). In the waterbelt state,
the equatorial ocean can be ice-free, although the middle and high
latitudes are fully glaciated (e.g.,
\citealt{hydeNeoproterozoicSnowballEarth2000};
\citealt{abbotJormungandGlobalClimate2011}). The current Earth is in
the temperate state. When the surface temperature further increases,
the planet could enter the moist greenhouse state in which the
atmosphere is warm enough so that water is no longer trapped by the
cold tropopause, resulting in a moist stratosphere (e.g.,
\citealt{kastingHabitableZonesMain1993}). In this scenario, hydrogen
loss to space is efficient due to the photolysis of water vapor. The
specific rate of water photolysis likely depends on the stellar
activity (e.g.,
\citealt{chenHabitabilitySpectroscopicObservability2019}). Also,
there is a strong hysteresis (or bistability) between the snowball
and temperate climate states, meaning that there exist two stable
climate solutions given the same stellar flux (e.g.,
\citealt{budykoEffectSolarRadiation1969}), although the stellar flux
range allowing the hysteresis is sensitive to other parameters such
as rotation rate and surface pressure. For example, it decreases or
even disappears if we increase the rotation rate of the planet
(\citealt{abbotDecreaseHysteresisPlanetary2018}) or on tidally locked exoplanets (\citealt{checlairNoSnowballHabitable2017}).

According to the definition of habitability, {a}
habitable climate includes waterbelt, temperate and cooler
moist greenhouse states. If the stellar flux is small, the climate
enters the snowball state. Snowball state can be habitable in some
conditions. A good example is the Earth{'s} climate in the
Neoproterozoic era. For exoplanets, in local regions such as high
geothermal heat flux or thin ice, photosynthesis life can still
exist. Another possibility is that although the ocean is covered
entirely by ice and snow, some continents are ice-free and are warm
enough to maintain liquid water (e.g.,
\citealt{paradiseHabitableSnowballsTemperate2019}).

If the stellar flux is large, the entire atmosphere will enter the
thermal runaway process, and there is no stable climate solution
until the ocean water is all evaporated into the atmosphere. The
underlying mechanism of the so-called runaway greenhouse climate is
owing to the limit of thermal emission in the moist atmosphere.
There are two limits (\citealt{nakajimaStudyRunawayGreenhouse1992}).
The moist stratosphere limit is the ``Komabayashi-Ingersoll limit"
(\citealt{komabayasiDiscreteEquilibriumTemperatures1967};
\citealt{ingersollRunawayGreenhouseHistory1969}) where the
stratospheric emissivity is set by 100\% relative humidity at the
tropopause.  Before the stratospheric limit is reached, the thermal
emission of the troposphere itself can be limited because the moist
adiabat in the troposphere has to follow the saturation vapor
pressure of water when the entire atmosphere becomes
water-dominated. This lower limit---
\cite{goldblattRunawayGreenhouseImplications2012} termed it
``Simpson-Nakajima limit"
(\citealt{simpsonStudiesTerrestrialRadiation1927};
\citealt{nakajimaStudyRunawayGreenhouse1992})---is more relevant to
runaway greenhouse effect. Venus might have experienced the runaway
greenhouse process (\citealt{ingersollRunawayGreenhouseHistory1969},
or at least moist greenhouse,
\citealt{kastingRunawayMoistGreenhouse1988}), resulting in very
little water and a high D/H ratio in the atmosphere.

The climate state classification might depend on specific models
because of the complexity of the 3D climate models. For example, the
simulations in \citealt{wolfConstraintsClimateHabitability2017}
found that there is no stable climate solution in the temperature
range of 250\,K$<T_s<$275\,K and 315\,K$<T_s<$330\,K. However, other
models could produce stable climate solutions between 250 and 275\,K
with oceanic dynamics (e.g.,
\citealt{yangInitiationModernSoft2011}). Also, using a different
cloud scheme, other models found stable climate solutions between
315 and 330{\,}K (e.g.,
\citealt{leconteIncreasedInsolationThreshold2013}). Moreover, a
moist greenhouse state does not even exist in some other Earth-like
models. For example,
\citealt{leconteIncreasedInsolationThreshold2013}) found an abrupt
transition from dry stratosphere to runaway greenhouse state. Future
model intercomparison is needed to understand the differences among
3D climate models.

Planets in orbits close to low-mass stars (such as M dwarfs) could
also be habitable. They are also important targets in future
observational surveys due to their proximity to the host stars so
that transits are more common and the {S/N}
is larger. These planets are likely to be synchronously rotating or
in spin-orbit resonances (such as Mercury) due to stellar tides. The
tidal locking semi-major axis is empirically found to scale with the
stellar mass to the 1/3 power (e.g.,
\citealt{pealeOriginEvolutionNatural1999};
\citealt{kastingHabitableZonesMain1993};
\citealt{dobrovolskisInsolationPatternsSynchronous2009};
\citealt{edsonAtmosphericCirculationsTerrestrial2011};
\citealt{haqq-misraGeothermalHeatingEnhances2015}). Assuming the
stellar mass-luminosity relationship $L\propto M^{\alpha}$ and the
mass-radius relationship $R\propto M^{\beta}$, the insolation $I$ at
the tidal locking distance is dependent on the stellar effective
temperature in the following relation
 \begin{equation}
I\propto T_{eff}^{4(3\alpha-2)/3(\alpha-2\beta)}.
\end{equation}
Taking $\alpha\sim 3.7$ and $\beta\sim 0.724$
(\citealt{demircanStellarMassluminosityMassradius1991}), we get the
``Tidal Locking Radius" curve: $I\propto T_{eff}^{5.4}$
(Fig.~\ref{fighz}). This implies that the incoming stellar flux at
the tidal locking distance depends strongly on stellar
temperature. Basically{,} any star hotter than the M (or late K)
type is not likely to have Earth-like planets (in terms of
instellation) in synchronous rotation
(\citealt{haqq-misraGeothermalHeatingEnhances2015}).

For planets in this tidal locking regime, a number of 3D
climate models have also been adapted from the Earth to study circulation patterns (e.g.,
\citealt{joshiSimulationsAtmospheresSynchronously1997};
\citealt{joshiClimateModelStudies2003};
\citealt{merlisAtmosphericDynamicsEarthlike2010};
\citealt{showmanAtmosphericCirculationExoplanets2010};
\citealt{wordsworthGLIESE581DFIRST2011};
\citealt{pierrehumbertPaletteClimatesGliese2011};
\citealt{edsonAtmosphericCirculationsTerrestrial2011};
\citealt{showmanAtmosphericCirculationTerrestrial2013};
\citealt{leconte3DClimateModeling2013};
\citealt{yangStabilizingCloudFeedback2013};
\citealt{huRoleOceanHeat2014};
\citealt{yangWaterTrappingTidally2014};
\citealt{wangCLIMATEPATTERNSHABITABLE2014};
\citealt{caroneConnectingDotsVersatile2014,
caroneConnectingDotsII2015,caroneConnectingDotsIII2016};
\citealt{wordsworthAtmosphericHeatRedistribution2015};
\citealt{kollDecipheringThermalPhase2015};
\citealt{kollTemperatureStructureAtmospheric2016};
\citealt{haqq-misraGeothermalHeatingEnhances2015};
\citealt{turbetHabitabilityProximaCentauri2016};
\citealt{pierrehumbertDynamicsAtmospheresNondilute2016};
\citealt{kopparapuINNEREDGEHABITABLE2016};
\citealt{kopparapuHabitableMoistAtmospheres2017};
\citealt{boutleExploringClimateProxima2017};
\citealt{wayResolvingOrbitalClimate2017};
\citealt{wolfConstraintsClimateHabitability2017};
\citealt{wolfAssessingHabitabilityTRAPPIST12017};
\citealt{fujiiNIRdrivenMoistUpper2017};
\citealt{nodaCirculationPatternDaynight2017};
\citealt{fauchezImpactCloudsHazes2019};
\citealt{wolfSimulatedPhasedependentSpectra2019}
\citealt{delgenioHabitableClimateScenarios2018};
\citealt{kopparapuHabitableMoistAtmospheres2017};
\citealt{haqq-misraDemarcatingCirculationRegimes2018};
\citealt{lewisInfluenceSubstellarContinent2018};
\citealt{chenHabitabilitySpectroscopicObservability2019};
\citealt{komacekAtmosphericCirculationClimate2019};
\citealt{yangOceanDynamicsInner2019};
\citealt{yangSimulationsWaterVapor2019};
\citealt{dingStabilizationDaysideSurface2020};
\citealt{yangTransitionEyeballSnowball2020};
\citealt{suissaFirstHabitableZone2020}, also see a recent review in
\citealt{shieldsHabitabilityPlanetsOrbiting2016},
\citealt{pierrehumbertAtmosphericCirculationTideLocked2019} and
\citealt{kopparapuCharacterizingExoplanetHabitability2019} and the
white paper by \citealt{wolfImportance3DGeneral2019}). Some of those
models also investigated the planets in other possible spin-orbit
resonance states (e.g., \citealt{wordsworthGLIESE581DFIRST2011};
\citealt{yangStabilizingCloudFeedback2013};
\citealt{wangCLIMATEPATTERNSHABITABLE2014};
\citealt{wayResolvingOrbitalClimate2017};
\citealt{boutleExploringClimateProxima2017};
\citealt{yangTransitionEyeballSnowball2020};
\citealt{suissaFirstHabitableZone2020}).
\cite{pierrehumbertAtmosphericCirculationTideLocked2019} provided a
detailed review of the atmospheric dynamics on tidally locked
terrestrial planets. Because of{ the} tidally
locked configuration, the climate behaves differently from{ that
on} non-tidally locked planets. In particular, interesting day-night
differences have been suggested in this regime compared with the
zonally homogenous fast rotators like Earth.  One of the noticeable
behaviors is that strong convective clouds form on the dayside and
significantly increase the planet's albedo, thus allowing the planet
to be habitable under higher insolation levels (e.g.,
\citealt{yangStabilizingCloudFeedback2013};
\citealt{kopparapuINNEREDGEHABITABLE2016};
\citealt{wayWasVenusFirst2016}). Also, a relatively thin atmosphere
could lead to atmospheric collapse on the nightside on
tidally locked planets (e.g.,
\citealt{kiteClimateInstabilityTidally2011};
\citealt{wordsworthAtmosphericHeatRedistribution2015}). In the
nearly snowball state, the climate of tidally locked
planets could show a peculiar ``eyeball" state with an open ocean
near the substellar point
(\citealt{pierrehumbertPaletteClimatesGliese2011}) but a recent
study {indicates} that the open ocean might also be
closed by sea-ice drift under certain circumstances
(\citealt{yangTransitionEyeballSnowball2020}).

In terms of classification, several regimes have been demarcated
using 3D simulations with Earth's atmospheric compositions. A series
of papers by
\cite{caroneConnectingDotsVersatile2014,caroneConnectingDotsII2015,caroneConnectingDotsIII2016,caroneStratosphereCirculationTidally2018}
{has} investigated the terrestrial troposphere
circulation regimes as functions of planetary radius and orbital
period (while still{ being} tidally locked). The wind field is
influenced by tropical Rossby waves that lead to equatorial
superrotation and by extratropical Rossby waves for two
high-latitude wind jets. They demarcated four circulation regimes,
including the troposphere and stratosphere, in terms of the
rotational period $P$ (see fig.~1 in
\citealt{caroneStratosphereCirculationTidally2018}). (1) For $P
<$3{\,}d, there is a strong mixture of
equatorial superrotation and high-latitude wind jets in the
troposphere. The stratosphere circulation from the equator to the
pole is inefficient due to an anti-Brewer-Dobson-circulation induced
by tropical Rossby waves. (2) For 3{\,}d$< P
<$6{\,}d, the tropospheric wind pattern
is either equatorial or high-latitude wind jets. The stratospheric
equator-to-pole transport is a bit more efficient because
extratropical Rossby waves counterbalance the effect of tropical
Rossby waves. (3) For 6{\,}d$< P <${\,}25{\,}d, there is weak superrotation in
the troposphere, and stratospheric transport could be
efficient if there is stratospheric wind breaking. (4) If $P
>$25{\,}d, the tropospheric wind {exhibits} radial flow structures. The
stratospheric transport is efficient because of thermally driven
circulation.

Similarly, in an idealized GCM study
\cite{nodaCirculationPatternDaynight2017} found that the
tropospheric circulation patterns change as the planetary rotation
rate increases. They {identified} four regimes. Type
I is a day-night thermally direct circulation, Type II shows a zonal
wavenumber one resonant Rossby wave on an equatorial westerly jet,
Type III exhibits long timescale north-south asymmetric
variation and Type IV {manifests} a pair of
mid-latitude westerly jets.

\cite{haqq-misraDemarcatingCirculationRegimes2018} further
characterized the circulation patterns in more realistic 3D
simulations into three dynamical regimes using the non-dimensional
equatorial Rossby deformation radius and the non-dimensional Rhines
length (to the planetary radius). The ``slow rotation regime" occurs
when both the Rossby deformation radius and the Rhines length exceed
the planetary radius, and the circulation has a mean zonal
circulation from the dayside to the nightside. In the ``rapid
rotation regime" with Rossby deformation radius less than the
planetary radius, the circulation {displays} a mean
zonal circulation that partially spans a hemisphere but with banded
clouds over the substellar point. {In
t}he third regime, ``Rhines rotation regime{,}"
which occurs when the Rhines length is greater than the radius but
the Rossby deformation radius is less than the radius, a thermally
direct circulation emerges from the dayside to the nightside, but
midlatitude jets also exist. These dynamical regimes can be characterized by
thermal emission phase curves from future observations.

As mentioned in Section \ref{sect:dynfund}, the ``WTG parameter"
$\Lambda$ in
\cite{pierrehumbertAtmosphericCirculationTideLocked2019} can be used
to characterize the dynamical regimes for tidally locked terrestrial
planets. Their 3D simulations revealed that if $\Lambda > 5$, the
horizontal temperature distributions are homogeneous (e.g., WTG
behavior) due to energy-transporting circulations. However,
nonlinearity could occur {that} break{s} the WTG
behavior if the circulations become very strong. As for
time variability, the circulation usually exhibits small temporal
fluctuation in the large $\Lambda$ regime (e.g., $\Lambda > 5$). The
planetary-scale transients such as strong eddies and wave
disturbances appear to be important if $\Lambda \leqslant 2$. A
likely source of these transients is the baroclinic instability.

The parameter space of climate on habitable planets is vast. There
is still much to explore to characterize all the climate regimes.
Other than the important parameters such as stellar flux and
planetary rotation rate that we have discussed above, other
parameters such as total atmospheric pressure, atmospheric
composition (especially the radiatively active species such as
\ch{CO2}, \ch{CH4} and clouds, or condensable atmospheres,
\citealt{dingConvectionCondensiblerichAtmospheres2016}), and the
ocean and ice dynamics (\citealt{huRoleOceanHeat2014};
\citealt{yangTransitionEyeballSnowball2020}) have also been shown to
be important. Furthermore, other less explored parameters might{
be crucial} as well, for example, planetary
eccentricity (e.g., \citealt{wangCLIMATEPATTERNSHABITABLE2014};
\citealt{adamsAquaplanetModelsEccentric2019}),  obliquity (e.g.,
\citealt{dobrovolskisInsolationPatternsSynchronous2009};
\citealt{wangEFFECTSOBLIQUITYHABITABILITY2016};
\citealt{kilicMultipleClimateStates2017};
\citealt{kangTropicalExtratropicalGeneral2019};
\citealt{kangMechanismsLeadingWarmer2019,kangWetterStratospheresHighobliquity2019}),
planetary gravity and radius (e.g.,
\citealt{yangEffectsRadiusGravity2019};
\citealt{yangHowPlanetaryRadius2019}), tidal heating (e.g.,
\citealt{barnesTidalVenusesTriggering2013}), magnetic field (e.g.,
\citealt{dongProximaCentauriHabitable2017},
\citealt{dongAtmosphericEscapeTRAPPIST12018}), stellar activity
(e.g., \citealt{badhanStellarActivityEffects2019};
\citealt{chenHabitabilitySpectroscopicObservability2019};
\citealt{airapetianImpactSpaceWeather2020}), initial water inventory
(e.g., \citealt{dingStabilizationDaysideSurface2020}), interior and
surface processes (e.g.,
\citealt{walkerNegativeFeedbackMechanism1981};
\citealt{charnayWarmColdEarly2017}) and so on. Check the
recent review by \cite{shieldsHabitabilityPlanetsOrbiting2016} and
\cite{kopparapuCharacterizingExoplanetHabitability2019} for detailed
summary of these parameters and discussions.

\section{Future Prospects}
\label{sect:prospects}

Looking back at the dawn of exoplanet and brown dwarf science,
we are impressed by the amount of information we have been able to retrieve
from these distant, unresolved faint pixels. Although it is
difficult to characterize their atmospheres without any bias due to
limited data quality and our a priori knowledge from the Solar
System bodies, preliminary analysis of the relationship between the
observed atmospheric characteristics and fundamental planetary
parameters has identified interesting yet somewhat arguable regimes
and trends in the current substellar atmosphere sample. The stellar
and planetary parameters considered here include mass, radius,
gravity, rotation rate, metallicity, surface albedo, internal
luminosity, stellar luminosity, stellar spectra and orbital
parameters. From these parameters, one can derive several
fundamental scales in the atmospheres. For length scales, there are
the pressure scale height, planetary radius, Rossby deformation
scale and Rhines scale. For timescales, there are
radiative timescale, conductive timescale, wind transport timescale,
eddy transport timescale, chemical timescale and mass loss
timescale. For velocity scales, there are light speed, sound speed,
rotational velocity, jet velocity, eddy velocity, escape
velocity and thermal velocity. For energy scales, there are
thermal energy, photon energy, latent heat, gravitational potential
energy, {KE} and convective potential
energy. In this review, we have shown that atmospheric trends and regimes
can be linked back to these scales in the atmospheres and thus to
the stellar and planetary parameters. Based on these scales, we can
derive several dimensionless numbers such as the Jeans parameter, Rossby
number, Froude number, Mach number, Alfv\'{e}n Mach number,
Richardson number, Rayleigh number, Ekman number, Taylor number,
Prandtl number, WTG parameter, $\Omega\tau_{\mathrm{dyn}},
\tau_{\mathrm{dyn}}/\tau_{\mathrm{drag}},
\tau_{\mathrm{dyn}}/\tau_{\mathrm{rad}}$, inverse Bowen ratio
($q/c_pT_{\rm eq}$), $\tau_{\mathrm{vis}}/\tau_{\mathrm{IR}}$,
$\tau_{\mathrm{dyn}}/\tau_{\mathrm{chem}}$ and
$F_{\mathrm{ext}}/F_{\mathrm{int}}$. We
{demonstrated} that these numbers are important to
understand the behaviors of various planetary climate. Simple
scaling laws shed light on the underlying mechanisms.

There are two key aspects of linking the atmospheric characteristics to bulk
planetary parameters. First, in this review, we have mainly
focused on the influence of the planetary parameters on the
atmospheric behaviors and resulting signals. Second, atmospheric processes
could also significantly impact the planetary parameters, which were
not discussed in this review. Here we briefly mentioned some
important aspects. As the outer boundary of a planetary body,
planetary atmospheres directly exchange mass and energy
with space. A number of studies have {signified} that
radiative processes in the outer gaseous envelopes (i.e.,
atmospheres) could greatly influence planetary-mass accretion rates
(e.g., \citealt{leeOpticallyThinCore2018};
\citealt{ginzburgEndgameGasGiant2019}) and subsequent radius
evolution (e.g., \citealt{chabrierTheoryLowMassStars2000};
\citealt{burrowsTheoryRadiusTransiting2003};
\citealt{fortneyGiantPlanetInterior2010}). Photoevaporative and
core-powered mass loss processes have been proposed to explain the
observed ``radius gap" in low-mass planets (e.g.,
\citealt{lopezRoleCoreMass2013};
\citealt{owenKeplerPlanetsTale2013};
\citealt{lopezRoleCoreMass2013};
\citealt{ginzburgCorepoweredMasslossRadius2018}) as discussed in
Section \ref{sect:radgap}. A significant mass loss would shrink the
planetary size and change the rotational and orbital angular
momentum and, thus, the planetary rotation rate and orbital
parameters. For example, if the planetary mass loss is larger than
10\%, \cite{matsumotoBreakingResonantChains2020} showed that
planetary orbits in resonant chains could be destabilized. Several
mechanisms related to atmospheric dynamics have been proposed to
explain the unexpected inflated sizes of some hot Jupiters (e.g.,
\citealt{showmanAtmosphericCirculationTides2002};
\citealt{youdinMechanicalGreenhouseBurial2010};
\citealt{tremblinAdvectionPotentialTemperature2017};
\citealt{sainsbury-martinezIdealisedSimulationsDeep2019}). The
ionization rate in the atmosphere is vital for the proposed Ohmic
heating inflation mechanisms on hot Jupiters (e.g.,
\citealt{batyginInflatingHotJupiters2010};
\citealt{huangOhmicDissipationInteriors2012};
\citealt{wuOHMICHEATINGSUSPENDS2013}) and the terminal rotation
rates of giant planets (e.g.,
\citealt{batyginTerminalRotationRates2018}). For smaller planets,
distribution of atmospheric compositions such as dusty outflow or
high-altitude photochemical haze might greatly enlarge the apparent
planetary radii in short wavelengths (e.g.,
\citealt{sekiyaDissipationPrimordialTerrestrial1980};
\citealt{cubillosOverabundanceLowdensityNeptunelike2017};
\citealt{lammerIdentifyingTrueRadius2016};
\citealt{wangDustyOutflowsPlanetary2019};
\citealt{gaoDeflatingSuperpuffsImpact2020}). Atmospheric escaping
flux from ablating planets might pollute the stellar emission (e.g.,
\citealt{haswellDispersedMatterPlanet2020}). This pollution might
not only change the received stellar flux in the atmosphere but also
create a possible new way to detect the atmospheric compositions
even without a single photon from the planet itself. Taking into
account the feedback of atmospheric processes on the basic stellar
and planetary parameters during planetary formation and
evolution would further complicate the characterization of
exoplanets and brown dwarfs. New trends and regimes might also
emerge in these interesting climate systems.

The future is challenging, but promising. For a long time we are likely to lack spatially-resolved
images or in-situ information from entry probes or flyby missions. That
will put a cap on our understanding of the detailed weather and
climate on specific bodies, as we have learned lessons from Solar System science in the spacecraft age. However, the
explosion of empirical data from dedicated telescopic
observations and large surveys would unveil, at least to
first order, the nature of diversity of planetary climate
and the main regimes in large parameter space. The population of exoplanets continues rapidly growing from past and
future space-based transit surveys such as Kepler, K2, TESS and
CHEOPS as well as ground-based surveys like WASP, KELT, MASCARA, HAT
and TrES for hot and warm gas giants, and NGTS, MEarth, Trappist,
SPECULOOS, and ExTrA for smaller and cooler planets. Most data
about exoplanet atmosphere observations will come
from current and future space telescopes like HST, CHEOPS, TESS,
JWST, PLATO, ARIEL, WFIRST, OST, HabEx, LUVOIR and high-precision
ground-based facilities such as VLT/FORS2, VLT/ESPRESSO,
VLT/CARMENES, TNG/HARPS, GTC/OSIRIS, VLT/SPHERE, Gemini/GPI,
Subaru/SCExAO, Magallan/MagAO(-X) and VLTI-GRAVITY. Several Chinese
space missions on exoplanets such as CSST, Miyin and CHES are
also in preparation. In the upcoming decade, JWST and ARIEL might be
the two most important {IR} telescopes for
characterizing substellar atmospheres (see discussions in
\citealt{greeneCharacterizingTransitingExoplanet2016};
\citealt{tinettiScienceARIELAtmospheric2016}). The detailed,
high-contrast observations of atmospheres on habitable terrestrial
planets and possible biosignatures might need to await the next
generation of large ground-based facilities such as the ELT,
GMT and TMT, and space-based observatories such as HabEx and
LUVOIR.

To aid the interpretation of telescopic observations,
laboratory experiments are critically important to
improve the accuracy of input parameters in atmospheric modeling
(\citealt{fortneyNeedLaboratoryMeasurements2019}). In terms of
atmospheric radiative properties, spectroscopic data include spectral line intensities, line broadening parameters, line
mixing parameters, collisional deactivation parameters for different
quantum states for Non-LTE calculations, collision-induced
absorption for various gas mixtures, aerosol optical properties such
as absorption and scattering coefficients, single scattering albedo
and the scattering phase function. These spectroscopic data need to be
improved to cover more wavelengths, higher resolutions, and various
temperature and pressure environment{s} to reveal diverse
substellar atmospheric conditions. In particular,
high-resolution spectroscopy might require a precise measurement of
the line core location and line shape that might deviate from the
conventionally assumed Voigt profile. For the chemical properties,
chemical kinetic data such as reaction rates and their
dependence on temperature, pressure and quantum states are
largely uncertain. To model the detailed formation process of clouds
and chemical hazes, one needs to measure the mechanical,
thermal and electronic properties of the particulate matters
in the atmosphere, such as surface tension, coagulation properties,
surface reaction rates of dusty grains, heat capacity, latent
heat and electronic charge. Laboratory chamber simulations of
the gas chemistry, photochemical haze formation and cloud
condensation are crucial to understanding these very complicated
processes. The electrical properties of the gases are also important
in understanding ion chemistry, lightning, and possible energy
processes such as Ohmic heating. {The e}quation of state
of gas mixtures under high temperature and pressure are also
important to model the deep atmospheres of giant planets. In the
situation where lab experiments are not available, \textit{ab
initio} calculations could provide alternatively useful information
for the model input, such as the molecular line information from
molecular physics simulations (e.g., Exomol,
\citealt{tennysonExoMolMolecularLine2012}) and the equation of state
of hydrogen-helium mixtures {in the} high-temperature
and high-pressure regimes from density functional molecular dynamics
simulations (e.g., \citealt{militzerInitioEquationState2013}).

Theoretical advancement is also required to synthesize new knowledge
to improve the current framework. It might proceed on three fronts.
First, refining the data retrieval techniques for the most robust
information to be derived from observations. This task
includes both designing the most efficient observational mode using
available facilities and improving numerical techniques for inverse
modeling. Second, exploring the parameter space and unveiling new
mechanisms. Some of the new physics might come from testing the
current theory in extreme conditions, such as super-fast rotators,
atmospheric collapses and strong star-planet interactions.
Some might come from the analysis of interactions among
different processes such as dynamics, radiation and chemistry
(including grain chemistry and cloud microphysics), such as the
thermal dissociation and recombination of hydrogen on ultra-hot
Jupiters. Could other exothermic and endothermic chemical reactions
also significantly affect the temperature and dynamical structures?
Is radiation pressure important for dynamics on very hot exoplanets?
Third, improving detailed numerical simulations. There are still
many challenging technical problems. For example, how to
appropriately represent the radiative transfer, chemistry and
cloud processes in 3D dynamical simulations while not significantly
slow down the computational efficiency? What are the proper upper
and lower boundary conditions in atmospheric simulations? Also, as
the model becomes more and more complicated, a hierarchical approach
to understanding the detailed mechanisms using models with different
levels of complexity would be appreciated.

The community should sustain intensive interaction to broaden and
deepen our understanding of the substellar atmospheres. We encourage
the observers, experimentalists and theorists work together to
collaborate and participate in long-term workshops about exoplanets and
brown dwarfs such as Exoplanets, Exoclimes, Cool Stars, Cloud
Academy, UCSC OWL Exoplanet Summer Workshop and many more (e.g.,
check \href{http://exoplanet.eu/meetings/}{Future Meetings on
Extrasolar Planets}). We also encourage data sharing activities such
as open-source data and software (e.g.,
\href{https://github.com/exosports/BART}{BART};
\href{https://github.com/elizakempton/Exo\textunderscore
Transmit}{Exo\textunderscore Transmit};
\href{https://github.com/exoclime}{Exoclimes Simulation Platform};
\href{https://github.com/VirtualPlanetaryLaboratory/vplanet}{VPLanet}).
Because atmospheric models are highly technical, model
intercomparison projects should be promoted to understand the
theoretical consistency and differences in the field (e.g.,
\citealt{yangSimulationsWaterVapor2019};
\citealt{barstowComparisonExoplanetSpectroscopic2020};
\citealt{fauchezTRAPPIST1HabitableAtmosphere2020}).

In the end, we should not forget to connect the new knowledge
learned from those exotic substellar atmospheres back to the Solar
System bodies where in-situ data are available. With rapidly evolving data and theory---
the two prongs in Hume's fork---a unified,
first-principle climate theory for diverse exoplanets and brown
dwarfs can be established. It will place the Earth and
Solar System in the large charts of atmospheric regimes and trends
across the entire multi-dimensional parameter space, which are yet
to be explored, in the universe.

\normalem
\begin{acknowledgements}
The stellar and planetary parameters were obtained from the
\href{https://exoplanetarchive.ipac.caltech.edu/}{NASA Exoplanet
Archive}, \href{http://exoplanet.eu/}{the Extrasolar Planets
Encyclopaedia} and
\href{http://www.exoplanetkyoto.org/?lang=en}{ExoKyoto}. I thank
Yunwen Duan for help with the data compilation and analysis. I thank
editor Jianghui Ji for inviting and motivating me to complete this
review. Thanks to Elena Manjavacas, Kevin Zahnle, Erik Petigura,
Mike Line, Hannah Wakeford, Mark Marley, Didier Saumon, Peter Gao,
Xianyu Tan, and Eric Wolf for the figure data. Special thanks to Kevin Zahnle, Xianyu Tan, Thaddeus
Komacek, Julianne Moses, Mark Marley, Pete \& Dinah Showman, Eric Wolf, and Jun Yang
for their detailed comments. I also thank Justin
Erwin, Jonathan Fortney, Vivien Parmentier, Michael Liu, Katelyn
Allers and Linfeng Wan for helpful discussions. Lastly, I
specifically dedicate this review article to my postdoc advisor Dr.
Adam Showman (1968--2020) for his help in these years. Brown dwarfs
never die, they simply fade away.
\end{acknowledgements}


\end{document}